**Title: The insertion of Neomycin cassette impairs maternal and social behaviors in *Arc/Arg3.1* knock-out mice**

Ana DUDAS[1], Emilia CAIRE[1], Abdurahman HASSAN A KUKU[2,3], Nicolas AZZOPARDI[1], Anil ANNAMNEEDI[1,3,4], Heba ELSEEDY[5], Gaëlle LEFORT[1], Benoît PIEGU[1], Romain YVINEC[1], Emmanuel PECNARD[1], Lucile DROBECQ[1], Anne-Charlotte TROUILLET[1], Angela SIRIGU[5], Dietmar KUHL[2], Pablo CHAMERO[1], Ora OHANA[2,3], Lucie P. PELLISSIER[1#]

[1]INRAE, CNRS, Université de Tours, PRC, 37380, Nouzilly, France

[2]Institute for Molecular and Cellular Cognition, Center for Molecular Neurobiology, University Medical Center Hamburg–Eppendorf, 20251 Hamburg, Germany

[3]Department of Neurosurgery, University Medical Center Hamburg–Eppendorf, 20251 Hamburg, Germany

[3]LE STUDIUM Loire Valley Institute for Advanced Studies, 45000, Orléans, France

[4]School of Arts and Sciences, Sai University, Chennai, India

[5]Institute of Neurosciences of la Timone, UMR7289 CNRS & Aix-Marseille Université, Marseille, France; iMIND Center of Excellence for Autism, Le Vinatier Hospital, Bron, France

#Corresponding author: Lucie P. Pellissier, PhD, Team biology of GPCR Signaling systems (BIOS), INRAE, CNRS, Université de Tours, PRC, 37380, Nouzilly, France. Phone: +33 4 47 42 79 62. Email: lucie.pellissier@inrae.fr

## ABSTRACT

The Neomycin resistance cassette ($Neo^+$) is commonly inserted in the genome of mice to generate knock-out (KO) models. The effect of gene deletion on social behaviors in mice is controversial between studies using different $Neo^+$ and $Neo^-$ mouse lines, particularly *Arc/Arg3.1* KO lines. In this study, we identified severe maternal behavior impairments in $Neo^+$, but not $Neo^-$ *Arc/Arg3.1* KO dams. These deficits resulted from reduced sociability and abnormal social information processing in $Neo^+$ *Arc/Arg3.1* KO dams, exacerbated by social communication impairments in pups. The expression of the Neo cassette product did not cause cytotoxicity, but led to altered ERK signaling, gene expression, and oxytocin system. However, oxytocin administration did not improve social impairments in $Neo^+$ *Arc/Arg3.1* KO animals. Interestingly, early social environment enrichment enhanced social interaction with familiar, but not unfamiliar conspecifics or maternal behavior. Overall, our findings reveal a major impact of the Neo cassette on behaviors, particularly social behaviors, in *Arc/Arg3.1* KO mice, underscoring the need to re-examine phenotypes of animal models carrying the Neo cassette in neuroscience research.

## SIGNIFICANCE STATEMENT

Gene deletion in mice is essential for studying gene functions, genetic diseases and developing potential treatments. To create these models, gene resistance to the antibiotic Neomycin is inserted into the genome, replacing the deleted gene. However, our findings show that the Neomycin resistance gene product itself can alter social behaviors in mice by disrupting cellular signaling and gene expression. This discovery has significant implications for mouse models of disorders characterized by social behavior impairments, such as autism spectrum disorders and schizophrenia. Therefore, this study emphasizes the need to remove the Neomycin cassette when investigating human diseases and developing new therapies.

## INTRODUCTION

The ability of mammalian brains to learn and adapt to new environments relies on synaptic plasticity (1). The activity-regulated cytoskeleton-associated protein (Arc), also known as activity-regulated gene of 3.1 kilobases (Arg3.1), is a key regulator of long-lasting synaptic plasticity, such as long-term potentiation and depression (2–4). Arc/Arg3.1 is an immediate early gene (IEG) predominantly expressed in excitatory glutamatergic neurons of the hippocampus, cortex, and striatum (5, 6). In response to various stimuli, it is transcribed within minutes, and its mRNA undergoes activity-dependent translation at post-synaptic sites in dendrites (7). Disruption of *Arc/Arg3.1* impairs learning, long-term memory consolidation, and sensory-motor gating, while its dysregulation affects cognitive flexibility, highlighting its essential role in synaptic plasticity (2, 8, 9).

In humans, *ARC/ARG3.1* gene has been linked to alcohol use disorder (10) and several neuropsychiatric and neurodevelopmental disorders characterized by social deficits (11). Notably, 43 *ARC/ARG3.1* single nucleotide polymorphisms have been associated with schizophrenia (12). Elevated Arc/Arg3.1 levels have been reported in mouse models of Fragile X and Angelman syndromes, both associated with autism spectrum disorders (ASD) (13, 14). Additionally, increased plasma levels of ARC/ARG3.1 have been reported in children with ASD (15). However, studies using *Arc/Arg3.1* knockout (KO) mice have yielded conflicting results: one study reported reduced sociability and social novelty in the first KO1 mouse line (16), while another found no effect on sociability in a second KO2 line (17). As pointed out by Gao et al. (17), a key difference between these lines is the presence of the Neomycin resistance cassette ($Neo^+$), which remained in the genome of the $Neo^+$ *Arc/Arg3.1* KO1 mouse line (16).

Whether the presence of $Neo^+$ in the mouse genome has an impact on the behavioral phenotypes is currently unknown.

In this study, we aimed to clarify these discrepancies about the role of Arc/Arg3.1 in social behaviors. Beyond studying sociability and social novelty, we examined various types of social behaviors, olfactory responses, and non-social behaviors in *Arc/Arg3.1* KO mouse lines. Our findings revealed significant behavior impairments in $Neo^+$, but not in $Neo^-$ *Arc/Arg3.1* KO lines. The expression of the Neo cassette product induced changes in cellular signaling and gene expression, particularly affecting the oxytocin system. Interestingly, the early social environment enrichment improved social interaction impairments with cage mates in $Neo^+$ *Arc/Arg3.1* KO mice. Overall, our findings reveal the strong effect of the Neo cassette on behaviors, particularly social behaviors.

## RESULTS

### Neo$^+$ *Arc/Arg3.1* KO dams display maternal behavior impairments

To investigate the impact of Arc/Arg3.1 on social behaviors, we examined maternal behavior in primiparous dams from three different *Arc/Arg3.1* KO lines. These included Neo$^+$ *Arc/Arg3.1* KO (Neo$^+$ KO1) and Neo$^-$ *Arc/Arg3.1* KO mice with the Neo cassette excised from the genome (Neo$^-$ KO1) derived from the first line, which previously showed sociability impairments (16), and Neo$^-$ *Arc/Arg3.1* KO (Neo$^-$ KO2) mice from the second line showing no sociability impairments (17).

We found no difference in parturition delay, litter size, lactation, or percentage of pup survival compared to corresponding WT controls in all three lines (**Table S1**). Remarkably, 93% of Neo$^+$ KO1 dams failed to retrieve their Neo$^+$ KO1 pups within 30 min of the pup retrieval test, resulting in a significantly longer pup retrieval latency compared to WT dams (**Figure 1A-B**). These dams displayed maternal care impairment, measured by reduced time crouching over the pups, rather than increased self-oriented behaviors (e.g., being outside the nest), while their ability to detect the pups was intact (**Figure S1A, Table S1**). In contrast, Neo$^-$ KO1 and KO2 dams showed no differences in maternal behavior compared to WT dams (**Figure 1C-F, Figure S1B-C, Table S1**). Notably, the differences in maternal behavior between WT or Neo$^-$ KO1 and KO2 dams from the first and second lines originated from the distinct genetic backgrounds, hybrid C57BL/6J;129S2 and pure C57BL/6J, respectively (**Figure 1**). Altogether, these results suggest that the presence of the Neo cassette in the genome of *Arc/Arg3.1* KO mice may induce maternal behavior impairments.

As these impairments might result from social impairments in KO pups (18), we tested Neo$^+$ KO1 dams with heterozygous Neo$^+$ *Arc*$^{+/-}$ pups. Neo$^+$ KO1 dams still displayed impaired maternal behavior (higher latency to retrieve their pups) compared to WT dams with heterozygous $Neo^+$ $Arc^{+/-}$ pups (**Figures 1G-H, S1D**), but to a lower extent than when tested with $Neo^+$ KO1 pups (**Figure 1A-B**). Additionally, $Neo^+$ $Arc^{+/-}$ dams showed maternal deficits comparable to $Neo^+$ KO1 with heterozygous $Neo^+$ $Arc^{+/-}$ pups.

In conclusion, the presence of the Neo cassette in the *Arc/Arg3.1* gene led to maternal behavior impairments in dams, which was also influenced by the genotype of pups.

**Neo cassette affects sociability in both dams and pups and non-social behaviors**

Subsequently, we verified if maternal impairments resulted from broader social deficits using the automatic Live Mouse Tracker (LMT) scoring device (19). Indeed, $Neo^+$ KO1 females displayed reduced social interaction (nose contact) and motivation (social approach) with both familiar (cage mates) and unfamiliar conspecifics in subsequent trials, as previously used (20), compared to WT and $Neo^-$ KO1 mice, and particularly in females (**Figures 2A-B, S2A-B, Table S2**). Notably, poor maternal care provided by $Neo^+$ KO1 dams did not cause these sociability impairments, since $Neo^+$ $Arc^{+/-}$ mice raised by either WT or $Neo^+$ KO1 dams did not exhibit social interaction impairments (**Figures 2C, S2C-D**). Sociability impairments in $Neo^+$ KO1 mice were confirmed in the sociability phase of the three-chambered test, and were not attributable to social memory deficits (**Figure S3, Table S1**).

Since maternal and social behaviors in rodents heavily rely on olfaction (21), we evaluated potential olfactory dysfunction in $Neo^+$ KO1 mice. These mice showed no impairments in general olfactory detection of neutral, food, or predator odors (**Figure S4A-C, Table S1**). In contrast, $Neo^+$ KO1 mice spent less time sniffing social odors (**Figure S4D**). To further evaluate potential alterations in social odor processing, we performed a social dominance test.

Interestingly, both dominant and subordinate $Neo^+$ KO1 mice displayed increased social dominance facing dominant WT mice (**Figure S4E**). This evidence indicates that the Neo cassette caused impairments in social odor processing rather than general odor recognition in $Neo^+$ KO1 mice.

As $Neo^+$ KO1 dams displayed worse maternal impairment with $Neo^+$ KO1 pups, we also assessed social deficits in pups. These pups exhibited shorter total time vocalizing, as well as delayed latency to vocalize, fewer and shorter ultrasound vocalizations following maternal separation compared to WT and $Neo^+$ $Arc^{+/-}$ pups (**Figure 2D-E, Table S3**). Additionally, in the olfactory-based homing test (22), both $Neo^+$ KO1 and $Arc^{+/-}$ pups exhibited delayed latency to reach the nest area and spent less time there than WT pups (**Figure S5A-B**), suggesting social impairments in both genotypes. These impairments did not result from motor impairments, as confirmed in three reflex-testing assays (**Figure S5C**).

Finally, we evaluated non-social behaviors, such as distinct short-term and mid-term memory, anxious-like, perseverative or compulsive behaviors and motor coordination (**Figures 2F-G, S6-7, Table S1**). $Neo^+$ KO1 mice displayed mild spatial memory impairment as previously reported (16), absent in $Neo^-$ KO1 mice, with no difference in short- and mid-term social or object recognition memory (**Figure S3B and S6**). Interestingly, while $Neo^+$ KO1 mice did not show anxious-like behavior, increased immobility during exploration of a novel environment was observed exclusively in $Neo^-$ KO1 mice (**Figures 2F, S7A**), suggesting that the Neo cassette may have masked this phenotype associated with *Arc/Arg3.1* deletion. Furthermore, both $Neo^+$ and $Neo^-$ *Arc* KO1 mice displayed increased self-grooming, indicating motor stereotypies due to *Arc/Arg3.1* deletion, with no difference in restrained or compulsive behaviors or motor coordination (**Figures 2G, S7B-C**).

In conclusion, the Neo cassette induced both social deficits in females and communication impairments in pups, worsening maternal behavior impairments in dams.

**Neo cassette perturbs gene expression, signaling, and oxytocin system**

Next, we studied the potential mechanisms underlying the effect of the Neo cassette on behaviors. While mainly reported in glutamatergic neurons of the hippocampus, cortex, and striatum (5, 6), we found that Arc/Arg3.1 was expressed in oxytocinergic (OT) neurons of the paraventricular nucleus of the hypothalamus (PVN) (**Figure 3A**), a critical region for regulating maternal and social behaviors (21). We next investigated the expression of the Neo cassette product in the brains of $Neo^{+}$ KO1 dams. Expression of the Neo protein product was only detected in the PVN (**Figures 3B, S8**). Consequently, we examined OT expression in the same $Neo^{+}$ KO1 dams and found comparable number of OT neurons in the PVN and supraoptic nucleus (SON), but OT fiber density was reduced in the nucleus accumbens (NAC) and prefrontal cortex (PFC) compared to WT dams (**Figures 3C, S9**). These results suggest that the Neo cassette affected OT connectivity, rather than inducing neuronal cytoxicity.

Neo product is a bacterial aminoglycoside phosphotransferase structurally resembling eukaryotic protein kinases (23). Given this similarity, we hypothesized that the Neo product might interfere with ERK1/2 kinase and ribosomal protein S6 (rpS6) signaling pathways, in a controlled cellular context. Phosphorylated ERK1/2, but not rpS6, was significantly increased in $Neo^{+}$-expressing HEK293 cells, while total ERK1/2, rpS6, and GAPDH levels remained unchanged (**Figure 3D, Table S4**), indicating that the Neo product selectively activates ERK1/2 without cytotoxic effects.

Since the ERK pathway, plays an important role in neuronal development and gene expression (24, 25), we analyzed the expression of oxytocin, vasopressin, and galanin systems regulating maternal and social behaviors (26), alongside ASD-associated genes, and other IEGs as markers of synaptic plasticity. These analyses were conducted in the PVN, SON, PFC and NAC, as well as the caudate putamen (CPU), a region implicated in motor stereotypies associated with *Arc/Arg3.1* deletion, of WT, $Neo^+$ and $Neo^-$ KO1 dams. To identify gene dysregulations associated with specific behavioral impairments, we performed integrated correlation analyses of gene expression with maternal behavior and motor stereotypies. $Neo^+$ KO1 dams segregated apart from $Neo^-$ KO1 and WT dams in both behavior and gene expression (**Figure 3E**). *Oxtr* in the PFC and *Shank3* in the SON were the most dysregulated genes correlated with pup retrieval latency, along with *Cck* and *Oxtr* in the PVN (**Figure 3E**), suggesting a potential link between maternal impairments and reduced OT signaling in the brain. Furthermore, we also identified *Oxtr* dysregulation across brain structures in naïve $Neo^+$ KO1 mice (**Figure S10, Table S5**). Social interaction deficits were correlated with IEG dysregulation, particularly *Homer1a* (**Figure S10**), suggesting impaired synaptic plasticity potentially driven by disrupted OT signaling. Focusing on the impact of *Arc/Arg3.1* deletion, the ASD-associated gene *Cntnap2* emerged as the most dysregulated gene in the PVN, correlating with increased self-grooming (**Figure 3F**), a direct behavioral consequence of *Arc/Arg3.1* deletion that was also masked by the Neo cassette.

Given the OT system dysregulation, we investigated whether acute (day 0) or subchronic (day 1-28) intranasal OT administration at 20 μg $kg^{-1}$ could alleviate sociability impairments of $Neo^+$ KO1 mice. OT treatment did not improve social interaction with unfamiliar conspecifics assessed in the LMT or reduce stereotyped behaviors in $Neo^+$ KO1 mice (**Figures 4A-B and S11A-D, Table S2**), suggesting that the Neo cassette-induced OT dysregulation cannot be counteracted by OT administration. By day 28, OT treatment significantly increased social interaction compared to saline (SAL) treatment, but only in WT mice (**Figure 4B**).

In summary, the Neo cassette induced changes in gene expression, oxytocin system dysregulation and, ultimately, maternal and social deficits that cannot be reversed by OT administration.

**Social enrichment partially improves sociability deficits in Neo$^{+}$ *Arc/Arg3.1* KO1 mice**

Interestingly, while SAL-treated Neo$^{+}$ KO1 mice showed sociability impairments over the first week, repeated exposure to unfamiliar conspecifics reduced genotype differences by day 14 (**Figure 4C**). This effect did not result from changes in anxious-like behavior, as no difference in stretch-attended posture (SAP) was observed over the trials (**Figure S11E-F, Table S2**), suggesting a potential effect of the social environment on sociability in Neo$^{+}$ KO1 mice.

To further investigate this, we tested whether an enriched early social environment within the critical post-natal period (20) would reverse the sociability and maternal impairments in Neo$^{+}$ KO1 mice by raising them with socially-competent WT and Neo$^{+}$ *Arc*$^{+/-}$ cage mates. Remarkably, social enrichment improved sociability impairment in Neo$^{+}$ KO1 when interacting with cage mates but not unfamiliar conspecifics (**Figure 4D-E, Table S2**). However, consistent with the Neo cassette's negative effect in heterozygous dams, an enriched early social environment did not reverse maternal impairments in Neo$^{+}$ KO1 dams with Neo$^{+}$ *Arc*$^{+/-}$ pups (**Figure 4F**), indicating that this behavior cannot be improved by social enrichment.

In conclusion, repeated social interactions and enriched early social environment partially rescued sociability impairments, but not maternal behavior in $Neo^+$ KO1 mice.

## DISCUSSION

In this study, we demonstrated that the presence of the Neo cassette in the *Arc/Arg3.1* gene led to significant impairments in maternal and social behaviors, as well as non-social behaviors. These results were confirmed by the absence of these impairments in two different lines of $Neo^-$ *Arc/Arg3.1* KO mice. Notably, no impairments were observed in $Neo^-$ KO1 derived from the $Neo^+$ KO1 line (retaining all other genomic elements, such as EGFP under the control of *Arc/Arg3.1* promoter). We identified that the severe maternal impairments caused by the Neo cassette resulted from sociability deficits, particularly in females. Although maternal care is vital in shaping social behaviors (27, 28), impaired maternal care of $Neo^+$ KO1 dams had no impact on the sociability of their adult offspring, ruling it out as a confounding factor in social deficits. The presence of the Neo cassette in *Arc/Arg3.1* KO mice also induced core social autistic-like traits – impairments in social communication and interaction.

Additionally, the short-term spatial memory impairment reported here and previously in $Neo^+$ KO1 mice (16), was not found after Neo cassette excision. This finding indicates that the Neo cassette may contribute to deficits in hippocampus-dependent short-term memory (29). In contrast, *Arc/Arg3.1* deletion in both KO1 and KO2 lines impairs long-term memory consolidation (2, 30). Nevertheless, *Arc/Arg3.1* deletion alone induced exacerbated stereotyped behaviors (self-grooming) and novelty-induced immobility, whereas the Neo cassette masked these anxious-like behaviors. The correlation between increased self-grooming and the ASD-associated gene *Cntnap2* due to *Arc/Arg3.1* deletion, was supported by previous findings showing that *Cntnap2* deletion in mice induces stereotyped behaviors (31).

Maternal impairments induced by the Neo cassette likely stem from social impairments and potential abnormal social cue processing. The impaired responses to social odors observed in Neo$^+$ *Arc* KO mice may result from altered organization of the olfactory bulb, where *Arc/Arg3.1* is expressed (32), and gene expression changes as previously reported in the presence of the Neo cassette (33), and/or disrupted olfactory cue processing in the central brain regions. The PVN receives indirect projections from the main olfactory system and projects to brain regions of the socio-emotional circuit, including the NAC and PFC (21). The expression of the Neo cassette in the PVN may impair olfactory cue processing and induce social interaction and maternal impairments.

The underlying molecular mechanisms through which Neo cassette affects social behaviors were linked to dysregulation in signaling and gene expression, particularly *Oxtr* and *Homer1a*. While the Neo cassette in the genome can alter the expression of nearby genes (34, 35), no major genes were located around the *Arc/Arg3.1* locus. However, the Neo cassette's protein product enhanced the ERK1/2 signaling pathway in cells. ERK dysregulations are reported in various neuropsychiatric conditions, including autism and schizophrenia (36, 37). Enhanced ERK signaling has been linked with impaired synaptic plasticity, learning, memory, and social interactions in mice (25) and autistic-like behavioral phenotypes (38), and leads to abnormal gene expression and neuronal development (24, 25). We observed in Neo$^+$ KO1 mice both abnormal gene expression and connectivity of OT neurons to the PFC and NAC, structures controlling social reward and interactions (39, 40). Although OT system is closely linked to maternal behavior (26), OT treatment did not reverse the behavioral impairments in Neo$^+$ KO1 mice, further suggesting an underlying OT signaling deficiency. In a neonatal maternal separation model of autism in rats, OT treatment rescued both social phenotype and ERK activation (41). Those results suggest that OT treatment may be ineffective when the ERK pathway is overactivated as observed in $Neo^+$ cells.

Interestingly, social interaction impairments caused by the Neo cassette were partially improved by repeated social interactions with unfamiliar conspecifics and early social enrichment. In addition, these impairments in naïve $Neo^+$ KO1 mice were linked to dysregulated IEG expression, particularly *Homer1a* as previously reported (42). IEGs play a critical role in brain plasticity and adaptation to novel environments (43), and their dysregulation has been observed in multiple mouse models of ASD showing social interaction impairments (42, 44). Therefore, repeated and long-lasting social stimulation might partially restore synaptic plasticity by modulating IEG expression, thereby relieving the Neo cassette's negative impact on social interactions.

Overall, the Neo cassette exerts off-target effects through multiple mechanisms, leading to altered behavioral phenotypes in *Arc/Arg3.1* KO mice and potentially misleading the interpretation of gene functions in other $Neo^+$ mouse models. Our findings have major implications for other KO mouse models that retain the Neo cassette. For instance, $Neo^+$ *Shank3* KO mouse lines display impaired sociability and social novelty (45), while $Neo^-$ *Shank3* KO exhibit intact sociability (46, 47). Similarly, $Neo^+$ *Fmr1* KO mice display inconsistent social behavior phenotype, ranging from enhanced social behavior (48) to intact sociability (49), whereas $Neo^-$ *Fmr1* KO show robust sociability impairments (42). These conflicting results suggest that the Neo cassette may contribute to the behavioral phenotype variability of mouse models. Additionally, the Neo cassette might contribute to the maternal behavior impairments of $Neo^+$ *Fosb* (50) and *Cd38* KO mice (51) or social interaction and/or communication phenotype of numerous KO mouse models of ASD, including $Neo^+$ *Shank3*,

*Cntnap2*, *Ube3a*, *Nlgn1-3*, *Nrxn1*, *Pten*, and *Cd38* KO mice (45, 51–56). Furthermore, through expression in the olfactory epithelia and their projecting areas (33), Neo cassette may also induce olfactory impairments reported in several mouse models of ASD (21), impacting social behaviors. Though driven by a ubiquitous phosphoglycerate kinase I (PGK) promoter, the local genomic environment may influence the expression of the Neo cassette product. Thus, the effect of the Neo cassette on behavior might vary between $Neo^{+}$ KO lines.

In conclusion, the Neo cassette exerts detrimental effects on behaviors in *Arc/Arg3.1* KO mice. The present data advocate for re-evaluating behavior phenotypes in KO mouse models carrying the Neo cassette, especially for those mimicking conditions characterized by social impairments. This re-evaluation could facilitate the predictive validity of KO models, enhancing their relevance in pre-clinical studies and improve translatability to humans.

## MATERIALS AND METHODS

For a detailed description of materials and methods, please see **SI Appendix, SI Materials and Methods**.

### *Animals*

All mouse breeding, care, and experimental procedures followed European, French and German Directives and were approved by the local ethics committees – CEEA Val de Loire N°19 and the French Ministry of Teaching, Research and Innovation (APAFIS #18035-2018121213436249), and the animal welfare office of the city of Hamburg and the animal care committee at the University Medical Center Hamburg-Eppendorf. The first line used in this study included WT and *Arc/Arg3.1* KO1 mice with the Neo cassette ($Neo^+$ KO1) (57) (JAX stock #007662), and $Neo^-$ KO1 mice, generated by excising the Neo cassette from the genome through the flox sites, bred on a hybrid 50% C57BL/6J – 50% 129S2 background. The second line comprised WT and *Arc/Arg3.1* KO mice exempt from the Neo cassette ($Neo^-$ KO2), bred on a pure C57BL/6J background (2). The two lines were housed in separate animal facilities under the same conditions: a 12-hour light/dark cycle, with food and water ad libitum, controlled temperature (21°C) and humidity (50%), housed in social groups of 2-4 animals. All experimental mice were two months old sexually naïve males and females at the start of the behavioral experiments.

### *Behaviors*

All behavioral tests were conducted at the start of the light phase in a dedicated quiet room or within the breeding facility for KO1 and KO2 lines, respectively, with one test per day. Maternal behavior was assessed in the pup retrieval and nest-building tests. Sociability was tested in the three-chambered test and social interaction in the Live Mouse Tracker (LMT) (19), while social memory was evaluated in the social recognition test. Olfaction was assessed over several assays with neutral, food, and social odors. Perseverative, stereotyped, and compulsive behaviors were tested in the spatial Y-maze task, motor stereotypy, and marble burying tests, respectively. Locomotion was assessed in the open-field test and motor coordination in the string test. Anxious-like behavior was evaluated in the elevated-plus maze, novelty-suppressed feeding, and open field tests. Spatial memory of 2M-old adult mice was assessed in object location recognition and Y-maze tasks, while the recognition memory was evaluated in the novel object recognition task. Ultrasound vocalizations following maternal separation were recorded in pups, along with the assessment of their social skills in the homing test, and their gripping, cliff avoidance, and righting reflexes in the respective tests. All animal research has been performed according to the ARRIVE guidelines (58).

***Quantitative PCR (qPCR)***

The expression of oxytocin family genes (*Oxt*, *Oxtr*, *Avp*, *Avpr1a*), immediate early genes (*Homer1a*, *Fos*, *Egr1*, *Foxp1*), ASD-associated genes (*Shank3*, *Fmr1*, *Cpeb4*, *Cntnap2*), neuropeptides genes (*Cck*, *Gal*, *Cartpt*, *Esr1*), and housekeeping gene *Gapdh* (list of primers is provided in **Table S6**) was investigated in the prefrontal cortex (PFC), nucleus accumbens (NAC), caudate putamen (CPU), paraventricular (PVN), and supraoptic (SON) nuclei of three-months-old WT, Neo$^+$ KO1, and Neo$^-$ KO1 dams, and two-month-old naïve WT and Neo$^+$ KO1 mice.

***Western blot***

Human embryonal kidney cells (HEK293) were transfected with equimolar quantities of Neo$^+$ pcDNA3.1, or control Neo$^-$ pUC19 plasmids. Protein levels of ERK1/2, phospho-ERK1/2, S6,

phospho-S6, and GAPDH in serum-deprived or serum-stimulated cells were determined by western blot.

### *Immunohistochemistry*

The brains for immunostaining were collected from WT and $Neo^+$ KO1 dams on pups' postnatal day 14 to immunolabel oxytocin pre-pro-peptide and Neo product (aminoglycoside phosphotransferase) in different brain regions, or from WT mice to co-label oxytocin pre-pro-peptide and Arc/Arg3.1.

### *Chronic oxytocin treatment*

WT and $Neo^+$ KO1 mice were injected daily for 28 days with a solution of 20 µg $kg^{-1}$ (0.03 IU) oxytocin (OT) or saline (NaCl 0.9%) as control. On days 0, 7, 14, 21, and 28 of treatment mice were subjected to social interaction in the LMT with up to four unfamiliar, age-, sex-, and treatment-matched conspecifics to evaluate the effect of OT treatment on social interaction and on day 25 for motor stereotypies.

### *Statistical analysis*

All data and statistical analyses were performed using R software (version 4.3.1). Raw data are reported at https://doi.org/10.82233/QIG5AS, while mean ± standard deviation (sd), n and statistics are reported in **Tables S1-S5**. For a detailed description of statistical methods, please see **legends** and **SI Materials and Methods**.

## ACKNOWLEDGEMENTS AND FUNDING SOURCES

Mouse breeding and care were performed at PAO, the rodent INRAE Animal Physiology Facility (https://doi.org/10.15454/1.5573896321728955E12). We thank Eva Kronberg from the UKE mouse facility at the ZMNH for breeding Neo$^-$ *Arc/Arg3.1* KO2 mice and assistance with the experiments. This work has benefited from the equipment and expertise of the Imaging facility "Plateau d'Imagerie Cellulaire" (PIC) of UMR-PRC (http://doi.org/10.17180/arap-gj59). We sincerely thank Dr. Emmanuel Valjent and Pascale Crepieux for her valuable assistance and insightful suggestions on the manuscript. We acknowledge the use of ChatGPT, developed by OpenAI, for assistance only with English editing during the preparation of this manuscript.

This project has received funding from the European Research Council (ERC) under the European Union's Horizon 2020 research and innovation program (grant agreement No. 428 851231). OO, DK and AK were supported by a grant from the Schaller-Nikolic foundation to OO and DK. LPP and AD acknowledge the LabEx MabImprove (grant ANR-10-LABX-53-01) for the financial support of AD PhD's co-fund. The funders played no role in the design, analysis, or interpretation of the study's data.

**Author contributions:** AD, ACT, HE, DK, PC, OO, and LPP designed experiments; AD, AHAK, AA, HE, EP, LD, and LPP performed experiments; AD, AHAK, AA, EP, LD, and LPP contributed to the data collection; AD, EC, NA, GL, BP, RY, OO, and LPP contributed to data analysis; AD performed all statistical analysis and data integration; AD and LPP wrote the original drafts; LPP, AS, OO and DK contributed to the funding acquisition; LPP contributed to project conceptualization and supervision.

**Declaration of interests:** Authors claim no competing interests.

**Abbreviations**

KO knock-out

Neo neomycin

IEG immediate early gene

ASD autism spectrum disorder

WT wild-type

LMT Live Mouse Tracker

PVN paraventricular nucleus of the hypothalamus

OT oxytocin

PFC prefrontal cortex

NAC nucleus accumbens

CPU caudate putamen

SON supraoptic nucleus

SAL saline

OB olfactory bulb

BNST bed nucleus of stria terminalis

MPOA medial preoptic area

## FIGURE LEGENDS

**Figure 1. The Neomycin cassette strongly impairs maternal behavior in Neo$^+$ *Arc/Arg3.1* KO dams**

In the first *Arc/Arg3.1* KO1 line raised in hybrid C57BL/6J;129S2 background, in the pup retrieval test with genotype-matched pups, a higher percentage of Neo$^+$ KO1 dams (brown; n = 15) failed to retrieve their pups compared to WT (dark gray; n = 14) (**A**), resulting in increased pup retrieval latency (**B**). In contrast, no difference in the percentage of females retrieving their pups (**C**) or pup retrieval latency (**D**) was observed between Neo$^-$ KO1 dams (yellow; n = 8) and WT (dark gray; n = 7). In the second Neo$^-$ *Arc/Arg3.1* KO2 line raised in pure C57BL/6J background, no difference in the percentage of females retrieving their pups (**E**) or pup retrieval latency (**F**) was observed between WT (pale gray; n = 6), and Neo$^-$ KO2 dams (turquoise; n = 7). In the pup retrieval test with heterozygous Neo$^+$ *Arc/Arg3.1*$^{+/-}$ pups, the percentage of females retrieving their pups was not different for Neo$^+$ *Arc/Arg3.1*$^{+/-}$ (beige; n = 15) and KO1 dams (brown; n = 14) compared to WT (dark gray; n = 10) (**G**), with increased pup retrieval latency (**H**). Data are presented as mean ± sd in **Table S1**. Groups in **A**, **C, E**, and **G** were compared by Fisher's exact test, and groups in **B**, **D, F**, and **H** by Kaplan-Meier test, with an asterisk indicating genotype effects. * $p < 0.05$; ** $p < 0.01$; *** $p < 0.001$; **** $p < 0.0001$; ns, not significant.

**Figure 2. Neo cassette affects sociability in both dams and pups and non-social behaviors**

Neo$^+$ KO1 mice (brown; n = 12, 6 females and 6 males) displayed social interaction impairments with cage mates (**A**) and unfamiliar conspecifics (**B**) compared to WT (dark gray; n = 48, 23 females and 25 males), which was not the case for Neo$^-$ KO1 mice (yellow; n = 31, 15 females and 16 males). Neo$^+$ *Arc/Arg3.1*$^{+/-}$ mice raised either by WT (beige with dark gray outline; n = 32, 16 females and 16 males) or Neo$^+$ *Arc/Arg3.1* KO1 dams (beige with brown outline; n = 27, 15 females and 12 males) displayed no impairments in social interaction with unfamiliar conspecifics compared to WT mice raised by WT dams (dark gray; n = 18, 8 females and 10 males) unlike Neo$^+$ *Arc/Arg3.1* KO1 mice raised by Neo$^+$ *Arc/Arg3.1* KO1 dams (brown; n = 12, 6 females and 6 males) (**C**). Upon maternal separation, Neo$^+$ KO1 pups (brown; n = 17) displayed increased latency to vocalize as shown by representative ultrasound vocalization traces and quantitative data (**D**) and reduced ultrasound vocalizations (**E**) compared to Neo$^+$ *Arc/Arg3.1*$^{+/-}$ (beige; n = 17) and WT pups (dark gray; n = 24). In the habituation phase of the three-chambered test, unlike Neo$^+$ KO1 mice (brown; n = 12, 6 females and 6 males), Neo$^-$ KO1 mice (yellow; n = 14, 7 females and 7 males) displayed increased novelty-induced immobility compared to WT (dark gray; n = 26, 12 females and 14 males; **F**). In the motor stereotypy test, both Neo$^+$ (brown; n = 20, 12 females and 8 males) and Neo$^-$ KO1 mice (yellow; n = 50, 26 females and 24 males) showed increased time self-grooming, but not digging compared to WT (dark gray; n = 62, 30 females and 32 males; **G**). Data are presented as mean ± sd in **Tables S1-3**. All groups were compared by Kruskal-Wallis test followed by Dunn's post-hoc test, with an asterisk indicating genotype effects (p = P adjusted). * $p < 0.05$; ** $p < 0.01$; *** $p < 0.001$.

**Figure 3. The Neomycin cassette perturbs gene expression, signaling, and oxytocin signaling**

Arc/Arg3.1 protein is expressed in oxytocinergic (OT) neurons, located in the paraventricular nucleus of the hypothalamus (PVN) (**A**). The protein product of the Neo cassette (Neo product) was detected in the PVN of Neo$^+$ *Arc/Arg3.1* KO1, but not WT dams (**B**). The density of the OT fibers was reduced in the NAC (bregma 0.86) and PFC (bregma 1.78) of Neo$^+$ KO1 dams (brown;

n = 3) compared to WT (dark gray; n = 3), as shown by representative images and fiber quantification (**C**). The expression of the Neo product in HEK293 cells transfected with $Neo^+$ pcDNA3.1 plasmid (dark blue; n = 5) induced increased phosphorylation of ERK1/2, but not rpS6, compared to starved HEK293 cells transfected with $Neo^-$ pUC19 plasmid (yellow; n = 5), as shown by a representative blot and protein quantification (**D**). Integration of maternal behavior and motor stereotypies with gene expression from the oxytocin family, immediate early genes, ASD-associated and neuropeptide genes revealed a clear separation of $Neo^+$ KO1 dams (brown triangles; n = 6) from WT (dark gray circles; n = 10) and $Neo^-$ KO1 (yellow crosses; n = 8) dams, while network analysis revealed strong positive correlations between pup retrieval latency and genes that were significantly dysregulated in $Neo^+$ KO1 dams compared to WT and $Neo^-$ KO1 (**E**). When focusing on *Arc/Arg3.1* deletion as the main discrimination factor, separation of WT and $Neo^-$ *Arc/Arg3.1* KO1 dams was less clear, with network analysis revealing a correlation between time spent self-grooming and significantly upregulated *Cntnap2* expression in the PVN of $Neo^-$ *Arc/Arg3.1* KO1 dams compared to WT (**F**). Behavior, protein quantification, and gene expression data are presented as mean ± sd in **Tables S1, S4, and S5**, respectively. Groups in **C**-**F** were compared by the Kruskal-Wallis test followed by Dunn's post-hoc test, with an asterisk indicating genotype effects (p = P adjusted). * $p < 0.05$; ** $p < 0.01$; *** $p < 0.001$. 3V, third ventricle; PFC, prefrontal cortex; NAC, nucleus accumbens; OT, oxytocin; PVN, paraventricular nucleus of the hypothalamus; SON, supraoptic nucleus; scale bars = 150 µm.

**Figure 4. Early social environment enrichment, but not intranasal oxytocin administration partially improves sociability deficits in Neo$^+$ *Arc/Arg3.1* KO1 mice**

On the first day of treatment (Day 0), acute saline (SAL) (light gray; n = 12, 8 females and 4 males) or oxytocin at 20 µg $kg^{-1}$ (OT20) administration (red; n = 16, 10 females and 6 males) Neo$^+$ KO1 mice displayed reduced social interaction with unfamiliar treatment-matched conspecifics, compared to SAL-treated (light gray; n = 15, 7 females and 8 males) and OT20-treated WT controls (red; n = 19, 8 females and 11 males) (**A**). On day 28, OT20-treated WT mice displayed increased social interaction compared to SAL-treated controls, while Neo$^+$ KO1 mice show no difference after 28 days of sub-chronic SAL or OT20 treatment (**B**). Over the trials, SAL-treated Neo$^+$ KO1 mice showed increased social interaction, while WT mice displayed constant interaction, except at day 28 (**C**). Neo$^+$ KO1 (pink; n = 12, 5 females and 7 males) and WT (light gray; n = 7, 5 females and 2 males) mice raised in early enriched social environment with WT and Neo$^+$ *Arc/Arg3.1*$^{+/-}$ littermates showed increased time in nose contact with cage mates compared to Neo$^+$ KO1 (brown; n = 24, 12 females and 12 males) and WT controls (dark gray; n = 43, 26 females and 17 males) raised with their sex-, genotype-matched littermates (**D**). However, Neo$^+$ KO1 mice raised in enriched environment displayed reduced social interaction with unfamiliar conspecifics compared to WT-enriched control (**E**). In the pup retrieval test, Neo$^+$ KO1 dams raised in the enriched environment (pink; n = 5) displayed increased latency to retrieve their Neo$^+$ *Arc/Arg3.1*$^{+/-}$ pups compared to WT control (dark gray; n = 10), similar to Neo$^+$ KO1 control (brown; n = 9) (**F**). Data are presented as mean ± sd in **Tables S1 and S2**. Groups in **A-E** were compared by the Kruskal-Wallis test followed by Dunn's post-hoc test and groups in **F** by Kaplan-Meier test, with an asterisk indicating genotype effects and ladder for day or social environment effect (p = P adjusted). # and * $p < 0.05$; ## and ** $p < 0.01$; ### and *** $p < 0.001$.

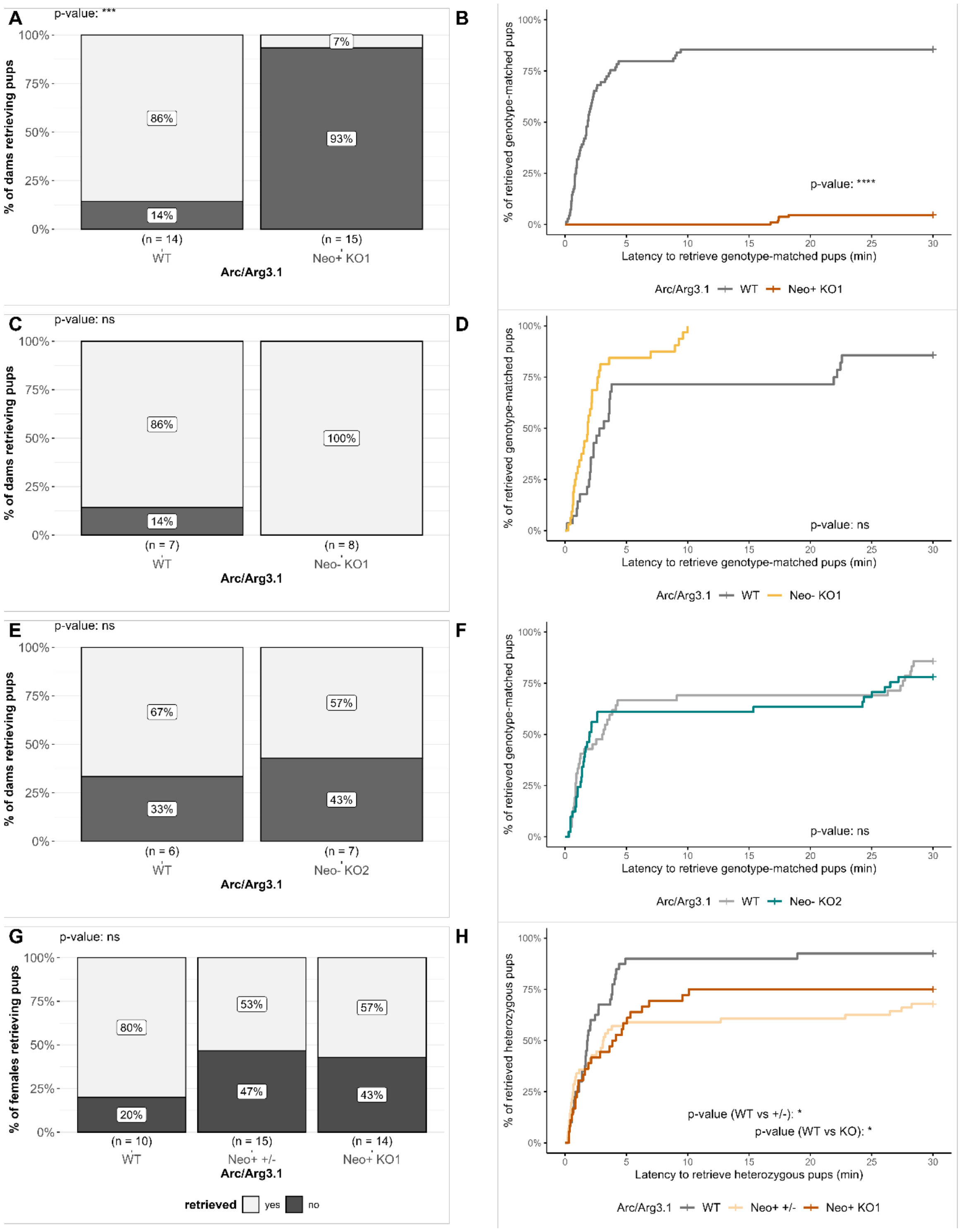

**Figure 1.** **The Neomycin cassette strongly impairs maternal behavior in Neo$^{+}$ *Arc/Arg3.1* KO dams**

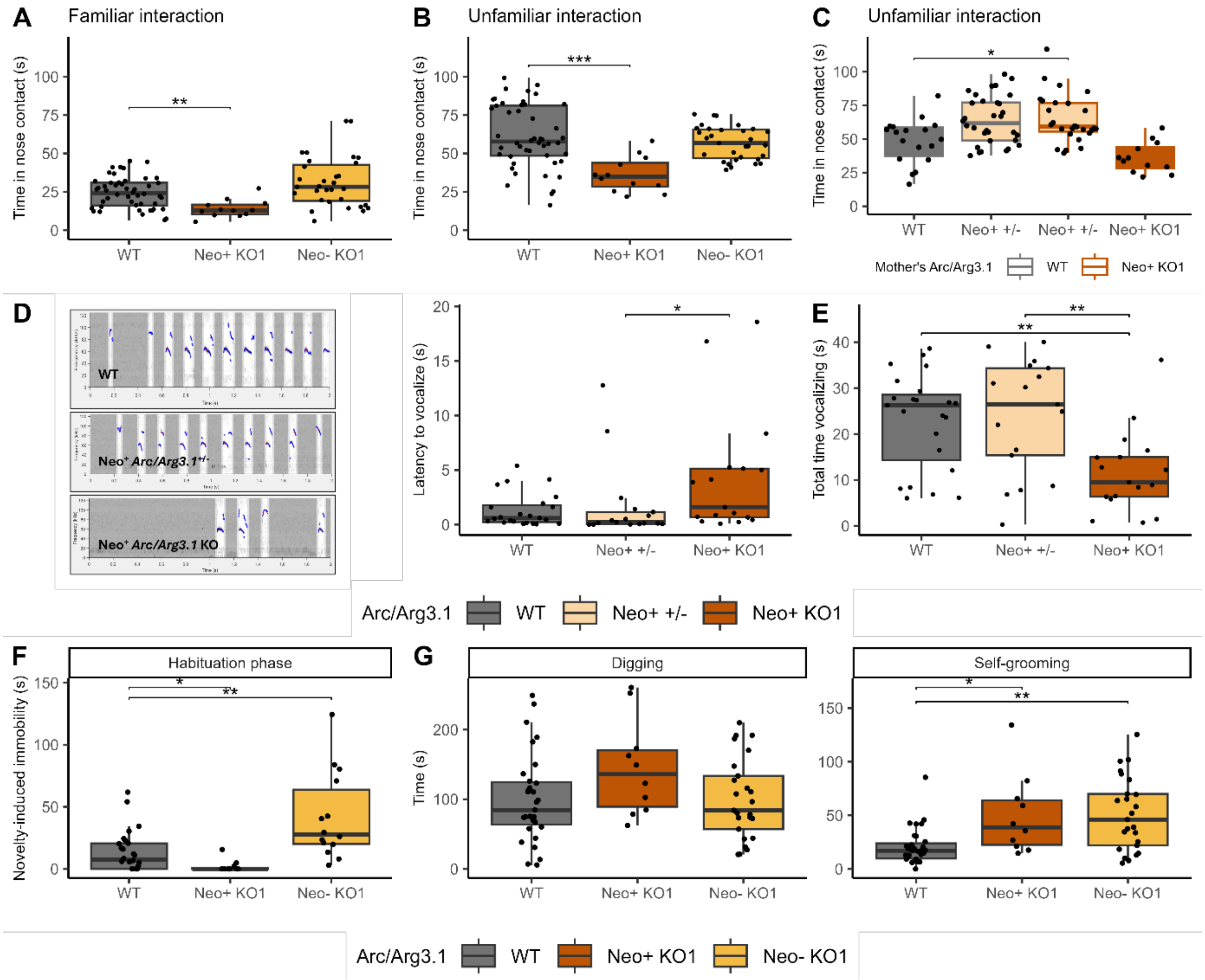

**Figure 2.** **Neo cassette affects sociability in both dams and pups and non-social behaviors**

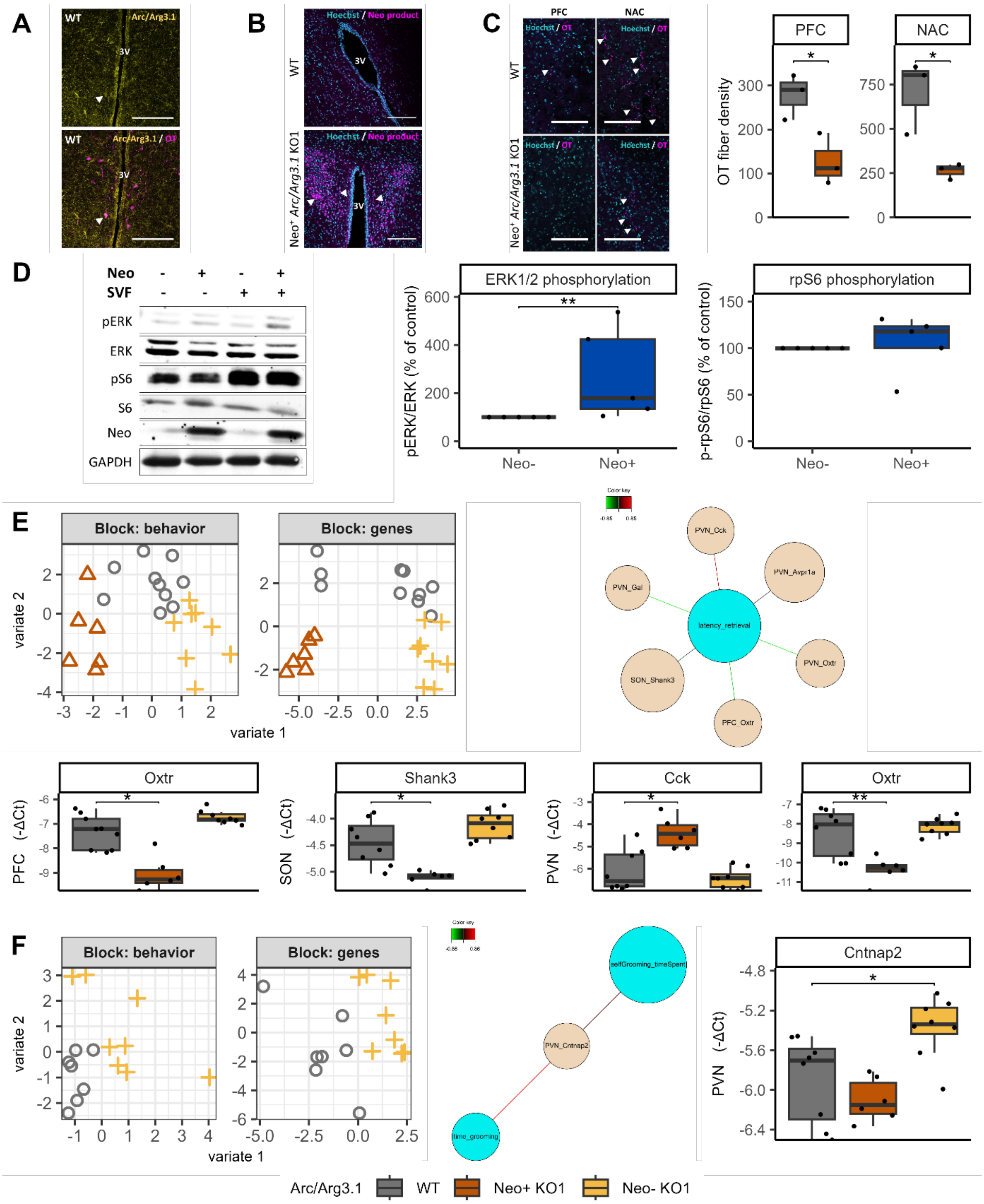

**Figure 3.** **The Neomycin cassette perturbs gene expression, signaling, and oxytocin signaling**

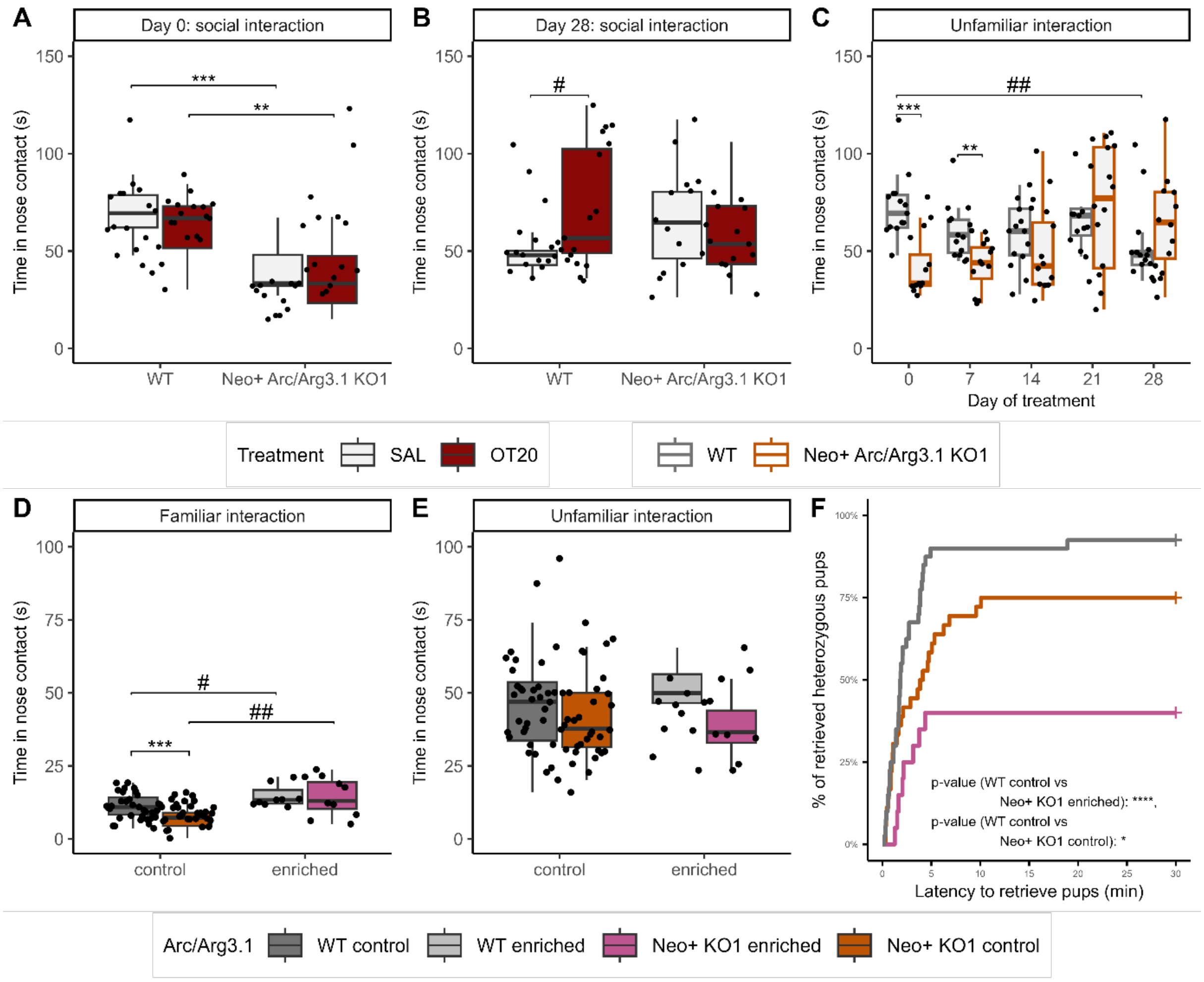

**Figure 4. Early social environment enrichment, but not intranasal oxytocin administration partially improves sociability deficits in Neo+ *Arc/Arg3.1* KO1 mice**

# Supporting Information for
The insertion of Neomycin cassette impairs maternal and social behaviors in *Arc/Arg3.1* knock-out mice.

Ana DUDAS[1], Emilia CAIRE[1], Abdurahman HASSAN A KUKU[2,3], Nicolas AZZOPARDI[1], Anil ANNAMNEEDI[1,3,4], Heba ELSEEDY[5], Gaëlle LEFORT[1], Benoît PIEGU[1], Romain YVINEC[1], Emmanuel PECNARD[1], Lucile DROBECQ[1], Anne-Charlotte TROUILLET[1], Angela SIRIGU[5], Dietmar KUHL[2], Pablo CHAMERO[1], Ora OHANA[2,3], Lucie P. PELLISSIER[1#]

[1]INRAE, CNRS, Université de Tours, PRC, 37380, Nouzilly, France

[2]Institute for Molecular and Cellular Cognition, Center for Molecular Neurobiology, University Medical Center Hamburg–Eppendorf, 20251 Hamburg, Germany

[3]Department of Neurosurgery, University Medical Center Hamburg–Eppendorf, 20251 Hamburg, Germany

[3]LE STUDIUM Loire Valley Institute for Advanced Studies, 45000, Orléans, France

[4]School of Arts and Sciences, Sai University, Chennai, India

[5]Institute of Neurosciences of la Timone, UMR7289 CNRS & Aix-Marseille Université, Marseille, France; iMIND Center of Excellence for Autism, Le Vinatier Hospital, Bron, France

#Corresponding author: Lucie P. Pellissier, PhD, Team biology of GPCR Signaling systems (BIOS), INRAE, CNRS, Université de Tours, PRC, 37380, Nouzilly, France. Phone: +33 4 47 42 79 62. Email: lucie.pellissier@inrae.fr

**This PDF file includes:**

Supporting text
Figures S1 to S11
Tables S1 to S6
SI References

**Animals**

All mice used in this study were raised in conventional type 2 cages with the same bedding and enrichment material (carton house and nesting material) in a dedicated breeding room with controlled temperature (21°C) and humidity (50%) under a 12-hour light/dark regular cycle with light on from 7 am till 7 pm for mice for $Neo^+$ and $Neo^-$ *Arc/Arg3.1* KO1 lines and from 8 pm till 8 am for mice for $Neo^-$ *Arc/Arg3.1* KO2 line. All the experimental mice were placed in the same breeding room of the animal facility. KO1 mice are raised in conventional health housing status, exempt from any monitored viral, bacterial, mycoplasma, fungi, parasites, or pathological lesions, except detected mouse Norovirus and Helicobacter spp. All $Neo^+$ and $Neo^-$ *Arc/Arg3.1* KO1 mice were bred on a mixed 50% C57BL/6J – 50% 129S2 background (Charles River, France), while $Neo^-$ *Arc/Arg3.1* KO2 mice were bred on a pure C57BL/6J background. Following the heterozygous breeding scheme, independent cohorts were generated from a minimum of three different homozygous non-inbred couples. All experimental mice were sexually naïve and between 7 and 9 weeks of age (2 months) when starting the experiments, while maternal behavior was tested in 3-4-month-old females, once they reached optimal reproductivity.

***Generation of $Neo^-$* Arc*/Arg3.1* KO1 *mouse line***

$Neo^+$ *Arc/Arg3.1* KO1 mice (1) (JAX stock #007662) contain a neomycin (Neo) resistance cassette flanked by loxP sites within the *Arc/Arg3.1* gene locus ($Neo^+$ KO1 line). To examine the potential effects of the Neo cassette on social behavior, a new line without the Neo cassette ($Neo^-$ KO1) was generated by site-specific recombination. $Neo^+$ *Arc/Arg3.1*$^{+/-}$ and KO1 males were bred with transgenic females expressing Cre recombinase (Cre/Cre genotype) under the control of the aromatase promoter (Aro) which lead to Cre expression in the gonads and therefore allows the recombination in the germline. PCR was used to confirm successful recombination in the first generation (F1), with primers binding to the *Arc/Arg3.1* promoter and Neo cassette (f1 and r2, shown in the figure below), and Neo cassette (f2 and r2). Additional primers were used to verify the *Arc/Arg3.1* locus genotype (f1 and r1), and the presence of Cre recombinase in the genome (not shown in the figure). In the second breeding round, non-inbred $Neo^{+}$ $Arc^{+/-}$ Aro-$Cre^{Tg/+}$ males and females from the F1 generation were bred and result in Neo cassette removel from one *Arc* locus. $Neo^{+}$ $Arc^{+/-}$ Aro-$Cre^{Tg/+}$ and $Neo^{-}$ $Arc^{+/-}$ Aro-$Cre^{Tg/+}$ mice from the F2 generation were then bred with wild-type (WT) mice to eliminate the Cre recombinase from the genome. Finally, $Neo^{-}$ $Arc^{+/-}$ mice without Cre recombinase in the genome were bred to obtain WT and $Neo^{-}$ KO1 mice. Following the heterozygous breeding scheme, independent experimental cohorts were generated from a minimum of three different homozygous non-inbred couples. The primers used for genotyping are reported in **Table S6**.

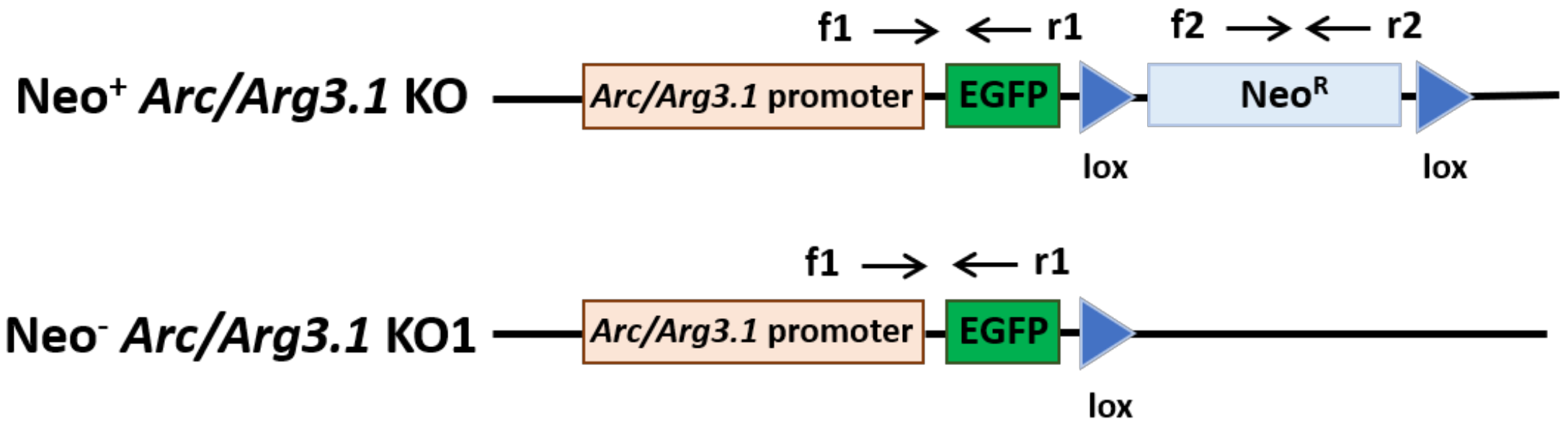

## Behavior experiments

All behavioral tests were carried out starting at the beginning of the light phase of the cycle to avoid any circadian cycle effect on behavior. One behavioral test was performed per day in a dedicated quiet room with controlled temperature (21°C) and humidity (50%), under a dim light intensity of 15 lux (except 400 lux for motor stereotypies and 60 lux for novelty-suppressed feeding test) by using the standard behavioral equipment. All floors were covered with an aluminum foil coated with a textured, non-reflective gray epoxy neutral paint to allow normal locomotion, well-being, and reduced anxious-like behaviors. Each type of behavior was tested in at least two behavioral paradigms. The experiments were performed by trained experimenters not blinded to the genotypes and treatments as treatment was provided per cage and genotypes were labelled on cages. In all except one cohort, maternal behavior was tested in females that had undergone other behavioral tests approximately 4 weeks before setting the breeding for maternal behavior assays. All behavior experiments were conducted in several cohorts of mice.

In the three-chambered, open-field, spatial Y-maze, novelty-suppressed feeding, elevated-plus maze, and all the memory tests the videos were recorded from above using a USB black and white camera with 2.8-12 mm varifocal optic and ANY-maze software (Stoelting, Ireland). In the motor stereotypies, social recognition, pup retrieval, olfactory habituation/dishabituation, food localization, innate olfactory preference, olfactory avoidance, resident-intruder aggression, string, and homing tests the videos were recorded from the side using a Sony HD FDR-AX33 4K Camescope. In the dominance tube, marble-burying tests, nesting quality evaluation, and pup reflexes tests no videos were recorded as the scoring was done in real-time by an experienced experimenter.

***Maternal behavior***

Two genotype- and age-matched primiparous dams were bred with a male of the same or opposite genotype for two weeks. After that period, the males were removed, and the dams were isolated in type 2 cages with extra enrichment (including a carton shelter, standard nesting material, and two cotton pads). The litter in their home cages was not changed until the end of the pup retrieval test to minimize disturbance. Fertility parameters such as delivery latency, litter size, percentage of dead pups in the litter, and percentage of pups with milk in their stomach were measured for each dam on postnatal days (PND) 0, 1, and 2. The maternal behavior experimental design, depicted in the figure below, included evaluation of nesting quality approximately five days before and five days after parturition (after the pup retrieval test), and the pup retrieval test conducted on PND 3-5.

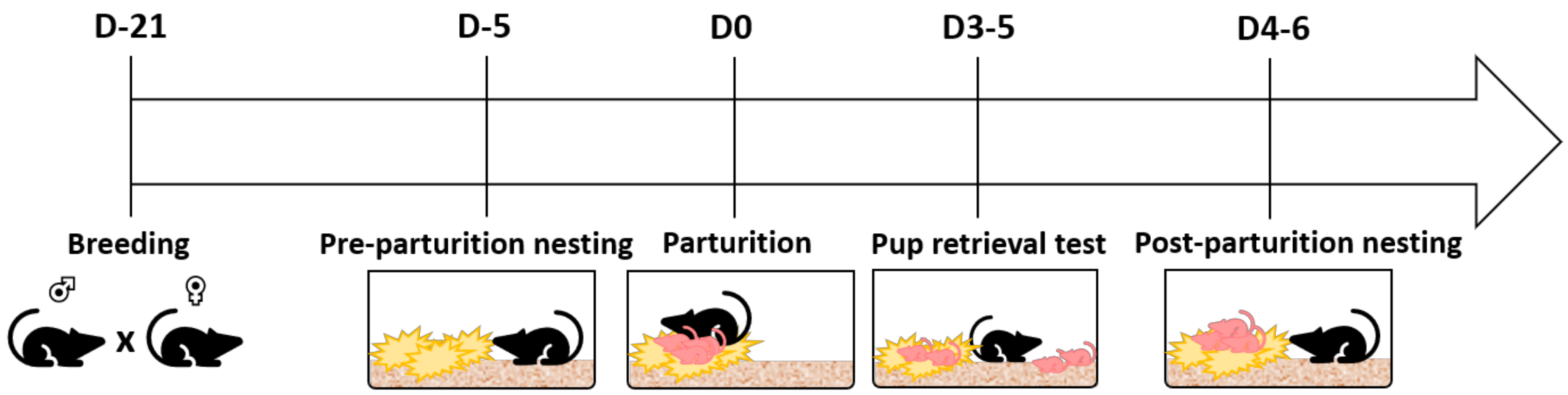

**Evaluation of nesting quality.** In a clean home cage of the experimental dams, two cotton pads were placed on the opposite side of the cage from the carton shelter in the afternoon. The next morning, the nest was observed, and nesting quality was ranked according to the criteria published in (2).

**Pup retrieval test.** On pup post-natal days (PND) 3-5, experimental dams with more than four pups in the litter were habituated in their home cages to the experimental room one to two hours before testing. After habituation, dams were briefly removed from their home cages (<1 min), and four pups from the litter were scattered on the opposite side of the cage from the nest. The dams were returned to their home cages by placing them in the nest with the remaining pups, the cage was closed with a clean grid without water or food and the behavior of the dams was recorded for 30 min. The latency of the dams to retrieve each pup to the nest, the total amount of time, number, and duration spent crouching, nesting, licking or sniffing the pups, self-grooming, digging outside the nest, and being outside the nest, as well as the latency to reach full maternal behavior (spending >1 min in the nest over the pups after all the pups have been retrieved and grouped) were measured. Retrieval was defined as a dam placing a pup into the nest. Crouching was defined as the dam being immobile in a hunched-back position over the pups in the nest and it was scored once the pups were retrieved and grouped. Licking or sniffing the pups was defined as the dam touching the pups with her nose or mouth and it was scored before retrieval when pups were visible outside the nest. Nesting was defined as collecting nesting material and placing it on the nest or around the nest to reduce dispersity in the cage.

***Social interactions***

Three-chambered test and complex social interaction in the Live Mouse Tracker (LMT) were performed as previously described (3, 4). Briefly, in the three-chambered test, experimental mice were subjected to four phases of 10 min separated by 5 min intervals: the habituation phase with empty grid cages, the sociability phase with unfamiliar sex- and age-matched WT interactor “mouse” versus Lego toy “object”, the social novelty phase with the “familiar” WT interactor from the sociability phase and a “novel” unfamiliar sex- and age-matched WT interactor replacing the toy, and the mate preference phase with the novel “familiar” WT interactor from the social novelty phase and a genotype-matched cage mate. The total amount of time, number, and mean duration of chamber entries, nose contacts to interactor, object or empty grid cages, the total distance traveled, self-grooming episodes, and immobility were measured.

For the complex social interaction in the LMT, up to four sex- and age-matched experimental mice were subjected to social interaction in the open field of the LMT across three consecutive trials which consisted of three interactions on separate days. The trials included interaction with up to four cage mates, followed by interaction with 2 unfamiliar WT and 2 unfamilair *Arc* KO mice, and finally, interaction with 4 genotype-matched mice.

Different parameters of the social interaction were automatically scored from the SQLite database using Python scripts developed by de Chaumont et al. (5).

***Social memory***

**Social recognition test.** Experimental mice were habituated to a clean conventional type 2 cage with fresh bedding for five minutes, after which they were subjected to four testing trials of 5 min separated by one-hour or 24-hour intervals. In the first trial, each experimental mouse met an unfamiliar age- and sex-matched WT interactor. One hour later, in the second trial, the experimental mice met the same interactor from the first trial. 24 hours later, in the third trial, the experimental mice met the same interactor mouse from the first and second trials. Finally, one hour later in the fourth trial, they met another unfamiliar age- and sex-matched WT interactor. The total amount of time, the number of visits, and mean duration spent in nose contact, self-grooming, self-grooming within 5 seconds following social contact, paw contacts, attacks, following, rearing, circling, huddling episodes, and the total time immobile were measured.

***Social dominance and aggression***

**Dominance tube test.** A dominance tube test adapted from Pallé et al. (6) was performed to test the inter- and intra-cage hierarchy of the experimental mice. The test consisted of five days of training and three days of testing in a conventional type 3 cage for rats using a plexiglass tube (35 cm long, 3 cm diameter) with moving transparent plexiglass gates in the middle. During the first three days of training, testing mice were placed with their cage mates for 30 minutes in a testing cage with the tube. In the remaining two training days the mice were individually placed in a testing cage with a tube and trained to cross the tube eight times, four times from each side without going back. On each testing day, each experimental mouse met an unfamiliar age-, sex-, and size-matched mouse of the opposite genotype to determine their dominance. This type of trial was performed four times for each experimental mouse. Each pair of animals was placed at the same time at each end of the tube allowing them to meet in the middle with the tube gates closed. The trial started by opening the gates and ended when one of the two animals retreated all four paws from the tube. Animals that retreated were considered losers from the trial and the other winners. Fifteen minutes after the dominance testing, the mice were tested to determine their intra-cage hierarchy. The experimental mice met each of their cage mates individually and the testing was performed as previously described. The number of wins on each testing day in dominance trials was measured, and the hierarchical status (dominant, intermediate, and subordinate) was determined for each experimental mouse by considering animals at the top, middle, and bottom of the hierarchy in the cage, respectively.

***Stereotyped and compulsive behavior***

**Motor stereotypies test.** Experimental mice were recorded individually in a clear conventional type 2 cage filled with a 3-4 cm thick layer of fresh litter for 10 minutes. The total time, number, and mean duration of self-grooming and digging for repetitive behaviors, the number of vertical jumping, circling, head shakes, and scratching episodes for stereotyped behaviors, as well as the total time spent immobile, and the number of rearing events, were measured.

**Marble-burying test.** Experimental mice were individually placed in a clean type 2 cage containing 20 glass marbles (diameter: 1.5 cm) evenly spaced on 4-cm thick fresh litter. After 15 min, the animals were removed from the testing cages, and the number of marbles not buried, buried more than half, more than two-thirds, or completely buried in the litter was determined by a trained experimenter. Marbles buried more than half were defined as marbles for which approximately 50% of the surface could be seen and those buried more than two-thirds as marbles for which only a small part of the surface could be seen in the litter. The number of marbles completely buried was determined by subtraction of the total number of marbles and the number of marbles buried more than half and more than two-thirds.

**Spatial Y-maze test.** Spatial Y-maze test was performed as previously described (3). Briefly, each experimental mouse was placed in one random arm of the Y-maze and was allowed to freely explore the three arms for 5 min. The percentage of spontaneous alternation (SPA), alternative arm returns (AAR), and same arm returns (SAR) determined from a triplet of arm entries was measured using ANY-maze software as well as the total distance traveled, the mean speed, the number of entries (entry of the 4 paws inside an arm), the total time, number and mean duration spent in each arm and analyzed using a custom-made Python script.

***Anxious-like behavior***

**Open-field test.** Experimental mice were placed for 10 minutes in an open-field arena (Ugo Basile, Italy; 100 x 100 cm) divided by 4 dark gray opaque partitions and walls in 4 open fields (46 x 46 cm). The total distance traveled, the total amount of time, the number of entries, the mean duration of center and corner entries, and the total time spent immobile were measured.

**Novelty-suppressed feeding test.** Novelty-suppressed feeding (NSF) test was performed on 16h-food-deprived mice, isolated in a standard housing cage for 30 min before individual testing. Three pellets of ordinary lab chow were placed on a white tissue in the center of each open-field arena filled with a thin layer of fresh litter and lit at 60 lux. Each mouse was placed in a corner of an arena and allowed to explore for a maximum of 10 min. Latency to feed was measured as the time necessary to bite a food pellet and was scored in real-time by a trained experimenter. Immediately after an eating event, each mouse was transferred to a clean conventional type 2 cage and allowed to feed on lab chow up to 60 min after the test. Food consumption in the cage 15 min and 60 min after the test was measured.

**Elevated plus maze test.** An elevated plus maze test was performed in a labyrinth of 4 arms 5 cm wide and located 80 cm above the ground. Two opposite arms were opened (without walls) while the other two arms were closed by side walls without the top cover. The experimental mice were placed in a closed arm of the maze and their behavior was recorded for 5 min. The total distance traveled, the total amount of time, the number of entries, and the mean duration of open-arm and closed-arm entries, as well as the total time spent immobile, were measured.

***Motor coordination***

**String test.** The testing apparatus consisted of two strings, one with and one without ridges, 38 cm long with 2 mm diameter, placed on the wooden support 49 cm above the bench surface. An empty clean cage was placed under the string for the mice to fall into. The test started by placing each experimental mouse with all four paws on the center of the string. Their latency to fall and duration of hind-limb grasp were measured in real-time by an experienced experimenter during 2 min of the test. If the mouse did not fall, its latency to fall was measured as the maximal duration of the test. The testing was first performed on a string with ridges and one and a half hours later on a string without ridges.

***Olfaction***

**Olfactory habituation/dishabituation test.** Experimental mice were first habituated in a clean conventional type 2 cage where the test took place with a clean piece of 2 cm x 2 cm Whatman paper glued to the side of the cage with adhesive paste for 5 min. 20 µL of blossom flower odor diluted 1:2 in saline was applied to a clean paper three times for 2 min, with one-minute intertrial intervals. The adhesive paste was changed after three applications of the blossom flower odor. Next, 20 µL of sex-matched urine collected from several 4-7 weeks old WT males or females not included in the test and pooled from different cages to avoid any olfactory bias (estrous state) was applied three times for 2 min in the same way. The total time, number, and duration of sniffing each odor, as well as time spent immobile and self-grooming were measured.

**Food localization test.** A day before the test, pieces of a sweet butter cookie were placed in the cages of experimental mice to familiarize them with the odor of the cookie. The experimental mice were then starved overnight (approximately 16 hours). A piece of cookie (approximately 2 g) was buried underneath a 3 cm thick layer of litter in a clean conventional type 2 cage and each mouse was introduced into the testing cage for 5 min. The latency to feed, defined as the time taken to dig out and bite a cookie, was measured.

**Innate olfactory preference test.** Experimental mice were tested in their home cage with two neural odors (2-phenoxyethanol diluted 1:1000 in mineral oil and 3-nonanone diluted 1:10000 in mineral oil), one food odor (10% milk powder diluted in distilled water), and pure same-sex and opposite-sex urine that was collected from 10 WT mice not included in the test and pooled to avoid any olfactory bias. A piece of Whatman paper (2 cm x 2 cm) was glued to the home cage of the mice using the adhesive paste on the opposite side of the carton shelter and the mice were habituated to the paper for 5 min. The odors were then applied to the Whatman paper for 2 min as follows: 15 μL of diluted 2-phenoxyethanol, 15 μL of diluted 3-nonanone, 15 μL of diluted milk, 5 μL of male, and finally 5 μL of female urine. The time between each trial was one minute and the adhesive paste was changed between odors. The total time, number, and duration of sniffing each odor were measured.

**Olfactory avoidance test.** Experimental mice were tested in their home cage with predator odors. A piece of Whatman paper (2 cm x 2 cm) was glued to the cage using the adhesive paste on the opposite side of the carton shelter, and the mice were habituated to the paper for 5 min. After habituation, 10 μL of predator odor (a mixture of mountain lion's, bobcat's, fox's, wolf's, coyote's, and marten's urines) was then applied to the Whatman paper, and total time, number, and duration of freezing, immobility and sniffing predator odors were measured for 15 min.

***Memory***

**Spatial object location recognition test.** Experimental mice were habituated two times for 10 min in the open field arena to two identical Lego Duplo toys taped at the border of the center area on the same side of the arena using adhesive paste. The interval between the two habituations was one and a half hours. Two hours after the last habituation, mice were tested for their spatial memory by placing them for 10 min in the open field arena where one Lego Duplo toy (called "old object") was placed in the same position as in habituation, while the other toy (called "spatial object") was moved to the other side of the center area, diagonal from the first object. The total time, mean duration of contact with the "old" and "spatial" object, and number of visits to each object were measured in the habituation and testing phases. From the total time in contact with the "old" and "spatial" object, the discrimination index was calculated as:

*discrimination index = (spatialObject_timeSpent – oldObject_timeSpent) / (spatialObject_timeSpent + oldObject_timeSpent)*

**Novel object recognition test.** Experimental mice were habituated two times for 10 min in the open field arena to the two identical Lego Duplo toys taped at the border of the center area on the same side of the arena using adhesive paste. The interval between the two habituations was one and a half hours. 24 hours later, experimental mice were placed in the open field arena for 10 min with one Lego Duplo toy from the habituation phase and a bigger Lego Duplo toy with a different color as a "novel" object. Both objects were placed in the same location like in the habituation phase. Two hours later, experimental mice were again placed in the open field arena for 10 min with a "novel" object from the previous phase and a Falcon tube replacing the Lego Duplo toy from the habituation phase as the second "novel" object. Both objects were placed in the same location as in the previous testing phase. The total time, mean duration of contact, and number of visits to both objects in the habituation and testing phases were measured.

**Spatial Y-maze memory test.** One arm of the Y-maze was closed with the doors provided by the manufacturer. The experimental mice were placed in one opened arm of the maze and were habituated to the apparatus for 15 minutes. One hour after the habituation, all the doors of the Y-maze were opened, mice were placed again in the same arm of the maze as in the habituation phase and their spontaneous alternation was recorded for 5 minutes. The total amount of time, mean duration, number of entries to the arm of the maze closed in the habituation ("hidden arm"), and the total time spent immobile were measured.

***Pup testing***

**Recording pups' ultrasound vocalizations (USVs).** On PND 6, each pup from the litter was taken from its home cage and placed into an empty plastic container (11 x 7 x 3.5 cm), which was then positioned inside a homemade sound-attenuating isolation box (32 x 21 x 14 cm) made of cardboard. USVs were recorded for 4 min using an ultrasonic microphone (Ultramic UM250K, Dodotronic, Italy), placed 20 cm above the pup in its plastic container, and connected to the SeaWave software (SeaPro Package, http://www-9.unipv.it/cibra/seawave.html). The total amount of time, number, duration, frequency, and amplitude of USVs were automatically measured by USVSeg software (7).

**Homing test.** The homing test was performed as previously described (8). Briefly, on PND 13, all pups from the litter were separated from their mother in a clean conventional type 2 cage and placed on a heating pad set at 35°C for 30 min. Each experimental pup was then moved to a type 2 cage which contained one-third of the litter from the pup's home cage (nest area) and two-thirds of the fresh litter (unfamiliar area), by placing it at the edge of the fresh litter, and its behavior was recorded for 5 min. The latency to reach the nest area by placing the front paws into it, the total amount of time, the number of visits, and the mean duration of visits to the nest area, as well as the time spent immobile, were measured.

**Grasping reflex test.** The grasping reflex was tested on PND7 by placing a cotton-tipped applicator close to each forepaw and hind paw of a pup. The score was given 0 if there was no grasping of the applicator with both forepaws and both hind paws, 1 if there was only one forepaw or one hind paw grasp, and 2 if the pup grasped the applicator with both forepaws and hind paws.

**Righting reflex test.** The righting reflex of the pups was assessed on PND7 by placing them on their backs and observing their ability to flip onto all four paws within 2 minutes. The time taken to fully right and achieve the correct posture was measured. If a pup did not fully right itself within the 2-minute test period, the full duration of the test was noted as the time spent to right.

**Cliff avoidance test.** Cliff avoidance was tested on PND7 by placing each pup near the edge of a table, gently nudging it toward the edge, and scoring avoidance. The score was given 0 if there were no movement or falls, 1 if there were attempts to move from the cliff but with hanging limbs, and 2 if the pup successfully moved from the cliff.

***Scoring***

Scoring was conducted manually for the pup retrieval test, sociability, social novelty, and mate interaction phases of the three-chambered test, social recognition, homing, olfactory habituation/dishabituation, innate olfactory preference, olfactory avoidance, food localization, and motor stereotypy tests. Manual scoring was done *a posteriori* by a trained experimenter blinded to the genotypes by using the Behavioral Observation Research Interactive Software (BORIS) (9). For the spatial Y-maze, open-field, elevated plus maze, object location recognition, spatial Y-maze memory, and novel object recognition tests, scoring was conducted automatically using the automatic animal tracking ANY-maze software. The automatic ANY-maze software was configured to detect animal immobility using a 95% threshold sensitivity, animal entries in compartment or arms using 80% of the animal body (e.g., when 4 paws of the animals entered the chamber or arms in the three-chambered and Y-maze tests), and the mouse head for object interaction. For the pup reflexes testing, novelty-suppressed feeding, marble burying, and string tests, scoring was done in real-time by a trained experimenter not blinded to the genotypes. Behavioral criteria to exclude animals from behavioral tests were: over 30% of the time spent immobile and/or lack of exploration of all arms or compartments of the testing apparatus.

**Quantitative PCR (qPCR)**

Three-month-old WT, Neo$^+$ *Arc* KO, and Neo$^-$ *Arc* KO1 dams and two-month-old naïve WT and Neo$^+$ *Arc* KO mice that underwent social interaction with unfamiliar conspecifics (SI condition) or stayed in their home cages as basal controls (basal condition) were euthanized by cervical dislocation. Their brains were rapidly dissected, and 1 mm thick brain slices were prepared using a coronal mouse brain matrix. The prefrontal cortex (PFC), nucleus accumbens (NAC), caudate putamen (CPU), paraventricular (PVN), and supraoptic (SON) nuclei were collected using a 2 mm diameter puncher (two punches for lateralized regions) as described in (3) and immediately frozen until further use. After tissue homogenization using a polytron (Grosseron, France, PT1200E), total RNAs were extracted according to the Direct-zol™ RNA Microprep and Miniprep kit (ZymoResearch, Ozyme, France, R2063 and R2050, respectively) and quantified using ND2000 nanodrop (Thermo Fisher Scientific, France). The cDNAs were generated from 0.5 (2 punches) or 0.25 µg (1 punch) of total RNAs using the SensiFast reverse transcriptase kit (Ozyme-Bioline, BIO-65054) and quantitative PCR (qPCR) was performed in triplicates according to 2X ONE Green Fast qPCR premix (Ozyme, France, OZYA008-1000) with 1 µL of cDNA and 1 µM of each validated couple of primers (list of primers in **Table S6**). The following qPCR protocol was applied for 40 cycles: 95 °C for 5 s, 60 °C for 15 s, and 60 °C for 30 s.

***Arc/Arg3.1* genomic environment exploration**

The three-dimensional genomic environment of the murine *Arc/Arg3.1* gene was explored using the 3D Genome Browser (10). The chromatin interaction data used was Hi-C, mouse assembly mm10, selected tissue was the brain (tissue "Jiang_2017-raw") and the interactions on chromosome 15 were visualized in the region 74000000-75000000. The Neo cassette in *Arc/Arg3.1* gene is located in a small topologically associating domain (TAD) on murine chromosome 15, and thus unlikely to cause major gene expression effects.

**Western blot**

The Human Embryonic Kidney 293 cell line (HEK293A) was cultivated in DMEM without phenol red supplemented with 10% heat-inactivated fetal bovine serum and 1% penicillin. The cells were plated in 6-well cell culture plates at concentration 5 x $10^5$ cells/well and transfected with 5.5 ng/1 kb of $Neo^+$ pcDNA3.1 plasmid and $Neo^-$ pUC19 control plasmid using Metafectene PRO (Biontex, Germany, T020-1.0) following the manufacturer's protocol. Two days after the transfections, cells were starved overnight by changing their culture medium to DMEM without phenol red and serum, or their medium was changed to fresh. Five independent experiments were performed with two biological replicates in each experiment. The cellular lysates were harvested in 150 µL of Radio-Immunoprecipitation Assay buffer (RIPA buffer, 50 mM Tris HCl pH=7.4, 150 mM NaCl, 1% Igepal CA-630 (Sigma-Aldrich, USA, I3021), 1% HaltTM protease and phosphatase inhibitor cocktail (Thermo Fisher Scientific, France, 78429), 1 µM PMSF), frozen at -20°C, thawed, homogenized by sonication 2 times for 20 seconds, and centrifuged at 10000 g for 10 minutes. Supernatants were collected and protein concentration was determined using the Bradford assay (Bio-Rad Protein Assay Dye Reagent Concentrate, Bio-Rad Laboratories, USA, 5000006). 40 µg of each sample were mixed with Laemli SDS Reducing Sample Buffer (Thermo Fisher Scientific, France, J61337.AC), heated at 85°C for 10 minutes, separated by SDS-PAGE electrophoresis and transferred to a nitrocellulose membrane (Trans-Blot Turbo RTA Transfer Kit Nitrocellulose, Bio-Rad Laboratories, USA, 1704271) using the Trans-Blot Turbo Transfer System (Bio-Rad Laboratories, USA, 1704150). Membranes were briefly stained with 0.1% Ponceau stain, de-colorized in distilled water, blocked in 5% milk in Tris Buffered Saline with 0.1% Tween® 20 (TBS-T; Sigma-Aldrich, USA, P1379) for one hour at room temperature (RT) and then incubated with mouse anti-p44/42 MAPK (Erk1/2) (1:1000; monoclonal to p44/42 MAPK, Cell Signaling Technology, USA, #9107) and rabbit anti-phospho-p44/42 MAPK (Erk1/2) (1:2000; monoclonal to phospho-p44/42 MAPK (Thr202/Tyr204), Cell Signaling Technology, USA, #4370), or mouse anti-S6 (1:1000; monoclonal to S6, Cell Signaling Technology, USA, #2317) and rabbit anti-phosho-S6 (1:2000; monoclonal to phosphor-S6 (Ser235/236), Cell Signaling Technology, USA, #4858) primary antibodies overnight at 4°C. Membranes were washed (3 x 10 minutes) in TBS-T and incubated for one hour with goat anti-mouse IRDye®680CW (1:15000; LI-COR®, Lincoln, USA, 926-68070) and anti-rabbit IRDye®800CW (1:15000; LI-COR®, Lincoln, USA, 926-3221) secondary antibodies in TBS-T with 5% milk at RT. Finally, blots were washed (3 x 10 minutes) in TBS-T and visualized by the Odyssey CLx Imager (LI-COR Biosciences, USA, 9140). After visualization, membranes were briefly washed in distilled water, stripped in 0.2 M NaOH for 10 min at RT, washed in distilled water until the pH returned to neutral, blocked again as previously described, and incubated with rabbit anti-GAPDH (1:15000; polyclonal, Proteintech, USA, 10494-1-AP) primary antibody overnight at 4°C. Finally, membranes were washed as described and incubated with goat anti-rabbit

IRDye®680CW (1:15000; LI-COR®, Lincoln, USA, 926-68070) secondary antibody following the previously described procedure. All the blots were analyzed and quantified in ImageJ using the Gel Analysis plugin (11). ERK1/2 and S6 protein levels were normalized to GAPDH protein levels, while phosho-ERK1/2 and phospho-S6 levels were normalized to total ERK1/2 or S6 protein levels. Protein levels in biological replicates from each experiment were averaged and all the data from the same experiment were normalized to the samples from pUC19 starved conditions as a control for interblot normalization.

The images of all blots used for quantification and the raw quantification data are available at https://doi.org/10.82233/QIG5AS.

**Immunostaining**

The brains for immunostaining were collected from dams on postnatal day 14 that were anesthetized by a mix of 100 mg/kg ketamine and 5 mg/kg xylazine while their pups were euthanized by guillotine. Dams were perfused transcardially with 0.1 M PBS followed by cold 0.1 M PBS containing 4% paraformaldehyde (PFA). Brains were removed, post-fixed overnight in 4% PFA, and cryoprotected in 0.1 M PBS containing 30% sucrose. After the brains sank in sucrose solution, they were snap-frozen in cold isopentane (temperature between -20°C and -30°C) and kept at -80°C until further processing. The frozen brains were embedded separately in the Tissue-Tek O.C.T. compound and cut into 30 μm-thick slices on a CryoStar NX70 cryostat (MM France, France, 957060). The brain sections were kept in PBS containing 0.1% sodium azide until immunostaining. Sections were washed (1 × 15 min) in PBS with shaking (150 rpm) and then blocked for one hour at room temperature in PBS containing 10% natural goat serum, 0.4% Triton, and 10 mg/mL bovine serum albumin (BSA). After blocking, the sections were incubated in the blocking solution supplemented with oxytocin primary antibody (1:500, rabbit monoclonal [EPR20973] to oxytocin-neurophysin 1, Abcam, USA, ab212193), aminoglycoside phosphotransferase (Neo product) primary antibody (1:150, rabbit polyclonal to Neomycin phosphotransferase II, Sigma-Aldrich, USA, #06-747), or oxytocin and Arc/Arg3.1 primary antibodies(rabbit polyclonal to oxytocin, Phoenix Pharmaceutical, USA, #G-051-01; mouse monoclonal [1G4B5] to Arc/Arg3.1, Proteintech, USA, #66550-1-Ig) and incubated overnight at 4°C. Sections were washed (3 x 15 min) in PBS with shaking (150 rpm) after which they were incubated with PBS containing 10 mg/mL BSA supplemented with secondary antibody (1:500, Cy3-conjugated goat anti-rabbit IgG, Jackson ImmunoResearch, USA, lot #150960) and Hoechst B2261 (1:1000, Sigma-Aldrich, USA, 14533) for one hour at room temperature with shaking (150 rpm). Samples were washed (3 x 15 min) in PBS with shaking (150 rpm) and mounted on Superfrost Plus adhesion microscope slides (Epredia, USA, J1800AMNZ) with Immu-Mount mounting medium (Epredia, USA, 9990412).

***Image acquisition and analysis***

For the oxytocin immunolabeling, images of all fluorescent-labeled sections were acquired using a 20x objective and Zeiss Axioscan Z1 (Carl Zeiss Microscopy, Germany). The lamp settings were kept the same during the acquisition of all images, with lamp intensity set at 30% and exposure time at 11.5 ms for Hoechst, lamp intensity at 15% and exposure time at 40 ms for Cy3. To improve the quality of the image, three images were taken along the z-axis with 11 μm range and averaged to obtain the final images used for quantitative analyses. The image analysis of oxytocin immunolabeling was performed using QuPath software (12). The number of oxytocinergic neurons was determined in the PVN and SON along the anterior-posterior axis, while the number, fluorescence intensity, and density of oxytocin fibers were determined in the projection areas of the oxytocinergic neurons (PFC, NAC, CPU, MPOA, BNST). The number of oxytocinergic neurons for each dam was determined by averaging the number in the two hemispheres and, as the number of neurons did not show differences between genotypes along the anterior-posterior axis of the PVN and SON, one specific coordinate with the highest number of slices for each dam was chosen (bregma -0.94) and the number of oxytocinergic neurons for this coordinate was averaged to get one final value per dam. For the fibre quantification, one specific coordinate with the lowest variability of fiber density was chosen in the NAC (bregma 0.86) and PFC (1.78), and the values were first averaged for the two hemispheres on the same slice, and then the values for all the slices were averaged to get one value per dam.

For Neo product immunolabeling, images for all fluorescent-labelled sections were acquired using a 20x objective with 0.8 numerical aperture and Zeiss inverted Axio Observer Z1 confocal microscope (Carl Zeiss Microscopy, Germany). The ranges of lasers were adjusted to avoid overlapping of signals from different fluorophores, with 405 nm laser detecting Hoechst B2261 signal in the range 410-471 nm and 561 nm laser detecting Cy3 signal in the range 565-620 nm. The laser settings were kept the same during the acquisition of all images, with the laser power at 2.0, master gain at 750, and pinhole opening of one airy unit for both lasers. Qualitative image analysis of Neo product immunolabeling was performed in Fiji (11).

Representative images used for quantitative analysis of oxytocin and qualitative analysis of Neo product immunolabelling are available at https://doi.org/10.82233/QIG5AS.

### Chronic oxytocin treatment

One week before starting the treatment, WT and Neo$^{+}$ *Arc* KO mice were implanted with RFID transponders for automatic chip detection in the Live Mouse Tracker (LMT), as detailed in (4). A solution of 20 µg/kg (0.3 IU) oxytocin was prepared by diluting 1 mg of ≥95% oxytocin corresponding to 600 IU (OT; Tocris Bioscience, USA, batch 16A/276657) in sterile saline (NaCl 0.9%). Aliquots of 250 µL of diluted oxytocin (OT) and saline (SAL) were stored at -20°C and on each treatment day, a fresh aliquot of OT and SAL was thawed. Mice were weighed on the first day of treatment and every 3-4 days thereafter to adjust the OT or SAL dosage accordingly. OT or SAL was administered intranasally every day for 28 days. Half of the dose was delivered into one nostril, allowing the mouse to inhale it, followed by the other half into the opposite nostril. Injections were given in the morning in a quiet room separate from the testing room, except on days 0, 7, 14, 21, and 28. On these days, injections were given in the experimental room, and 15 minutes later, the mice were subjected to social interaction in the LMT with up to four unfamiliar, age-, sex-, and treatment-matched congeners as previously described to evaluate the effect of OT treatment on social interaction.

**Statistical analysis**

All data and statistical analyses were performed using R software (version 4.3.1). Animal outliers (±2 standard deviations) were removed from the qPCR dataset, while for behavior, females from one cohort tested in maternal behavior assays were excluded due to the abnormal behavior of the WT (increased jumping while manipulating, increased self-grooming and immobility displayed in different behavior assays compared to WT from other experimental cohorts). For qPCR after maternal behavior assays, batch corrections were applied using the ComBat sva package (13). The behavior of WT from distinct experimental cohorts tested in the three-chambered, object location recognition, motor stereotypy tests, and the Live Mouse Tracker remained consistent and did not require correction, so the data from WT mice from different cohorts were pooled together. For comparisons between groups, the Kruskal-Wallis tests with Dunn's posthoc tests were performed using the rstatix package (14). P-values were adjusted with Benjamini-Hochberg correction (15). For the pup retrieval test, comparisons of retrieval latencies between groups were conducted by the Kaplan-Meyer test from the survminer package (16), and the percentage of females retrieving pups, as well as pup reflex scores between groups was compared by the Fisher's test using the ggstatsplot package (17). Furthermore, the integration of qPCR and maternal behavior data was performed using DIABLO (Multiblock (s)PLS-DA) implemented in the mixOmics package (18). The compromise parameter was set at 0.8 for data obtained in dams and at 0.75 for data obtained in naïve mice to maximize the correlation between qPCR and behavioral datasets. The probability of error level (alpha) was set at 0.05. Raw data are available at https://doi.org/10.82233/QIG5AS, while mean ± standard deviation (sd), and statistics are represented in **Tables S1-S5**.

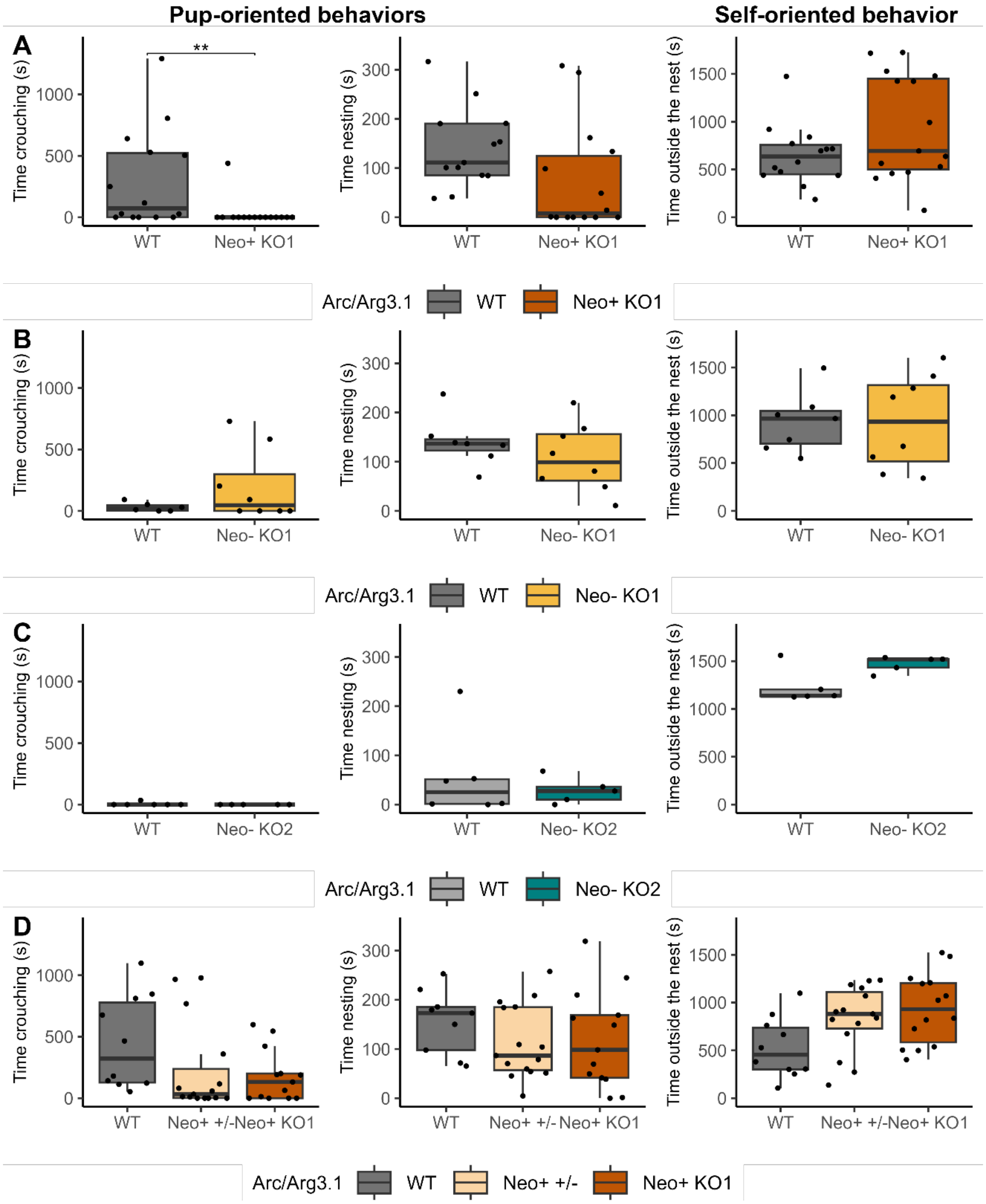

**Fig. S1. Neo⁺ *Arc/Arg3.1*-deficient dams display maternal care impairments**

In the pup retrieval test with their genotype-matched pups, Neo⁺ KO1 dams of the first *Arc/Arg3.1* line (brown; n = 15) spent significantly less time crouching over the pups, and nesting compared to WT (dark gray; n = 14), with comparable time spent outside of the nest between genotypes (**A**). Neo⁻ KO1 dams (yellow; n = 8) showed no difference in the time spent crouching over the pups, nesting, or outside the nest compared to WT (dark gray; n = 7) (**B**). Similarly, $Neo^-$ KO2 dams from the second *Arc/Arg3.1* line (turquoise; n = 5) showed no difference compared to WT (pale gray; n = 6) (**C**). In the pup retrieval test with $Neo^+$ *Arc/Arg3.1*$^{+/-}$ pups, both $Neo^+$ *Arc/Arg3.1*$^{+/-}$ (beige; n = 15) and $Neo^+$ KO1 dams (brown; n = 14) spent less time crouching over the pups and nesting, and more time outside the nest compared to WT (dark gray; n = 10) (**D**). Data are presented as mean ± sd in **Table S1**. All groups were compared by Kruskal-Wallis test followed by Dunn's post-hoc test, with an asterisk indicating genotype effects (p = P adjusted). * $p < 0.05$; ** $p < 0.01$; *** $p < 0.001$.

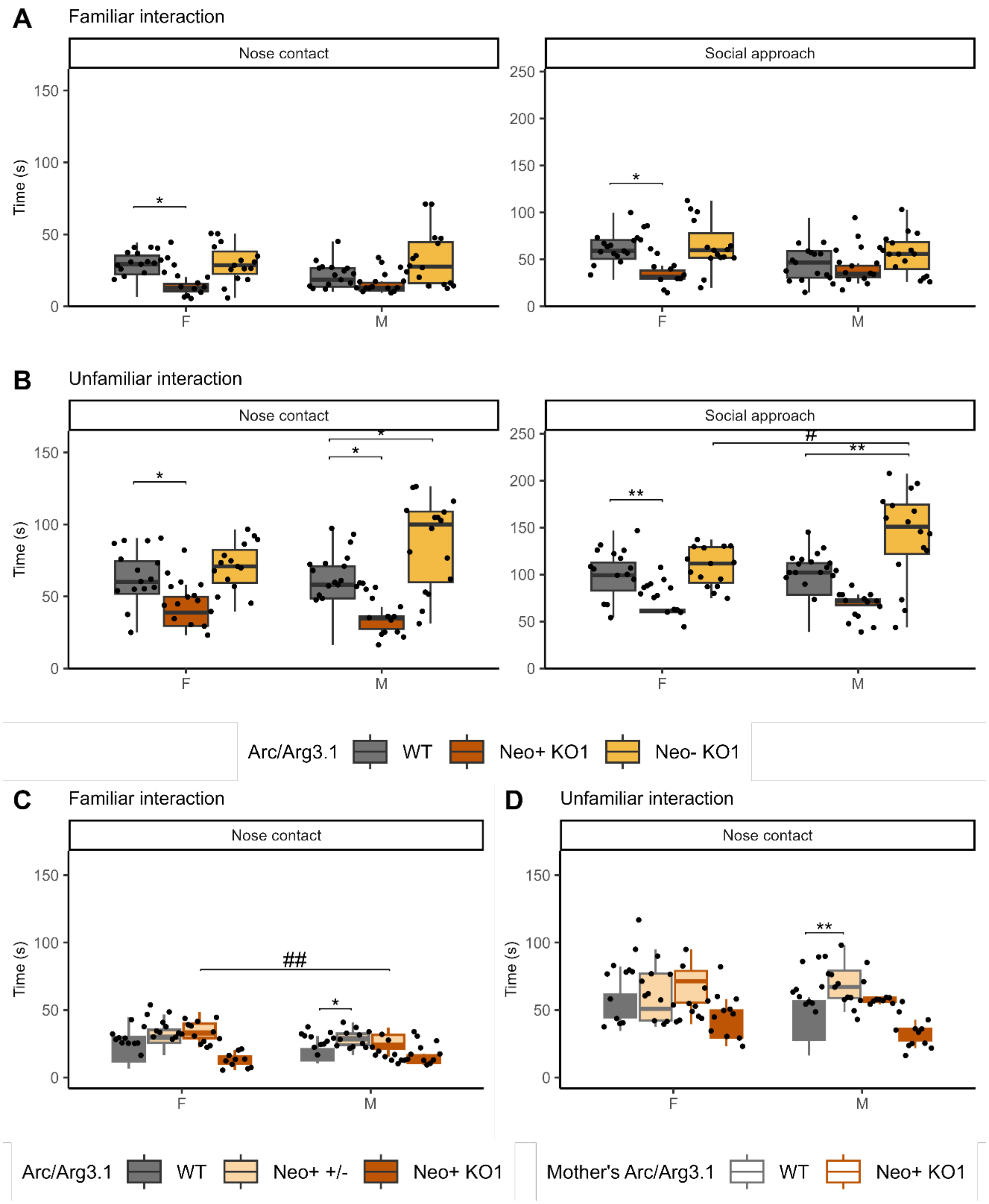

**Fig. S2. Social interaction impairments are more pronounced in Neo$^{+}$ *Arc/Arg3.1* KO1 females than males**

In in the Live Mouse Tracker (LMT), when interacting with genotype-matched cage mates, Neo$^{+}$ KO1 females (brown; n = 6) displayed significantly reduced time in nose contact and social approach compared to WT females (dark gray; n = 23), while Neo$^{-}$ KO1 females (yellow; n = 15), $Neo^+$ *Arc/Arg3.1* KO1 (brown; n = 6) and $Neo^-$ KO1 males (yellow; n = 16) displayed no difference compared to WT (dark gray; n = 25 males) (**A**). When interacting with unfamiliar conspecifics, both $Neo^+$ KO1 females and males showed significantly decreased time in nose contact, while only females spent less time in social approach compared to WT. In constrast, $Neo^-$ KO1 males showed increased time in nose contact and social approach compared to WT males, and increased time in social approach compared to $Neo^-$ KO1 females (**B**). $Neo^+$ *Arc/Arg3.1*$^{+/-}$ males or females raised by either WT (beige with dark gray outline; n = 32, 16 females and 16 males) or $Neo^+$ KO1 dams (beige with brown outline; n = 27, 15 females and 12 males) showed no impairments in social interaction with their genotype-matched cage mates compared to WT (dark gray; n = 18, 8 females and 10 males), while $Neo^+$ KO1 control mice (brown; n = 12, 6 females and 6 males) showed reduced social interaction compared to WT (**C**). No difference in social interaction with unfamiliar conspecifics was observed between $Neo^+$ *Arc/Arg3.1*$^{+/-}$ mice raised by $Neo^+$ KO1 or WT dams, while $Neo^+$ KO1 controls showed reduced social interaction compared to WT (**D**). Data are presented as mean ± sd in **Table S2**. All groups were compared by the Kruskal-Wallis test followed by Dunn's post-hoc test, with an asterisk indicating genotype effects and ladder sex effects (p = P adjusted). * or # $p < 0.05$; ** or ## $p < 0.01$; *** or ### $p < 0.001$.

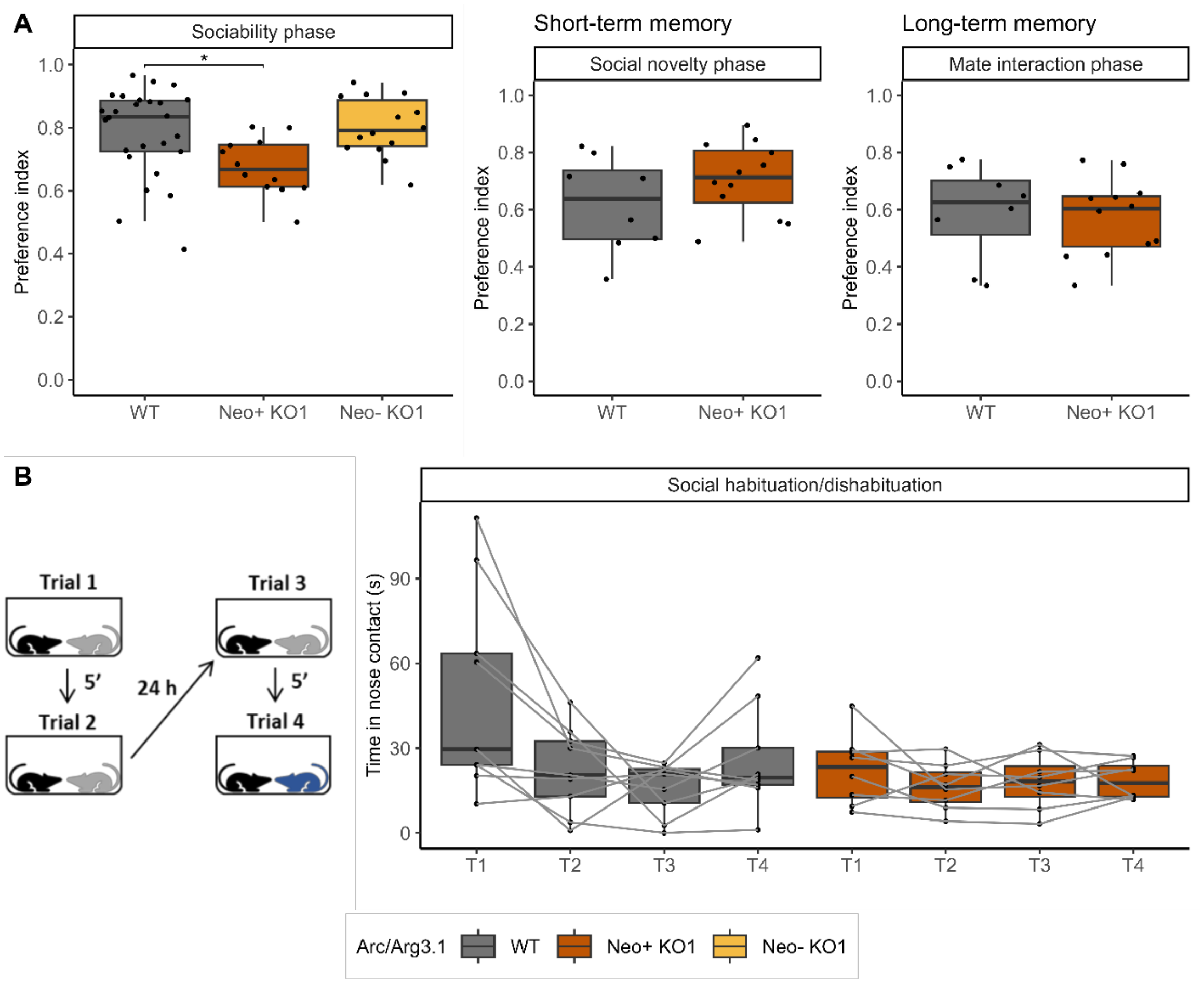

**Fig. S3. Neo⁺ *Arc/Arg3.1* KO1 mice display social deficits, rather than social memory impairments.**

In the three-chambered test, Neo⁺ *Arc/Arg3.1* KO1 mice (brown; n = 12, 6 females and 6 males) showed reduced sociability preference index and no difference in the social novelty or cage mate preference index compared to WT (dark gray; n = 8, 4 females and 4 males) (**A**). In the social recognition test depicted in the scheme, WT mice (n = 9, 4 females and 5 males) displayed habituation to the unfamiliar WT interactor over the first three trials, whereas Neo⁺ KO1 mice (n = 8, 4 females and 4 males) did not, due to low time in nose contact with the interactor already in the first trial (**B**). Data are presented as mean ± sd in **Table S1**. All groups were compared by the Kruskal-Wallis test followed by Dunn's post-hoc test, with an asterisk indicating genotype effects (p = P adjusted). * $p < 0.05$; ** $p < 0.01$; *** $p < 0.001$.

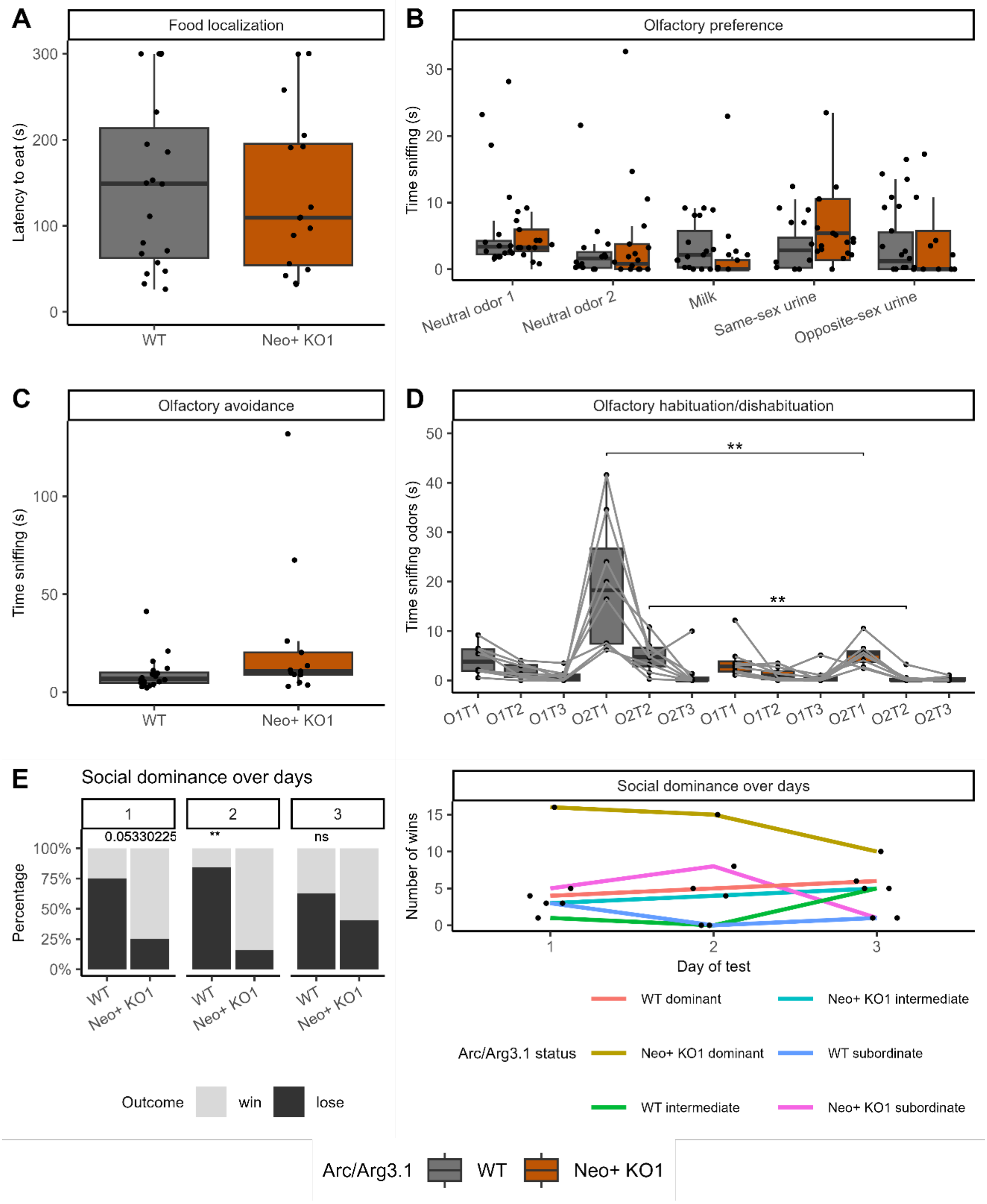

**Fig. S4. Neo$^{+}$ *Arc/Arg3.1* KO1 mice display olfactory impairments only with social odors.** In the food localization test, Neo$^{+}$ KO1 mice (brown; n = 16, 12 females and 4 males) displayed no difference in the latency to eat compared to WT (dark gray; n = 19, 7 females and 12 males) (**A**). In the innate olfactory preference test, Neo$^{+}$ KO1 mice (brown; n = 13, 9 females and 4 males) showed no difference in the time sniffing both neutral odors, milk as a food odor, or social

odors (same-sex and opposite-sex urine) compared to WT (dark gray; n = 20, 10 females and 10 males) (**B**). In the olfactory avoidance test, $Neo^+$ KO1 mice (brown; n = 13, 9 females and 4 males) displayed no difference in the time sniffing predator odors compared to WT (dark gray; n = 20, 10 females and 10 males) (**C**). In the olfactory habituation/dishabituation test, $Neo^+$ KO1 mice (brown; n = 8, 4 females and 4 males) showed reduced time sniffing social odor (same-sex urine in O2T1, O2T2, and O2T3), while the time sniffing non-social odor (blossom flower odor in O1T1, O1T2, and O1T3) and habituation over the three trials to both odors was comparable to WT (dark gray; n = 8, 4 females and 4 males) (**D**). In the social dominance test, $Neo^+$ KO1 mice (n = 16; 8 females and 8 males) won in more trials than WT mice (n = 16, 8 females and 8 males) over the three days of the test since $Neo^+$ KO1 mice dominant (ochre), intermediate (turquoise) and subordinate (pink) in their home cage were winning more over dominant (salmon), intermediate (green) and subordinate (blue) WT mice (**E**). Data are presented as mean ± sd in **Table S1**. Groups in **A**-**D** were compared by the Kruskal-Wallis test followed by Dunn’s post-hoc test, or by Fisher’s exact test in **E** (left) with an asterisk indicating genotype effects (p = P adjusted). * $p < 0.05$; ** $p < 0.01$; *** $p < 0.001$; ns, not significant.

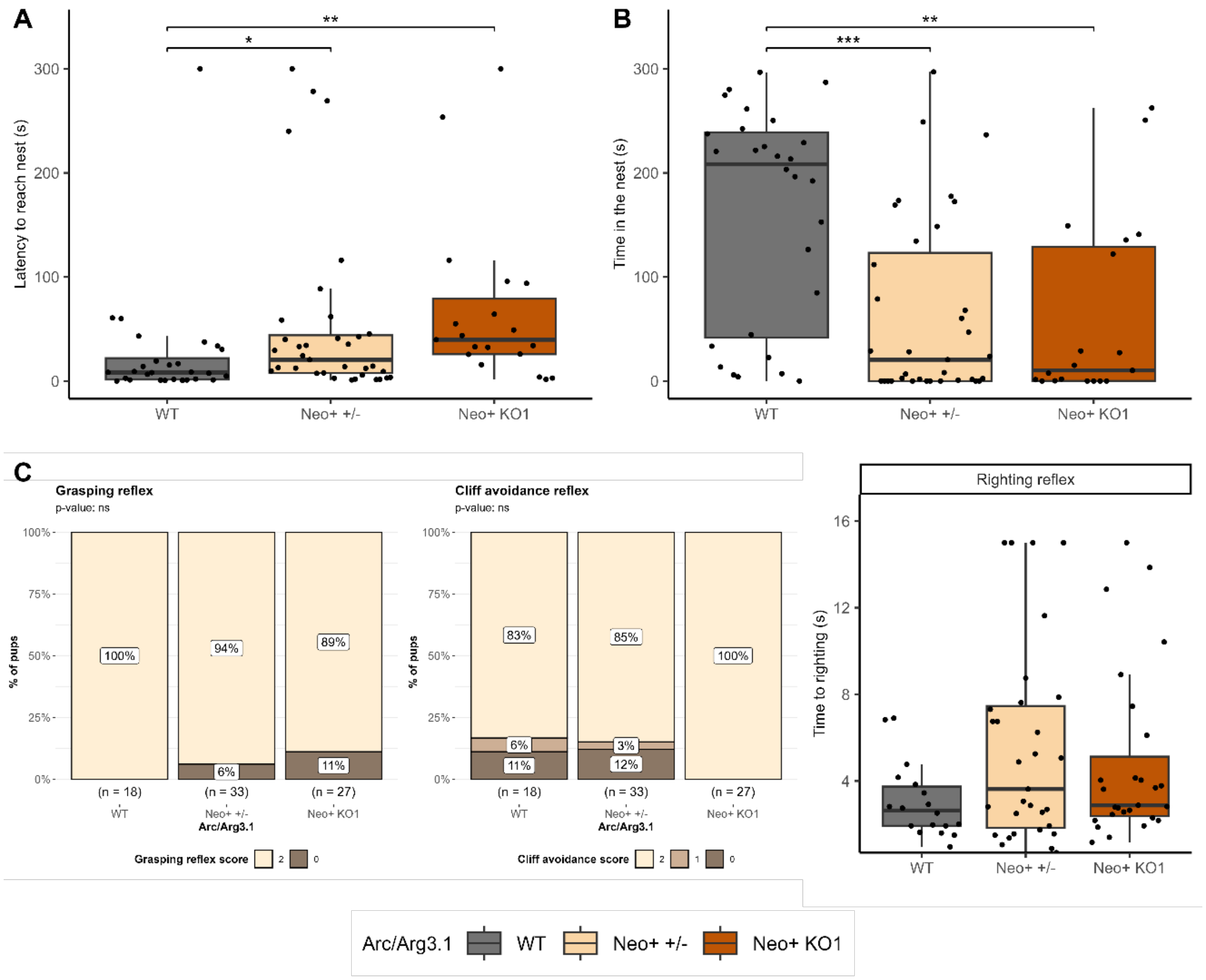

**Fig. S5. Neo⁺ *Arc/Arg3.1* KO1 pups exhibit social communication impairments, but not motor impairments.**

In the olfactory-based homing test, Neo⁺ KO1 (brown; n = 19) and Neo⁺ *Arc/Arg3.1*$^{+/-}$ pups (beige; n = 35) showed significantly increased latency to reach the "nest area" (litter from their home cage) (**A**) and spend less time in it compared to WT (dark gray; n = 28) pups (**B**). When testing their reflexes, Neo⁺ KO1 (n = 27) and Neo⁺ *Arc/Arg3.1*$^{+/-}$ pups (n = 33) displayed no difference in the grasping reflex, cliff avoidance reflex, or time to righting compared to WT pups (n = 18) (**C**). Data are presented as mean ± sd in Table S3. Groups in **A**, **B**, and **C** (right) were compared by Kruskal-Wallis test followed by Dunn's post-hoc test or by Fisher's exact test in **C** (left and center), with an asterisk indicating genotype effects (p = P adjusted). * p < 0.05; ** p < 0.01; *** p < 0.001; ns, not significant.

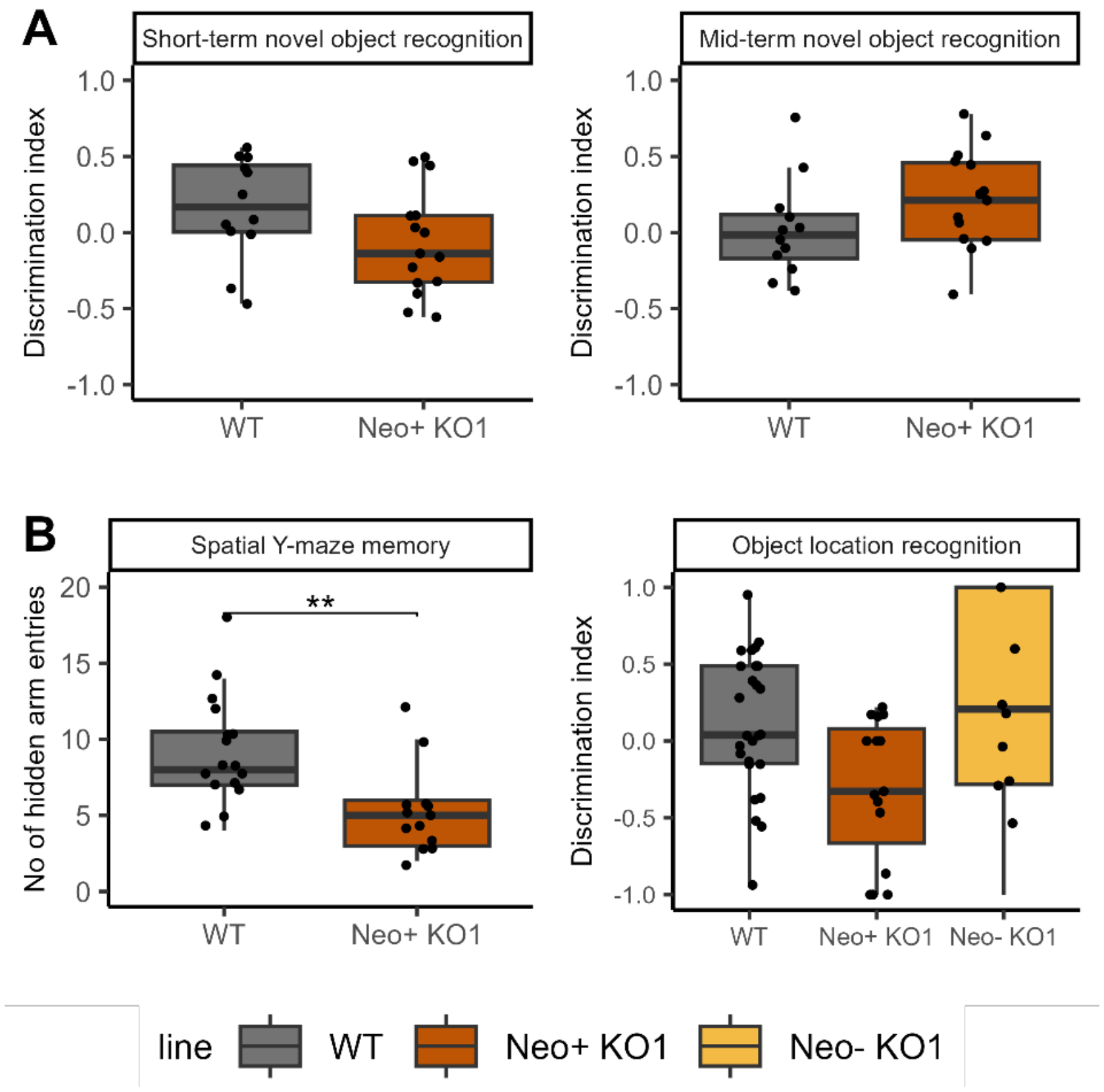

**Fig.S6. Neo⁺ KO1 mice display mild spatial short-term memory impairments which are absent in Neo⁻ KO1 mice.**

In the novel object recognition test, Neo⁺ *1* KO1 mice (brown; n = 15, 8 females and 7 males) showed no difference in the index to discriminate between a novel and an old object when testing their short-term or mid-term recognition memory compared to WT (dark gray; n = 12, 8 females and 4 males) (**A**). In the spatial Y-maze memory assay, Neo⁺ KO1 mice (brown; n = 13, 6 females and 7 males) showed a reduced number of entries in the arm hidden in the habituation phase compared to WT (dark gray; n = 16, 8 females and 8 males). In the object location recognition test, Neo⁺ KO1 mice (brown; n = 15, 8 females and 7 males) displayed a reduced index to discriminate between an object that changed its location and the one that did not, compared to WT (dark gray; n = 28, 16 females and 12 males), which was not the case for Neo⁻ KO1 mice (yellow; n = 16, 8 females and 8 males) (**B**). Data are presented as mean ± sd in **Table S1**. All groups were compared by the Kruskal-Wallis test followed by Dunn's post-hoc test, with an asterisk indicating genotype effects ($p$ = P adjusted). * $p < 0.05$; ** $p < 0.01$; *** $p < 0.001$.

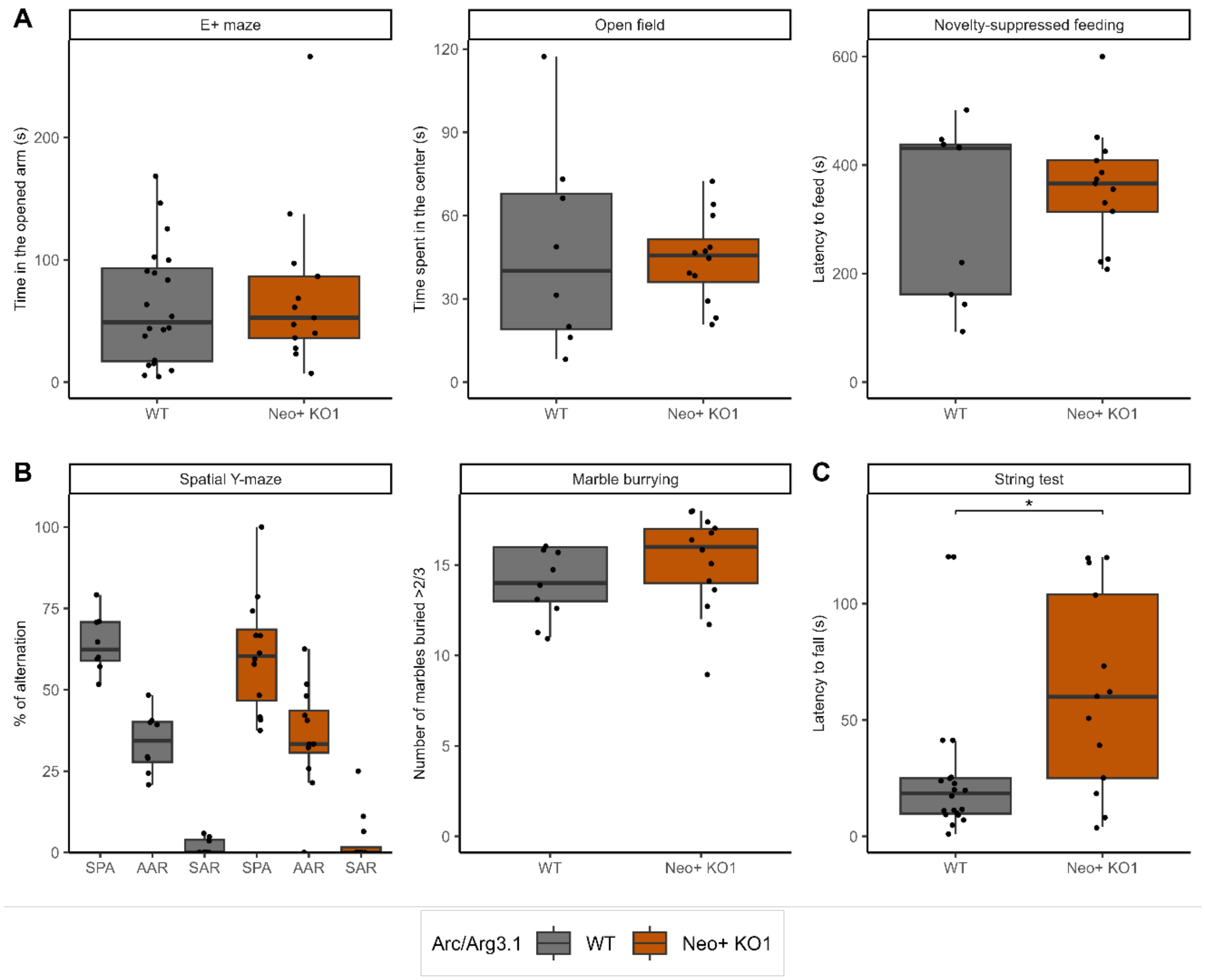

**Fig. S7. Social behavior impairments of Neo⁺ *Arc/Arg3.1* KO1 are not due to anxious-like behavior or motor impairments.**

Anxious-like behavior was evaluated in the elevated plus maze, open-field, and novelty-suppressed feeding tests (**A**). In the elevated plus maze test, Neo⁺ KO1 mice (brown; n = 13, 9 females and 4 males) showed no difference in the time spent in the opened arms of the maze compared to WT (dark gray; n = 20, 10 females and 10 males) (**A, left**). In the open-field test, Neo⁺ KO1 mice (brown; n = 25, 15 females and 10 males) displayed no difference in the time spent in the center of the open-field arena compared to WT (dark gray; n = 16, 8 females and 8 males) (**A, center**). In the novelty-suppressed feeding test, Neo⁺ KO1 mice (brown; n = 13, 7 females and 6 males) displayed comparable food latency to WT mice (dark gray; n = 9, 4 females and 5 males) (**A, right**). Perseverative behavior was tested in the spatial Y-maze task, where Neo⁺ KO1 mice (brown; n = 12, 6 females and 6 males) showed no difference in the percentage of alternation compared to WT (dark gray; n = 8, 4 females and 4 males) (**B, left**), and compulsive behavior in the marble-burying test, in which $Neo^+$ KO1 mice (brown; n = 13, 7 females and 6 males) showed no difference in the number of marbles buried (>2/3) compared to WT (dark gray; n = 9, 4 females and 5 males) (**B, right**). In the string test assessing motor coordination, $Neo^+$ KO1 mice (brown; n = 13, 9 females and 4 males) displayed increased latency to fall from the string compared to WT (dark gray; n = 20, 10 males and 10 females) (**C**). Data are presented as mean ± sd in **Table S1**. All groups were compared by the Kruskal-Wallis test followed by Dunn's post-hoc test, with an asterisk indicating genotype effects (p = P adjusted). * $p < 0.05$; ** $p < 0.01$; *** $p < 0.001$.

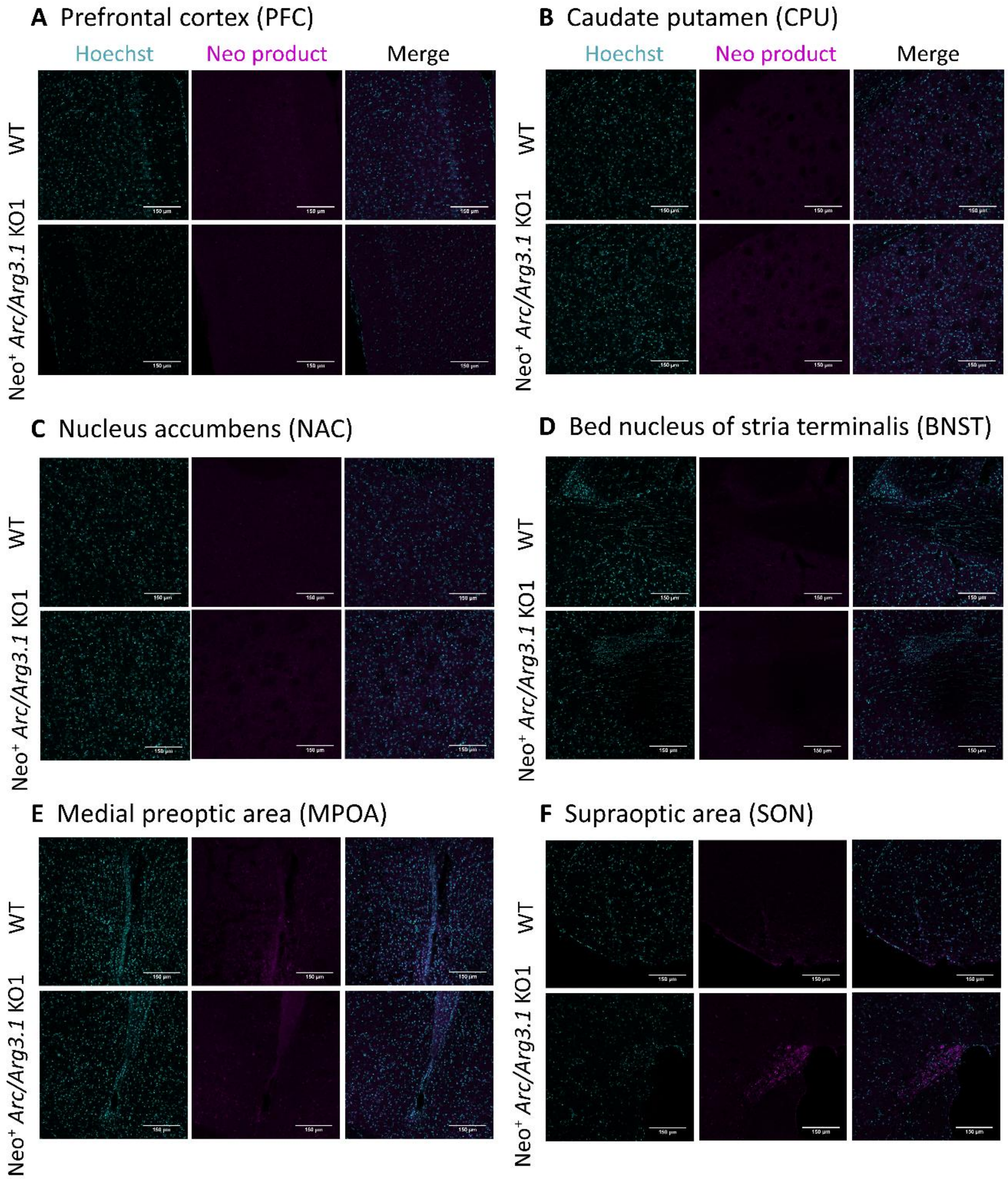

**Fig. S8. Neo protein product is not specifically expressed in other brain region in Neo$^+$ *Arc/Arg3.1* KO1 dams**

Aminoglycoside phosphotransferase coded by the Neo cassette (Neo product) was not detected at the protein level in the PFC (**A**), CPU (**B**), NAC (**C**), or BNST (**D**) of WT or Neo$^+$ KO1 dams, while a non-specific signal in WT and Neo$^+$ KO1 dams was observed in the MPOA (**E**) and SON (**F**).

Scale bars = 150 μm. PFC, prefrontal cortex; NAC, nucleus accumbens; CPU, caudate putamen; BNST, bed nucleus of stria terminalis; MPOA, medial preoptic area; PVN, paraventricular nucleus of the hypothalamus; SON, supraoptic nucleus.

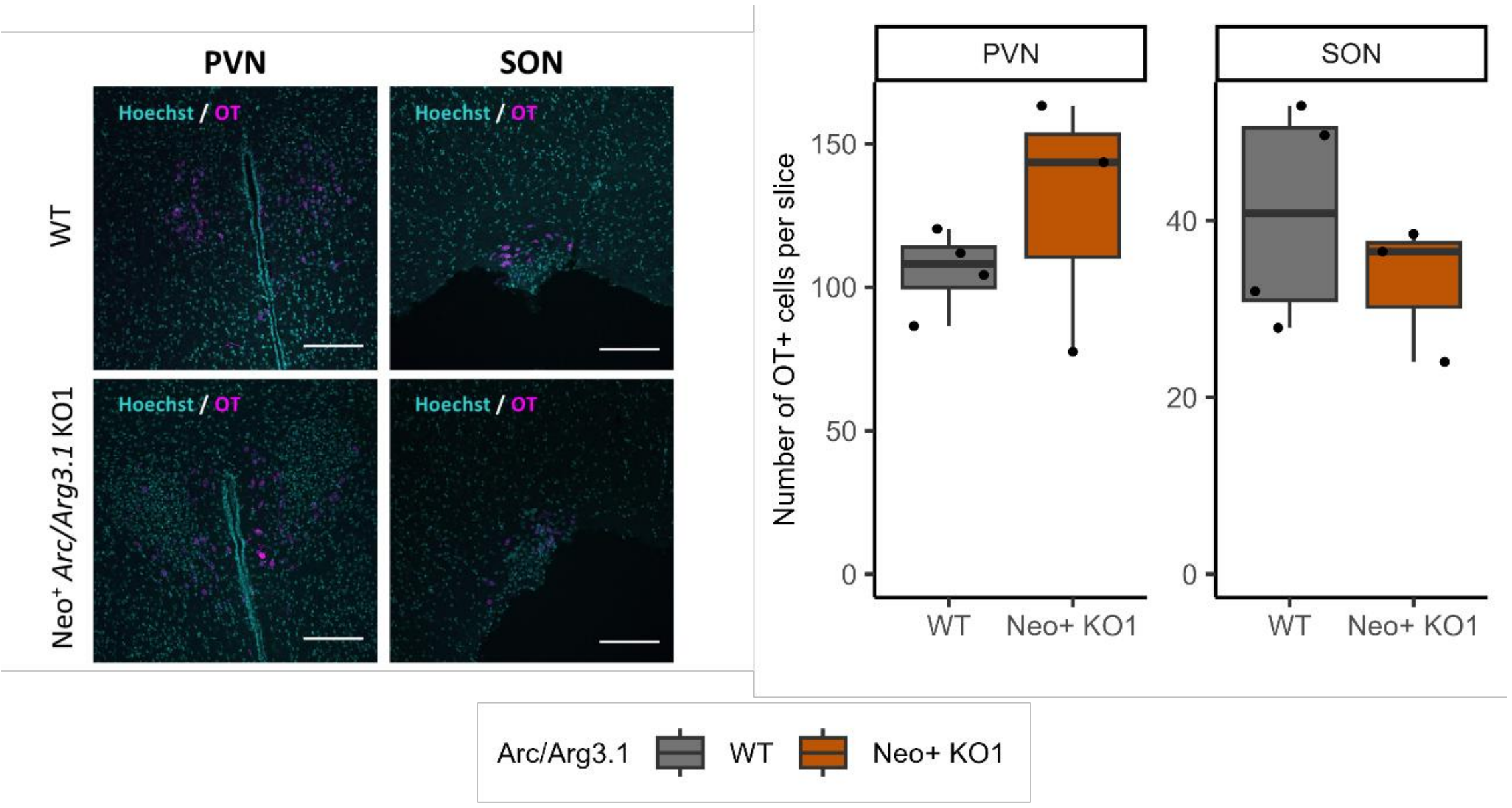

**Fig. S9. Neomycin cassette does not affect the number of oxytocinergic neurons in the PVN and SON**

The number of oxytocinergic (OT) neurons in the PVN and SON was comparable between $Neo^+$ KO1 (brown; n = 3) and WT dams (dark gray; n = 4) as shown by representative images (scale bars = 150 μm) and cellular quantification. PVN, paraventricular nucleus of the hypothalamus; SON, supraoptic nucleus. Raw data are available at https://doi.org/10.82233/QIG5AS. Groups were compared by the Kruskal-Wallis test, followed by Dunn's post-hoc test, with an asterisk indicating genotype effects (p = P adjusted). * $p < 0.05$; ** $p < 0.01$; *** $p < 0.001$.

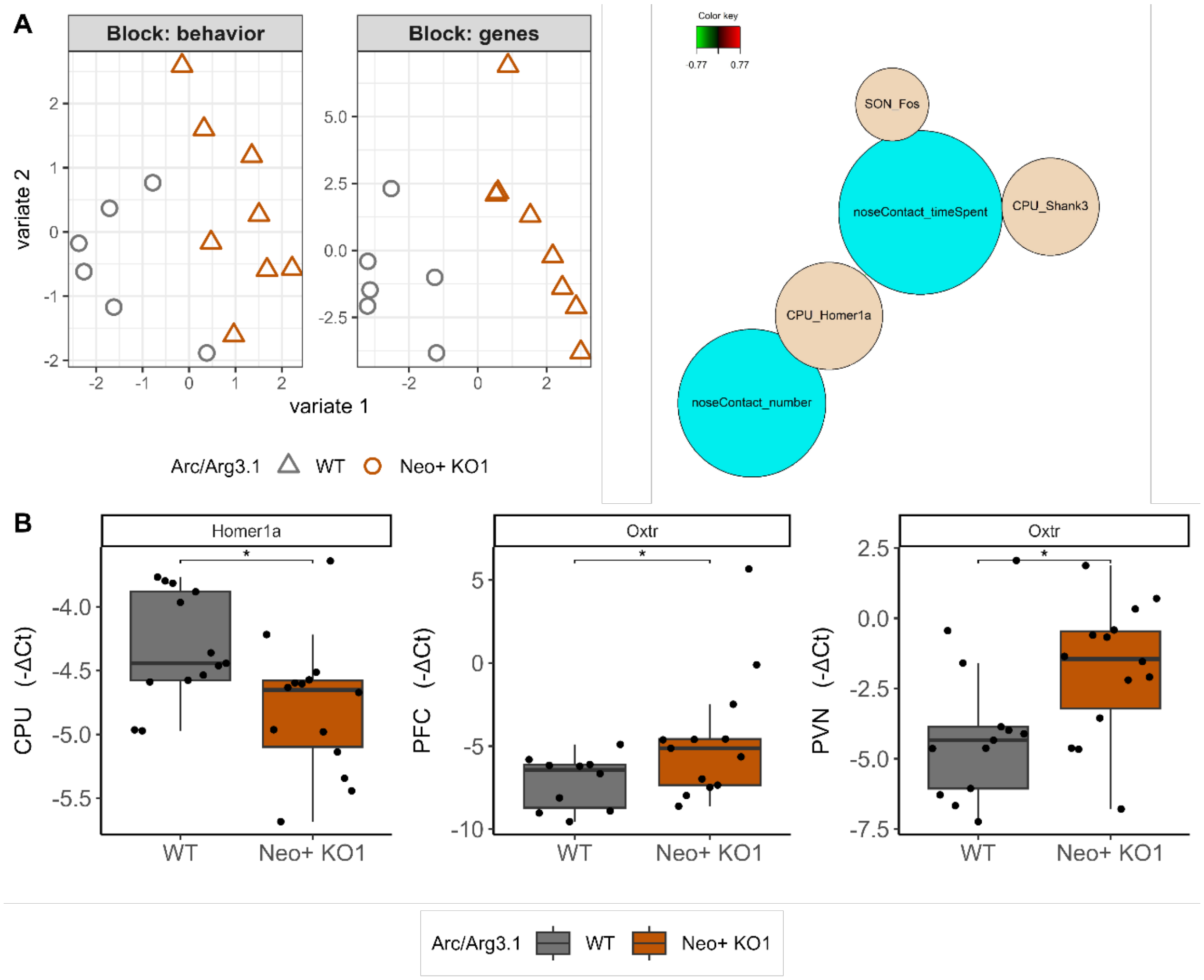

**Fig. S10. Neomycin cassette-induced social interaction impairments in Neo⁺ *Arc/Arg3.1* KO1 mice were linked with dysregulation of immediate early gene expression**

Integration of social interaction with unfamiliar conspecifics with genes expression from the oxytocin family, immediate early genes, ASD-associated and neuropeptide genes revealed a clear separation of $Neo^{+}$ KO1 sexually naïve mice (brown triangles; n = 8, 4 females and 4 males) from WT mice (dark gray circles; n = 6, 4 females and 2 males) on both behavior and gene expression. While network analysis revealed strong correlations between time and number of nose contacts, including *Homer1a* in the CPU (**A**). The expression of *Homer1a* in the CPU showed dysregulation in the Neo+ KO1 naïve mice compared to WT, along with significant dysregulation of *Oxtr* expression in the PFC and PVN (**B**). Data are presented as mean ± standard deviation in **Table S5**. Groups from gene expression analysis in **B** were compared by the Kruskal-Wallis test followed by Dunn's post-hoc test, with an asterisk indicating genotype effects (p = P adjusted). * $p < 0.05$;

** $p < 0.01$; *** $p < 0.001$. PFC, prefrontal cortex; CPU, caudate putamen; PVN, paraventricular nucleus of the hypothalamus; SON, supraoptic nucleus.

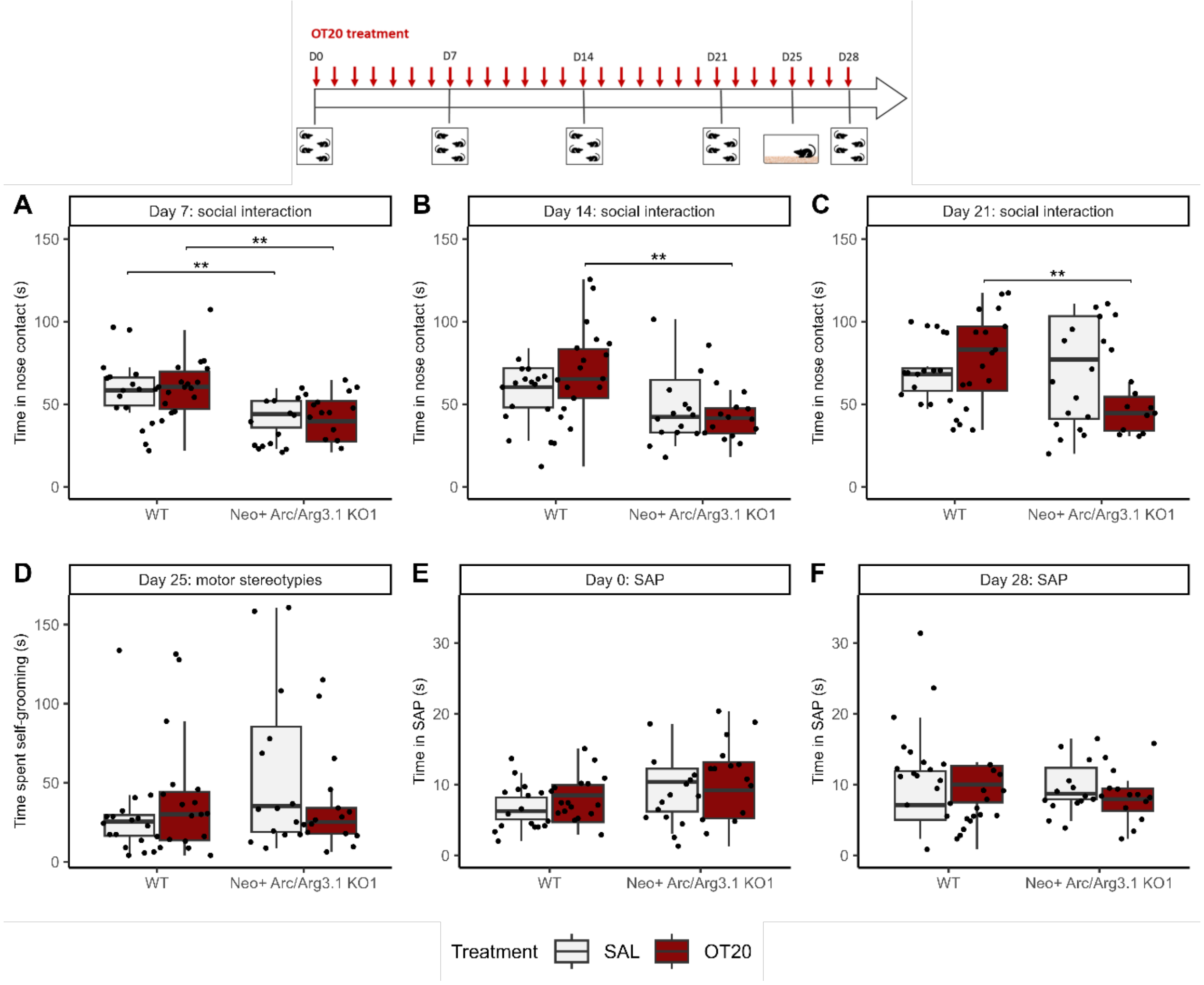

**Fig. S11. Subchronic oxytocin administration does not improve social interaction impairments, stereotyped or anxious-like behaviors.**

On day 7 of oxytocin at 20 µg kg$^{-1}$ (OT20) administration, both saline (SAL)- (light gray; n = 12, 8 females and 4 males) and OT20-treated Neo$^+$ KO1 mice (red; n = 16, 10 females and 6 males) displayed reduced social interaction with unfamiliar treatment-matched conspecifics, compared to the SAL-treated (light gray; n = 15, 7 females and 8 males) and OT20-treated WT controls (red; n = 19, 8 females and 11 males) (**A**). On day 14, OT20-treated Neo$^+$ KO1 mice showed significantly reduced social interaction compared to OT20-treated WT control (**B**). On day 21, OT20-treated Neo$^+$ KO1 mice continued to display reduced social interaction compared to OT20-treated WT (**C**). In the motor stereotypy test, performed on day 25, no difference in the time spent self-grooming was observed between Neo$^+$ KO1 (light gray for saline treatment, n = 12, 8 females and 4 males; red for oxytocin treatment, n = 17, 11 females and 6 males) compared to WT (light gray for saline treatment, n = 16, 8 females and 8 males; red for oxytocin treatment, n = 19, 8 females and 11 males) (**D**). No difference in the time spent in stretch-attend posture (SAP), a measure of anxious-like behavior, was observed between genotypes or treatments either on the first day of treatment (**E**) or day 28 (**F**). Data are presented as mean ± sd in **Table S2**. All groups were compared by the Kruskal-Wallis test followed by Dunn's post-hoc test, with an asterisk indicating genotype effects and ladder treatment effects (p = P adjusted). * or # $p < 0.05$; ** or ## $p < 0.01$; *** or ### $p < 0.001$.

**Table S1. All raw data, mean, and statistics from standardized behavior tests performed in adult mice.**

| var_s1 | genotype | line | pup_genotype | N | n_females | n_males | mean | sd |
|---|---|---|---|---|---|---|---|---|
| delivery_latency | Neo+ Arc KO1 | Neo+ Arc | +/- | 14 | 14 | 0 | 20.7142857 | 1.38278267 |
| delivery_latency | Neo+ Arc KO1 | Neo+ Arc | KO | 15 | 15 | 0 | 21.6666667 | 1.29099445 |
| delivery_latency | Neo- Arc KO2 | Neo- Arc Bl6J | KO | 7 | 7 | 0 | NA | NA |
| delivery_latency | Neo- Arc KO1 | Neo- Arc | +/- | 7 | 7 | 0 | 20.7142857 | 1.38013112 |
| delivery_latency | Neo- Arc KO1 | Neo- Arc | KO | 8 | 8 | 0 | 21.375 | 0.91612538 |
| delivery_latency | WT 1 | Neo+ Arc | +/- | 10 | 10 | 0 | 21.2 | 1.75119007 |
| delivery_latency | WT 1 | Neo+ Arc | WT | 14 | 14 | 0 | 21.9285714 | 2.09263491 |
| delivery_latency | WT 2 | Neo- Arc | +/- | 8 | 8 | 0 | 22.75 | 4.68279526 |
| delivery_latency | WT 2 | Neo- Arc | WT | 7 | 7 | 0 | 21.4285714 | 0.97590007 |
| delivery_latency | WT Bl6J | Neo- Arc Bl6J | WT | 6 | 6 | 0 | NA | NA |
| delivery_latency | WT 1 | Neo+ Arc | +/- | 15 | 15 | 0 | 22.8666667 | 4.12079512 |
| duration_crouching | Neo+ Arc KO1 | Neo+ Arc | +/- | 14 | 14 | 0 | 34.3364286 | 42.2739889 |
| duration_crouching | Neo+ Arc KO1 | Neo+ Arc | KO | 15 | 15 | 0 | 14.6365333 | 56.6870498 |
| duration_crouching | Neo- Arc KO2 | Neo- Arc Bl6J | KO | 7 | 7 | 0 | 0 | 0 |
| duration_crouching | Neo- Arc KO1 | Neo- Arc | +/- | 7 | 7 | 0 | 14.8748571 | 19.2493451 |
| duration_crouching | Neo- Arc KO1 | Neo- Arc | KO | 8 | 8 | 0 | 21.528 | 33.4219929 |
| duration_crouching | WT 1 | Neo+ Arc | +/- | 10 | 10 | 0 | 70.0606 | 49.4364414 |
| duration_crouching | WT 1 | Neo+ Arc | WT | 14 | 14 | 0 | 63.948 | 93.1498319 |
| duration_crouching | WT 2 | Neo- Arc | +/- | 8 | 8 | 0 | 19.044 | 26.5788724 |
| duration_crouching | WT 2 | Neo- Arc | WT | 7 | 7 | 0 | 9.11628571 | 10.7489169 |
| duration_crouching | WT Bl6J | Neo- Arc Bl6J | WT | 6 | 6 | 0 | 2.85433333 | 6.99166022 |
| duration_crouching | WT 1 | Neo+ Arc | +/- | 15 | 15 | 0 | 23.4942667 | 30.5824361 |
| duration_digging | Neo+ Arc KO1 | Neo+ Arc | +/- | 14 | 14 | 0 | 4.55928571 | 4.13536711 |
| duration_digging | Neo+ Arc KO1 | Neo+ Arc | KO | 15 | 15 | 0 | 2.9964 | 2.11139486 |
| duration_digging | Neo- Arc KO2 | Neo- Arc Bl6J | KO | 7 | 7 | 0 | 3.3994 | 1.16500657 |
| duration_digging | Neo- Arc KO1 | Neo- Arc | +/- | 7 | 7 | 0 | 3.51742857 | 1.66514643 |
| duration_digging | Neo- Arc KO1 | Neo- Arc | KO | 8 | 8 | 0 | 2.032625 | 0.78458523 |
| duration_digging | WT 1 | Neo+ Arc | +/- | 10 | 10 | 0 | 1.708 | 1.13136771 |
| duration_digging | WT 1 | Neo+ Arc | WT | 14 | 14 | 0 | 2.17857143 | 1.61671227 |
| duration_digging | WT 2 | Neo- Arc | +/- | 8 | 8 | 0 | 1.899625 | 1.25466808 |
| duration_digging | WT 2 | Neo- Arc | WT | 7 | 7 | 0 | 1.98014286 | 1.51027563 |
| duration_digging | WT Bl6J | Neo- Arc Bl6J | WT | 6 | 6 | 0 | 3.57666667 | 1.37732068 |
| duration_digging | WT 1 | Neo+ Arc | +/- | 15 | 15 | 0 | 3.02953333 | 1.34882583 |
| duration_grooming | Neo+ Arc KO1 | Neo+ Arc | +/- | 14 | 14 | 0 | 24.5305 | 26.612681 |
| duration_grooming | Neo+ Arc KO1 | Neo+ Arc | KO | 15 | 15 | 0 | 13.8824 | 16.445013 |
| duration_grooming | Neo- Arc KO2 | Neo- Arc Bl6J | KO | 7 | 7 | 0 | 6.2406 | 2.42848014 |
| duration_grooming | Neo- Arc KO1 | Neo- Arc | +/- | 7 | 7 | 0 | 16.3387143 | 9.4119908 |
| duration_grooming | Neo- Arc KO1 | Neo- Arc | KO | 8 | 8 | 0 | 29.280875 | 21.0492997 |
| duration_grooming | WT 1 | Neo+ Arc | +/- | 10 | 10 | 0 | 18.3617 | 17.7182309 |
| duration_grooming | WT 1 | Neo+ Arc | WT | 14 | 14 | 0 | 20.3437857 | 16.617421 |
| duration_grooming | WT 2 | Neo- Arc | +/- | 8 | 8 | 0 | 15.812125 | 20.9977994 |
| duration_grooming | WT 2 | Neo- Arc | WT | 7 | 7 | 0 | 14.4734286 | 16.6236992 |
| duration_grooming | WT Bl6J | Neo- Arc Bl6J | WT | 6 | 6 | 0 | 7.85766667 | 9.84429079 |
| duration_grooming | WT 1 | Neo+ Arc | +/- | 15 | 15 | 0 | 12.7756667 | 11.4954445 |
| duration_inside_nest | Neo+ Arc KO1 | Neo+ Arc | +/- | 14 | 14 | 0 | 45.5151429 | 28.8607264 |
| duration_inside_nest | Neo+ Arc KO1 | Neo+ Arc | KO | 15 | 15 | 0 | 81.0774667 | 140.390426 |
| duration_inside_nest | Neo- Arc KO2 | Neo- Arc Bl6J | KO | 7 | 7 | 0 | 8.0426 | 4.39620709 |
| duration_inside_nest | Neo- Arc KO1 | Neo- Arc | +/- | 7 | 7 | 0 | 32.269 | 23.0278692 |
| duration_inside_nest | Neo- Arc KO1 | Neo- Arc | KO | 8 | 8 | 0 | 47.76925 | 35.4378983 |
| duration_inside_nest | WT 1 | Neo+ Arc | +/- | 10 | 10 | 0 | 90.032 | 63.8720853 |
| duration_inside_nest | WT 1 | Neo+ Arc | WT | 14 | 14 | 0 | 59.081292 | 39.2301924 |
| duration_inside_nest | WT 2 | Neo- Arc | +/- | 8 | 8 | 0 | 34.14325 | 24.4164517 |
| duration_inside_nest | WT 2 | Neo- Arc | WT | 7 | 7 | 0 | 39.5987143 | 28.851265 |
| duration_inside_nest | WT Bl6J | Neo- Arc Bl6J | WT | 6 | 6 | 0 | 15.2618333 | 12.1646901 |
| duration_inside_nest | WT 1 | Neo+ Arc | +/- | 15 | 15 | 0 | 58.624 | 51.4952065 |
| duration_nesting | Neo+ Arc KO1 | Neo+ Arc | +/- | 14 | 14 | 0 | 14.2894286 | 12.1753351 |
| duration_nesting | Neo+ Arc KO1 | Neo+ Arc | KO | 15 | 15 | 0 | 6.94493333 | 8.93768888 |
| duration_nesting | Neo- Arc KO2 | Neo- Arc Bl6J | KO | 7 | 7 | 0 | 5.858 | 7.11837734 |
| duration_nesting | Neo- Arc KO1 | Neo- Arc | +/- | 7 | 7 | 0 | 3.507 | 5.15731396 |
| duration_nesting | Neo- Arc KO1 | Neo- Arc | KO | 8 | 8 | 0 | 9.739125 | 5.78894645 |
| duration_nesting | WT 1 | Neo+ Arc | +/- | 10 | 10 | 0 | 19.2613 | 8.05691686 |
| duration_nesting | WT 1 | Neo+ Arc | WT | 14 | 14 | 0 | 14.2789286 | 5.65803102 |
| duration_nesting | WT 2 | Neo- Arc | +/- | 8 | 8 | 0 | 11.278625 | 8.56057408 |
| duration_nesting | WT 2 | Neo- Arc | WT | 7 | 7 | 0 | 10.6685714 | 5.47197283 |
| duration_nesting | WT Bl6J | Neo- Arc Bl6J | WT | 6 | 6 | 0 | 4.871 | 6.01696016 |
| duration_nesting | WT 1 | Neo+ Arc | +/- | 15 | 15 | 0 | 12.3451333 | 5.06830338 |
| duration_outside_nest | Neo+ Arc KO1 | Neo+ Arc | +/- | 14 | 14 | 0 | 50.1244235 | 48.5494187 |
| duration_outside_nest | Neo+ Arc KO1 | Neo+ Arc | KO | 15 | 15 | 0 | 69.5014667 | 56.1736868 |
| duration_outside_nest | Neo- Arc KO2 | Neo- Arc Bl6J | KO | 7 | 7 | 0 | 41.2322 | 16.0848155 |
| duration_outside_nest | Neo- Arc KO1 | Neo- Arc | +/- | 7 | 7 | 0 | 54.8928571 | 50.5049868 |
| duration_outside_nest | Neo- Arc KO1 | Neo- Arc | KO | 8 | 8 | 0 | 43.60725 | 21.5138352 |
| duration_outside_nest | WT 1 | Neo+ Arc | +/- | 10 | 10 | 0 | 25.1518 | 6.01815278 |
| duration_outside_nest | WT 1 | Neo+ Arc | WT | 14 | 14 | 0 | 28.1217143 | 12.8984107 |
| duration_outside_nest | WT 2 | Neo- Arc | +/- | 8 | 8 | 0 | 40.310625 | 14.414404 |
| duration_outside_nest | WT 2 | Neo- Arc | WT | 7 | 7 | 0 | 35.4047143 | 12.8151177 |
| duration_outside_nest | WT Bl6J | Neo- Arc Bl6J | WT | 6 | 6 | 0 | 93.837 | 75.9179541 |
| duration_outside_nest | WT 1 | Neo+ Arc | +/- | 15 | 15 | 0 | 34.8614 | 12.1649883 |
| duration_sniffing_pups | Neo+ Arc KO1 | Neo+ Arc | +/- | 14 | 14 | 0 | 2.81728571 | 1.64846639 |
| duration_sniffing_pups | Neo+ Arc KO1 | Neo+ Arc | KO | 15 | 15 | 0 | 1.5186 | 0.66922693 |
| duration_sniffing_pups | Neo- Arc KO2 | Neo- Arc Bl6J | KO | 7 | 7 | 0 | 1.1234 | 0.19610533 |
| duration_sniffing_pups | Neo- Arc KO1 | Neo- Arc | +/- | 7 | 7 | 0 | 1.31785714 | 0.32918102 |
| duration_sniffing_pups | Neo- Arc KO1 | Neo- Arc | KO | 8 | 8 | 0 | 1.658375 | 0.5092347 |
| duration_sniffing_pups | WT 1 | Neo+ Arc | +/- | 10 | 10 | 0 | 3.3172 | 2.45504609 |
| duration_sniffing_pups | WT 1 | Neo+ Arc | WT | 14 | 14 | 0 | 1.89957143 | 0.96987031 |
| duration_sniffing_pups | WT 2 | Neo- Arc | +/- | 8 | 8 | 0 | 1.474125 | 0.45435683 |
| duration_sniffing_pups | WT 2 | Neo- Arc | WT | 7 | 7 | 0 | 1.69742857 | 0.42087166 |
| duration_sniffing_pups | WT Bl6J | Neo- Arc Bl6J | WT | 6 | 6 | 0 | 1.2175 | 0.33614922 |
| duration_sniffing_pups | WT 1 | Neo+ Arc | +/- | 15 | 15 | 0 | 1.89973333 | 0.82459067 |
| latency_first | Neo+ Arc KO1 | Neo+ Arc | +/- | 14 | 14 | 0 | 508.259 | 710.438842 |
| latency_first | Neo+ Arc KO1 | Neo+ Arc | KO | 15 | 15 | 0 | 1746.99173 | 205.300134 |
| latency_first | Neo- Arc KO2 | Neo- Arc Bl6J | KO | 7 | 7 | 0 | 614.532 | 770.842646 |
| latency_first | Neo- Arc KO1 | Neo- Arc | +/- | 7 | 7 | 0 | 1047.49943 | 938.631766 |
| latency_first | Neo- Arc KO1 | Neo- Arc | KO | 8 | 8 | 0 | 108.31825 | 175.520585 |
| latency_first | WT 1 | Neo+ Arc | +/- | 10 | 10 | 0 | 76.0816 | 65.3032504 |
| latency_first | WT 1 | Neo+ Arc | WT | 14 | 14 | 0 | 337.608143 | 633.190722 |
| latency_first | WT 2 | Neo- Arc | +/- | 8 | 8 | 0 | 84.4545 | 76.8883205 |
| latency_first | WT 2 | Neo- Arc | WT | 7 | 7 | 0 | 508.515857 | 731.566559 |
| latency_first | WT Bl6J | Neo- Arc Bl6J | WT | 6 | 6 | 0 | 433.2825 | 698.507788 |
| latency_first | WT 1 | Neo+ Arc | +/- | 15 | 15 | 0 | 370.4274 | 559.191086 |
| latency_fourth | Neo+ Arc KO1 | Neo+ Arc | +/- | 14 | 14 | 0 | 900.420857 | 819.314783 |
| latency_fourth | Neo+ Arc KO1 | Neo+ Arc | KO | 15 | 15 | 0 | 1749.72833 | 194.701328 |
| latency_fourth | Neo- Arc KO2 | Neo- Arc Bl6J | KO | 7 | 7 | 0 | 625.1195 | 832.135994 |
| latency_fourth | Neo- Arc KO1 | Neo- Arc | +/- | 7 | 7 | 0 | 1080.084 | 897.939897 |
| latency_fourth | Neo- Arc KO1 | Neo- Arc | KO | 8 | 8 | 0 | 223.62625 | 189.101967 |
| latency_fourth | WT 1 | Neo+ Arc | +/- | 10 | 10 | 0 | 487.0805 | 696.33888 |
| latency_fourth | WT 1 | Neo+ Arc | WT | 14 | 14 | 0 | 623.678286 | 782.324132 |
| latency_fourth | WT 2 | Neo- Arc | +/- | 8 | 8 | 0 | 420.361286 | 614.582918 |
| latency_fourth | WT 2 | Neo- Arc | WT | 7 | 7 | 0 | 582.737571 | 693.300279 |
| latency_fourth | WT Bl6J | Neo- Arc Bl6J | WT | 6 | 6 | 0 | 643.721 | 842.29681 |
| latency_fourth | WT 1 | Neo+ Arc | +/- | 15 | 15 | 0 | 1066.63933 | 816.520903 |
| latency_full_maternal_behaviour | Neo+ Arc KO1 | Neo+ Arc | +/- | 14 | 14 | 0 | 1118.21429 | 646.301927 |
| latency_full_maternal_behaviour | Neo+ Arc KO1 | Neo+ Arc | KO | 15 | 15 | 0 | 1764 | 139.4274 |
| latency_full_maternal_behaviour | Neo- Arc KO2 | Neo- Arc Bl6J | KO | 7 | 7 | 0 | 1484.16667 | 574.929706 |
| latency_full_maternal_behaviour | Neo- Arc KO1 | Neo- Arc | +/- | 7 | 7 | 0 | 1425.71429 | 528.76766 |
| latency_full_maternal_behaviour | Neo- Arc KO1 | Neo- Arc | KO | 8 | 8 | 0 | 866.875 | 717.997898 |
| latency_full_maternal_behaviour | WT 1 | Neo+ Arc | +/- | 10 | 10 | 0 | 687 | 593.109508 |
| latency_full_maternal_behaviour | WT 1 | Neo+ Arc | WT | 14 | 14 | 0 | 913.076923 | 713.25702 |
| latency_full_maternal_behaviour | WT 2 | Neo- Arc | +/- | 8 | 8 | 0 | 999.375 | 631.480333 |
| latency_full_maternal_behaviour | WT 2 | Neo- Arc | WT | 7 | 7 | 0 | 1046.14286 | 598.647762 |
| latency_full_maternal_behaviour | WT Bl6J | Neo- Arc Bl6J | WT | 6 | 6 | 0 | 1225.5 | 567.414751 |
| latency_full_maternal_behaviour | WT 1 | Neo+ Arc | +/- | 15 | 15 | 0 | 1159.66667 | 734.203812 |
| latency_second | Neo+ Arc KO1 | Neo+ Arc | +/- | 14 | 14 | 0 | 661.2565 | 770.062331 |
| latency_second | Neo+ Arc KO1 | Neo+ Arc | KO | 15 | 15 | 0 | 1749.33227 | 196.235287 |
| latency_second | Neo- Arc KO2 | Neo- Arc Bl6J | KO | 7 | 7 | 0 | 714.953429 | 823.194034 |
| latency_second | Neo- Arc KO1 | Neo- Arc | +/- | 7 | 7 | 0 | 1054.88014 | 929.401609 |
| latency_second | Neo- Arc KO1 | Neo- Arc | KO | 8 | 8 | 0 | 151.929375 | 168.142365 |
| latency_second | WT 1 | Neo+ Arc | +/- | 10 | 10 | 0 | 214.3223 | 330.597659 |
| latency_second | WT 1 | Neo+ Arc | WT | 14 | 14 | 0 | 351.894429 | 627.091266 |
| latency_second | WT 2 | Neo- Arc | +/- | 8 | 8 | 0 | 135.808125 | 98.8939187 |
| latency_second | WT 2 | Neo- Arc | WT | 7 | 7 | 0 | 530.143429 | 721.39132 |
| latency_second | WT Bl6J | Neo- Arc Bl6J | WT | 6 | 6 | 0 | 617.1535 | 835.435756 |
| latency_second | WT 1 | Neo+ Arc | +/- | 15 | 15 | 0 | 411.556467 | 575.704536 |
| latency_sniffing_pups | Neo+ Arc KO1 | Neo+ Arc | +/- | 14 | 14 | 0 | 8.07521429 | 9.53074327 |
| latency_sniffing_pups | Neo+ Arc KO1 | Neo+ Arc | KO | 15 | 15 | 0 | 9.1578 | 7.68663469 |
| latency_sniffing_pups | Neo- Arc KO2 | Neo- Arc Bl6J | KO | 7 | 7 | 0 | 4.5836 | 5.82663735 |
| latency_sniffing_pups | Neo- Arc KO1 | Neo- Arc | +/- | 7 | 7 | 0 | 12.3441429 | 7.28079534 |
| latency_sniffing_pups | Neo- Arc KO1 | Neo- Arc | KO | 8 | 8 | 0 | 6.705625 | 3.17736912 |
| latency_sniffing_pups | WT 1 | Neo+ Arc | +/- | 10 | 10 | 0 | 3.6092 | 1.17839163 |
| latency_sniffing_pups | WT 1 | Neo+ Arc | WT | 14 | 14 | 0 | 7.651 | 9.53321395 |
| latency_sniffing_pups | WT 2 | Neo- Arc | +/- | 8 | 8 | 0 | 5.817125 | 4.26205495 |
| latency_sniffing_pups | WT 2 | Neo- Arc | WT | 7 | 7 | 0 | 5.68357143 | 5.2716977 |
| latency_sniffing_pups | WT Bl6J | Neo- Arc Bl6J | WT | 6 | 6 | 0 | 4.58316667 | 4.92603294 |
| latency_sniffing_pups | WT 1 | Neo+ Arc | +/- | 15 | 15 | 0 | 4.83666667 | 2.3071591 |
| latency_third | Neo+ Arc KO1 | Neo+ Arc | +/- | 14 | 14 | 0 | 812.215071 | 788.277613 |
| latency_third | Neo+ Arc KO1 | Neo+ Arc | KO | 15 | 15 | 0 | 1749.67427 | 194.910727 |
| latency_third | Neo- Arc KO2 | Neo- Arc Bl6J | KO | 7 | 7 | 0 | 613.700167 | 831.762498 |
| latency_third | Neo- Arc KO1 | Neo- Arc | +/- | 7 | 7 | 0 | 1061.43871 | 921.194414 |
| latency_third | Neo- Arc KO1 | Neo- Arc | KO | 8 | 8 | 0 | 172.082 | 168.769756 |
| latency_third | WT 1 | Neo+ Arc | +/- | 10 | 10 | 0 | 314.2894 | 529.117571 |
| latency_third | WT 1 | Neo+ Arc | WT | 14 | 14 | 0 | 370.637643 | 618.982511 |
| latency_third | WT 2 | Neo- Arc | +/- | 8 | 8 | 0 | 158.731429 | 93.0009793 |
| latency_third | WT 2 | Neo- Arc | WT | 7 | 7 | 0 | 560.028143 | 706.613229 |
| latency_third | WT Bl6J | Neo- Arc Bl6J | WT | 6 | 6 | 0 | 634.428833 | 844.750444 |
| latency_third | WT 1 | Neo+ Arc | +/- | 15 | 15 | 0 | 812.538667 | 824.784859 |
| litter_size | Neo+ Arc KO1 | Neo+ Arc | +/- | 14 | 14 | 0 | 6.57142857 | 1.28388148 |
| litter_size | Neo+ Arc KO1 | Neo+ Arc | KO | 15 | 15 | 0 | 7.8 | 1.93464652 |
| litter_size | Neo- Arc KO2 | Neo- Arc Bl6J | KO | 7 | 7 | 0 | 5.85714286 | 2.41029538 |
| litter_size | Neo- Arc KO1 | Neo- Arc | +/- | 7 | 7 | 0 | 7.28571429 | 1.11269728 |
| litter_size | Neo- Arc KO1 | Neo- Arc | KO | 8 | 8 | 0 | 8 | 1.30930734 |
| litter_size | WT 1 | Neo+ Arc | +/- | 10 | 10 | 0 | 8.7 | 1.82878223 |
| litter_size | WT 1 | Neo+ Arc | WT | 14 | 14 | 0 | 6.64285714 | 1.69193303 |
| litter_size | WT 2 | Neo- Arc | +/- | 8 | 8 | 0 | 6.625 | 2.82526863 |
| litter_size | WT 2 | Neo- Arc | WT | 7 | 7 | 0 | 8.57142857 | 1.27241802 |
| litter_size | WT Bl6J | Neo- Arc Bl6J | WT | 6 | 6 | 0 | 7.33333333 | 2.25092574 |
| litter_size | WT 1 | Neo+ Arc | +/- | 15 | 15 | 0 | 8.13333333 | 1.7674302 |
| number_crouching | Neo+ Arc KO1 | Neo+ Arc | +/- | 14 | 14 | 0 | 3.07142857 | 2.61546542 |
| number_crouching | Neo+ Arc KO1 | Neo+ Arc | KO | 15 | 15 | 0 | 0.13333333 | 0.51639778 |
| number_crouching | Neo- Arc KO2 | Neo- Arc Bl6J | KO | 7 | 7 | 0 | 0 | 0 |
| number_crouching | Neo- Arc KO1 | Neo- Arc | +/- | 7 | 7 | 0 | 3.57142857 | 5.68205194 |
| number_crouching | Neo- Arc KO1 | Neo- Arc | KO | 8 | 8 | 0 | 5.125 | 6.17454452 |
| number_crouching | WT 1 | Neo+ Arc | +/- | 10 | 10 | 0 | 6.7 | 3.6224608 |
| number_crouching | WT 1 | Neo+ Arc | WT | 14 | 14 | 0 | 3.42857143 | 3.7357789 |
| number_crouching | WT 2 | Neo- Arc | +/- | 8 | 8 | 0 | 1.625 | 2.55999442 |
| number_crouching | WT 2 | Neo- Arc | WT | 7 | 7 | 0 | 2 | 2.82842712 |
| number_crouching | WT Bl6J | Neo- Arc Bl6J | WT | 6 | 6 | 0 | 0.33333333 | 0.81649658 |
| number_crouching | NA | Neo+ Arc | +/- | 15 | 15 | 0 | 4.8 | 5.87002069 |
| number_digging | Neo+ Arc KO1 | Neo+ Arc | +/- | 14 | 14 | 0 | 27.5 | 17.0101779 |
| number_digging | Neo+ Arc KO1 | Neo+ Arc | KO | 15 | 15 | 0 | 23.3333333 | 23.8317515 |
| number_digging | Neo- Arc KO2 | Neo- Arc Bl6J | KO | 7 | 7 | 0 | 86.6 | 37.520661 |
| number_digging | Neo- Arc KO1 | Neo- Arc | +/- | 7 | 7 | 0 | 25.4285714 | 20.9113776 |
| number_digging | Neo- Arc KO1 | Neo- Arc | KO | 8 | 8 | 0 | 13 | 6.86606562 |
| number_digging | WT 1 | Neo+ Arc | +/- | 10 | 10 | 0 | 13.7 | 17.2243368 |
| number_digging | WT 1 | Neo+ Arc | WT | 14 | 14 | 0 | 14.5714286 | 17.1541172 |
| number_digging | WT 2 | Neo- Arc | +/- | 8 | 8 | 0 | 17.375 | 15.1179883 |
| number_digging | WT 2 | Neo- Arc | WT | 7 | 7 | 0 | 10.5714286 | 12.5014285 |
| number_digging | WT Bl6J | Neo- Arc Bl6J | WT | 6 | 6 | 0 | 76.3333333 | 32.3645897 |
| number_digging | WT 1 | Neo+ Arc | +/- | 15 | 15 | 0 | 31.6 | 24.0380651 |
| number_grooming | Neo+ Arc KO1 | Neo+ Arc | +/- | 14 | 14 | 0 | 2.35714286 | 1.44686094 |
| number_grooming | Neo+ Arc KO1 | Neo+ Arc | KO | 15 | 15 | 0 | 2.26666667 | 1.9808608 |
| number_grooming | Neo- Arc KO2 | Neo- Arc Bl6J | KO | 7 | 7 | 0 | 2.6 | 2.19089023 |
| number_grooming | Neo- Arc KO1 | Neo- Arc | +/- | 7 | 7 | 0 | 3.14285714 | 2.11570094 |
| number_grooming | Neo- Arc KO1 | Neo- Arc | KO | 8 | 8 | 0 | 2.875 | 2.35660167 |
| number_grooming | WT 1 | Neo+ Arc | +/- | 10 | 10 | 0 | 1.6 | 2.11869981 |
| number_grooming | WT 1 | Neo+ Arc | WT | 14 | 14 | 0 | 3 | 3.30500786 |
| number_grooming | WT 2 | Neo- Arc | +/- | 8 | 8 | 0 | 1.625 | 1.92260983 |
| number_grooming | WT 2 | Neo- Arc | WT | 7 | 7 | 0 | 1.28571429 | 1.11269728 |
| number_grooming | WT Bl6J | Neo- Arc Bl6J | WT | 6 | 6 | 0 | 3.16666667 | 2.56255081 |
| number_grooming | WT 1 | Neo+ Arc | +/- | 15 | 15 | 0 | 2.33333333 | 2.43975018 |
| number_inside_nest | Neo+ Arc KO1 | Neo+ Arc | +/- | 14 | 14 | 0 | 23.7857143 | 9.84634701 |
| number_inside_nest | Neo+ Arc KO1 | Neo+ Arc | KO | 15 | 15 | 0 | 17.8666667 | 11.9514892 |
| number_inside_nest | Neo- Arc KO2 | Neo- Arc Bl6J | KO | 7 | 7 | 0 | 38.2 | 11.1669154 |
| number_inside_nest | Neo- Arc KO1 | Neo- Arc | +/- | 7 | 7 | 0 | 24.2857143 | 9.26848218 |
| number_inside_nest | Neo- Arc KO1 | Neo- Arc | KO | 8 | 8 | 0 | 20.875 | 6.17454452 |
| number_inside_nest | WT 1 | Neo+ Arc | +/- | 10 | 10 | 0 | 20.2 | 10.6332811 |
| number_inside_nest | WT 1 | Neo+ Arc | WT | 14 | 14 | 0 | 24.0714286 | 9.47437261 |
| number_inside_nest | WT 2 | Neo- Arc | +/- | 8 | 8 | 0 | 26.625 | 9.14857522 |
| number_inside_nest | WT 2 | Neo- Arc | WT | 7 | 7 | 0 | 29.4285714 | 14.2811898 |
| number_inside_nest | WT Bl6J | Neo- Arc Bl6J | WT | 6 | 6 | 0 | 21.5 | 13.8094171 |
| number_inside_nest | WT 1 | Neo+ Arc | +/- | 15 | 15 | 0 | 24.6666667 | 12.2396701 |
| number_nesting | Neo+ Arc KO1 | Neo+ Arc | +/- | 14 | 14 | 0 | 9.07142857 | 6.46248685 |
| number_nesting | Neo+ Arc KO1 | Neo+ Arc | KO | 15 | 15 | 0 | 7.93333333 | 10.2153013 |
| number_nesting | Neo- Arc KO2 | Neo- Arc Bl6J | KO | 7 | 7 | 0 | 5.6 | 5.77061522 |
| number_nesting | Neo- Arc KO1 | Neo- Arc | +/- | 7 | 7 | 0 | 3 | 3.65148372 |
| number_nesting | Neo- Arc KO1 | Neo- Arc | KO | 8 | 8 | 0 | 11.25 | 6.27352715 |
| number_nesting | WT 1 | Neo+ Arc | +/- | 10 | 10 | 0 | 9.6 | 4.52646539 |
| number_nesting | WT 1 | Neo+ Arc | WT | 14 | 14 | 0 | 11.5 | 6.24807663 |
| number_nesting | WT 2 | Neo- Arc | +/- | 8 | 8 | 0 | 10.125 | 9.04650682 |
| number_nesting | WT 2 | Neo- Arc | WT | 7 | 7 | 0 | 16.2857143 | 8.4402663 |
| number_nesting | WT Bl6J | Neo- Arc Bl6J | WT | 6 | 6 | 0 | 6.66666667 | 7.08989892 |
| number_nesting | WT 1 | Neo+ Arc | +/- | 15 | 15 | 0 | 9.66666667 | 7.0979541 |
| number_outside_nest | Neo+ Arc KO1 | Neo+ Arc | +/- | 14 | 14 | 0 | 24.2142857 | 9.86975622 |
| number_outside_nest | Neo+ Arc KO1 | Neo+ Arc | KO | 15 | 15 | 0 | 18.2 | 11.8212883 |
| number_outside_nest | Neo- Arc KO2 | Neo- Arc Bl6J | KO | 7 | 7 | 0 | 39 | 11.2915898 |
| number_outside_nest | Neo- Arc KO1 | Neo- Arc | +/- | 7 | 7 | 0 | 24 | 8.86942313 |
| number_outside_nest | Neo- Arc KO1 | Neo- Arc | KO | 8 | 8 | 0 | 21.125 | 6.7281392 |
| number_outside_nest | WT 1 | Neo+ Arc | +/- | 10 | 10 | 0 | 20.3 | 10.6671875 |
| number_outside_nest | WT 1 | Neo+ Arc | WT | 14 | 14 | 0 | 24.4285714 | 9.81924551 |
| number_outside_nest | WT 2 | Neo- Arc | +/- | 8 | 8 | 0 | 26.875 | 9.17196816 |
| number_outside_nest | WT 2 | Neo- Arc | WT | 7 | 7 | 0 | 29.5714286 | 14.6612544 |
| number_outside_nest | WT Bl6J | Neo- Arc Bl6J | WT | 6 | 6 | 0 | 21.8333333 | 13.8768392 |
| number_outside_nest | WT 1 | Neo+ Arc | +/- | 15 | 15 | 0 | 25.1333333 | 12.0999803 |
| number_rearing | Neo+ Arc KO1 | Neo+ Arc | +/- | 14 | 14 | 0 | 60.0714286 | 47.1274284 |
| number_rearing | Neo+ Arc KO1 | Neo+ Arc | KO | 15 | 15 | 0 | 62.7333333 | 43.30204 |
| number_rearing | Neo- Arc KO2 | Neo- Arc Bl6J | KO | 7 | 7 | 0 | 103.2 | 36.2932501 |
| number_rearing | Neo- Arc KO1 | Neo- Arc | +/- | 7 | 7 | 0 | 87 | 49.0815648 |
| number_rearing | Neo- Arc KO1 | Neo- Arc | KO | 8 | 8 | 0 | 90.5 | 67.8485709 |
| number_rearing | WT 1 | Neo+ Arc | +/- | 10 | 10 | 0 | 46 | 39.8636565 |
| number_rearing | WT 1 | Neo+ Arc | WT | 14 | 14 | 0 | 51.1428571 | 36.1468312 |
| number_rearing | WT 2 | Neo- Arc | +/- | 8 | 8 | 0 | 98.375 | 41.9555462 |
| number_rearing | WT 2 | Neo- Arc | WT | 7 | 7 | 0 | 73 | 46.921921 |
| number_rearing | WT Bl6J | Neo- Arc Bl6J | WT | 6 | 6 | 0 | 76 | 21.2790977 |
| number_rearing | WT 1 | Neo+ Arc | +/- | 15 | 15 | 0 | 59.5333333 | 40.0211849 |
| number_sniffing_pups | Neo+ Arc KO1 | Neo+ Arc | +/- | 14 | 14 | 0 | 27.8571429 | 19.9300976 |
| number_sniffing_pups | Neo+ Arc KO1 | Neo+ Arc | KO | 15 | 15 | 0 | 18.4666667 | 11.7099876 |
| number_sniffing_pups | Neo- Arc KO2 | Neo- Arc Bl6J | KO | 7 | 7 | 0 | 63.6 | 48.0343627 |
| number_sniffing_pups | Neo- Arc KO1 | Neo- Arc | +/- | 7 | 7 | 0 | 21.1428571 | 17.6203346 |
| number_sniffing_pups | Neo- Arc KO1 | Neo- Arc | KO | 8 | 8 | 0 | 14.125 | 13.6532309 |
| number_sniffing_pups | WT 1 | Neo+ Arc | +/- | 10 | 10 | 0 | 20.7 | 14.4764176 |
| number_sniffing_pups | WT 1 | Neo+ Arc | WT | 14 | 14 | 0 | 12.0714286 | 10.8448094 |
| number_sniffing_pups | WT 2 | Neo- Arc | +/- | 8 | 8 | 0 | 15 | 7.50238057 |
| number_sniffing_pups | WT 2 | Neo- Arc | WT | 7 | 7 | 0 | 21.8571429 | 11.8100038 |
| number_sniffing_pups | WT Bl6J | Neo- Arc Bl6J | WT | 6 | 6 | 0 | 29.5 | 26.281172 |
| number_sniffing_pups | WT 1 | Neo+ Arc | +/- | 15 | 15 | 0 | 19.2666667 | 13.6510945 |
| percentage_dead_pups | Neo+ Arc KO1 | Neo+ Arc | +/- | 14 | 14 | 0 | 3.23129252 | 6.44362251 |
| percentage_dead_pups | Neo+ Arc KO1 | Neo+ Arc | KO | 15 | 15 | 0 | 35.7282251 | 44.4757563 |
| percentage_dead_pups | Neo- Arc KO2 | Neo- Arc Bl6J | KO | 7 | 7 | 0 | NA | NA |
| percentage_dead_pups | Neo- Arc KO1 | Neo- Arc | +/- | 7 | 7 | 0 | 0 | 0 |
| percentage_dead_pups | Neo- Arc KO1 | Neo- Arc | KO | 8 | 8 | 0 | 1.78571429 | 5.05076272 |
| percentage_dead_pups | WT 1 | Neo+ Arc | +/- | 10 | 10 | 0 | 0 | 0 |
| percentage_dead_pups | WT 1 | Neo+ Arc | WT | 14 | 14 | 0 | 7.67857143 | 23.5842248 |
| percentage_dead_pups | WT 2 | Neo- Arc | +/- | 8 | 8 | 0 | 5.41666667 | 11.8103398 |
| percentage_dead_pups | WT 2 | Neo- Arc | WT | 7 | 7 | 0 | 1.58730159 | 4.19960526 |
| percentage_dead_pups | WT Bl6J | Neo- Arc Bl6J | WT | 6 | 6 | 0 | NA | NA |
| percentage_dead_pups | WT 1 | Neo+ Arc | +/- | 15 | 15 | 0 | 2.06466667 | 5.46701912 |
| percentage_pups_milk_PND0 | Neo+ Arc KO1 | Neo+ Arc | +/- | 14 | 14 | 0 | 100 | 0 |
| percentage_pups_milk_PND0 | Neo+ Arc KO1 | Neo+ Arc | KO | 15 | 15 | 0 | 92.6346801 | 12.1648807 |
| percentage_pups_milk_PND0 | Neo- Arc KO2 | Neo- Arc Bl6J | KO | 7 | 7 | 0 | NA | NA |
| percentage_pups_milk_PND0 | Neo- Arc KO1 | Neo- Arc | +/- | 7 | 7 | 0 | 100 | 0 |
| percentage_pups_milk_PND0 | Neo- Arc KO1 | Neo- Arc | KO | 8 | 8 | 0 | 100 | 0 |
| percentage_pups_milk_PND0 | WT 1 | Neo+ Arc | +/- | 10 | 10 | 0 | 99 | 3.16227766 |
| percentage_pups_milk_PND0 | WT 1 | Neo+ Arc | WT | 14 | 14 | 0 | 98.8092857 | 4.4552449 |
| percentage_pups_milk_PND0 | WT 2 | Neo- Arc | +/- | 8 | 8 | 0 | 75 | 46.291005 |
| percentage_pups_milk_PND0 | WT 2 | Neo- Arc | WT | 7 | 7 | 0 | 98.4126984 | 4.19960526 |
| percentage_pups_milk_PND0 | WT Bl6J | Neo- Arc Bl6J | WT | 6 | 6 | 0 | NA | NA |
| percentage_pups_milk_PND0 | WT 1 | Neo+ Arc | +/- | 15 | 15 | 0 | 92.3806667 | 25.819968 |
| percentage_time_outside_nest | Neo+ Arc KO1 | Neo+ Arc | +/- | 14 | 14 | 0 | 52.0373795 | 20.5989795 |
| percentage_time_outside_nest | Neo+ Arc KO1 | Neo+ Arc | KO | 15 | 15 | 0 | 52.5920724 | 30.967325 |
| percentage_time_outside_nest | Neo- Arc KO2 | Neo- Arc Bl6J | KO | 7 | 7 | 0 | 82.6988567 | 8.50006559 |
| percentage_time_outside_nest | Neo- Arc KO1 | Neo- Arc | +/- | 7 | 7 | 0 | 58.5937782 | 25.1044376 |
| percentage_time_outside_nest | Neo- Arc KO1 | Neo- Arc | KO | 8 | 8 | 0 | 51.9056607 | 27.6977326 |
| percentage_time_outside_nest | WT 1 | Neo+ Arc | +/- | 10 | 10 | 0 | 29.4513196 | 17.6884464 |
| percentage_time_outside_nest | WT 1 | Neo+ Arc | WT | 14 | 14 | 0 | 36.2849224 | 17.5593306 |
| percentage_time_outside_nest | WT 2 | Neo- Arc | +/- | 8 | 8 | 0 | 57.7365088 | 21.7495433 |
| percentage_time_outside_nest | WT 2 | Neo- Arc | WT | 7 | 7 | 0 | 51.7592796 | 17.7912376 |
| percentage_time_outside_nest | WT Bl6J | Neo- Arc Bl6J | WT | 6 | 6 | 0 | 80.5875005 | 15.4066659 |
| percentage_time_outside_nest | WT 1 | Neo+ Arc | +/- | 15 | 15 | 0 | 46.383957 | 19.1711538 |
| postparturition_nesting_score | Neo+ Arc KO1 | Neo+ Arc | +/- | 14 | 14 | 0 | 2.14285714 | 0.88640526 |
| postparturition_nesting_score | Neo+ Arc KO1 | Neo+ Arc | KO | 15 | 15 | 0 | 2.53333333 | 1.06009883 |
| postparturition_nesting_score | Neo- Arc KO2 | Neo- Arc Bl6J | KO | 7 | 7 | 0 | NA | NA |
| postparturition_nesting_score | Neo- Arc KO1 | Neo- Arc | +/- | 7 | 7 | 0 | 2.5 | 0.76376262 |
| postparturition_nesting_score | Neo- Arc KO1 | Neo- Arc | KO | 8 | 8 | 0 | 2.75 | 1.28173989 |
| postparturition_nesting_score | WT 1 | Neo+ Arc | +/- | 10 | 10 | 0 | 2.6 | 0.77459667 |
| postparturition_nesting_score | WT 1 | Neo+ Arc | WT | 14 | 14 | 0 | 3.42857143 | 0.75592895 |
| postparturition_nesting_score | WT 2 | Neo- Arc | +/- | 8 | 8 | 0 | 3.28571429 | 0.75592895 |
| postparturition_nesting_score | WT 2 | Neo- Arc | WT | 7 | 7 | 0 | 3.35714286 | 0.74801324 |
| postparturition_nesting_score | WT Bl6J | Neo- Arc Bl6J | WT | 6 | 6 | 0 | NA | NA |
| postparturition_nesting_score | WT 1 | Neo+ Arc | +/- | 15 | 15 | 0 | 2.6 | 0.89042526 |
| preparturition_nesting_score | Neo+ Arc KO1 | Neo+ Arc | +/- | 14 | 14 | 0 | 3.03571429 | 1.21630408 |
| preparturition_nesting_score | Neo+ Arc KO1 | Neo+ Arc | KO | 15 | 15 | 0 | 2.86666667 | 1.30201309 |
| preparturition_nesting_score | Neo- Arc KO2 | Neo- Arc Bl6J | KO | 7 | 7 | 0 | NA | NA |
| preparturition_nesting_score | Neo- Arc KO1 | Neo- Arc | +/- | 7 | 7 | 0 | 2.71428571 | 1.46791073 |
| preparturition_nesting_score | Neo- Arc KO1 | Neo- Arc | KO | 8 | 8 | 0 | 2.625 | 1.06066017 |
| preparturition_nesting_score | WT 1 | Neo+ Arc | +/- | 10 | 10 | 0 | 3.25 | 0.79056942 |
| preparturition_nesting_score | WT 1 | Neo+ Arc | WT | 14 | 14 | 0 | 2.71428571 | 0.91387353 |
| preparturition_nesting_score | WT 2 | Neo- Arc | +/- | 8 | 8 | 0 | 2.6875 | 1.27999721 |
| preparturition_nesting_score | WT 2 | Neo- Arc | WT | 7 | 7 | 0 | 2.92857143 | 1.01770049 |
| preparturition_nesting_score | WT Bl6J | Neo- Arc Bl6J | WT | 6 | 6 | 0 | NA | NA |
| preparturition_nesting_score | WT 1 | Neo+ Arc | +/- | 15 | 15 | 0 | 2.8 | 1.26491106 |
| test_duration | Neo+ Arc KO1 | Neo+ Arc | +/- | 14 | 14 | 0 | 1793.288 | 13.9254579 |
| test_duration | Neo+ Arc KO1 | Neo+ Arc | KO | 15 | 15 | 0 | 1792.67407 | 11.1898965 |
| test_duration | Neo- Arc KO2 | Neo- Arc Bl6J | KO | 7 | 7 | 0 | 1786.5096 | 113.97156 |
| test_duration | Neo- Arc KO1 | Neo- Arc | +/- | 7 | 7 | 0 | 1798.963 | 5.53988505 |
| test_duration | Neo- Arc KO1 | Neo- Arc | KO | 8 | 8 | 0 | 1794.878 | 7.95484278 |
| test_duration | WT 1 | Neo+ Arc | +/- | 10 | 10 | 0 | 1793.0529 | 7.82694221 |
| test_duration | WT 1 | Neo+ Arc | WT | 14 | 14 | 0 | 1788.34164 | 13.7355234 |
| test_duration | WT 2 | Neo- Arc | +/- | 8 | 8 | 0 | 1791.23863 | 11.9172563 |
| test_duration | WT 2 | Neo- Arc | WT | 7 | 7 | 0 | 1796.88957 | 6.86272538 |
| test_duration | WT Bl6J | Neo- Arc Bl6J | WT | 6 | 6 | 0 | 1664.874 | 257.908687 |
| test_duration | WT 1 | Neo+ Arc | +/- | 15 | 15 | 0 | 1793.65953 | 9.83233613 |
| time_crouching | Neo+ Arc KO1 | Neo+ Arc | +/- | 14 | 14 | 0 | 168.268643 | 209.823538 |
| time_crouching | Neo+ Arc KO1 | Neo+ Arc | KO | 15 | 15 | 0 | 29.2730667 | 113.3741 |
| time_crouching | Neo- Arc KO2 | Neo- Arc Bl6J | KO | 7 | 7 | 0 | 0 | 0 |
| time_crouching | Neo- Arc KO1 | Neo- Arc | +/- | 7 | 7 | 0 | 132.858429 | 217.755124 |
| time_crouching | Neo- Arc KO1 | Neo- Arc | KO | 8 | 8 | 0 | 200.9555 | 292.65913 |
| time_crouching | WT 1 | Neo+ Arc | +/- | 10 | 10 | 0 | 450.9109 | 381.030295 |
| time_crouching | WT 1 | Neo+ Arc | WT | 14 | 14 | 0 | 299.284357 | 399.945156 |
| time_crouching | WT 2 | Neo- Arc | +/- | 8 | 8 | 0 | 61.695 | 101.939343 |
| time_crouching | WT 2 | Neo- Arc | WT | 7 | 7 | 0 | 26.3481429 | 34.9510163 |
| time_crouching | WT Bl6J | Neo- Arc Bl6J | WT | 6 | 6 | 0 | 5.70866667 | 13.9833204 |
| time_crouching | WT 1 | Neo+ Arc | +/- | 15 | 15 | 0 | 225.811067 | 365.71656 |
| time_digging | Neo+ Arc KO1 | Neo+ Arc | +/- | 14 | 14 | 0 | 119.316714 | 153.231467 |
| time_digging | Neo+ Arc KO1 | Neo+ Arc | KO | 15 | 15 | 0 | 82.8230667 | 104.795831 |
| time_digging | Neo- Arc KO2 | Neo- Arc Bl6J | KO | 7 | 7 | 0 | 305.2876 | 154.546119 |
| time_digging | Neo- Arc KO1 | Neo- Arc | +/- | 7 | 7 | 0 | 100.412571 | 114.531969 |
| time_digging | Neo- Arc KO1 | Neo- Arc | KO | 8 | 8 | 0 | 26.01575 | 16.2692953 |
| time_digging | WT 1 | Neo+ Arc | +/- | 10 | 10 | 0 | 36.1397 | 56.8772032 |
| time_digging | WT 1 | Neo+ Arc | WT | 14 | 14 | 0 | 46.9025714 | 56.7814651 |
| time_digging | WT 2 | Neo- Arc | +/- | 8 | 8 | 0 | 48.62825 | 60.2416729 |
| time_digging | WT 2 | Neo- Arc | WT | 7 | 7 | 0 | 29.0751429 | 47.5853811 |
| time_digging | WT Bl6J | Neo- Arc Bl6J | WT | 6 | 6 | 0 | 300.173667 | 194.845622 |
| time_digging | WT 1 | Neo+ Arc | +/- | 15 | 15 | 0 | 110.2996 | 103.259278 |
| time_grooming | Neo+ Arc KO1 | Neo+ Arc | +/- | 14 | 14 | 0 | 73.9040714 | 133.967413 |
| time_grooming | Neo+ Arc KO1 | Neo+ Arc | KO | 15 | 15 | 0 | 44.9331333 | 51.89352 |
| time_grooming | Neo- Arc KO2 | Neo- Arc Bl6J | KO | 7 | 7 | 0 | 18.9998 | 19.1482017 |
| time_grooming | Neo- Arc KO1 | Neo- Arc | +/- | 7 | 7 | 0 | 58.984 | 59.3233946 |
| time_grooming | Neo- Arc KO1 | Neo- Arc | KO | 8 | 8 | 0 | 92.996875 | 89.7317631 |
| time_grooming | WT 1 | Neo+ Arc | +/- | 10 | 10 | 0 | 45.4745 | 80.5778013 |
| time_grooming | WT 1 | Neo+ Arc | WT | 14 | 14 | 0 | 95.3552857 | 128.931017 |
| time_grooming | WT 2 | Neo- Arc | +/- | 8 | 8 | 0 | 29.70425 | 30.2521688 |
| time_grooming | WT 2 | Neo- Arc | WT | 7 | 7 | 0 | 21.3591429 | 18.3318345 |
| time_grooming | WT Bl6J | Neo- Arc Bl6J | WT | 6 | 6 | 0 | 42.7485 | 71.6716109 |
| time_grooming | WT 1 | Neo+ Arc | +/- | 15 | 15 | 0 | 33.5198667 | 37.1080596 |
| time_inside_nest | Neo+ Arc KO1 | Neo+ Arc | +/- | 14 | 14 | 0 | 857.892 | 367.185023 |
| time_inside_nest | Neo+ Arc KO1 | Neo+ Arc | KO | 15 | 15 | 0 | 849.082467 | 557.038593 |
| time_inside_nest | Neo- Arc KO2 | Neo- Arc Bl6J | KO | 7 | 7 | 0 | 315.1008 | 160.723426 |
| time_inside_nest | Neo- Arc KO1 | Neo- Arc | +/- | 7 | 7 | 0 | 743.815286 | 452.46838 |
| time_inside_nest | Neo- Arc KO1 | Neo- Arc | KO | 8 | 8 | 0 | 859.804125 | 500.393393 |
| time_inside_nest | WT 1 | Neo+ Arc | +/- | 10 | 10 | 0 | 1263.1901 | 318.879552 |
| time_inside_nest | WT 1 | Neo+ Arc | WT | 14 | 14 | 0 | 1137.35114 | 313.651534 |
| time_inside_nest | WT 2 | Neo- Arc | +/- | 8 | 8 | 0 | 755.14725 | 390.655808 |
| time_inside_nest | WT 2 | Neo- Arc | WT | 7 | 7 | 0 | 865.293 | 320.372202 |
| time_inside_nest | WT Bl6J | Neo- Arc Bl6J | WT | 6 | 6 | 0 | 334.88 | 293.109732 |
| time_inside_nest | WT 1 | Neo+ Arc | +/- | 15 | 15 | 0 | 958.731467 | 344.031059 |
| time_nesting | Neo+ Arc KO1 | Neo+ Arc | +/- | 14 | 14 | 0 | 151.743857 | 153.711521 |
| time_nesting | Neo+ Arc KO1 | Neo+ Arc | KO | 15 | 15 | 0 | 95.6044 | 130.955286 |
| time_nesting | Neo- Arc KO2 | Neo- Arc Bl6J | KO | 7 | 7 | 0 | 28.3984 | 26.301024 |
| time_nesting | Neo- Arc KO1 | Neo- Arc | +/- | 7 | 7 | 0 | 21.0245714 | 30.6735796 |
| time_nesting | Neo- Arc KO1 | Neo- Arc | KO | 8 | 8 | 0 | 107.88925 | 69.1701428 |
| time_nesting | WT 1 | Neo+ Arc | +/- | 10 | 10 | 0 | 178.2241 | 95.2225992 |
| time_nesting | WT 1 | Neo+ Arc | WT | 14 | 14 | 0 | 155.080286 | 96.9977882 |
| time_nesting | WT 2 | Neo- Arc | +/- | 8 | 8 | 0 | 135.042 | 144.778457 |
| time_nesting | WT 2 | Neo- Arc | WT | 7 | 7 | 0 | 139.889286 | 51.0005527 |
| time_nesting | WT Bl6J | Neo- Arc Bl6J | WT | 6 | 6 | 0 | 55.708 | 88.7368592 |
| time_nesting | WT 1 | Neo+ Arc | +/- | 15 | 15 | 0 | 113.060933 | 74.433013 |
| time_outside_nest | Neo+ Arc KO1 | Neo+ Arc | +/- | 14 | 14 | 0 | 934.429714 | 372.197735 |
| time_outside_nest | Neo+ Arc KO1 | Neo+ Arc | KO | 15 | 15 | 0 | 941.2982 | 552.376557 |
| time_outside_nest | Neo- Arc KO2 | Neo- Arc Bl6J | KO | 7 | 7 | 0 | 1470.7102 | 81.696861 |
| time_outside_nest | Neo- Arc KO1 | Neo- Arc | +/- | 7 | 7 | 0 | 1053.55229 | 449.5673 |
| time_outside_nest | Neo- Arc KO1 | Neo- Arc | KO | 8 | 8 | 0 | 931.169125 | 495.219099 |
| time_outside_nest | WT 1 | Neo+ Arc | +/- | 10 | 10 | 0 | 527.5569 | 315.697251 |
| time_outside_nest | WT 1 | Neo+ Arc | WT | 14 | 14 | 0 | 648.8145 | 313.317069 |
| time_outside_nest | WT 2 | Neo- Arc | +/- | 8 | 8 | 0 | 1034.15675 | 390.490804 |
| time_outside_nest | WT 2 | Neo- Arc | WT | 7 | 7 | 0 | 929.396143 | 316.855192 |
| time_outside_nest | WT Bl6J | Neo- Arc Bl6J | WT | 6 | 6 | 0 | 1329.66317 | 290.054989 |
| time_outside_nest | WT 1 | Neo+ Arc | +/- | 15 | 15 | 0 | 832.070667 | 343.793901 |
| time_sniffing_pups | Neo+ Arc KO1 | Neo+ Arc | +/- | 14 | 14 | 0 | 82.1684286 | 83.5484936 |
| time_sniffing_pups | Neo+ Arc KO1 | Neo+ Arc | KO | 15 | 15 | 0 | 29.5408 | 21.6663307 |
| time_sniffing_pups | Neo- Arc KO2 | Neo- Arc Bl6J | KO | 7 | 7 | 0 | 74.7364 | 65.9633894 |
| time_sniffing_pups | Neo- Arc KO1 | Neo- Arc | +/- | 7 | 7 | 0 | 30.5788571 | 27.3876273 |
| time_sniffing_pups | Neo- Arc KO1 | Neo- Arc | KO | 8 | 8 | 0 | 26.282875 | 30.908998 |
| time_sniffing_pups | WT 1 | Neo+ Arc | +/- | 10 | 10 | 0 | 61.7031 | 54.7627198 |
| time_sniffing_pups | WT 1 | Neo+ Arc | WT | 14 | 14 | 0 | 23.326 | 25.7467344 |
| time_sniffing_pups | WT 2 | Neo- Arc | +/- | 8 | 8 | 0 | 23.5615 | 17.0002915 |
| time_sniffing_pups | WT 2 | Neo- Arc | WT | 7 | 7 | 0 | 34.2325714 | 16.8882572 |
| time_sniffing_pups | WT Bl6J | Neo- Arc Bl6J | WT | 6 | 6 | 0 | 39.1606667 | 37.2526238 |
| time_sniffing_pups | WT 1 | Neo+ Arc | +/- | 15 | 15 | 0 | 33.8155333 | 21.89279 |
| E+Maze_center_numberEntry | Neo+ Arc KO1 | Neo+ Arc KO1 | NA | 13 | 9 | 4 | 7.07692308 | 4.05095747 |
| E+Maze_center_numberEntry | WT | WT | NA | 20 | 10 | 10 | 20.1 | 9.17031825 |
| E+Maze_center_timeSpent | Neo+ Arc KO1 | Neo+ Arc KO1 | NA | 13 | 9 | 4 | 111.730769 | 92.7270671 |
| E+Maze_center_timeSpent | WT | WT | NA | 20 | 10 | 10 | 106.795 | 75.1075893 |
| E+Maze_closedArms_numberEntry | Neo+ Arc KO1 | Neo+ Arc KO1 | NA | 13 | 9 | 4 | 4.61538462 | 4.95880465 |
| E+Maze_closedArms_numberEntry | WT | WT | NA | 20 | 10 | 10 | 13.1 | 5.88396572 |
| E+Maze_closedArms_timeSpent | Neo+ Arc KO1 | Neo+ Arc KO1 | NA | 13 | 9 | 4 | 115.161538 | 94.1910606 |
| E+Maze_closedArms_timeSpent | WT | WT | NA | 20 | 10 | 10 | 130.255 | 80.1451018 |
| E+Maze_distanceTraveled | Neo+ Arc KO1 | Neo+ Arc KO1 | NA | 13 | 9 | 4 | 2.14846154 | 1.32914519 |
| E+Maze_distanceTraveled | WT | WT | NA | 20 | 10 | 10 | 3.6197 | 1.32550559 |
| E+Maze_openedArms_numberEntry | Neo+ Arc KO1 | Neo+ Arc KO1 | NA | 13 | 9 | 4 | 8.69230769 | 5.80781966 |
| E+Maze_openedArms_numberEntry | WT | WT | NA | 20 | 10 | 10 | 15.4 | 8.71417478 |
| E+Maze_openedArms_timeSpent | Neo+ Arc KO1 | Neo+ Arc KO1 | NA | 13 | 9 | 4 | 73.1153846 | 67.5918985 |
| E+Maze_openedArms_timeSpent | WT | WT | NA | 20 | 10 | 10 | 62.94 | 48.6363563 |
| E+maze_immobility_timeSpent | Neo+ Arc KO1 | Neo+ Arc KO1 | NA | 13 | 9 | 4 | 72.7307692 | 62.5122306 |
| E+maze_immobility_timeSpent | WT | WT | NA | 20 | 10 | 10 | 66.075 | 80.8955264 |
| foodLocalization_latencyFeed | Neo+ Arc KO1 | Neo+ Arc KO1 | NA | 16 | 12 | 4 | 136.5625 | 92.7915001 |
| foodLocalization_latencyFeed | WT | WT | NA | 19 | 7 | 12 | 147.473684 | 99.6880515 |
| innateOlfactoryPreference_odor1_numberVisits | Neo+ Arc KO1 | Neo+ Arc KO1 | NA | 13 | 9 | 4 | 3.30769231 | 1.84321347 |
| innateOlfactoryPreference_odor1_numberVisits | WT | WT | NA | 18 | 8 | 10 | 2.44444444 | 1.29352333 |
| innateOlfactoryPreference_odor1_timeSpent | Neo+ Arc KO1 | Neo+ Arc KO1 | NA | 13 | 9 | 4 | 6.21815385 | 6.96389407 |
| innateOlfactoryPreference_odor1_timeSpent | WT | WT | NA | 18 | 8 | 10 | 5.23583333 | 6.29623596 |
| innateOlfactoryPreference_odor2_numberVisits | Neo+ Arc KO1 | Neo+ Arc KO1 | NA | 13 | 9 | 4 | 1.92307692 | 2.56455124 |
| innateOlfactoryPreference_odor2_numberVisits | WT | WT | NA | 20 | 10 | 10 | 1.9 | 1.94395148 |
| innateOlfactoryPreference_odor2_timeSpent | Neo+ Arc KO1 | Neo+ Arc KO1 | NA | 13 | 9 | 4 | 4.17546154 | 8.86076585 |
| innateOlfactoryPreference_odor2_timeSpent | WT | WT | NA | 20 | 10 | 10 | 3.37495 | 5.68422242 |
| innateOlfactoryPreference_odor3_numberVisits | Neo+ Arc KO1 | Neo+ Arc KO1 | NA | 13 | 9 | 4 | 0.92307692 | 1.32045058 |
| innateOlfactoryPreference_odor3_numberVisits | WT | WT | NA | 20 | 10 | 10 | 1.9 | 1.44732057 |
| innateOlfactoryPreference_odor3_timeSpent | Neo+ Arc KO1 | Neo+ Arc KO1 | NA | 13 | 9 | 4 | 1.22569231 | 2.38112912 |
| innateOlfactoryPreference_odor3_timeSpent | WT | WT | NA | 20 | 10 | 10 | 4.06835 | 5.50200747 |
| innateOlfactoryPreference_odor4_numberVisits | Neo+ Arc KO1 | Neo+ Arc KO1 | NA | 13 | 9 | 4 | 3.38461538 | 4.01120226 |
| innateOlfactoryPreference_odor4_numberVisits | WT | WT | NA | 20 | 10 | 10 | 2.25 | 2.67296406 |
| innateOlfactoryPreference_odor4_timeSpent | Neo+ Arc KO1 | Neo+ Arc KO1 | NA | 13 | 9 | 4 | 6.72338462 | 6.81767309 |
| innateOlfactoryPreference_odor4_timeSpent | WT | WT | NA | 20 | 10 | 10 | 3.2673 | 3.16592285 |
| innateOlfactoryPreference_odor5_numberVisits | Neo+ Arc KO1 | Neo+ Arc KO1 | NA | 13 | 9 | 4 | 2.07692308 | 2.56455124 |
| innateOlfactoryPreference_odor5_numberVisits | WT | WT | NA | 20 | 10 | 10 | 1.75 | 2.42519668 |
| innateOlfactoryPreference_odor5_timeSpent | Neo+ Arc KO1 | Neo+ Arc KO1 | NA | 13 | 9 | 4 | 3.65576923 | 5.40905419 |
| innateOlfactoryPreference_odor5_timeSpent | WT | WT | NA | 20 | 10 | 10 | 4.0415 | 5.66390073 |
| marble_half_burried | Neo+ Arc KO1 | Neo+ Arc KO1 | NA | 13 | 7 | 6 | 1.92307692 | 0.8623165 |
| marble_half_burried | WT | WT | NA | 9 | 4 | 5 | 2.55555556 | 1.50923086 |
| marble_more_half_burried | Neo+ Arc KO1 | Neo+ Arc KO1 | NA | 13 | 7 | 6 | 17 | 2.41522946 |
| marble_more_half_burried | WT | WT | NA | 9 | 4 | 5 | 16.4444444 | 1.50923086 |
| marble_more_two_thirds_burried | Neo+ Arc KO1 | Neo+ Arc KO1 | NA | 13 | 7 | 6 | 15.0769231 | 2.62873666 |
| marble_more_two_thirds_burried | WT | WT | NA | 9 | 4 | 5 | 13.8888889 | 2.02758751 |
| marble_not_burried | Neo+ Arc KO1 | Neo+ Arc KO1 | NA | 13 | 7 | 6 | 3 | 2.41522946 |
| marble_not_burried | WT | WT | NA | 9 | 4 | 5 | 3.55555556 | 1.50923086 |
| marble_totally_burried | Neo+ Arc KO1 | Neo+ Arc KO1 | NA | 13 | 7 | 6 | 11.3846154 | 3.33012666 |
| marble_totally_burried | WT | WT | NA | 9 | 4 | 5 | 10.6666667 | 2.6925824 |
| marble_two_thirds_burried | Neo+ Arc KO1 | Neo+ Arc KO1 | NA | 13 | 7 | 6 | 3.69230769 | 2.05688338 |
| marble_two_thirds_burried | WT | WT | NA | 9 | 4 | 5 | 3.22222222 | 1.30170828 |
| ms_digging_duration | Neo+ Arc KO1 | Neo+ Arc KO1 | NA | 10 | 6 | 4 | 4.8952 | 1.46306845 |
| ms_digging_duration | Neo- Arc KO1 | Neo- Arc KO1 | NA | 25 | 13 | 12 | 2.66816 | 0.91034388 |
| ms_digging_duration | WT | WT | NA | 31 | 15 | 16 | 2.38357766 | 1.19723475 |
| ms_digging_number | Neo+ Arc KO1 | Neo+ Arc KO1 | NA | 10 | 6 | 4 | 30.7 | 13.2417689 |
| ms_digging_number | Neo- Arc KO1 | Neo- Arc KO1 | NA | 25 | 13 | 12 | 36.68 | 17.1918585 |
| ms_digging_number | WT | WT | NA | 31 | 15 | 16 | 41.3548387 | 20.6825988 |
| ms_digging_timeSpent | Neo+ Arc KO1 | Neo+ Arc KO1 | NA | 10 | 6 | 4 | 144.8339 | 69.1796005 |
| ms_digging_timeSpent | Neo- Arc KO1 | Neo- Arc KO1 | NA | 25 | 13 | 12 | 99.86768 | 57.6930366 |
| ms_digging_timeSpent | WT | WT | NA | 31 | 15 | 16 | 99.925871 | 63.0786932 |
| ms_headShakes | Neo+ Arc KO1 | Neo+ Arc KO1 | NA | 10 | 6 | 4 | 5.2 | 2.65832027 |
| ms_headShakes | Neo- Arc KO1 | Neo- Arc KO1 | NA | 25 | 13 | 12 | 3.24 | 2.50466232 |
| ms_headShakes | WT | WT | NA | 31 | 15 | 16 | 2.90322581 | 1.90359027 |
| ms_immobility_duration | Neo+ Arc KO1 | Neo+ Arc KO1 | NA | 10 | 6 | 4 | 1.6477 | 3.5706541 |
| ms_immobility_duration | Neo- Arc KO1 | Neo- Arc KO1 | NA | 25 | 13 | 12 | 0.15664 | 0.7832 |
| ms_immobility_duration | WT | WT | NA | 31 | 15 | 16 | 0 | 0 |
| ms_immobility_number | Neo+ Arc KO1 | Neo+ Arc KO1 | NA | 10 | 6 | 4 | 0.2 | 0.42163702 |
| ms_immobility_number | Neo- Arc KO1 | Neo- Arc KO1 | NA | 25 | 13 | 12 | 0.08 | 0.4 |
| ms_immobility_number | WT | WT | NA | 31 | 15 | 16 | 0 | 0 |
| ms_immobility_timeSpent | Neo+ Arc KO1 | Neo+ Arc KO1 | NA | 10 | 6 | 4 | 1.6477 | 3.5706541 |
| ms_immobility_timeSpent | Neo- Arc KO1 | Neo- Arc KO1 | NA | 25 | 13 | 12 | 0.31332 | 1.5666 |
| ms_immobility_timeSpent | WT | WT | NA | 31 | 15 | 16 | 0 | 0 |
| ms_rearing | Neo+ Arc KO1 | Neo+ Arc KO1 | NA | 10 | 6 | 4 | 54.3 | 21.5924369 |
| ms_rearing | Neo- Arc KO1 | Neo- Arc KO1 | NA | 25 | 13 | 12 | 59.84 | 18.8540005 |
| ms_rearing | WT | WT | NA | 31 | 15 | 16 | 66.9032258 | 24.9537637 |
| ms_scratching | Neo+ Arc KO1 | Neo+ Arc KO1 | NA | 10 | 6 | 4 | 1.7 | 1.56702124 |
| ms_scratching | Neo- Arc KO1 | Neo- Arc KO1 | NA | 25 | 13 | 12 | 1.72 | 1.96892526 |
| ms_scratching | WT | WT | NA | 31 | 15 | 16 | 0.77419355 | 1.02338254 |
| ms_selfGrooming_duration | Neo+ Arc KO1 | Neo+ Arc KO1 | NA | 10 | 6 | 4 | 15.36295 | 4.85735472 |
| ms_selfGrooming_duration | Neo- Arc KO1 | Neo- Arc KO1 | NA | 25 | 13 | 12 | 12.2116 | 6.82963256 |
| ms_selfGrooming_duration | WT | WT | NA | 31 | 15 | 16 | 8.42258065 | 3.60491297 |
| ms_selfGrooming_number | Neo+ Arc KO1 | Neo+ Arc KO1 | NA | 10 | 6 | 4 | 3 | 1.41421356 |
| ms_selfGrooming_number | Neo- Arc KO1 | Neo- Arc KO1 | NA | 25 | 13 | 12 | 4.24 | 2.4200551 |
| ms_selfGrooming_number | WT | WT | NA | 31 | 15 | 16 | 2.51612903 | 1.91260673 |
| ms_selfGrooming_timeSpent | Neo+ Arc KO1 | Neo+ Arc KO1 | NA | 10 | 6 | 4 | 49.8724 | 37.1359942 |
| ms_selfGrooming_timeSpent | Neo- Arc KO1 | Neo- Arc KO1 | NA | 25 | 13 | 12 | 51.04108 | 33.9308879 |
| ms_selfGrooming_timeSpent | WT | WT | NA | 31 | 15 | 16 | 20.9134839 | 16.8280434 |
| novelObject_hab1_discriminationIndex | Neo+ Arc KO1 | Neo+ Arc KO1 | NA | 15 | 8 | 7 | -0.08421771 | 0.40678097 |
| novelObject_hab1_discriminationIndex | WT | WT | NA | 12 | 8 | 4 | 0.01384455 | 0.31045292 |
| novelObject_hab1_interaction_object1_number | Neo+ Arc KO1 | Neo+ Arc KO1 | NA | 15 | 8 | 7 | 13.1333333 | 14.3967589 |
| novelObject_hab1_interaction_object1_number | WT | WT | NA | 12 | 8 | 4 | 10.9166667 | 9.23883437 |
| novelObject_hab1_interaction_object1_timeSpent | Neo+ Arc KO1 | Neo+ Arc KO1 | NA | 15 | 8 | 7 | 21.3 | 31.8479199 |
| novelObject_hab1_interaction_object1_timeSpent | WT | WT | NA | 12 | 8 | 4 | 12.1833333 | 10.5395216 |
| novelObject_hab1_interaction_object2_number | Neo+ Arc KO1 | Neo+ Arc KO1 | NA | 15 | 8 | 7 | 12 | 9.84160266 |
| novelObject_hab1_interaction_object2_number | WT | WT | NA | 12 | 8 | 4 | 7.58333333 | 4.29499356 |
| novelObject_hab1_interaction_object2_timeSpent | Neo+ Arc KO1 | Neo+ Arc KO1 | NA | 15 | 8 | 7 | 15.2 | 12.7433456 |
| novelObject_hab1_interaction_object2_timeSpent | WT | WT | NA | 12 | 8 | 4 | 13.025 | 13.0703777 |
| novelObject_hab2_discriminationIndex | Neo+ Arc KO1 | Neo+ Arc KO1 | NA | 15 | 8 | 7 | 0.03050903 | 0.57370162 |
| novelObject_hab2_discriminationIndex | WT | WT | NA | 12 | 8 | 4 | -0.20364806 | 0.39272387 |
| novelObject_hab2_interaction_object1_number | Neo+ Arc KO1 | Neo+ Arc KO1 | NA | 15 | 8 | 7 | 7.06666667 | 6.19292992 |
| novelObject_hab2_interaction_object1_number | WT | WT | NA | 12 | 8 | 4 | 12.1666667 | 10.0257245 |
| novelObject_hab2_interaction_object1_timeSpent | Neo+ Arc KO1 | Neo+ Arc KO1 | NA | 15 | 8 | 7 | 12.2733333 | 13.3431345 |
| novelObject_hab2_interaction_object1_timeSpent | WT | WT | NA | 12 | 8 | 4 | 17.5166667 | 21.620522 |
| novelObject_hab2_interaction_object2_number | Neo+ Arc KO1 | Neo+ Arc KO1 | NA | 15 | 8 | 7 | 7.73333333 | 8.89194392 |
| novelObject_hab2_interaction_object2_number | WT | WT | NA | 12 | 8 | 4 | 7.75 | 6.19567298 |
| novelObject_hab2_interaction_object2_timeSpent | Neo+ Arc KO1 | Neo+ Arc KO1 | NA | 15 | 8 | 7 | 14.0933333 | 20.252459 |
| novelObject_hab2_interaction_object2_timeSpent | WT | WT | NA | 12 | 8 | 4 | 10 | 11.5502853 |
| novelObject_ltm_discriminationIndex | Neo+ Arc KO1 | Neo+ Arc KO1 | NA | 15 | 8 | 7 | 0.14252863 | 0.4443839 |
| novelObject_ltm_discriminationIndex | WT | WT | NA | 12 | 8 | 4 | 0.02074363 | 0.32147898 |
| novelObject_ltm_interaction_object1_number | Neo+ Arc KO1 | Neo+ Arc KO1 | NA | 15 | 8 | 7 | 13.6 | 10.7158094 |
| novelObject_ltm_interaction_object1_number | WT | WT | NA | 12 | 8 | 4 | 13.25 | 8.22551463 |
| novelObject_ltm_interaction_object1_timeSpent | Neo+ Arc KO1 | Neo+ Arc KO1 | NA | 15 | 8 | 7 | 27.2533333 | 38.3711362 |
| novelObject_ltm_interaction_object1_timeSpent | WT | WT | NA | 12 | 8 | 4 | 26.0833333 | 18.3189834 |
| novelObject_ltm_interaction_object2_number | Neo+ Arc KO1 | Neo+ Arc KO1 | NA | 15 | 8 | 7 | 14.0666667 | 12.3315314 |
| novelObject_ltm_interaction_object2_number | WT | WT | NA | 12 | 8 | 4 | 13.5 | 6.92163932 |
| novelObject_ltm_interaction_object2_timeSpent | Neo+ Arc KO1 | Neo+ Arc KO1 | NA | 15 | 8 | 7 | 40.3866667 | 38.1126501 |
| novelObject_ltm_interaction_object2_timeSpent | WT | WT | NA | 12 | 8 | 4 | 27.5666667 | 18.820846 |
| novelObject_stm_discriminationIndex | Neo+ Arc KO1 | Neo+ Arc KO1 | NA | 15 | 8 | 7 | -0.06661015 | 0.34563392 |
| novelObject_stm_discriminationIndex | WT | WT | NA | 12 | 8 | 4 | 0.16023413 | 0.33895987 |
| novelObject_stm_interaction_object1_number | Neo+ Arc KO1 | Neo+ Arc KO1 | NA | 15 | 8 | 7 | 15.5333333 | 13.1304807 |
| novelObject_stm_interaction_object1_number | WT | WT | NA | 12 | 8 | 4 | 9.91666667 | 4.66043957 |
| novelObject_stm_interaction_object1_timeSpent | Neo+ Arc KO1 | Neo+ Arc KO1 | NA | 15 | 8 | 7 | 39.5866667 | 41.4758254 |
| novelObject_stm_interaction_object1_timeSpent | WT | WT | NA | 12 | 8 | 4 | 16.7583333 | 10.8085369 |
| novelObject_stm_interaction_object2_number | Neo+ Arc KO1 | Neo+ Arc KO1 | NA | 15 | 8 | 7 | 17.4 | 13.5319727 |
| novelObject_stm_interaction_object2_number | WT | WT | NA | 12 | 8 | 4 | 16 | 7.83929496 |
| novelObject_stm_interaction_object2_timeSpent | Neo+ Arc KO1 | Neo+ Arc KO1 | NA | 15 | 8 | 7 | 25.08 | 16.5799879 |
| novelObject_stm_interaction_object2_timeSpent | WT | WT | NA | 12 | 8 | 4 | 25.8416667 | 20.8057011 |
| nsf_food_intake_15min | Neo+ Arc KO1 | Neo+ Arc KO1 | NA | 13 | 7 | 6 | 0.25384615 | 0.05637747 |
| nsf_food_intake_15min | WT | WT | NA | 8 | 4 | 4 | 0.5125 | 0.16201852 |
| nsf_food_intake_60min | Neo+ Arc KO1 | Neo+ Arc KO1 | NA | 9 | 3 | 6 | 0.53333333 | 0.1040833 |
| nsf_food_intake_60min | WT | WT | NA | 8 | 4 | 4 | 1.1125 | 0.24016363 |
| nsf_latency_feed | Neo+ Arc KO1 | Neo+ Arc KO1 | NA | 13 | 7 | 6 | 359.076923 | 106.646192 |
| nsf_latency_feed | WT | WT | NA | 9 | 4 | 5 | 318.777778 | 160.218427 |
| olfactoryAvoidance_immobility_timeSpent | Neo+ Arc KO1 | Neo+ Arc KO1 | NA | 13 | 9 | 4 | 0.69230769 | 2.49615088 |
| olfactoryAvoidance_immobility_timeSpent | WT | WT | NA | 20 | 10 | 10 | 0.72005 | 3.22016149 |
| olfactoryAvoidance_predatorOdor_numberVisits | Neo+ Arc KO1 | Neo+ Arc KO1 | NA | 13 | 9 | 4 | 11.8461538 | 5.32049737 |
| olfactoryAvoidance_predatorOdor_numberVisits | WT | WT | NA | 20 | 10 | 10 | 6.3 | 4.48506293 |
| olfactoryAvoidance_predatorOdor_timeSpent | Neo+ Arc KO1 | Neo+ Arc KO1 | NA | 13 | 9 | 4 | 24.6436154 | 36.3518573 |
| olfactoryAvoidance_predatorOdor_timeSpent | WT | WT | NA | 20 | 10 | 10 | 9.4794 | 8.78092642 |
| olfactoryAvoidance_selfgrooming_numberVisits | Neo+ Arc KO1 | Neo+ Arc KO1 | NA | 13 | 9 | 4 | 2.53846154 | 1.94145069 |
| olfactoryAvoidance_selfgrooming_numberVisits | WT | WT | NA | 20 | 10 | 10 | 2.25 | 1.40955387 |
| olfactoryAvoidance_selfgrooming_timeSpent | Neo+ Arc KO1 | Neo+ Arc KO1 | NA | 13 | 9 | 4 | 36.2353846 | 43.2549753 |
| olfactoryAvoidance_selfgrooming_timeSpent | WT | WT | NA | 20 | 10 | 10 | 44.9874 | 46.3837376 |
| openField_center_distanceTraveled | Neo+ Arc KO1 | Neo+ Arc KO1 | NA | 12 | 6 | 6 | 2.80441667 | 1.30885949 |
| openField_center_distanceTraveled | WT | WT | NA | 8 | 4 | 4 | 2.9885 | 1.99286233 |
| openField_center_duration | Neo+ Arc KO1 | Neo+ Arc KO1 | NA | 12 | 6 | 6 | 1.3001553 | 0.21797768 |
| openField_center_duration | WT | WT | NA | 8 | 4 | 4 | 1.57410886 | 0.50132334 |
| openField_center_numberEntry | Neo+ Arc KO1 | Neo+ Arc KO1 | NA | 12 | 6 | 6 | 34.3333333 | 11.4123803 |
| openField_center_numberEntry | WT | WT | NA | 8 | 4 | 4 | 31.125 | 22.9374148 |
| openField_center_timeSpent | Neo+ Arc KO1 | Neo+ Arc KO1 | NA | 12 | 6 | 6 | 44.5083333 | 15.791048 |
| openField_center_timeSpent | WT | WT | NA | 8 | 4 | 4 | 47.65 | 36.6774746 |
| openField_corners_distanceTraveled | Neo+ Arc KO1 | Neo+ Arc KO1 | NA | 12 | 6 | 6 | 13.2575 | 2.79897411 |
| openField_corners_distanceTraveled | WT | WT | NA | 8 | 4 | 4 | 16.48425 | 5.57230393 |
| openField_corners_duration | Neo+ Arc KO1 | Neo+ Arc KO1 | NA | 12 | 6 | 6 | 11.573491 | 10.1765015 |
| openField_corners_duration | WT | WT | NA | 8 | 4 | 4 | 10.7238292 | 5.14135144 |
| openField_corners_numberEntry | Neo+ Arc KO1 | Neo+ Arc KO1 | NA | 12 | 6 | 6 | 74.25 | 40.6048027 |
| openField_corners_numberEntry | WT | WT | NA | 8 | 4 | 4 | 61.625 | 31.231566 |
| openField_corners_timeSpent | Neo+ Arc KO1 | Neo+ Arc KO1 | NA | 12 | 6 | 6 | 517 | 45.7752613 |
| openField_corners_timeSpent | WT | WT | NA | 8 | 4 | 4 | 528.1875 | 35.0610666 |
| openField_distanceTraveled | Neo+ Arc KO1 | Neo+ Arc KO1 | NA | 12 | 6 | 6 | 16.5464167 | 4.05943958 |
| openField_distanceTraveled | WT | WT | NA | 8 | 4 | 4 | 20.0355 | 7.34477669 |
| openField_immobility_timeSpent | Neo+ Arc KO1 | Neo+ Arc KO1 | NA | 12 | 6 | 6 | 344.233333 | 51.5705577 |
| openField_immobility_timeSpent | WT | WT | NA | 8 | 4 | 4 | 343.9 | 76.5661432 |
| openField_lineCrossings | Neo+ Arc KO1 | Neo+ Arc KO1 | NA | 12 | 6 | 6 | 293.75 | 77.3752662 |
| openField_lineCrossings | WT | WT | NA | 8 | 4 | 4 | 382.375 | 109.821852 |
| openField_meanSpeed | Neo+ Arc KO1 | Neo+ Arc KO1 | NA | 12 | 6 | 6 | 0.02758333 | 0.00673469 |
| openField_meanSpeed | WT | WT | NA | 8 | 4 | 4 | 0.0335 | 0.01236354 |
| p1_chA_numberEntry | Neo+ Arc KO1 | Neo+ Arc KO1 | NA | 12 | 6 | 6 | 20.6666667 | 7.07535201 |
| p1_chA_numberEntry | Neo- Arc KO1 | Neo- Arc KO1 | NA | 14 | 7 | 7 | 14.8571429 | 7.20957849 |
| p1_chA_numberEntry | WT | WT | NA | 26 | 12 | 14 | 21.7692308 | 11.4028337 |
| p1_chA_timeSpent | Neo+ Arc KO1 | Neo+ Arc KO1 | NA | 12 | 6 | 6 | 187.233333 | 42.9298229 |
| p1_chA_timeSpent | Neo- Arc KO1 | Neo- Arc KO1 | NA | 14 | 7 | 7 | 157.042857 | 72.4925112 |
| p1_chA_timeSpent | WT | WT | NA | 26 | 12 | 14 | 162.561538 | 64.7004487 |
| p1_chB_numberEntry | Neo+ Arc KO1 | Neo+ Arc KO1 | NA | 12 | 6 | 6 | 19.5833333 | 6.68047812 |
| p1_chB_numberEntry | Neo- Arc KO1 | Neo- Arc KO1 | NA | 14 | 7 | 7 | 11.9285714 | 4.85900094 |
| p1_chB_numberEntry | WT | WT | NA | 26 | 12 | 14 | 19 | 9.20869155 |
| p1_chB_timeSpent | Neo+ Arc KO1 | Neo+ Arc KO1 | NA | 12 | 6 | 6 | 163.691667 | 60.1697819 |
| p1_chB_timeSpent | Neo- Arc KO1 | Neo- Arc KO1 | NA | 14 | 7 | 7 | 159.242857 | 99.4196136 |
| p1_chB_timeSpent | WT | WT | NA | 26 | 12 | 14 | 149.796154 | 65.4787201 |
| p1_chamberCenter_timeSpent | Neo+ Arc KO1 | Neo+ Arc KO1 | NA | 12 | 6 | 6 | 109.416667 | 23.3272308 |
| p1_chamberCenter_timeSpent | Neo- Arc KO1 | Neo- Arc KO1 | NA | 14 | 7 | 7 | 283.714286 | 87.4636879 |
| p1_chamberCenter_timeSpent | WT | WT | NA | 26 | 12 | 14 | 256.25 | 120.292887 |
| p1_cyA_numberVisits | Neo+ Arc KO1 | Neo+ Arc KO1 | NA | 12 | 6 | 6 | 40.0833333 | 20.5976977 |
| p1_cyA_numberVisits | Neo- Arc KO1 | Neo- Arc KO1 | NA | 14 | 7 | 7 | 47.7142857 | 31.6651436 |
| p1_cyA_numberVisits | WT | WT | NA | 26 | 12 | 14 | 52.8461538 | 24.5319258 |
| p1_cyA_timeSpent | Neo+ Arc KO1 | Neo+ Arc KO1 | NA | 12 | 6 | 6 | 93.8333333 | 40.8314347 |
| p1_cyA_timeSpent | Neo- Arc KO1 | Neo- Arc KO1 | NA | 14 | 7 | 7 | 82.7428571 | 49.8476867 |
| p1_cyA_timeSpent | WT | WT | NA | 26 | 12 | 14 | 83.3730769 | 42.4466965 |
| p1_cyB_numberVisits | Neo+ Arc KO1 | Neo+ Arc KO1 | NA | 12 | 6 | 6 | 30.25 | 16.1421245 |
| p1_cyB_numberVisits | Neo- Arc KO1 | Neo- Arc KO1 | NA | 14 | 7 | 7 | 52.3571429 | 30.8560109 |
| p1_cyB_numberVisits | WT | WT | NA | 26 | 12 | 14 | 46.7692308 | 27.4157731 |
| p1_cyB_timeSpent | Neo+ Arc KO1 | Neo+ Arc KO1 | NA | 12 | 6 | 6 | 61.7666667 | 30.675259 |
| p1_cyB_timeSpent | Neo- Arc KO1 | Neo- Arc KO1 | NA | 14 | 7 | 7 | 86.0357143 | 62.739763 |
| p1_cyB_timeSpent | WT | WT | NA | 26 | 12 | 14 | 85.9307692 | 63.968941 |
| p1_immobility_timeSpent | Neo+ Arc KO1 | Neo+ Arc KO1 | NA | 12 | 6 | 6 | 1.88333333 | 4.59185709 |
| p1_immobility_timeSpent | Neo- Arc KO1 | Neo- Arc KO1 | NA | 14 | 7 | 7 | 41.9142857 | 35.1276323 |
| p1_immobility_timeSpent | WT | WT | NA | 26 | 12 | 14 | 13.8192308 | 16.576152 |
| p1_preference | Neo+ Arc KO1 | Neo+ Arc KO1 | NA | 12 | 6 | 6 | 0.41556325 | 0.18678553 |
| p1_preference | Neo- Arc KO1 | Neo- Arc KO1 | NA | 14 | 7 | 7 | 0.49741508 | 0.26749472 |
| p1_preference | WT | WT | NA | 26 | 12 | 14 | 0.48814505 | 0.20813313 |
| p2_chM_numberEntry | Neo+ Arc KO1 | Neo+ Arc KO1 | NA | 12 | 6 | 6 | 8.25 | 2.13733054 |
| p2_chM_numberEntry | Neo- Arc KO1 | Neo- Arc KO1 | NA | 14 | 7 | 7 | 4.64285714 | 1.78054196 |
| p2_chM_numberEntry | WT | WT | NA | 26 | 12 | 14 | 10.2307692 | 4.2923904 |
| p2_chM_timeSpent | Neo+ Arc KO1 | Neo+ Arc KO1 | NA | 12 | 6 | 6 | 307.426917 | 81.7753029 |
| p2_chM_timeSpent | Neo- Arc KO1 | Neo- Arc KO1 | NA | 14 | 7 | 7 | 377.828786 | 115.577672 |
| p2_chM_timeSpent | WT | WT | NA | 26 | 12 | 14 | 317.603077 | 95.0501966 |
| p2_chT_numberEntry | Neo+ Arc KO1 | Neo+ Arc KO1 | NA | 12 | 6 | 6 | 7.58333333 | 2.3915888 |
| p2_chT_numberEntry | Neo- Arc KO1 | Neo- Arc KO1 | NA | 14 | 7 | 7 | 3.85714286 | 1.40642169 |
| p2_chT_numberEntry | WT | WT | NA | 26 | 12 | 14 | 8.76923077 | 4.35713385 |
| p2_chT_timeSpent | Neo+ Arc KO1 | Neo+ Arc KO1 | NA | 12 | 6 | 6 | 197.23425 | 63.7334182 |
| p2_chT_timeSpent | Neo- Arc KO1 | Neo- Arc KO1 | NA | 14 | 7 | 7 | 177.820857 | 108.945452 |
| p2_chT_timeSpent | WT | WT | NA | 26 | 12 | 14 | 185.986385 | 91.2791965 |
| p2_chamberCenter_timeSpent | Neo+ Arc KO1 | Neo+ Arc KO1 | NA | 12 | 6 | 6 | 95.3388333 | 33.881078 |
| p2_chamberCenter_timeSpent | Neo- Arc KO1 | Neo- Arc KO1 | NA | 14 | 7 | 7 | 44.3503571 | 39.1838792 |
| p2_chamberCenter_timeSpent | WT | WT | NA | 26 | 12 | 14 | 96.4105385 | 33.6495444 |
| p2_cyM_numberVisits | Neo+ Arc KO1 | Neo+ Arc KO1 | NA | 12 | 6 | 6 | 30.5833333 | 13.2490709 |
| p2_cyM_numberVisits | Neo- Arc KO1 | Neo- Arc KO1 | NA | 14 | 7 | 7 | 31.2857143 | 13.1992341 |
| p2_cyM_numberVisits | WT | WT | NA | 26 | 12 | 14 | 48.1538462 | 25.0354825 |
| p2_cyM_timeSpent | Neo+ Arc KO1 | Neo+ Arc KO1 | NA | 12 | 6 | 6 | 52.7860833 | 23.280035 |
| p2_cyM_timeSpent | Neo- Arc KO1 | Neo- Arc KO1 | NA | 14 | 7 | 7 | 66.2522857 | 38.1345816 |
| p2_cyM_timeSpent | WT | WT | NA | 26 | 12 | 14 | 81.5787308 | 37.9786413 |
| p2_cyT_numberVisits | Neo+ Arc KO1 | Neo+ Arc KO1 | NA | 12 | 6 | 6 | 17.0833333 | 6.24439142 |
| p2_cyT_numberVisits | Neo- Arc KO1 | Neo- Arc KO1 | NA | 14 | 7 | 7 | 11.8571429 | 5.06658955 |
| p2_cyT_numberVisits | WT | WT | NA | 26 | 12 | 14 | 16.2307692 | 6.94151391 |
| p2_cyT_timeSpent | Neo+ Arc KO1 | Neo+ Arc KO1 | NA | 12 | 6 | 6 | 24.2569167 | 10.9995936 |
| p2_cyT_timeSpent | Neo- Arc KO1 | Neo- Arc KO1 | NA | 14 | 7 | 7 | 14.3639286 | 5.65092276 |
| p2_cyT_timeSpent | WT | WT | NA | 26 | 12 | 14 | 18.7379231 | 9.63939461 |
| p2_immobility_timeSpent | Neo+ Arc KO1 | Neo+ Arc KO1 | NA | 12 | 6 | 6 | 5.58541667 | 5.7685244 |
| p2_immobility_timeSpent | Neo- Arc KO1 | Neo- Arc KO1 | NA | 14 | 7 | 7 | 28.4383571 | 36.3144801 |
| p2_immobility_timeSpent | WT | WT | NA | 26 | 12 | 14 | 8.87523077 | 17.4551709 |
| p2_preference | Neo+ Arc KO1 | Neo+ Arc KO1 | NA | 12 | 6 | 6 | 0.67678233 | 0.09093937 |
| p2_preference | Neo- Arc KO1 | Neo- Arc KO1 | NA | 14 | 7 | 7 | 0.80197228 | 0.0935761 |
| p2_preference | WT | WT | NA | 26 | 12 | 14 | 0.78627622 | 0.14090102 |
| p2_selfGrooming_number | Neo+ Arc KO1 | Neo+ Arc KO1 | NA | 12 | 6 | 6 | 2 | 0.95346259 |
| p2_selfGrooming_number | Neo- Arc KO1 | Neo- Arc KO1 | NA | 14 | 7 | 7 | 1.92857143 | 1.49173547 |
| p2_selfGrooming_number | WT | WT | NA | 26 | 12 | 14 | 1.38461538 | 0.94135745 |
| p2_selfGrooming_timeSpent | Neo+ Arc KO1 | Neo+ Arc KO1 | NA | 12 | 6 | 6 | 50.9426667 | 33.3634133 |
| p2_selfGrooming_timeSpent | Neo- Arc KO1 | Neo- Arc KO1 | NA | 14 | 7 | 7 | 39.787 | 41.8913099 |
| p2_selfGrooming_timeSpent | WT | WT | NA | 26 | 12 | 14 | 31.9846154 | 38.5960713 |
| p3_chM_numberEntry | Neo+ Arc KO1 | Neo+ Arc KO1 | NA | 12 | 6 | 6 | 7.5 | 2.39317211 |
| p3_chM_numberEntry | WT | WT | NA | 8 | 4 | 4 | 13.75 | 2.12132034 |
| p3_chM_timeSpent | Neo+ Arc KO1 | Neo+ Arc KO1 | NA | 12 | 6 | 6 | 234.263083 | 100.154595 |
| p3_chM_timeSpent | WT | WT | NA | 8 | 4 | 4 | 241.914 | 71.6900533 |
| p3_chN_numberEntry | Neo+ Arc KO1 | Neo+ Arc KO1 | NA | 12 | 6 | 6 | 8.41666667 | 2.96826651 |
| p3_chN_numberEntry | WT | WT | NA | 8 | 4 | 4 | 13.375 | 2.61520281 |
| p3_chN_timeSpent | Neo+ Arc KO1 | Neo+ Arc KO1 | NA | 12 | 6 | 6 | 255.547167 | 103.829752 |
| p3_chN_timeSpent | WT | WT | NA | 8 | 4 | 4 | 241.086 | 61.8533486 |
| p3_chamberCenter_timeSpent | Neo+ Arc KO1 | Neo+ Arc KO1 | NA | 12 | 6 | 6 | 110.18975 | 43.2169856 |
| p3_chamberCenter_timeSpent | WT | WT | NA | 8 | 4 | 4 | 117 | 53.5606509 |
| p3_cyM_duration | Neo+ Arc KO1 | Neo+ Arc KO1 | NA | 12 | 6 | 6 | 1.238 | 0.37499139 |
| p3_cyM_duration | WT | WT | NA | 8 | 4 | 4 | 1.32975 | 0.37258585 |
| p3_cyM_numberVisits | Neo+ Arc KO1 | Neo+ Arc KO1 | NA | 12 | 6 | 6 | 17.8333333 | 9.20309566 |
| p3_cyM_numberVisits | WT | WT | NA | 8 | 4 | 4 | 26.375 | 15.8288841 |
| p3_cyM_timeSpent | Neo+ Arc KO1 | Neo+ Arc KO1 | NA | 12 | 6 | 6 | 22.4191667 | 13.9191487 |
| p3_cyM_timeSpent | WT | WT | NA | 8 | 4 | 4 | 37.8675 | 30.4516552 |
| p3_cyN_duration | Neo+ Arc KO1 | Neo+ Arc KO1 | NA | 12 | 6 | 6 | 2.1405 | 0.63840917 |
| p3_cyN_duration | WT | WT | NA | 8 | 4 | 4 | 1.86975 | 0.41958066 |
| p3_cyN_numberVisits | Neo+ Arc KO1 | Neo+ Arc KO1 | NA | 12 | 6 | 6 | 26.25 | 12.2409521 |
| p3_cyN_numberVisits | WT | WT | NA | 8 | 4 | 4 | 31.5 | 14.1824842 |
| p3_cyN_timeSpent | Neo+ Arc KO1 | Neo+ Arc KO1 | NA | 12 | 6 | 6 | 55.25125 | 26.0374167 |
| p3_cyN_timeSpent | WT | WT | NA | 8 | 4 | 4 | 56.122125 | 21.9152501 |
| p3_immobility_timeSpent | Neo+ Arc KO1 | Neo+ Arc KO1 | NA | 12 | 6 | 6 | 2.12441667 | 4.12314173 |
| p3_immobility_timeSpent | WT | WT | NA | 8 | 4 | 4 | 0.186375 | 0.52714811 |
| p3_preference | Neo+ Arc KO1 | Neo+ Arc KO1 | NA | 12 | 6 | 6 | 0.70674137 | 0.12762723 |
| p3_preference | WT | WT | NA | 8 | 4 | 4 | 0.61920076 | 0.16714509 |
| p3_selfGrooming_duration | Neo+ Arc KO1 | Neo+ Arc KO1 | NA | 12 | 6 | 6 | 39.8575833 | 39.3479704 |
| p3_selfGrooming_duration | WT | WT | NA | 8 | 4 | 4 | 20.3435 | 14.1046986 |
| p3_selfGrooming_number | Neo+ Arc KO1 | Neo+ Arc KO1 | NA | 12 | 6 | 6 | 2.33333333 | 1.23091491 |
| p3_selfGrooming_number | WT | WT | NA | 8 | 4 | 4 | 1.5 | 0.9258201 |
| p3_selfGrooming_timeSpent | Neo+ Arc KO1 | Neo+ Arc KO1 | NA | 12 | 6 | 6 | 88.7108333 | 81.5491727 |
| p3_selfGrooming_timeSpent | WT | WT | NA | 8 | 4 | 4 | 35.561125 | 28.1194503 |
| p4_chMa_numerEntry | Neo+ Arc KO1 | Neo+ Arc KO1 | NA | 12 | 6 | 6 | 8.58333333 | 2.50302847 |
| p4_chMa_numerEntry | WT | WT | NA | 8 | 4 | 4 | 12.75 | 2.54950976 |
| p4_chMa_timeSpent | Neo+ Arc KO1 | Neo+ Arc KO1 | NA | 12 | 6 | 6 | 189.6565 | 77.2508742 |
| p4_chMa_timeSpent | WT | WT | NA | 8 | 4 | 4 | 272.138 | 97.0659564 |
| p4_chN_numberEntry | Neo+ Arc KO1 | Neo+ Arc KO1 | NA | 12 | 6 | 6 | 8.33333333 | 2.80691786 |
| p4_chN_numberEntry | WT | WT | NA | 8 | 4 | 4 | 11.625 | 3.5831949 |
| p4_chN_timeSpent | Neo+ Arc KO1 | Neo+ Arc KO1 | NA | 12 | 6 | 6 | 275.9305 | 99.9902224 |
| p4_chN_timeSpent | WT | WT | NA | 8 | 4 | 4 | 219.26 | 86.8499808 |
| p4_chamberCenter_timeSpent | Neo+ Arc KO1 | Neo+ Arc KO1 | NA | 12 | 6 | 6 | 134.413 | 80.3076197 |
| p4_chamberCenter_timeSpent | WT | WT | NA | 8 | 4 | 4 | 108.602 | 23.882021 |
| p4_cyMa_duration | Neo+ Arc KO1 | Neo+ Arc KO1 | NA | 12 | 6 | 6 | 1.41033333 | 0.30219059 |
| p4_cyMa_duration | WT | WT | NA | 8 | 4 | 4 | 1.372125 | 0.40866488 |
| p4_cyMa_numberVisits | Neo+ Arc KO1 | Neo+ Arc KO1 | NA | 12 | 6 | 6 | 22 | 9.31274786 |
| p4_cyMa_numberVisits | WT | WT | NA | 8 | 4 | 4 | 26.625 | 7.46300399 |
| p4_cyMa_timeSpent | Neo+ Arc KO1 | Neo+ Arc KO1 | NA | 12 | 6 | 6 | 31.7540833 | 15.8247388 |
| p4_cyMa_timeSpent | WT | WT | NA | 8 | 4 | 4 | 38.1835 | 23.6502661 |
| p4_cyN_duration | Neo+ Arc KO1 | Neo+ Arc KO1 | NA | 12 | 6 | 6 | 1.33133333 | 0.53610402 |
| p4_cyN_duration | WT | WT | NA | 8 | 4 | 4 | 1.156625 | 0.3359111 |
| p4_cyN_numberVisits | Neo+ Arc KO1 | Neo+ Arc KO1 | NA | 12 | 6 | 6 | 17.1666667 | 5.55686853 |
| p4_cyN_numberVisits | WT | WT | NA | 8 | 4 | 4 | 22.5 | 10.743769 |
| p4_cyN_timeSpent | Neo+ Arc KO1 | Neo+ Arc KO1 | NA | 12 | 6 | 6 | 22.9630833 | 11.7866741 |
| p4_cyN_timeSpent | WT | WT | NA | 8 | 4 | 4 | 27.008 | 18.1754493 |
| p4_immobility_timeSpent | Neo+ Arc KO1 | Neo+ Arc KO1 | NA | 12 | 6 | 6 | 6.2925 | 9.52547995 |
| p4_immobility_timeSpent | WT | WT | NA | 8 | 4 | 4 | 0.314 | 0.88812612 |
| p4_preference | Neo+ Arc KO1 | Neo+ Arc KO1 | NA | 12 | 6 | 6 | 0.57228236 | 0.1351576 |
| p4_preference | WT | WT | NA | 8 | 4 | 4 | 0.58973517 | 0.16658549 |
| p4_selfGrooming_duration | Neo+ Arc KO1 | Neo+ Arc KO1 | NA | 12 | 6 | 6 | 51.4800833 | 41.4492212 |
| p4_selfGrooming_duration | WT | WT | NA | 8 | 4 | 4 | 21.969375 | 17.7338007 |
| p4_selfGrooming_number | Neo+ Arc KO1 | Neo+ Arc KO1 | NA | 12 | 6 | 6 | 2.25 | 0.75377836 |
| p4_selfGrooming_number | WT | WT | NA | 8 | 4 | 4 | 2 | 0.9258201 |
| p4_selfGrooming_timeSpent | Neo+ Arc KO1 | Neo+ Arc KO1 | NA | 12 | 6 | 6 | 112.908417 | 80.8269002 |
| p4_selfGrooming_timeSpent | WT | WT | NA | 8 | 4 | 4 | 48.937875 | 51.6385518 |
| residentIntruder_day1_attack_duration | Neo+ Arc KO1 | Neo+ Arc KO1 | NA | 7 | 0 | 7 | 3.07728571 | 3.63722819 |
| residentIntruder_day1_attack_duration | WT | WT | NA | 8 | 0 | 8 | 3.977875 | 6.35061033 |
| residentIntruder_day1_attack_latency | Neo+ Arc KO1 | Neo+ Arc KO1 | NA | 7 | 0 | 7 | 224.629857 | 72.8783265 |
| residentIntruder_day1_attack_latency | WT | WT | NA | 8 | 0 | 8 | 258.434 | 70.8521517 |
| residentIntruder_day1_attack_number | Neo+ Arc KO1 | Neo+ Arc KO1 | NA | 7 | 0 | 7 | 3.42857143 | 3.82348632 |
| residentIntruder_day1_attack_number | WT | WT | NA | 8 | 0 | 8 | 1.875 | 2.58774585 |
| residentIntruder_day1_attack_timeSpent | Neo+ Arc KO1 | Neo+ Arc KO1 | NA | 7 | 0 | 7 | 20.8702857 | 26.8849629 |
| residentIntruder_day1_attack_timeSpent | WT | WT | NA | 8 | 0 | 8 | 9.350625 | 11.9786161 |
| residentIntruder_day2_attack_duration | Neo+ Arc KO1 | Neo+ Arc KO1 | NA | 7 | 0 | 7 | 0.74757143 | 1.10771234 |
| residentIntruder_day2_attack_duration | WT | WT | NA | 8 | 0 | 8 | 3.462875 | 2.22038545 |
| residentIntruder_day2_attack_latency | Neo+ Arc KO1 | Neo+ Arc KO1 | NA | 7 | 0 | 7 | 231.725 | 104.978982 |
| residentIntruder_day2_attack_latency | WT | WT | NA | 8 | 0 | 8 | 155.639375 | 104.882015 |
| residentIntruder_day2_attack_number | Neo+ Arc KO1 | Neo+ Arc KO1 | NA | 7 | 0 | 7 | 3.28571429 | 6.23736819 |
| residentIntruder_day2_attack_number | WT | WT | NA | 8 | 0 | 8 | 6.5 | 5.23722937 |
| residentIntruder_day2_attack_timeSpent | Neo+ Arc KO1 | Neo+ Arc KO1 | NA | 7 | 0 | 7 | 8.13885714 | 18.5287161 |
| residentIntruder_day2_attack_timeSpent | WT | WT | NA | 8 | 0 | 8 | 30.31375 | 25.9908027 |
| socialMemory_trial1_circling_number | Neo+ Arc KO1 | Neo+ Arc KO1 | NA | 8 | 4 | 4 | 0 | 0 |
| socialMemory_trial1_circling_number | WT | WT | NA | 9 | 4 | 5 | 0.11111111 | 0.33333333 |
| socialMemory_trial1_following_duration | Neo+ Arc KO1 | Neo+ Arc KO1 | NA | 8 | 4 | 4 | 1.0115 | 0.74833873 |
| socialMemory_trial1_following_duration | WT | WT | NA | 9 | 4 | 5 | 1.08733333 | 0.60724727 |
| socialMemory_trial1_following_number | Neo+ Arc KO1 | Neo+ Arc KO1 | NA | 8 | 4 | 4 | 3.5 | 3.07059789 |
| socialMemory_trial1_following_number | WT | WT | NA | 9 | 4 | 5 | 2.22222222 | 2.10818511 |
| socialMemory_trial1_following_timeSpent | Neo+ Arc KO1 | Neo+ Arc KO1 | NA | 8 | 4 | 4 | 4.47675 | 3.96677418 |
| socialMemory_trial1_following_timeSpent | WT | WT | NA | 9 | 4 | 5 | 3.38066667 | 4.87652968 |
| socialMemory_trial1_noseContact_duration | Neo+ Arc KO1 | Neo+ Arc KO1 | NA | 8 | 4 | 4 | 1.065 | 0.29517646 |
| socialMemory_trial1_noseContact_duration | WT | WT | NA | 9 | 4 | 5 | 1.56666667 | 0.70000714 |
| socialMemory_trial1_noseContact_number | Neo+ Arc KO1 | Neo+ Arc KO1 | NA | 8 | 4 | 4 | 20.75 | 9.77971662 |
| socialMemory_trial1_noseContact_number | WT | WT | NA | 9 | 4 | 5 | 28.2222222 | 11.0428458 |
| socialMemory_trial1_noseContact_timeSpent | Neo+ Arc KO1 | Neo+ Arc KO1 | NA | 8 | 4 | 4 | 22.4925 | 12.4864461 |
| socialMemory_trial1_noseContact_timeSpent | WT | WT | NA | 9 | 4 | 5 | 48.9362222 | 36.1582821 |
| socialMemory_trial1_pawContact_duration | Neo+ Arc KO1 | Neo+ Arc KO1 | NA | 8 | 4 | 4 | 0.7665 | 0.6068217 |
| socialMemory_trial1_pawContact_duration | WT | WT | NA | 9 | 4 | 5 | 0.79344444 | 0.75421004 |
| socialMemory_trial1_pawContact_number | Neo+ Arc KO1 | Neo+ Arc KO1 | NA | 8 | 4 | 4 | 4.25 | 5.8002463 |
| socialMemory_trial1_pawContact_number | WT | WT | NA | 9 | 4 | 5 | 3.55555556 | 3.43187671 |
| socialMemory_trial1_pawContact_timeSpent | Neo+ Arc KO1 | Neo+ Arc KO1 | NA | 8 | 4 | 4 | 5.00625 | 8.01193144 |
| socialMemory_trial1_pawContact_timeSpent | WT | WT | NA | 9 | 4 | 5 | 4.13022222 | 5.2540062 |
| socialMemory_trial1_rearing_number | Neo+ Arc KO1 | Neo+ Arc KO1 | NA | 8 | 4 | 4 | 24 | 7.92824967 |
| socialMemory_trial1_rearing_number | WT | WT | NA | 9 | 4 | 5 | 16.7777778 | 11.47582 |
| socialMemory_trial1_selfGrooming_afterSI_duration | Neo+ Arc KO1 | Neo+ Arc KO1 | NA | 8 | 4 | 4 | 2.576 | 4.10117098 |
| socialMemory_trial1_selfGrooming_afterSI_duration | WT | WT | NA | 9 | 4 | 5 | 3.80755556 | 6.13259974 |
| socialMemory_trial1_selfGrooming_afterSI_number | Neo+ Arc KO1 | Neo+ Arc KO1 | NA | 8 | 4 | 4 | 0.375 | 0.51754917 |
| socialMemory_trial1_selfGrooming_afterSI_number | WT | WT | NA | 9 | 4 | 5 | 0.66666667 | 1 |
| socialMemory_trial1_selfGrooming_afterSI_timeSpent | Neo+ Arc KO1 | Neo+ Arc KO1 | NA | 8 | 4 | 4 | 2.576 | 4.10117098 |
| socialMemory_trial1_selfGrooming_afterSI_timeSpent | WT | WT | NA | 9 | 4 | 5 | 5.181 | 7.86875929 |
| socialMemory_trial1_selfGrooming_duration | Neo+ Arc KO1 | Neo+ Arc KO1 | NA | 8 | 4 | 4 | 6.91925 | 8.27267833 |
| socialMemory_trial1_selfGrooming_duration | WT | WT | NA | 9 | 4 | 5 | 3.16566667 | 5.46774188 |
| socialMemory_trial1_selfGrooming_number | Neo+ Arc KO1 | Neo+ Arc KO1 | NA | 8 | 4 | 4 | 0.5 | 0.53452248 |
| socialMemory_trial1_selfGrooming_number | WT | WT | NA | 9 | 4 | 5 | 0.44444444 | 0.72648316 |
| socialMemory_trial1_selfGrooming_timeSpent | Neo+ Arc KO1 | Neo+ Arc KO1 | NA | 8 | 4 | 4 | 6.91925 | 8.27267833 |
| socialMemory_trial1_selfGrooming_timeSpent | WT | WT | NA | 9 | 4 | 5 | 3.691 | 5.85025412 |
| socialMemory_trial2_attack_numer | Neo+ Arc KO1 | Neo+ Arc KO1 | NA | 8 | 4 | 4 | 2.5 | 4.53557368 |
| socialMemory_trial2_attack_numer | WT | WT | NA | 9 | 4 | 5 | 1.22222222 | 2.99072641 |
| socialMemory_trial2_following_duration | Neo+ Arc KO1 | Neo+ Arc KO1 | NA | 8 | 4 | 4 | 0.684375 | 0.51698216 |
| socialMemory_trial2_following_duration | WT | WT | NA | 9 | 4 | 5 | 0.80488889 | 1.8560065 |
| socialMemory_trial2_following_number | Neo+ Arc KO1 | Neo+ Arc KO1 | NA | 8 | 4 | 4 | 1.25 | 1.03509834 |
| socialMemory_trial2_following_number | WT | WT | NA | 9 | 4 | 5 | 0.66666667 | 1.11803399 |
| socialMemory_trial2_following_timeSpent | Neo+ Arc KO1 | Neo+ Arc KO1 | NA | 8 | 4 | 4 | 1.192625 | 1.28456929 |
| socialMemory_trial2_following_timeSpent | WT | WT | NA | 9 | 4 | 5 | 2.15 | 5.59872264 |
| socialMemory_trial2_noseContact_duration | Neo+ Arc KO1 | Neo+ Arc KO1 | NA | 8 | 4 | 4 | 1.093375 | 0.359734 |
| socialMemory_trial2_noseContact_duration | WT | WT | NA | 9 | 4 | 5 | 1.21644444 | 0.44289477 |
| socialMemory_trial2_noseContact_number | Neo+ Arc KO1 | Neo+ Arc KO1 | NA | 8 | 4 | 4 | 15.75 | 6.94365075 |
| socialMemory_trial2_noseContact_number | WT | WT | NA | 9 | 4 | 5 | 16 | 8.54400375 |
| socialMemory_trial2_noseContact_timeSpent | Neo+ Arc KO1 | Neo+ Arc KO1 | NA | 8 | 4 | 4 | 16.427125 | 8.29219456 |
| socialMemory_trial2_noseContact_timeSpent | WT | WT | NA | 9 | 4 | 5 | 22.3807778 | 15.0799589 |
| socialMemory_trial2_pawContact_duration | Neo+ Arc KO1 | Neo+ Arc KO1 | NA | 8 | 4 | 4 | 1.7365 | 0.86241173 |
| socialMemory_trial2_pawContact_duration | WT | WT | NA | 9 | 4 | 5 | 1.63444444 | 1.79712166 |
| socialMemory_trial2_pawContact_number | Neo+ Arc KO1 | Neo+ Arc KO1 | NA | 8 | 4 | 4 | 5.75 | 5.94618725 |
| socialMemory_trial2_pawContact_number | WT | WT | NA | 9 | 4 | 5 | 4 | 4.41588043 |
| socialMemory_trial2_pawContact_timeSpent | Neo+ Arc KO1 | Neo+ Arc KO1 | NA | 8 | 4 | 4 | 11.455375 | 14.0415795 |
| socialMemory_trial2_pawContact_timeSpent | WT | WT | NA | 9 | 4 | 5 | 9.13144444 | 11.3794784 |
| socialMemory_trial2_rearing_number | Neo+ Arc KO1 | Neo+ Arc KO1 | NA | 8 | 4 | 4 | 13.75 | 6.73477118 |
| socialMemory_trial2_rearing_number | WT | WT | NA | 9 | 4 | 5 | 13.5555556 | 9.01541889 |
| socialMemory_trial2_selfGrooming_afterSI_duration | Neo+ Arc KO1 | Neo+ Arc KO1 | NA | 8 | 4 | 4 | 0 | 0 |
| socialMemory_trial2_selfGrooming_afterSI_duration | WT | WT | NA | 9 | 4 | 5 | 4.02111111 | 6.62017582 |
| socialMemory_trial2_selfGrooming_afterSI_number | Neo+ Arc KO1 | Neo+ Arc KO1 | NA | 8 | 4 | 4 | 0 | 0 |
| socialMemory_trial2_selfGrooming_afterSI_number | WT | WT | NA | 9 | 4 | 5 | 0.55555556 | 1.01379376 |
| socialMemory_trial2_selfGrooming_afterSI_timeSpent | Neo+ Arc KO1 | Neo+ Arc KO1 | NA | 8 | 4 | 4 | 0 | 0 |
| socialMemory_trial2_selfGrooming_afterSI_timeSpent | WT | WT | NA | 9 | 4 | 5 | 5.68155556 | 9.02955871 |
| socialMemory_trial2_selfGrooming_duration | Neo+ Arc KO1 | Neo+ Arc KO1 | NA | 8 | 4 | 4 | 12.776125 | 22.5398887 |
| socialMemory_trial2_selfGrooming_duration | WT | WT | NA | 9 | 4 | 5 | 2.37011111 | 6.0764955 |
| socialMemory_trial2_selfGrooming_number | Neo+ Arc KO1 | Neo+ Arc KO1 | NA | 8 | 4 | 4 | 0.5 | 0.75592895 |
| socialMemory_trial2_selfGrooming_number | WT | WT | NA | 9 | 4 | 5 | 0.22222222 | 0.44095855 |
| socialMemory_trial2_selfGrooming_timeSpent | Neo+ Arc KO1 | Neo+ Arc KO1 | NA | 8 | 4 | 4 | 17.754875 | 31.8470513 |
| socialMemory_trial2_selfGrooming_timeSpent | WT | WT | NA | 9 | 4 | 5 | 2.37011111 | 6.0764955 |
| socialMemory_trial3_attack_numer | Neo+ Arc KO1 | Neo+ Arc KO1 | NA | 8 | 4 | 4 | 0.5 | 1.41421356 |
| socialMemory_trial3_attack_numer | WT | WT | NA | 9 | 4 | 5 | 0.44444444 | 1.01379376 |
| socialMemory_trial3_following_duration | Neo+ Arc KO1 | Neo+ Arc KO1 | NA | 8 | 4 | 4 | 0.498875 | 0.54904734 |
| socialMemory_trial3_following_duration | WT | WT | NA | 9 | 4 | 5 | 0.23655556 | 0.47936862 |
| socialMemory_trial3_following_number | Neo+ Arc KO1 | Neo+ Arc KO1 | NA | 8 | 4 | 4 | 1.5 | 1.41421356 |
| socialMemory_trial3_following_number | WT | WT | NA | 9 | 4 | 5 | 0.66666667 | 1.32287566 |
| socialMemory_trial3_following_timeSpent | Neo+ Arc KO1 | Neo+ Arc KO1 | NA | 8 | 4 | 4 | 1.26 | 1.43924117 |
| socialMemory_trial3_following_timeSpent | WT | WT | NA | 9 | 4 | 5 | 0.70988889 | 1.43853766 |
| socialMemory_trial3_noseContact_duration | Neo+ Arc KO1 | Neo+ Arc KO1 | NA | 8 | 4 | 4 | 0.980875 | 0.28247601 |
| socialMemory_trial3_noseContact_duration | WT | WT | NA | 9 | 4 | 5 | 0.98766667 | 0.52901323 |
| socialMemory_trial3_noseContact_number | Neo+ Arc KO1 | Neo+ Arc KO1 | NA | 8 | 4 | 4 | 17.5 | 6.92820323 |
| socialMemory_trial3_noseContact_number | WT | WT | NA | 9 | 4 | 5 | 13.8888889 | 7.57371185 |
| socialMemory_trial3_noseContact_timeSpent | Neo+ Arc KO1 | Neo+ Arc KO1 | NA | 8 | 4 | 4 | 18.079625 | 9.60734073 |
| socialMemory_trial3_noseContact_timeSpent | WT | WT | NA | 9 | 4 | 5 | 15.8122222 | 9.33984642 |
| socialMemory_trial3_pawContact_duration | Neo+ Arc KO1 | Neo+ Arc KO1 | NA | 8 | 4 | 4 | 0.497625 | 0.589595 |
| socialMemory_trial3_pawContact_duration | WT | WT | NA | 9 | 4 | 5 | 0.59055556 | 0.60584551 |
| socialMemory_trial3_pawContact_number | Neo+ Arc KO1 | Neo+ Arc KO1 | NA | 8 | 4 | 4 | 2.875 | 5.19443383 |
| socialMemory_trial3_pawContact_number | WT | WT | NA | 9 | 4 | 5 | 1.44444444 | 1.33333333 |
| socialMemory_trial3_pawContact_timeSpent | Neo+ Arc KO1 | Neo+ Arc KO1 | NA | 8 | 4 | 4 | 2.8235 | 4.8727113 |
| socialMemory_trial3_pawContact_timeSpent | WT | WT | NA | 9 | 4 | 5 | 1.24077778 | 1.87731882 |
| socialMemory_trial3_rearing_number | Neo+ Arc KO1 | Neo+ Arc KO1 | NA | 8 | 4 | 4 | 21.875 | 12.6878739 |
| socialMemory_trial3_rearing_number | WT | WT | NA | 9 | 4 | 5 | 16.5555556 | 8.70504324 |
| socialMemory_trial3_selfGrooming_afterSI_duration | Neo+ Arc KO1 | Neo+ Arc KO1 | NA | 8 | 4 | 4 | 0 | 0 |
| socialMemory_trial3_selfGrooming_afterSI_duration | WT | WT | NA | 9 | 4 | 5 | 0.26 | 0.78 |
| socialMemory_trial3_selfGrooming_afterSI_number | Neo+ Arc KO1 | Neo+ Arc KO1 | NA | 8 | 4 | 4 | 0 | 0 |
| socialMemory_trial3_selfGrooming_afterSI_number | WT | WT | NA | 9 | 4 | 5 | 0.11111111 | 0.33333333 |
| socialMemory_trial3_selfGrooming_afterSI_timeSpent | Neo+ Arc KO1 | Neo+ Arc KO1 | NA | 8 | 4 | 4 | 0 | 0 |
| socialMemory_trial3_selfGrooming_afterSI_timeSpent | WT | WT | NA | 9 | 4 | 5 | 0.26 | 0.78 |
| socialMemory_trial3_selfGrooming_duration | Neo+ Arc KO1 | Neo+ Arc KO1 | NA | 8 | 4 | 4 | 8.15725 | 16.6521849 |
| socialMemory_trial3_selfGrooming_duration | WT | WT | NA | 9 | 4 | 5 | 10.3206667 | 11.6255527 |
| socialMemory_trial3_selfGrooming_number | Neo+ Arc KO1 | Neo+ Arc KO1 | NA | 8 | 4 | 4 | 0.625 | 0.74402381 |
| socialMemory_trial3_selfGrooming_number | WT | WT | NA | 9 | 4 | 5 | 0.66666667 | 0.5 |
| socialMemory_trial3_selfGrooming_timeSpent | Neo+ Arc KO1 | Neo+ Arc KO1 | NA | 8 | 4 | 4 | 14.188125 | 33.4798203 |
| socialMemory_trial3_selfGrooming_timeSpent | WT | WT | NA | 9 | 4 | 5 | 10.3206667 | 11.6255527 |
| socialMemory_trial4_attack_numer | Neo+ Arc KO1 | Neo+ Arc KO1 | NA | 8 | 4 | 4 | 1.5 | 3.2071349 |
| socialMemory_trial4_attack_numer | WT | WT | NA | 9 | 4 | 5 | 2.55555556 | 5.41089436 |
| socialMemory_trial4_following_duration | Neo+ Arc KO1 | Neo+ Arc KO1 | NA | 8 | 4 | 4 | 0.4135 | 0.45686978 |
| socialMemory_trial4_following_duration | WT | WT | NA | 9 | 4 | 5 | 0.313 | 0.34506195 |
| socialMemory_trial4_following_number | Neo+ Arc KO1 | Neo+ Arc KO1 | NA | 8 | 4 | 4 | 1 | 1.41421356 |
| socialMemory_trial4_following_number | WT | WT | NA | 9 | 4 | 5 | 1.55555556 | 2.00693243 |
| socialMemory_trial4_following_timeSpent | Neo+ Arc KO1 | Neo+ Arc KO1 | NA | 8 | 4 | 4 | 0.776375 | 1.0448669 |
| socialMemory_trial4_following_timeSpent | WT | WT | NA | 9 | 4 | 5 | 0.92955556 | 1.23838939 |
| socialMemory_trial4_noseContact_duration | Neo+ Arc KO1 | Neo+ Arc KO1 | NA | 8 | 4 | 4 | 1.151125 | 0.28515331 |
| socialMemory_trial4_noseContact_duration | WT | WT | NA | 9 | 4 | 5 | 1.04988889 | 0.25943951 |
| socialMemory_trial4_noseContact_number | Neo+ Arc KO1 | Neo+ Arc KO1 | NA | 8 | 4 | 4 | 16.625 | 6.23211727 |
| socialMemory_trial4_noseContact_number | WT | WT | NA | 9 | 4 | 5 | 22.7777778 | 13.6269749 |
| socialMemory_trial4_noseContact_timeSpent | Neo+ Arc KO1 | Neo+ Arc KO1 | NA | 8 | 4 | 4 | 18.699875 | 6.66953308 |
| socialMemory_trial4_noseContact_timeSpent | WT | WT | NA | 9 | 4 | 5 | 25.9447778 | 18.4907492 |
| socialMemory_trial4_pawContact_duration | Neo+ Arc KO1 | Neo+ Arc KO1 | NA | 8 | 4 | 4 | 2.187875 | 2.74586913 |
| socialMemory_trial4_pawContact_duration | WT | WT | NA | 9 | 4 | 5 | 1.205 | 0.89621245 |
| socialMemory_trial4_pawContact_number | Neo+ Arc KO1 | Neo+ Arc KO1 | NA | 8 | 4 | 4 | 4.625 | 2.82526863 |
| socialMemory_trial4_pawContact_number | WT | WT | NA | 9 | 4 | 5 | 2.77777778 | 2.68224616 |
| socialMemory_trial4_pawContact_timeSpent | Neo+ Arc KO1 | Neo+ Arc KO1 | NA | 8 | 4 | 4 | 12.939125 | 17.8091822 |
| socialMemory_trial4_pawContact_timeSpent | WT | WT | NA | 9 | 4 | 5 | 4.56022222 | 4.87204715 |
| socialMemory_trial4_rearing_number | Neo+ Arc KO1 | Neo+ Arc KO1 | NA | 8 | 4 | 4 | 14.875 | 8.54295867 |
| socialMemory_trial4_rearing_number | WT | WT | NA | 9 | 4 | 5 | 10.6666667 | 5 |
| socialMemory_trial4_selfGrooming_afterSI_duration | Neo+ Arc KO1 | Neo+ Arc KO1 | NA | 8 | 4 | 4 | 2.047625 | 5.79155809 |
| socialMemory_trial4_selfGrooming_afterSI_duration | WT | WT | NA | 9 | 4 | 5 | 2.01077778 | 6.03233333 |
| socialMemory_trial4_selfGrooming_afterSI_number | Neo+ Arc KO1 | Neo+ Arc KO1 | NA | 8 | 4 | 4 | 0.25 | 0.70710678 |
| socialMemory_trial4_selfGrooming_afterSI_number | WT | WT | NA | 9 | 4 | 5 | 0.11111111 | 0.33333333 |
| socialMemory_trial4_selfGrooming_afterSI_timeSpent | Neo+ Arc KO1 | Neo+ Arc KO1 | NA | 8 | 4 | 4 | 4.095375 | 11.5834697 |
| socialMemory_trial4_selfGrooming_afterSI_timeSpent | WT | WT | NA | 9 | 4 | 5 | 2.01077778 | 6.03233333 |
| socialMemory_trial4_selfGrooming_duration | Neo+ Arc KO1 | Neo+ Arc KO1 | NA | 8 | 4 | 4 | 17.119125 | 11.479196 |
| socialMemory_trial4_selfGrooming_duration | WT | WT | NA | 9 | 4 | 5 | 3.48044444 | 5.49031634 |
| socialMemory_trial4_selfGrooming_number | Neo+ Arc KO1 | Neo+ Arc KO1 | NA | 8 | 4 | 4 | 1.25 | 0.88640526 |
| socialMemory_trial4_selfGrooming_number | WT | WT | NA | 9 | 4 | 5 | 0.33333333 | 0.5 |
| socialMemory_trial4_selfGrooming_timeSpent | Neo+ Arc KO1 | Neo+ Arc KO1 | NA | 8 | 4 | 4 | 24.70825 | 20.7319972 |
| socialMemory_trial4_selfGrooming_timeSpent | WT | WT | NA | 9 | 4 | 5 | 3.48044444 | 5.49031634 |
| socialOlfaction_O1T1_sniffing_duration | Neo+ Arc KO1 | Neo+ Arc KO1 | NA | 8 | 4 | 4 | 0.78775536 | 0.36695235 |
| socialOlfaction_O1T1_sniffing_duration | WT | WT | NA | 8 | 4 | 4 | 1.28433333 | 0.70646742 |
| socialOlfaction_O1T1_sniffing_number | Neo+ Arc KO1 | Neo+ Arc KO1 | NA | 8 | 4 | 4 | 4.875 | 2.64237447 |
| socialOlfaction_O1T1_sniffing_number | WT | WT | NA | 8 | 4 | 4 | 3.5 | 2.44948974 |
| socialOlfaction_O1T1_sniffing_timeSpent | Neo+ Arc KO1 | Neo+ Arc KO1 | NA | 8 | 4 | 4 | 3.848375 | 3.56343325 |
| socialOlfaction_O1T1_sniffing_timeSpent | WT | WT | NA | 8 | 4 | 4 | 4.222375 | 3.01845197 |
| socialOlfaction_O1T2_sniffing_duration | Neo+ Arc KO1 | Neo+ Arc KO1 | NA | 8 | 4 | 4 | 0.49533333 | 0.43363778 |
| socialOlfaction_O1T2_sniffing_duration | WT | WT | NA | 8 | 4 | 4 | 0.6558125 | 0.43898208 |
| socialOlfaction_O1T2_sniffing_number | Neo+ Arc KO1 | Neo+ Arc KO1 | NA | 8 | 4 | 4 | 2 | 2.13808994 |
| socialOlfaction_O1T2_sniffing_number | WT | WT | NA | 8 | 4 | 4 | 2.25 | 1.28173989 |
| socialOlfaction_O1T2_sniffing_timeSpent | Neo+ Arc KO1 | Neo+ Arc KO1 | NA | 8 | 4 | 4 | 1.248125 | 1.30432253 |
| socialOlfaction_O1T2_sniffing_timeSpent | WT | WT | NA | 8 | 4 | 4 | 1.790375 | 1.50732866 |
| socialOlfaction_O1T3_sniffing_duration | Neo+ Arc KO1 | Neo+ Arc KO1 | NA | 8 | 4 | 4 | 0.3824375 | 0.56878602 |
| socialOlfaction_O1T3_sniffing_duration | WT | WT | NA | 8 | 4 | 4 | 0.28127083 | 0.29678518 |
| socialOlfaction_O1T3_sniffing_number | Neo+ Arc KO1 | Neo+ Arc KO1 | NA | 8 | 4 | 4 | 1.125 | 1.12599163 |
| socialOlfaction_O1T3_sniffing_number | WT | WT | NA | 8 | 4 | 4 | 1.75 | 1.75254916 |
| socialOlfaction_O1T3_sniffing_timeSpent | Neo+ Arc KO1 | Neo+ Arc KO1 | NA | 8 | 4 | 4 | 0.9115 | 1.74356007 |
| socialOlfaction_O1T3_sniffing_timeSpent | WT | WT | NA | 8 | 4 | 4 | 0.87725 | 1.18715648 |
| socialOlfaction_O2T1_sniffing_duration | Neo+ Arc KO1 | Neo+ Arc KO1 | NA | 8 | 4 | 4 | 1.34229375 | 0.59509589 |
| socialOlfaction_O2T1_sniffing_duration | WT | WT | NA | 8 | 4 | 4 | 2.09599868 | 0.60232132 |
| socialOlfaction_O2T1_sniffing_number | Neo+ Arc KO1 | Neo+ Arc KO1 | NA | 8 | 4 | 4 | 4.75 | 2.71240536 |
| socialOlfaction_O2T1_sniffing_number | WT | WT | NA | 8 | 4 | 4 | 9 | 4.6291005 |
| socialOlfaction_O2T1_sniffing_timeSpent | Neo+ Arc KO1 | Neo+ Arc KO1 | NA | 8 | 4 | 4 | 5.3285 | 2.54845864 |
| socialOlfaction_O2T1_sniffing_timeSpent | WT | WT | NA | 8 | 4 | 4 | 19.683875 | 13.1992095 |
| socialOlfaction_O2T2_sniffing_duration | Neo+ Arc KO1 | Neo+ Arc KO1 | NA | 8 | 4 | 4 | 0.20158333 | 0.37405189 |
| socialOlfaction_O2T2_sniffing_duration | WT | WT | NA | 8 | 4 | 4 | 1.2946994 | 0.91665869 |
| socialOlfaction_O2T2_sniffing_number | Neo+ Arc KO1 | Neo+ Arc KO1 | NA | 8 | 4 | 4 | 0.75 | 1.16496475 |
| socialOlfaction_O2T2_sniffing_number | WT | WT | NA | 8 | 4 | 4 | 4.25 | 2.96407056 |
| socialOlfaction_O2T2_sniffing_timeSpent | Neo+ Arc KO1 | Neo+ Arc KO1 | NA | 8 | 4 | 4 | 0.50475 | 1.11968959 |
| socialOlfaction_O2T2_sniffing_timeSpent | WT | WT | NA | 8 | 4 | 4 | 4.860375 | 3.31456138 |
| socialOlfaction_O2T3_sniffing_duration | Neo+ Arc KO1 | Neo+ Arc KO1 | NA | 8 | 4 | 4 | 0.17983333 | 0.29154275 |
| socialOlfaction_O2T3_sniffing_duration | WT | WT | NA | 8 | 4 | 4 | 0.29814583 | 0.5779002 |
| socialOlfaction_O2T3_sniffing_number | Neo+ Arc KO1 | Neo+ Arc KO1 | NA | 8 | 4 | 4 | 0.625 | 1.06066017 |
| socialOlfaction_O2T3_sniffing_number | WT | WT | NA | 8 | 4 | 4 | 1.25 | 2.18762755 |
| socialOlfaction_O2T3_sniffing_timeSpent | Neo+ Arc KO1 | Neo+ Arc KO1 | NA | 8 | 4 | 4 | 0.27 | 0.43278896 |
| socialOlfaction_O2T3_sniffing_timeSpent | WT | WT | NA | 8 | 4 | 4 | 1.451875 | 3.48193528 |
| spatialObject_hab1_discriminationIndex | Neo+ Arc KO1 | Neo+ Arc KO1 | NA | 15 | 8 | 7 | -0.11766106 | 0.32406328 |
| spatialObject_hab1_discriminationIndex | Neo- Arc KO1 | Neo- Arc KO1 | NA | 16 | 8 | 8 | 0.23524871 | 0.30208091 |
| spatialObject_hab1_discriminationIndex | WT | WT | NA | 28 | 16 | 12 | 0.12228213 | 0.42789049 |
| spatialObject_hab1_interaction_oldObject_number | Neo+ Arc KO1 | Neo+ Arc KO1 | NA | 15 | 8 | 7 | 15.9333333 | 13.5249329 |
| spatialObject_hab1_interaction_oldObject_number | Neo- Arc KO1 | Neo- Arc KO1 | NA | 16 | 8 | 8 | 19.0625 | 10.4592463 |
| spatialObject_hab1_interaction_oldObject_number | WT | WT | NA | 28 | 16 | 12 | 21.3214286 | 26.0370869 |
| spatialObject_hab1_interaction_oldObject_timeSpent | Neo+ Arc KO1 | Neo+ Arc KO1 | NA | 15 | 8 | 7 | 20.3066667 | 18.0234952 |
| spatialObject_hab1_interaction_oldObject_timeSpent | Neo- Arc KO1 | Neo- Arc KO1 | NA | 16 | 8 | 8 | 17.075 | 12.1169028 |
| spatialObject_hab1_interaction_oldObject_timeSpent | WT | WT | NA | 28 | 16 | 12 | 21.8857143 | 31.2935293 |
| spatialObject_hab1_interaction_spatialObject_number | Neo+ Arc KO1 | Neo+ Arc KO1 | NA | 15 | 8 | 7 | 17.2666667 | 14.0023807 |
| spatialObject_hab1_interaction_spatialObject_number | Neo- Arc KO1 | Neo- Arc KO1 | NA | 16 | 8 | 8 | 26.375 | 14.4585615 |
| spatialObject_hab1_interaction_spatialObject_number | WT | WT | NA | 28 | 16 | 12 | 22.8928571 | 19.8407349 |
| spatialObject_hab1_interaction_spatialObject_timeSpent | Neo+ Arc KO1 | Neo+ Arc KO1 | NA | 15 | 8 | 7 | 18.98 | 18.4583392 |
| spatialObject_hab1_interaction_spatialObject_timeSpent | Neo- Arc KO1 | Neo- Arc KO1 | NA | 16 | 8 | 8 | 27.08125 | 15.4453539 |
| spatialObject_hab1_interaction_spatialObject_timeSpent | WT | WT | NA | 28 | 16 | 12 | 22.0321429 | 25.2448756 |
| spatialObject_hab2_discriminationIndex | Neo+ Arc KO1 | Neo+ Arc KO1 | NA | 15 | 8 | 7 | -0.00413148 | 0.48015381 |
| spatialObject_hab2_discriminationIndex | Neo- Arc KO1 | Neo- Arc KO1 | NA | 15 | 7 | 8 | 0.08595122 | 0.61364594 |
| spatialObject_hab2_discriminationIndex | WT | WT | NA | 27 | 15 | 12 | 0.13378325 | 0.4111905 |
| spatialObject_hab2_interaction_oldObject_number | Neo+ Arc KO1 | Neo+ Arc KO1 | NA | 15 | 8 | 7 | 12.8 | 17.2924426 |
| spatialObject_hab2_interaction_oldObject_number | Neo- Arc KO1 | Neo- Arc KO1 | NA | 16 | 8 | 8 | 6.4375 | 7.27524341 |
| spatialObject_hab2_interaction_oldObject_number | WT | WT | NA | 28 | 16 | 12 | 10.6785714 | 10.317384 |
| spatialObject_hab2_interaction_oldObject_timeSpent | Neo+ Arc KO1 | Neo+ Arc KO1 | NA | 15 | 8 | 7 | 18.7333333 | 25.3059752 |
| spatialObject_hab2_interaction_oldObject_timeSpent | Neo- Arc KO1 | Neo- Arc KO1 | NA | 16 | 8 | 8 | 9.2875 | 17.4650842 |
| spatialObject_hab2_interaction_oldObject_timeSpent | WT | WT | NA | 28 | 16 | 12 | 12.4892857 | 11.8594341 |
| spatialObject_hab2_interaction_spatialObject_number | Neo+ Arc KO1 | Neo+ Arc KO1 | NA | 15 | 8 | 7 | 10 | 12 |
| spatialObject_hab2_interaction_spatialObject_number | Neo- Arc KO1 | Neo- Arc KO1 | NA | 16 | 8 | 8 | 10.6875 | 18.5534139 |
| spatialObject_hab2_interaction_spatialObject_number | WT | WT | NA | 28 | 16 | 12 | 11.7857143 | 6.60166726 |
| spatialObject_hab2_interaction_spatialObject_timeSpent | Neo+ Arc KO1 | Neo+ Arc KO1 | NA | 15 | 8 | 7 | 15.8 | 20.8422648 |
| spatialObject_hab2_interaction_spatialObject_timeSpent | Neo- Arc KO1 | Neo- Arc KO1 | NA | 16 | 8 | 8 | 12.41875 | 21.1106361 |
| spatialObject_hab2_interaction_spatialObject_timeSpent | WT | WT | NA | 28 | 16 | 12 | 13.4642857 | 9.73322733 |
| spatialObject_test_discriminationIndex | Neo+ Arc KO1 | Neo+ Arc KO1 | NA | 15 | 8 | 7 | -0.31175169 | 0.46469935 |
| spatialObject_test_discriminationIndex | Neo- Arc KO1 | Neo- Arc KO1 | NA | 14 | 6 | 8 | 0.20661168 | 0.74883882 |
| spatialObject_test_discriminationIndex | WT | WT | NA | 26 | 14 | 12 | 0.11594323 | 0.44915643 |
| spatialObject_test_interaction_oldObject_number | Neo+ Arc KO1 | Neo+ Arc KO1 | NA | 15 | 8 | 7 | 10.6666667 | 23.3656239 |
| spatialObject_test_interaction_oldObject_number | Neo- Arc KO1 | Neo- Arc KO1 | NA | 14 | 6 | 8 | 2.64285714 | 3.27242895 |
| spatialObject_test_interaction_oldObject_number | WT | WT | NA | 26 | 14 | 12 | 9.03846154 | 6.90785506 |
| spatialObject_test_interaction_oldObject_timeSpent | Neo+ Arc KO1 | Neo+ Arc KO1 | NA | 15 | 8 | 7 | 21.8 | 55.484335 |
| spatialObject_test_interaction_oldObject_timeSpent | Neo- Arc KO1 | Neo- Arc KO1 | NA | 14 | 6 | 8 | 2.97857143 | 4.71595958 |
| spatialObject_test_interaction_oldObject_timeSpent | WT | WT | NA | 26 | 14 | 12 | 10.0653846 | 9.44821432 |
| spatialObject_test_interaction_spatialObject_number | Neo+ Arc KO1 | Neo+ Arc KO1 | NA | 15 | 8 | 7 | 4.2 | 4.26279586 |
| spatialObject_test_interaction_spatialObject_number | Neo- Arc KO1 | Neo- Arc KO1 | NA | 14 | 6 | 8 | 4.07142857 | 4.85900094 |
| spatialObject_test_interaction_spatialObject_number | WT | WT | NA | 26 | 14 | 12 | 10.2692308 | 7.13054103 |
| spatialObject_test_interaction_spatialObject_timeSpent | Neo+ Arc KO1 | Neo+ Arc KO1 | NA | 15 | 8 | 7 | 9.31333333 | 20.2218011 |
| spatialObject_test_interaction_spatialObject_timeSpent | Neo- Arc KO1 | Neo- Arc KO1 | NA | 14 | 6 | 8 | 9.64285714 | 21.0564339 |
| spatialObject_test_interaction_spatialObject_timeSpent | WT | WT | NA | 26 | 14 | 12 | 18.3538462 | 23.6641202 |
| stringTest_hindlimbGrasp_timeSpent | Neo+ Arc KO1 | Neo+ Arc KO1 | NA | 13 | 9 | 4 | 33.8461538 | 32.472671 |
| stringTest_hindlimbGrasp_timeSpent | WT | WT | NA | 20 | 10 | 10 | 21.8 | 28.5539194 |
| stringTest_latencyFall | Neo+ Arc KO1 | Neo+ Arc KO1 | NA | 13 | 9 | 4 | 61.6923077 | 42.8045648 |
| stringTest_latencyFall | WT | WT | NA | 20 | 10 | 10 | 27.55 | 33.3631928 |
| tubeTest_winsIntercage_day1_number | Neo+ Arc KO1 | Neo+ Arc KO1 | NA | 8 | 4 | 4 | 3 | 1.19522861 |
| tubeTest_winsIntercage_day1_number | WT | WT | NA | 8 | 4 | 4 | 1 | 0.75592895 |
| tubeTest_winsIntercage_day2_number | Neo+ Arc KO1 | Neo+ Arc KO1 | NA | 8 | 4 | 4 | 3.375 | 0.74402381 |
| tubeTest_winsIntercage_day2_number | WT | WT | NA | 8 | 4 | 4 | 0.625 | 0.91612538 |
| tubeTest_winsIntercage_day3_number | Neo+ Arc KO1 | Neo+ Arc KO1 | NA | 8 | 4 | 4 | 2.375 | 1.30247018 |
| tubeTest_winsIntercage_day3_number | WT | WT | NA | 8 | 4 | 4 | 1.5 | 1.19522861 |
| tubeTest_winsIntracage_day1_number | Neo+ Arc KO1 | Neo+ Arc KO1 | NA | 8 | 4 | 4 | 1.125 | 0.83452296 |
| tubeTest_winsIntracage_day1_number | WT | WT | NA | 8 | 4 | 4 | 1 | 1.06904497 |
| tubeTest_winsIntracage_day2_number | Neo+ Arc KO1 | Neo+ Arc KO1 | NA | 8 | 4 | 4 | 0.75 | 0.70710678 |
| tubeTest_winsIntracage_day2_number | WT | WT | NA | 8 | 4 | 4 | 0.875 | 0.83452296 |
| tubeTest_winsIntracage_day3_number | Neo+ Arc KO1 | Neo+ Arc KO1 | NA | 8 | 4 | 4 | 0.75 | 0.70710678 |
| tubeTest_winsIntracage_day3_number | WT | WT | NA | 8 | 4 | 4 | 1 | 1.06904497 |
| ymazeMemory_hiddenArm_numberVisits | Neo+ Arc KO1 | Neo+ Arc KO1 | NA | 13 | 6 | 7 | 5.30769231 | 2.86893172 |
| ymazeMemory_hiddenArm_numberVisits | WT | WT | NA | 16 | 8 | 8 | 9.3125 | 3.55375388 |
| ymazeMemory_hiddenArm_timeSpent | Neo+ Arc KO1 | Neo+ Arc KO1 | NA | 13 | 6 | 7 | 76.7153846 | 57.1758376 |
| ymazeMemory_hiddenArm_timeSpent | WT | WT | NA | 16 | 8 | 8 | 73.54375 | 39.4646165 |
| ymaze_alternation | Neo+ Arc KO1 | Neo+ Arc KO1 | NA | 12 | 6 | 6 | 22.8333333 | 6.88652615 |
| ymaze_alternation | WT | WT | NA | 8 | 4 | 4 | 32.25 | 6.27352715 |
| ymaze_percentage_aar | Neo+ Arc KO1 | Neo+ Arc KO1 | NA | 12 | 6 | 6 | 35.3829697 | 15.9508978 |
| ymaze_percentage_aar | WT | WT | NA | 8 | 4 | 4 | 33.9745077 | 9.45659493 |
| ymaze_percentage_sar | Neo+ Arc KO1 | Neo+ Arc KO1 | NA | 12 | 6 | 6 | 3.54689367 | 7.62242292 |
| ymaze_percentage_sar | WT | WT | NA | 8 | 4 | 4 | 1.79147879 | 2.54887561 |
| ymaze_percentage_spa | Neo+ Arc KO1 | Neo+ Arc KO1 | NA | 12 | 6 | 6 | 61.0701366 | 18.0371451 |
| ymaze_percentage_spa | WT | WT | NA | 8 | 4 | 4 | 64.2340135 | 8.96198741 |

| var_s1 | group1 | group2 | n1 | n2 | statistic | p | p.adj | p.adj.signif |
|---|---|---|---|---|---|---|---|---|
| duration_crouching | Neo+ Arc KO1 | WT 1 | 15 | 14 | 3.33172388 | 0.0008631 | 0.01294648 | * |
| duration_nesting | Neo+ Arc KO1 | WT 1 | 15 | 14 | 3.18581997 | 0.00144344 | 0.02165166 | * |
| latency_first | Neo+ Arc KO1 | WT 1 | 15 | 14 | -4.31658354 | 1.58E-05 | 0.00023769 | *** |
| latency_fourth | Neo+ Arc KO1 | WT 1 | 15 | 14 | -3.64378411 | 0.00026866 | 0.00376122 | ** |
| latency_full_maternal_behaviour | Neo+ Arc KO1 | WT 1 | 15 | 13 | -3.56872755 | 0.00035872 | 0.00538079 | ** |
| latency_second | Neo+ Arc KO1 | WT 1 | 15 | 14 | -4.48336242 | 7.35E-06 | 0.00011021 | *** |
| latency_third | Neo+ Arc KO1 | WT 1 | 15 | 14 | -4.38637226 | 1.15E-05 | 0.00017289 | *** |
| number_crouching | Neo+ Arc KO1 | WT 1 | 15 | 14 | 3.23821963 | 0.00120278 | 0.01804173 | * |
| time_crouching | Neo+ Arc KO1 | WT 1 | 15 | 14 | 3.35472984 | 0.00079443 | 0.01191638 | * |
| E+Maze_center_numberEntry | Neo+ Arc KO1 | WT | 13 | 20 | 3.80222617 | 0.0001434 | 0.0001434 | *** |
| E+Maze_closedArms_numberEntry | Neo+ Arc KO1 | WT | 13 | 20 | 3.47057508 | 0.00051935 | 0.00051935 | *** |
| ms_digging_duration | Neo+ Arc KO1 | WT | 10 | 31 | -4.53522442 | 5.75E-06 | 1.73E-05 | **** |
| ms_selfGrooming_duration | Neo+ Arc KO1 | WT | 10 | 31 | -3.68551478 | 0.00022824 | 0.00068472 | *** |
| ms_selfGrooming_timeSpent | Neo- Arc KO1 | WT | 25 | 31 | -3.56003997 | 0.0003708 | 0.0011124 | ** |
| ms_selfGrooming_timeSpent | Neo+ Arc KO1 | WT | 10 | 31 | -2.67723793 | 0.00742319 | 0.01484638 | * |
| nsf_food_intake_60min | Neo+ Arc KO1 | WT | 9 | 8 | 3.51619629 | 0.00043778 | 0.00043778 | *** |
| olfactoryAvoidance_predatorOdor_numberVisits | Neo+ Arc KO1 | WT | 13 | 20 | -3.04217095 | 0.00234878 | 0.00234878 | ** |
| p1_chamberCenter_timeSpent | Neo+ Arc KO1 | WT | 12 | 26 | 3.32102197 | 0.00089688 | 0.00179377 | ** |
| p1_immobility_timeSpent | Neo- Arc KO1 | WT | 14 | 26 | -2.9512693 | 0.00316471 | 0.00632942 | ** |
| p1_immobility_timeSpent | Neo+ Arc KO1 | WT | 12 | 26 | 2.50071647 | 0.01239424 | 0.01239424 | * |
| p2_chM_numberEntry | Neo- Arc KO1 | WT | 14 | 26 | 4.34913141 | 1.37E-05 | 4.10E-05 | **** |
| p2_chT_numberEntry | Neo- Arc KO1 | WT | 14 | 26 | 3.96852771 | 7.23E-05 | 0.00021695 | *** |
| p2_chamberCenter_timeSpent | Neo- Arc KO1 | WT | 14 | 26 | 3.61250867 | 0.00030325 | 0.00090975 | *** |
| p3_chM_numberEntry | Neo+ Arc KO1 | WT | 12 | 8 | 3.4191843 | 0.00062809 | 0.00062809 | *** |
| socialOlfaction_O2T1_sniffing_timeSpent | Neo+ Arc KO1 | WT | 8 | 8 | 2.94058818 | 0.0032759 | 0.0032759 | ** |
| socialOlfaction_O2T2_sniffing_timeSpent | Neo+ Arc KO1 | WT | 8 | 8 | 3.04041443 | 0.00236253 | 0.00236253 | ** |
| spatialObject_test_interaction_oldObject_number | Neo- Arc KO1 | WT | 14 | 26 | 3.42346782 | 0.00061828 | 0.00185483 | ** |
| spatialObject_test_interaction_spatialObject_number | Neo- Arc KO1 | WT | 14 | 26 | 2.88168591 | 0.00395554 | 0.01059403 | * |
| spatialObject_test_interaction_spatialObject_number | Neo+ Arc KO1 | WT | 15 | 26 | 2.9172485 | 0.00353134 | 0.01059403 | * |
| tubeTest_winsIntercage_day2_number | Neo+ Arc KO1 | WT | 8 | 8 | -3.35061718 | 0.00080632 | 0.00080632 | *** |
| ymazeMemory_hiddenArm_numberVisits | Neo+ Arc KO1 | WT | 13 | 16 | 3.12939721 | 0.00175165 | 0.00175165 | ** |

**Table S2. All raw data, mean, and statistics from the social interaction in the Live Mouse Tracker.**

| event | expe | day | genotype | line | genotype_mother | treatment | KO1_postweaning | N | n_females | n_males | mean | sd |
|---|---|---|---|---|---|---|---|---|---|---|---|---|
| Cuddling | OT treatment | 1 | KO1 | Neo+ Arc | KO1 | OT20 | KO1 control | 16 | 10 | 6 | 39.75 | 43.5943472 |
| Cuddling | OT treatment | 1 | KO1 | Neo+ Arc | KO1 | SAL | KO1 control | 12 | 8 | 4 | 24.3694444 | 7.58989837 |
| Cuddling | OT treatment | 1 | WT | Neo+ Arc | WT | OT20 | WT control | 19 | 8 | 11 | 39.6807018 | 12.7585716 |
| Cuddling | OT treatment | 1 | WT | Neo+ Arc | WT | SAL | WT control | 15 | 7 | 8 | 40.1866667 | 11.331476 |
| Cuddling | OT treatment | 7 | KO1 | Neo+ Arc | KO1 | OT20 | KO1 control | 16 | 10 | 6 | 27.4791667 | 15.7791699 |
| Cuddling | OT treatment | 7 | KO1 | Neo+ Arc | KO1 | SAL | KO1 control | 12 | 8 | 4 | 26.2944444 | 10.5308861 |
| Cuddling | OT treatment | 7 | WT | Neo+ Arc | WT | OT20 | WT control | 19 | 8 | 11 | 37.4315789 | 22.0398922 |
| Cuddling | OT treatment | 7 | WT | Neo+ Arc | WT | SAL | WT control | 15 | 7 | 8 | 35.6177778 | 19.5481086 |
| Cuddling | OT treatment | 14 | KO1 | Neo+ Arc | KO1 | OT20 | KO1 control | 16 | 10 | 6 | 22.5875 | 4.74061842 |
| Cuddling | OT treatment | 14 | KO1 | Neo+ Arc | KO1 | SAL | KO1 control | 12 | 8 | 4 | 27 | 9.00188532 |
| Cuddling | OT treatment | 14 | WT | Neo+ Arc | WT | OT20 | WT control | 19 | 8 | 11 | 54.4280702 | 37.1399211 |
| Cuddling | OT treatment | 14 | WT | Neo+ Arc | WT | SAL | WT control | 15 | 7 | 8 | 37.6311111 | 19.9937152 |
| Cuddling | OT treatment | 21 | KO1 | Neo+ Arc | KO1 | OT20 | KO1 control | 16 | 10 | 6 | 31.9395833 | 17.1466192 |
| Cuddling | OT treatment | 21 | KO1 | Neo+ Arc | KO1 | SAL | KO1 control | 12 | 8 | 4 | 65.8333333 | 38.3277624 |
| Cuddling | OT treatment | 21 | WT | Neo+ Arc | WT | OT20 | WT control | 19 | 8 | 11 | 52.1684211 | 49.5228809 |
| Cuddling | OT treatment | 21 | WT | Neo+ Arc | WT | SAL | WT control | 15 | 7 | 8 | 43.9511111 | 19.0957066 |
| Cuddling | OT treatment | 28 | KO1 | Neo+ Arc | KO1 | OT20 | KO1 control | 16 | 10 | 6 | 36.31875 | 15.014837 |
| Cuddling | OT treatment | 28 | KO1 | Neo+ Arc | KO1 | SAL | KO1 control | 12 | 8 | 4 | 45.6194444 | 21.8807071 |
| Cuddling | OT treatment | 28 | WT | Neo+ Arc | WT | OT20 | WT control | 19 | 8 | 11 | 46.6894737 | 38.5154868 |
| Cuddling | OT treatment | 28 | WT | Neo+ Arc | WT | SAL | WT control | 15 | 7 | 8 | 24.7377778 | 12.5670953 |
| Cuddling | genotype differences | 1 | KO1 | Neo- Arc | NA | NA | KO1 control | 31 | 15 | 16 | 23.8623656 | 20.0112963 |
| Cuddling | genotype differences | 1 | WT | Neo- Arc | NA | NA | WT control | 30 | 15 | 15 | 20.1888889 | 8.97866067 |
| Cuddling | genotype differences | 2 | KO1 | Neo- Arc | NA | NA | KO1 control | 31 | 15 | 16 | 45.8430108 | 14.9740224 |
| Cuddling | genotype differences | 2 | WT | Neo- Arc | NA | NA | WT control | 30 | 15 | 15 | 48.8133333 | 16.3810296 |
| Cuddling | genotype differences | 3 | KO1 | Neo- Arc | NA | NA | KO1 control | 31 | 15 | 16 | 83.2301075 | 39.1805029 |
| Cuddling | genotype differences | 3 | WT | Neo- Arc | NA | NA | WT control | 30 | 15 | 15 | 50.6555556 | 17.1749201 |
| Cuddling | social enrichment | 1 | +/- | Neo+ Arc | NA | NA | enriched | 18 | 11 | 7 | 11.9462963 | 4.69130943 |
| Cuddling | social enrichment | 1 | +/- | Neo+ Arc | NA | NA | social | 17 | 11 | 6 | 8 | 2.51749435 |
| Cuddling | social enrichment | 1 | KO1 | Neo+ Arc | NA | NA | enriched | 13 | 6 | 7 | 12.1102564 | 5.95752535 |
| Cuddling | social enrichment | 1 | KO1 | Neo+ Arc | NA | NA | non-social | 32 | 22 | 10 | 5.35416667 | 1.90156981 |
| Cuddling | social enrichment | 1 | WT | Neo+ Arc | NA | NA | enriched | 7 | 5 | 2 | 10.6 | 5.65240034 |
| Cuddling | social enrichment | 1 | WT | Neo+ Arc | NA | NA | social | 48 | 31 | 17 | 7.17777778 | 3.78683381 |
| Cuddling | social enrichment | 2 | +/- | Neo+ Arc | NA | NA | enriched | 16 | 9 | 7 | 30.2104167 | 9.46592109 |
| Cuddling | social enrichment | 2 | +/- | Neo+ Arc | NA | NA | social | 8 | 3 | 5 | 33.8791667 | 10.4523247 |
| Cuddling | social enrichment | 2 | KO1 | Neo+ Arc | NA | NA | enriched | 12 | 5 | 7 | 29.0972222 | 10.2583738 |
| Cuddling | social enrichment | 2 | KO1 | Neo+ Arc | NA | NA | non-social | 24 | 12 | 12 | 30.9555556 | 12.9829679 |
| Cuddling | social enrichment | 2 | WT | Neo+ Arc | NA | NA | enriched | 7 | 5 | 2 | 34.9952381 | 10.306029 |
| Cuddling | social enrichment | 2 | WT | Neo+ Arc | NA | NA | social | 43 | 26 | 17 | 31.820155 | 13.4404903 |
| Cuddling | social enrichment | 3 | +/- | Neo+ Arc | NA | NA | enriched | 17 | 10 | 7 | 30.8803922 | 24.0226486 |
| Cuddling | social enrichment | 3 | +/- | Neo+ Arc | NA | NA | social | 10 | 5 | 5 | 28.59 | 16.2945196 |
| Cuddling | social enrichment | 3 | KO1 | Neo+ Arc | NA | NA | enriched | 12 | 5 | 7 | 24.9333333 | 7.1055676 |
| Cuddling | social enrichment | 3 | KO1 | Neo+ Arc | NA | NA | non-social | 24 | 12 | 12 | 27.9319444 | 9.03356853 |
| Cuddling | social enrichment | 3 | WT | Neo+ Arc | NA | NA | enriched | 7 | 5 | 2 | 28.7809524 | 11.1194334 |
| Cuddling | social enrichment | 3 | WT | Neo+ Arc | NA | NA | social | 39 | 24 | 15 | 30.1213675 | 12.7332587 |
| FollowZone | OT treatment | 1 | KO1 | Neo+ Arc | KO1 | OT20 | KO1 control | 11 | 7 | 4 | 1.26363636 | 1.05392964 |
| FollowZone | OT treatment | 1 | KO1 | Neo+ Arc | KO1 | SAL | KO1 control | 8 | 7 | 1 | 1.375 | 1.26838855 |
| FollowZone | OT treatment | 1 | WT | Neo+ Arc | WT | OT20 | WT control | 19 | 8 | 11 | 2.97894737 | 2.36995626 |
| FollowZone | OT treatment | 1 | WT | Neo+ Arc | WT | SAL | WT control | 15 | 7 | 8 | 3.74222222 | 2.86690106 |
| FollowZone | OT treatment | 7 | KO1 | Neo+ Arc | KO1 | OT20 | KO1 control | 12 | 8 | 4 | 0.96388889 | 0.49019854 |
| FollowZone | OT treatment | 7 | KO1 | Neo+ Arc | KO1 | SAL | KO1 control | 11 | 8 | 3 | 1.4 | 0.8566083 |
| FollowZone | OT treatment | 7 | WT | Neo+ Arc | WT | OT20 | WT control | 15 | 7 | 8 | 1.12 | 1.06434271 |
| FollowZone | OT treatment | 7 | WT | Neo+ Arc | WT | SAL | WT control | 13 | 7 | 6 | 1.64615385 | 1.17980997 |
| FollowZone | OT treatment | 14 | KO1 | Neo+ Arc | KO1 | OT20 | KO1 control | 8 | 5 | 3 | 1.375 | 0.88402938 |
| FollowZone | OT treatment | 14 | KO1 | Neo+ Arc | KO1 | SAL | KO1 control | 9 | 6 | 3 | 1.64074074 | 1.46095267 |
| FollowZone | OT treatment | 14 | WT | Neo+ Arc | WT | OT20 | WT control | 15 | 8 | 7 | 1.67111111 | 1.96569252 |
| FollowZone | OT treatment | 14 | WT | Neo+ Arc | WT | SAL | WT control | 11 | 5 | 6 | 1.94848485 | 2.09031436 |
| FollowZone | OT treatment | 21 | KO1 | Neo+ Arc | KO1 | OT20 | KO1 control | 12 | 8 | 4 | 1.40555556 | 1.51369728 |
| FollowZone | OT treatment | 21 | KO1 | Neo+ Arc | KO1 | SAL | KO1 control | 10 | 7 | 3 | 1.54333333 | 1.01032326 |
| FollowZone | OT treatment | 21 | WT | Neo+ Arc | WT | OT20 | WT control | 15 | 7 | 8 | 2.54666667 | 3.05761604 |
| FollowZone | OT treatment | 21 | WT | Neo+ Arc | WT | SAL | WT control | 12 | 6 | 6 | 1.825 | 1.48031699 |
| FollowZone | OT treatment | 28 | KO1 | Neo+ Arc | KO1 | OT20 | KO1 control | 14 | 9 | 5 | 1.48571429 | 0.98403619 |
| FollowZone | OT treatment | 28 | KO1 | Neo+ Arc | KO1 | SAL | KO1 control | 10 | 7 | 3 | 2.35666667 | 1.70388445 |
| FollowZone | OT treatment | 28 | WT | Neo+ Arc | WT | OT20 | WT control | 18 | 8 | 10 | 2.44259259 | 1.87918789 |
| FollowZone | OT treatment | 28 | WT | Neo+ Arc | WT | SAL | WT control | 14 | 7 | 7 | 2.66428571 | 3.22793534 |
| FollowZone | genotype differences | 1 | +/- | Neo+ Arc | +/+ | NA | NA | 32 | 16 | 16 | 10.6104167 | 4.29345119 |
| FollowZone | genotype differences | 1 | +/- | Neo+ Arc | -/- | NA | NA | 28 | 16 | 12 | 9.96428571 | 3.57894999 |
| FollowZone | genotype differences | 1 | KO1 | Neo+ Arc | NA | NA | KO1 control | 13 | 6 | 7 | 3.92820513 | 2.45235511 |
| FollowZone | genotype differences | 1 | KO1 | Neo- Arc | NA | NA | KO1 control | 13 | 5 | 8 | 0.50769231 | 0.27893775 |
| FollowZone | genotype differences | 1 | WT | Neo+ Arc | NA | NA | WT control | 19 | 8 | 11 | 8.01578947 | 4.43737449 |
| FollowZone | genotype differences | 1 | WT | Neo- Arc | NA | NA | WT control | 21 | 12 | 9 | 1.03968254 | 0.77313353 |
| FollowZone | genotype differences | 2 | +/- | Neo+ Arc | +/+ | NA | NA | 32 | 16 | 16 | 16.2770833 | 10.1815859 |
| FollowZone | genotype differences | 2 | +/- | Neo+ Arc | -/- | NA | NA | 27 | 15 | 12 | 20.308642 | 7.49732484 |
| FollowZone | genotype differences | 2 | KO1 | Neo+ Arc | NA | NA | KO1 control | 12 | 6 | 6 | 8.99166667 | 6.41694608 |
| FollowZone | genotype differences | 2 | KO1 | Neo- Arc | NA | NA | KO1 control | 20 | 12 | 8 | 1.145 | 0.77695424 |
| FollowZone | genotype differences | 2 | WT | Neo+ Arc | NA | NA | WT control | 18 | 8 | 10 | 14.612963 | 8.86168324 |
| FollowZone | genotype differences | 2 | WT | Neo- Arc | NA | NA | WT control | 24 | 11 | 13 | 1.84027778 | 1.25393985 |
| FollowZone | genotype differences | 3 | KO1 | Neo- Arc | NA | NA | KO1 control | 20 | 7 | 13 | 1.06166667 | 1.09940308 |
| FollowZone | genotype differences | 3 | WT | Neo- Arc | NA | NA | WT control | 24 | 13 | 11 | 1.89861111 | 1.70397317 |
| FollowZone | social enrichment | 1 | +/- | Neo+ Arc | NA | NA | enriched | 13 | 8 | 5 | 1.3025641 | 0.89695592 |
| FollowZone | social enrichment | 1 | +/- | Neo+ Arc | NA | NA | social | 9 | 8 | 1 | 2.02592593 | 1.19335868 |
| FollowZone | social enrichment | 1 | KO1 | Neo+ Arc | NA | NA | enriched | 8 | 3 | 5 | 0.87916667 | 0.68566675 |
| FollowZone | social enrichment | 1 | KO1 | Neo+ Arc | NA | NA | non-social | 10 | 5 | 5 | 0.57333333 | 0.26703678 |
| FollowZone | social enrichment | 1 | WT | Neo+ Arc | NA | NA | enriched | 4 | 3 | 1 | 0.55 | 0.26736021 |
| FollowZone | social enrichment | 1 | WT | Neo+ Arc | NA | NA | social | 37 | 25 | 12 | 0.81261261 | 0.5400818 |
| FollowZone | social enrichment | 2 | +/- | Neo+ Arc | NA | NA | enriched | 14 | 8 | 6 | 2.70714286 | 2.0948524 |
| FollowZone | social enrichment | 2 | +/- | Neo+ Arc | NA | NA | social | 8 | 3 | 5 | 1.82083333 | 1.36113136 |
| FollowZone | social enrichment | 2 | KO1 | Neo+ Arc | NA | NA | enriched | 11 | 5 | 6 | 1.08484848 | 0.8949691 |
| FollowZone | social enrichment | 2 | KO1 | Neo+ Arc | NA | NA | non-social | 22 | 11 | 11 | 2.02727273 | 1.40763738 |
| FollowZone | social enrichment | 2 | WT | Neo+ Arc | NA | NA | enriched | 6 | 4 | 2 | 1.59444444 | 0.71784141 |
| FollowZone | social enrichment | 2 | WT | Neo+ Arc | NA | NA | social | 41 | 26 | 15 | 2.21138211 | 2.02922549 |
| FollowZone | social enrichment | 3 | +/- | Neo+ Arc | NA | NA | enriched | 17 | 10 | 7 | 2.18235294 | 1.35739321 |
| FollowZone | social enrichment | 3 | +/- | Neo+ Arc | NA | NA | social | 8 | 5 | 3 | 2.54583333 | 2.84431695 |
| FollowZone | social enrichment | 3 | KO1 | Neo+ Arc | NA | NA | enriched | 12 | 5 | 7 | 1.83333333 | 0.83714281 |
| FollowZone | social enrichment | 3 | KO1 | Neo+ Arc | NA | NA | non-social | 17 | 10 | 7 | 1.10392157 | 0.84966352 |
| FollowZone | social enrichment | 3 | WT | Neo+ Arc | NA | NA | enriched | 7 | 5 | 2 | 2.00952381 | 2.54999741 |
| FollowZone | social enrichment | 3 | WT | Neo+ Arc | NA | NA | social | 33 | 21 | 12 | 2.58888889 | 1.98556714 |
| Make contact | OT treatment | 1 | KO1 | Neo+ Arc | KO1 | OT20 | KO1 control | 16 | 10 | 6 | 25.7958333 | 11.6763825 |
| Make contact | OT treatment | 1 | KO1 | Neo+ Arc | KO1 | SAL | KO1 control | 12 | 8 | 4 | 30.2527778 | 12.0532443 |
| Make contact | OT treatment | 1 | WT | Neo+ Arc | WT | OT20 | WT control | 19 | 8 | 11 | 31.3929825 | 15.6790085 |
| Make contact | OT treatment | 1 | WT | Neo+ Arc | WT | SAL | WT control | 15 | 7 | 8 | 35.5666667 | 16.5797639 |
| Make contact | OT treatment | 7 | KO1 | Neo+ Arc | KO1 | OT20 | KO1 control | 16 | 10 | 6 | 31.7958333 | 14.2700301 |
| Make contact | OT treatment | 7 | KO1 | Neo+ Arc | KO1 | SAL | KO1 control | 12 | 8 | 4 | 36.3861111 | 9.50813005 |
| Make contact | OT treatment | 7 | WT | Neo+ Arc | WT | OT20 | WT control | 19 | 8 | 11 | 29.2929825 | 13.5387855 |
| Make contact | OT treatment | 7 | WT | Neo+ Arc | WT | SAL | WT control | 15 | 7 | 8 | 32.0333333 | 18.277152 |
| Make contact | OT treatment | 14 | KO1 | Neo+ Arc | KO1 | OT20 | KO1 control | 16 | 10 | 6 | 31.7208333 | 12.6628876 |
| Make contact | OT treatment | 14 | KO1 | Neo+ Arc | KO1 | SAL | KO1 control | 12 | 8 | 4 | 33.7388889 | 8.81244338 |
| Make contact | OT treatment | 14 | WT | Neo+ Arc | WT | OT20 | WT control | 19 | 8 | 11 | 29.7087719 | 17.2282349 |
| Make contact | OT treatment | 14 | WT | Neo+ Arc | WT | SAL | WT control | 15 | 7 | 8 | 28.4466667 | 14.2267678 |
| Make contact | OT treatment | 21 | KO1 | Neo+ Arc | KO1 | OT20 | KO1 control | 16 | 10 | 6 | 29.8 | 11.7939658 |
| Make contact | OT treatment | 21 | KO1 | Neo+ Arc | KO1 | SAL | KO1 control | 12 | 8 | 4 | 37.2222222 | 13.6469505 |
| Make contact | OT treatment | 21 | WT | Neo+ Arc | WT | OT20 | WT control | 19 | 8 | 11 | 36.5368421 | 22.1347052 |
| Make contact | OT treatment | 21 | WT | Neo+ Arc | WT | SAL | WT control | 15 | 7 | 8 | 36.7555556 | 20.5900449 |
| Make contact | OT treatment | 28 | KO1 | Neo+ Arc | KO1 | OT20 | KO1 control | 16 | 10 | 6 | 42.2416667 | 31.7500458 |
| Make contact | OT treatment | 28 | KO1 | Neo+ Arc | KO1 | SAL | KO1 control | 12 | 8 | 4 | 46.6638889 | 20.4603359 |
| Make contact | OT treatment | 28 | WT | Neo+ Arc | WT | OT20 | WT control | 19 | 8 | 11 | 38.5157895 | 25.7726225 |
| Make contact | OT treatment | 28 | WT | Neo+ Arc | WT | SAL | WT control | 15 | 7 | 8 | 29.7 | 25.0293447 |
| Make contact | genotype differences | 1 | KO1 | Neo- Arc | NA | NA | KO1 control | 31 | 15 | 16 | 20.5150538 | 10.0456766 |
| Make contact | genotype differences | 1 | WT | Neo- Arc | NA | NA | WT control | 30 | 15 | 15 | 22.7344444 | 15.149552 |
| Make contact | genotype differences | 2 | KO1 | Neo- Arc | NA | NA | KO1 control | 31 | 15 | 16 | 26.5817204 | 11.8094966 |
| Make contact | genotype differences | 2 | WT | Neo- Arc | NA | NA | WT control | 30 | 15 | 15 | 36.3488889 | 12.8847492 |
| Make contact | genotype differences | 3 | KO1 | Neo- Arc | NA | NA | KO1 control | 31 | 15 | 16 | 30.2516129 | 13.7200336 |
| Make contact | genotype differences | 3 | WT | Neo- Arc | NA | NA | WT control | 30 | 15 | 15 | 35.8722222 | 15.0663566 |
| Make contact | social enrichment | 1 | +/- | Neo+ Arc | NA | NA | enriched | 18 | 11 | 7 | 18.3592593 | 8.49054496 |
| Make contact | social enrichment | 1 | +/- | Neo+ Arc | NA | NA | social | 17 | 11 | 6 | 11.2254902 | 7.8255061 |
| Make contact | social enrichment | 1 | KO1 | Neo+ Arc | NA | NA | enriched | 13 | 6 | 7 | 10.6512821 | 7.25968849 |
| Make contact | social enrichment | 1 | KO1 | Neo+ Arc | NA | NA | non-social | 34 | 22 | 12 | 7.19019608 | 4.61120382 |
| Make contact | social enrichment | 1 | WT | Neo+ Arc | NA | NA | enriched | 7 | 5 | 2 | 11.4857143 | 3.32954547 |
| Make contact | social enrichment | 1 | WT | Neo+ Arc | NA | NA | social | 48 | 31 | 17 | 12.7763889 | 7.08898176 |
| Make contact | social enrichment | 2 | +/- | Neo+ Arc | NA | NA | enriched | 16 | 9 | 7 | 20.3666667 | 8.47702778 |
| Make contact | social enrichment | 2 | +/- | Neo+ Arc | NA | NA | social | 8 | 3 | 5 | 18.0416667 | 8.77038216 |
| Make contact | social enrichment | 2 | KO1 | Neo+ Arc | NA | NA | enriched | 12 | 5 | 7 | 17.7777778 | 6.68576557 |
| Make contact | social enrichment | 2 | KO1 | Neo+ Arc | NA | NA | non-social | 24 | 12 | 12 | 16.4152778 | 6.16913761 |
| Make contact | social enrichment | 2 | WT | Neo+ Arc | NA | NA | enriched | 7 | 5 | 2 | 18.4761905 | 8.20170972 |
| Make contact | social enrichment | 2 | WT | Neo+ Arc | NA | NA | social | 43 | 26 | 17 | 21.075969 | 8.05142968 |
| Make contact | social enrichment | 3 | +/- | Neo+ Arc | NA | NA | enriched | 17 | 10 | 7 | 21.6431373 | 8.10009229 |
| Make contact | social enrichment | 3 | +/- | Neo+ Arc | NA | NA | social | 10 | 5 | 5 | 18.7233333 | 8.14437699 |
| Make contact | social enrichment | 3 | KO1 | Neo+ Arc | NA | NA | enriched | 12 | 5 | 7 | 17.4472222 | 5.83572173 |
| Make contact | social enrichment | 3 | KO1 | Neo+ Arc | NA | NA | non-social | 24 | 12 | 12 | 19.1666667 | 8.35528702 |
| Make contact | social enrichment | 3 | WT | Neo+ Arc | NA | NA | enriched | 7 | 5 | 2 | 19.6857143 | 4.8196687 |
| Make contact | social enrichment | 3 | WT | Neo+ Arc | NA | NA | social | 39 | 24 | 15 | 21.0461538 | 10.370326 |
| Move in contact | OT treatment | 1 | KO1 | Neo+ Arc | KO1 | OT20 | KO1 control | 16 | 10 | 6 | 50.0375 | 42.495481 |
| Move in contact | OT treatment | 1 | KO1 | Neo+ Arc | KO1 | SAL | KO1 control | 12 | 8 | 4 | 39.3305556 | 9.33452869 |
| Move in contact | OT treatment | 1 | WT | Neo+ Arc | WT | OT20 | WT control | 19 | 8 | 11 | 77.0982456 | 35.0427155 |
| Move in contact | OT treatment | 1 | WT | Neo+ Arc | WT | SAL | WT control | 15 | 7 | 8 | 87.0466667 | 33.9724534 |
| Move in contact | OT treatment | 7 | KO1 | Neo+ Arc | KO1 | OT20 | KO1 control | 16 | 10 | 6 | 48.13125 | 21.5354951 |
| Move in contact | OT treatment | 7 | KO1 | Neo+ Arc | KO1 | SAL | KO1 control | 12 | 8 | 4 | 49 | 9.77110769 |
| Move in contact | OT treatment | 7 | WT | Neo+ Arc | WT | OT20 | WT control | 19 | 8 | 11 | 44.9210526 | 26.1966241 |
| Move in contact | OT treatment | 7 | WT | Neo+ Arc | WT | SAL | WT control | 15 | 7 | 8 | 50.9644444 | 24.4937908 |
| Move in contact | OT treatment | 14 | KO1 | Neo+ Arc | KO1 | OT20 | KO1 control | 16 | 10 | 6 | 47.4041667 | 15.8085297 |
| Move in contact | OT treatment | 14 | KO1 | Neo+ Arc | KO1 | SAL | KO1 control | 12 | 8 | 4 | 60.1111111 | 20.3677867 |
| Move in contact | OT treatment | 14 | WT | Neo+ Arc | WT | OT20 | WT control | 19 | 8 | 11 | 52.8052632 | 34.371996 |
| Move in contact | OT treatment | 14 | WT | Neo+ Arc | WT | SAL | WT control | 15 | 7 | 8 | 50.8466667 | 29.8483995 |
| Move in contact | OT treatment | 21 | KO1 | Neo+ Arc | KO1 | OT20 | KO1 control | 16 | 10 | 6 | 59.425 | 30.3735009 |
| Move in contact | OT treatment | 21 | KO1 | Neo+ Arc | KO1 | SAL | KO1 control | 12 | 8 | 4 | 90.5555556 | 40.5242846 |
| Move in contact | OT treatment | 21 | WT | Neo+ Arc | WT | OT20 | WT control | 15 | 8 | 7 | 58.76 | 36.2997752 |
| Move in contact | OT treatment | 21 | WT | Neo+ Arc | WT | SAL | WT control | 15 | 7 | 8 | 51.5333333 | 29.6976296 |
| Move in contact | OT treatment | 28 | KO1 | Neo+ Arc | KO1 | OT20 | KO1 control | 16 | 10 | 6 | 73.175 | 23.6063064 |
| Move in contact | OT treatment | 28 | KO1 | Neo+ Arc | KO1 | SAL | KO1 control | 12 | 8 | 4 | 76.3888889 | 27.8523935 |
| Move in contact | OT treatment | 28 | WT | Neo+ Arc | WT | OT20 | WT control | 18 | 7 | 11 | 62.9962963 | 43.4998233 |
| Move in contact | OT treatment | 28 | WT | Neo+ Arc | WT | SAL | WT control | 14 | 7 | 7 | 48.1928571 | 24.3080803 |
| Move in contact | genotype differences | 1 | +/- | Neo+ Arc | +/+ | NA | NA | 32 | 16 | 16 | 37.9385417 | 10.198412 |
| Move in contact | genotype differences | 1 | +/- | Neo+ Arc | -/- | NA | NA | 28 | 16 | 12 | 36.8166667 | 8.38011314 |
| Move in contact | genotype differences | 1 | KO1 | Neo+ Arc | NA | NA | KO1 control | 13 | 6 | 7 | 17.3615385 | 3.68525204 |
| Move in contact | genotype differences | 1 | KO1 | Neo- Arc | NA | NA | KO1 control | 31 | 15 | 16 | 28.7215054 | 12.2796898 |
| Move in contact | genotype differences | 1 | WT | Neo+ Arc | NA | NA | WT control | 19 | 8 | 11 | 29.2175439 | 14.0906298 |
| Move in contact | genotype differences | 1 | WT | Neo- Arc | NA | NA | WT control | 30 | 15 | 15 | 29.1955556 | 12.193682 |
| Move in contact | genotype differences | 2 | +/- | Neo+ Arc | +/+ | NA | NA | 32 | 16 | 16 | 56.8875 | 13.5864252 |
| Move in contact | genotype differences | 2 | +/- | Neo+ Arc | -/- | NA | NA | 27 | 15 | 12 | 67.3234568 | 14.8404069 |
| Move in contact | genotype differences | 2 | KO1 | Neo+ Arc | NA | NA | KO1 control | 12 | 6 | 6 | 34.3111111 | 8.70655545 |
| Move in contact | genotype differences | 2 | KO1 | Neo- Arc | NA | NA | KO1 control | 31 | 15 | 16 | 54.455914 | 19.2329489 |
| Move in contact | genotype differences | 2 | WT | Neo+ Arc | NA | NA | WT control | 18 | 8 | 10 | 48.0314815 | 16.8484381 |
| Move in contact | genotype differences | 2 | WT | Neo- Arc | NA | NA | WT control | 30 | 15 | 15 | 62.9966667 | 19.1139101 |
| Move in contact | genotype differences | 3 | KO1 | Neo- Arc | NA | NA | KO1 control | 31 | 15 | 16 | 64.8172043 | 25.280616 |
| Move in contact | genotype differences | 3 | WT | Neo- Arc | NA | NA | WT control | 30 | 15 | 15 | 75.7966667 | 31.0106153 |
| Move in contact | social enrichment | 1 | +/- | Neo+ Arc | NA | NA | enriched | 18 | 11 | 7 | 23.5518519 | 5.82521252 |
| Move in contact | social enrichment | 1 | +/- | Neo+ Arc | NA | NA | social | 17 | 11 | 6 | 17.9568627 | 6.47965582 |
| Move in contact | social enrichment | 1 | KO1 | Neo+ Arc | NA | NA | enriched | 13 | 6 | 7 | 18.9282051 | 6.12973696 |
| Move in contact | social enrichment | 1 | KO1 | Neo+ Arc | NA | NA | non-social | 34 | 22 | 12 | 10.2862745 | 4.93405071 |
| Move in contact | social enrichment | 1 | WT | Neo+ Arc | NA | NA | enriched | 7 | 5 | 2 | 18.1238095 | 4.07475127 |
| Move in contact | social enrichment | 1 | WT | Neo+ Arc | NA | NA | social | 48 | 31 | 17 | 16.5666667 | 6.77336949 |
| Move in contact | social enrichment | 2 | +/- | Neo+ Arc | NA | NA | enriched | 16 | 9 | 7 | 60.01875 | 19.8206027 |
| Move in contact | social enrichment | 2 | +/- | Neo+ Arc | NA | NA | social | 8 | 3 | 5 | 62.1416667 | 22.9235496 |
| Move in contact | social enrichment | 2 | KO1 | Neo+ Arc | NA | NA | enriched | 12 | 5 | 7 | 44.1833333 | 18.225005 |
| Move in contact | social enrichment | 2 | KO1 | Neo+ Arc | NA | NA | non-social | 24 | 12 | 12 | 50.7708333 | 19.7094776 |
| Move in contact | social enrichment | 2 | WT | Neo+ Arc | NA | NA | enriched | 7 | 5 | 2 | 53.5095238 | 28.547444 |
| Move in contact | social enrichment | 2 | WT | Neo+ Arc | NA | NA | social | 43 | 26 | 17 | 54.455814 | 19.6327158 |
| Move in contact | social enrichment | 3 | +/- | Neo+ Arc | NA | NA | enriched | 17 | 10 | 7 | 62.527451 | 27.2662317 |
| Move in contact | social enrichment | 3 | +/- | Neo+ Arc | NA | NA | social | 11 | 6 | 5 | 51.4333333 | 10.4199275 |
| Move in contact | social enrichment | 3 | KO1 | Neo+ Arc | NA | NA | enriched | 12 | 5 | 7 | 36.3666667 | 6.63962698 |
| Move in contact | social enrichment | 3 | KO1 | Neo+ Arc | NA | NA | non-social | 24 | 12 | 12 | 42.0236111 | 13.6700649 |
| Move in contact | social enrichment | 3 | WT | Neo+ Arc | NA | NA | enriched | 7 | 5 | 2 | 49.1047619 | 25.1565132 |
| Move in contact | social enrichment | 3 | WT | Neo+ Arc | NA | NA | social | 39 | 24 | 15 | 57.5401709 | 23.7200105 |
| MoveIsolated_time | genotype differences | 1 | +/- | Neo+ Arc | +/+ | NA | NA | 32 | 16 | 16 | 244.496875 | 42.5880453 |
| MoveIsolated_time | genotype differences | 1 | +/- | Neo+ Arc | -/- | NA | NA | 28 | 16 | 12 | 236.652381 | 52.2268008 |
| MoveIsolated_time | genotype differences | 1 | KO1 | Neo+ Arc | NA | NA | KO1 control | 13 | 6 | 7 | 213.792308 | 34.9786659 |
| MoveIsolated_time | genotype differences | 1 | WT | Neo+ Arc | NA | NA | WT control | 19 | 8 | 11 | 255.410526 | 51.0387532 |
| MoveIsolated_time | genotype differences | 2 | +/- | Neo+ Arc | +/+ | NA | NA | 32 | 16 | 16 | 182.773958 | 44.8407782 |
| MoveIsolated_time | genotype differences | 2 | +/- | Neo+ Arc | -/- | NA | NA | 27 | 15 | 12 | 194.485185 | 39.8808903 |
| MoveIsolated_time | genotype differences | 2 | KO1 | Neo+ Arc | NA | NA | KO1 control | 12 | 6 | 6 | 180.991667 | 47.560813 |
| MoveIsolated_time | genotype differences | 2 | WT | Neo+ Arc | NA | NA | WT control | 18 | 8 | 10 | 215.781481 | 60.6077523 |
| Nose contact | OT treatment | 1 | KO1 | Neo+ Arc | KO1 | OT20 | KO1 control | 16 | 10 | 6 | 43.4145833 | 31.5344608 |
| Nose contact | OT treatment | 1 | KO1 | Neo+ Arc | KO1 | SAL | KO1 control | 12 | 8 | 4 | 42.6666667 | 16.961543 |
| Nose contact | OT treatment | 1 | WT | Neo+ Arc | WT | OT20 | WT control | 19 | 8 | 11 | 61.7157895 | 15.736889 |
| Nose contact | OT treatment | 1 | WT | Neo+ Arc | WT | SAL | WT control | 15 | 7 | 8 | 72.0911111 | 16.3980906 |
| Nose contact | OT treatment | 7 | KO1 | Neo+ Arc | KO1 | OT20 | KO1 control | 16 | 10 | 6 | 40.0020833 | 14.704363 |
| Nose contact | OT treatment | 7 | KO1 | Neo+ Arc | KO1 | SAL | KO1 control | 12 | 8 | 4 | 42.7083333 | 12.6377751 |
| Nose contact | OT treatment | 7 | WT | Neo+ Arc | WT | OT20 | WT control | 19 | 8 | 11 | 59.645614 | 21.6133452 |
| Nose contact | OT treatment | 7 | WT | Neo+ Arc | WT | SAL | WT control | 15 | 7 | 8 | 60.2444444 | 13.5883217 |
| Nose contact | OT treatment | 14 | KO1 | Neo+ Arc | KO1 | OT20 | KO1 control | 16 | 10 | 6 | 40.0729167 | 11.2290972 |
| Nose contact | OT treatment | 14 | KO1 | Neo+ Arc | KO1 | SAL | KO1 control | 12 | 8 | 4 | 50.9083333 | 24.0062918 |
| Nose contact | OT treatment | 14 | WT | Neo+ Arc | WT | OT20 | WT control | 19 | 8 | 11 | 67.122807 | 29.6052945 |
| Nose contact | OT treatment | 14 | WT | Neo+ Arc | WT | SAL | WT control | 15 | 7 | 8 | 58.9733333 | 16.3511623 |
| Nose contact | OT treatment | 21 | KO1 | Neo+ Arc | KO1 | OT20 | KO1 control | 16 | 10 | 6 | 48.7958333 | 19.5415943 |
| Nose contact | OT treatment | 21 | KO1 | Neo+ Arc | KO1 | SAL | KO1 control | 12 | 8 | 4 | 71.7388889 | 32.9836655 |
| Nose contact | OT treatment | 21 | WT | Neo+ Arc | WT | OT20 | WT control | 19 | 8 | 11 | 78.8175439 | 26.3564488 |
| Nose contact | OT treatment | 21 | WT | Neo+ Arc | WT | SAL | WT control | 15 | 7 | 8 | 68.1066667 | 19.7243231 |
| Nose contact | OT treatment | 28 | KO1 | Neo+ Arc | KO1 | OT20 | KO1 control | 16 | 10 | 6 | 58.5208333 | 20.4642086 |
| Nose contact | OT treatment | 28 | KO1 | Neo+ Arc | KO1 | SAL | KO1 control | 12 | 8 | 4 | 64.2555556 | 25.4743616 |
| Nose contact | OT treatment | 28 | WT | Neo+ Arc | WT | OT20 | WT control | 19 | 8 | 11 | 71.9754386 | 29.5684785 |
| Nose contact | OT treatment | 28 | WT | Neo+ Arc | WT | SAL | WT control | 15 | 7 | 8 | 52.3711111 | 19.564573 |
| Nose contact | genotype differences | 1 | +/- | Neo+ Arc | +/+ | NA | NA | 32 | 16 | 16 | 30.43125 | 8.01673848 |
| Nose contact | genotype differences | 1 | +/- | Neo+ Arc | -/- | NA | NA | 28 | 16 | 12 | 30.2964286 | 8.4462003 |
| Nose contact | genotype differences | 1 | KO1 | Neo+ Arc | NA | NA | KO1 control | 13 | 6 | 7 | 15.0666667 | 6.68259209 |
| Nose contact | genotype differences | 1 | KO1 | Neo- Arc | NA | NA | KO1 control | 31 | 15 | 16 | 31.2903226 | 16.2529546 |
| Nose contact | genotype differences | 1 | WT | Neo+ Arc | NA | NA | WT control | 19 | 8 | 11 | 21.5596491 | 11.8505018 |
| Nose contact | genotype differences | 1 | WT | Neo- Arc | NA | NA | WT control | 30 | 15 | 15 | 27.0233333 | 8.92263105 |
| Nose contact | genotype differences | 2 | +/- | Neo+ Arc | +/+ | NA | NA | 32 | 16 | 16 | 63.890625 | 17.8339579 |
| Nose contact | genotype differences | 2 | +/- | Neo+ Arc | -/- | NA | NA | 27 | 15 | 12 | 64.8382716 | 18.0546599 |
| Nose contact | genotype differences | 2 | KO1 | Neo+ Arc | NA | NA | KO1 control | 12 | 6 | 6 | 36.2277778 | 11.4317901 |
| Nose contact | genotype differences | 2 | KO1 | Neo- Arc | NA | NA | KO1 control | 31 | 15 | 16 | 56.8397849 | 10.9005077 |
| Nose contact | genotype differences | 2 | WT | Neo+ Arc | NA | NA | WT control | 18 | 8 | 10 | 48.5685185 | 16.5476514 |
| Nose contact | genotype differences | 2 | WT | Neo- Arc | NA | NA | WT control | 30 | 15 | 15 | 68.2811111 | 19.4717374 |
| Nose contact | genotype differences | 3 | KO1 | Neo- Arc | NA | NA | KO1 control | 31 | 15 | 16 | 78.8387097 | 26.2839581 |
| Nose contact | genotype differences | 3 | WT | Neo- Arc | NA | NA | WT control | 30 | 15 | 15 | 66.67 | 17.3448555 |
| Nose contact | social enrichment | 1 | +/- | Neo+ Arc | NA | NA | enriched | 18 | 11 | 7 | 17.3425926 | 6.74319609 |
| Nose contact | social enrichment | 1 | +/- | Neo+ Arc | NA | NA | social | 17 | 11 | 6 | 12.327451 | 5.46368358 |
| Nose contact | social enrichment | 1 | KO1 | Neo+ Arc | NA | NA | enriched | 13 | 6 | 7 | 13.8410256 | 6.2049585 |
| Nose contact | social enrichment | 1 | KO1 | Neo+ Arc | NA | NA | non-social | 34 | 22 | 12 | 7.19509804 | 2.8613897 |
| Nose contact | social enrichment | 1 | WT | Neo+ Arc | NA | NA | enriched | 7 | 5 | 2 | 14.8857143 | 3.95197359 |
| Nose contact | social enrichment | 1 | WT | Neo+ Arc | NA | NA | social | 48 | 31 | 17 | 11.4055556 | 3.73857493 |
| Nose contact | social enrichment | 2 | +/- | Neo+ Arc | NA | NA | enriched | 16 | 9 | 7 | 43.3416667 | 13.1721058 |
| Nose contact | social enrichment | 2 | +/- | Neo+ Arc | NA | NA | social | 8 | 3 | 5 | 47.4416667 | 18.2950835 |
| Nose contact | social enrichment | 2 | KO1 | Neo+ Arc | NA | NA | enriched | 12 | 5 | 7 | 37.475 | 9.28735564 |
| Nose contact | social enrichment | 2 | KO1 | Neo+ Arc | NA | NA | non-social | 24 | 12 | 12 | 40.0666667 | 12.7299601 |
| Nose contact | social enrichment | 2 | WT | Neo+ Arc | NA | NA | enriched | 7 | 5 | 2 | 49.2333333 | 13.2188698 |
| Nose contact | social enrichment | 2 | WT | Neo+ Arc | NA | NA | social | 43 | 26 | 17 | 46.8682171 | 16.6001912 |
| Nose contact | social enrichment | 3 | +/- | Neo+ Arc | NA | NA | enriched | 17 | 10 | 7 | 41.6784314 | 18.6238607 |
| Nose contact | social enrichment | 3 | +/- | Neo+ Arc | NA | NA | social | 11 | 6 | 5 | 42.7272727 | 13.7657451 |
| Nose contact | social enrichment | 3 | KO1 | Neo+ Arc | NA | NA | enriched | 12 | 5 | 7 | 33.1638889 | 8.34736344 |
| Nose contact | social enrichment | 3 | KO1 | Neo+ Arc | NA | NA | non-social | 24 | 12 | 12 | 34.9861111 | 8.62809654 |
| Nose contact | social enrichment | 3 | WT | Neo+ Arc | NA | NA | enriched | 7 | 5 | 2 | 41.2047619 | 21.7062581 |
| Nose contact | social enrichment | 3 | WT | Neo+ Arc | NA | NA | social | 39 | 24 | 15 | 42.9264957 | 16.1508626 |
| Periphery Zone | OT treatment | 1 | KO1 | Neo+ Arc | KO1 | OT20 | KO1 control | 16 | 10 | 6 | 506.81875 | 57.0924448 |
| Periphery Zone | OT treatment | 1 | KO1 | Neo+ Arc | KO1 | SAL | KO1 control | 12 | 8 | 4 | 486.383333 | 41.5859327 |
| Periphery Zone | OT treatment | 1 | WT | Neo+ Arc | WT | OT20 | WT control | 19 | 8 | 11 | 506.561404 | 38.5894892 |
| Periphery Zone | OT treatment | 1 | WT | Neo+ Arc | WT | SAL | WT control | 15 | 7 | 8 | 515.006667 | 35.1001429 |
| Periphery Zone | OT treatment | 7 | KO1 | Neo+ Arc | KO1 | OT20 | KO1 control | 16 | 10 | 6 | 503.204167 | 54.3151011 |
| Periphery Zone | OT treatment | 7 | KO1 | Neo+ Arc | KO1 | SAL | KO1 control | 12 | 8 | 4 | 478.338889 | 61.854722 |
| Periphery Zone | OT treatment | 7 | WT | Neo+ Arc | WT | OT20 | WT control | 19 | 8 | 11 | 541.140351 | 25.6869679 |
| Periphery Zone | OT treatment | 7 | WT | Neo+ Arc | WT | SAL | WT control | 15 | 7 | 8 | 514.566667 | 52.9506794 |
| Periphery Zone | OT treatment | 14 | KO1 | Neo+ Arc | KO1 | OT20 | KO1 control | 16 | 10 | 6 | 470.64375 | 53.5972107 |
| Periphery Zone | OT treatment | 14 | KO1 | Neo+ Arc | KO1 | SAL | KO1 control | 12 | 8 | 4 | 485.802778 | 56.8661062 |
| Periphery Zone | OT treatment | 14 | WT | Neo+ Arc | WT | OT20 | WT control | 19 | 8 | 11 | 520.212281 | 58.8471458 |
| Periphery Zone | OT treatment | 14 | WT | Neo+ Arc | WT | SAL | WT control | 15 | 7 | 8 | 516.68 | 43.2995579 |
| Periphery Zone | OT treatment | 21 | KO1 | Neo+ Arc | KO1 | OT20 | KO1 control | 16 | 10 | 6 | 477.60625 | 58.257916 |
| Periphery Zone | OT treatment | 21 | KO1 | Neo+ Arc | KO1 | SAL | KO1 control | 12 | 8 | 4 | 530.177778 | 28.2172321 |
| Periphery Zone | OT treatment | 21 | NA | Neo+ Arc | NA | NA | NA | 3 | 0 | 0 | 0.03333333 | 0 |
| Periphery Zone | OT treatment | 21 | WT | Neo+ Arc | WT | OT20 | WT control | 19 | 8 | 11 | 552.04386 | 35.4006069 |
| Periphery Zone | OT treatment | 21 | WT | Neo+ Arc | WT | SAL | WT control | 15 | 7 | 8 | 529.868889 | 39.2695376 |
| Periphery Zone | OT treatment | 28 | KO1 | Neo+ Arc | KO1 | OT20 | KO1 control | 16 | 10 | 6 | 502.145833 | 51.0251077 |
| Periphery Zone | OT treatment | 28 | KO1 | Neo+ Arc | KO1 | SAL | KO1 control | 12 | 8 | 4 | 530.202778 | 39.9324998 |
| Periphery Zone | OT treatment | 28 | WT | Neo+ Arc | WT | OT20 | WT control | 19 | 8 | 11 | 543.731579 | 30.0531268 |
| Periphery Zone | OT treatment | 28 | WT | Neo+ Arc | WT | SAL | WT control | 15 | 7 | 8 | 459.18 | 103.457006 |
| Rearing | OT treatment | 1 | KO1 | Neo+ Arc | KO1 | OT20 | KO1 control | 15 | 9 | 6 | 5.68888889 | 5.49169166 |
| Rearing | OT treatment | 1 | KO1 | Neo+ Arc | KO1 | SAL | KO1 control | 12 | 8 | 4 | 3.65555556 | 3.22616617 |
| Rearing | OT treatment | 1 | WT | Neo+ Arc | WT | OT20 | WT control | 19 | 8 | 11 | 2.54385965 | 2.01258548 |
| Rearing | OT treatment | 1 | WT | Neo+ Arc | WT | SAL | WT control | 15 | 7 | 8 | 4.63777778 | 3.08991875 |
| Rearing | OT treatment | 7 | KO1 | Neo+ Arc | KO1 | OT20 | KO1 control | 16 | 10 | 6 | 4.88958333 | 2.67824483 |
| Rearing | OT treatment | 7 | KO1 | Neo+ Arc | KO1 | SAL | KO1 control | 12 | 8 | 4 | 4.575 | 2.77208761 |
| Rearing | OT treatment | 7 | WT | Neo+ Arc | WT | OT20 | WT control | 19 | 8 | 11 | 4.06140351 | 2.54973403 |
| Rearing | OT treatment | 7 | WT | Neo+ Arc | WT | SAL | WT control | 15 | 7 | 8 | 3.40888889 | 3.12949275 |
| Rearing | OT treatment | 14 | KO1 | Neo+ Arc | KO1 | OT20 | KO1 control | 16 | 10 | 6 | 5.17291667 | 3.39516473 |
| Rearing | OT treatment | 14 | KO1 | Neo+ Arc | KO1 | SAL | KO1 control | 12 | 8 | 4 | 4.18888889 | 2.66845899 |
| Rearing | OT treatment | 14 | WT | Neo+ Arc | WT | OT20 | WT control | 19 | 8 | 11 | 1.38947368 | 1.41176943 |
| Rearing | OT treatment | 14 | WT | Neo+ Arc | WT | SAL | WT control | 14 | 7 | 7 | 2.8 | 2.56411795 |
| Rearing | OT treatment | 21 | KO1 | Neo+ Arc | KO1 | OT20 | KO1 control | 15 | 9 | 6 | 2.90888889 | 2.24776668 |
| Rearing | OT treatment | 21 | KO1 | Neo+ Arc | KO1 | SAL | KO1 control | 12 | 8 | 4 | 2.96944444 | 2.3765461 |
| Rearing | OT treatment | 21 | WT | Neo+ Arc | WT | OT20 | WT control | 19 | 8 | 11 | 1.36666667 | 1.36522055 |
| Rearing | OT treatment | 21 | WT | Neo+ Arc | WT | SAL | WT control | 15 | 7 | 8 | 1.8 | 1.76841782 |
| Rearing | OT treatment | 28 | KO1 | Neo+ Arc | KO1 | OT20 | KO1 control | 16 | 10 | 6 | 2.88958333 | 2.64086739 |
| Rearing | OT treatment | 28 | KO1 | Neo+ Arc | KO1 | SAL | KO1 control | 12 | 8 | 4 | 2.31944444 | 2.09344616 |
| Rearing | OT treatment | 28 | WT | Neo+ Arc | WT | OT20 | WT control | 19 | 8 | 11 | 1.5245614 | 1.67788234 |
| Rearing | OT treatment | 28 | WT | Neo+ Arc | WT | SAL | WT control | 15 | 7 | 8 | 1.48 | 1.21510337 |
| Rearing | genotype differences | 1 | KO1 | Neo- Arc | NA | NA | KO1 control | 30 | 15 | 15 | 7.09111111 | 6.26281328 |
| Rearing | genotype differences | 1 | WT | Neo- Arc | NA | NA | WT control | 30 | 15 | 15 | 3.93444444 | 4.59728792 |
| Rearing | genotype differences | 2 | KO1 | Neo- Arc | NA | NA | KO1 control | 31 | 15 | 16 | 4.4172043 | 4.33061343 |
| Rearing | genotype differences | 2 | WT | Neo- Arc | NA | NA | WT control | 29 | 14 | 15 | 5.02183908 | 4.89392461 |
| Rearing | genotype differences | 3 | KO1 | Neo- Arc | NA | NA | KO1 control | 31 | 15 | 16 | 3.68602151 | 3.16811399 |
| Rearing | genotype differences | 3 | WT | Neo- Arc | NA | NA | WT control | 30 | 15 | 15 | 5.09888889 | 4.35714486 |
| Rearing | social enrichment | 1 | +/- | Neo+ Arc | NA | NA | enriched | 18 | 11 | 7 | 2.16481481 | 1.85456414 |
| Rearing | social enrichment | 1 | +/- | Neo+ Arc | NA | NA | social | 17 | 11 | 6 | 2.41176471 | 2.44699195 |
| Rearing | social enrichment | 1 | KO1 | Neo+ Arc | NA | NA | enriched | 13 | 6 | 7 | 2.26153846 | 2.11287263 |
| Rearing | social enrichment | 1 | KO1 | Neo+ Arc | NA | NA | non-social | 31 | 20 | 11 | 1.94623656 | 1.6950839 |
| Rearing | social enrichment | 1 | WT | Neo+ Arc | NA | NA | enriched | 7 | 5 | 2 | 2.6952381 | 2.26743684 |
| Rearing | social enrichment | 1 | WT | Neo+ Arc | NA | NA | social | 47 | 31 | 16 | 2.83049645 | 3.42907313 |
| Rearing | social enrichment | 2 | +/- | Neo+ Arc | NA | NA | enriched | 16 | 9 | 7 | 2.38958333 | 1.74736707 |
| Rearing | social enrichment | 2 | +/- | Neo+ Arc | NA | NA | social | 7 | 2 | 5 | 3.38571429 | 1.95380513 |
| Rearing | social enrichment | 2 | KO1 | Neo+ Arc | NA | NA | enriched | 12 | 5 | 7 | 2.51111111 | 3.04734802 |
| Rearing | social enrichment | 2 | KO1 | Neo+ Arc | NA | NA | non-social | 23 | 12 | 11 | 2.42173913 | 2.25692221 |
| Rearing | social enrichment | 2 | WT | Neo+ Arc | NA | NA | enriched | 7 | 5 | 2 | 2.23809524 | 1.23219596 |
| Rearing | social enrichment | 2 | WT | Neo+ Arc | NA | NA | social | 43 | 26 | 17 | 2.70077519 | 2.93196657 |
| Rearing | social enrichment | 3 | +/- | Neo+ Arc | NA | NA | enriched | 17 | 10 | 7 | 2.41568627 | 2.02831066 |
| Rearing | social enrichment | 3 | +/- | Neo+ Arc | NA | NA | social | 11 | 6 | 5 | 3.56666667 | 3.82973744 |
| Rearing | social enrichment | 3 | KO1 | Neo+ Arc | NA | NA | enriched | 12 | 5 | 7 | 2.425 | 1.62934733 |
| Rearing | social enrichment | 3 | KO1 | Neo+ Arc | NA | NA | non-social | 24 | 12 | 12 | 2.27638889 | 2.3385548 |
| Rearing | social enrichment | 3 | WT | Neo+ Arc | NA | NA | enriched | 7 | 5 | 2 | 4.5952381 | 2.94224836 |
| Rearing | social enrichment | 3 | WT | Neo+ Arc | NA | NA | social | 38 | 24 | 14 | 2.61842105 | 2.34414096 |
| SAP | OT treatment | 1 | KO1 | Neo+ Arc | KO1 | OT20 | KO1 control | 16 | 10 | 6 | 9.80416667 | 5.78145022 |
| SAP | OT treatment | 1 | KO1 | Neo+ Arc | KO1 | SAL | KO1 control | 12 | 8 | 4 | 9.71388889 | 4.31189237 |
| SAP | OT treatment | 1 | WT | Neo+ Arc | WT | OT20 | WT control | 19 | 8 | 11 | 7.85964912 | 3.38915443 |
| SAP | OT treatment | 1 | WT | Neo+ Arc | WT | SAL | WT control | 15 | 7 | 8 | 6.86 | 3.07328995 |
| SAP | OT treatment | 7 | KO1 | Neo+ Arc | KO1 | OT20 | KO1 control | 16 | 10 | 6 | 8.59583333 | 2.93201675 |
| SAP | OT treatment | 7 | KO1 | Neo+ Arc | KO1 | SAL | KO1 control | 12 | 8 | 4 | 8.52222222 | 3.05638373 |
| SAP | OT treatment | 7 | WT | Neo+ Arc | WT | OT20 | WT control | 19 | 8 | 11 | 12.5947368 | 7.59249356 |
| SAP | OT treatment | 7 | WT | Neo+ Arc | WT | SAL | WT control | 15 | 7 | 8 | 12.0155556 | 8.83980775 |
| SAP | OT treatment | 14 | KO1 | Neo+ Arc | KO1 | OT20 | KO1 control | 16 | 10 | 6 | 8.34791667 | 3.77816277 |
| SAP | OT treatment | 14 | KO1 | Neo+ Arc | KO1 | SAL | KO1 control | 12 | 8 | 4 | 7.79722222 | 2.77424877 |
| SAP | OT treatment | 14 | WT | Neo+ Arc | WT | OT20 | WT control | 19 | 8 | 11 | 14.0210526 | 8.71861362 |
| SAP | OT treatment | 14 | WT | Neo+ Arc | WT | SAL | WT control | 15 | 7 | 8 | 11.7577778 | 8.33734634 |
| SAP | OT treatment | 21 | KO1 | Neo+ Arc | KO1 | OT20 | KO1 control | 16 | 10 | 6 | 9.51666667 | 4.95571499 |
| SAP | OT treatment | 21 | KO1 | Neo+ Arc | KO1 | SAL | KO1 control | 12 | 8 | 4 | 11.5666667 | 3.58399923 |
| SAP | OT treatment | 21 | WT | Neo+ Arc | WT | OT20 | WT control | 19 | 8 | 11 | 11.0070175 | 5.89910622 |
| SAP | OT treatment | 21 | WT | Neo+ Arc | WT | SAL | WT control | 15 | 7 | 8 | 10.9888889 | 5.64063582 |
| SAP | OT treatment | 28 | KO1 | Neo+ Arc | KO1 | OT20 | KO1 control | 16 | 10 | 6 | 8.12708333 | 3.71984107 |
| SAP | OT treatment | 28 | KO1 | Neo+ Arc | KO1 | SAL | KO1 control | 12 | 8 | 4 | 9.88055556 | 3.3484644 |
| SAP | OT treatment | 28 | WT | Neo+ Arc | WT | OT20 | WT control | 19 | 8 | 11 | 12.422807 | 8.72075043 |
| SAP | OT treatment | 28 | WT | Neo+ Arc | WT | SAL | WT control | 15 | 7 | 8 | 8.91555556 | 5.15157345 |
| SAP | genotype differences | 1 | +/- | Neo+ Arc | +/+ | NA | NA | 32 | 16 | 16 | 10.846875 | 4.32393405 |
| SAP | genotype differences | 1 | +/- | Neo+ Arc | -/- | NA | NA | 28 | 16 | 12 | 12.3821429 | 4.41653722 |
| SAP | genotype differences | 1 | KO1 | Neo+ Arc | NA | NA | KO1 control | 13 | 6 | 7 | 18.2923077 | 6.71685489 |
| SAP | genotype differences | 1 | KO1 | Neo- Arc | NA | NA | KO1 control | 31 | 15 | 16 | 10.1709677 | 5.47815015 |
| SAP | genotype differences | 1 | WT | Neo+ Arc | NA | NA | WT control | 19 | 8 | 11 | 14.2684211 | 4.69523627 |
| SAP | genotype differences | 1 | WT | Neo- Arc | NA | NA | WT control | 30 | 15 | 15 | 9.37 | 5.27608898 |
| SAP | genotype differences | 2 | +/- | Neo+ Arc | +/+ | NA | NA | 32 | 16 | 16 | 12.5 | 4.9714093 |
| SAP | genotype differences | 2 | +/- | Neo+ Arc | -/- | NA | NA | 27 | 15 | 12 | 12.1493827 | 4.19738798 |
| SAP | genotype differences | 2 | KO1 | Neo+ Arc | NA | NA | KO1 control | 12 | 6 | 6 | 17.0388889 | 4.3578058 |
| SAP | genotype differences | 2 | KO1 | Neo- Arc | NA | NA | KO1 control | 31 | 15 | 16 | 12.3376344 | 5.38523181 |
| SAP | genotype differences | 2 | WT | Neo+ Arc | NA | NA | WT control | 18 | 8 | 10 | 15.1425926 | 5.1524955 |
| SAP | genotype differences | 2 | WT | Neo- Arc | NA | NA | WT control | 30 | 15 | 15 | 10.9611111 | 4.16999025 |
| SAP | genotype differences | 3 | KO1 | Neo- Arc | NA | NA | KO1 control | 31 | 15 | 16 | 15.9376344 | 5.18661912 |
| SAP | genotype differences | 3 | WT | Neo- Arc | NA | NA | WT control | 30 | 15 | 15 | 10.1788889 | 4.9185043 |
| SAP | social enrichment | 1 | +/- | Neo+ Arc | NA | NA | enriched | 18 | 11 | 7 | 4.43518519 | 3.06494991 |
| SAP | social enrichment | 1 | +/- | Neo+ Arc | NA | NA | social | 17 | 11 | 6 | 2.89607843 | 1.63842244 |
| SAP | social enrichment | 1 | KO1 | Neo+ Arc | NA | NA | enriched | 13 | 6 | 7 | 6.08205128 | 2.44455614 |
| SAP | social enrichment | 1 | KO1 | Neo+ Arc | NA | NA | non-social | 34 | 22 | 12 | 4.10294118 | 2.21495328 |
| SAP | social enrichment | 1 | WT | Neo+ Arc | NA | NA | enriched | 7 | 5 | 2 | 8.3 | 4.07644544 |
| SAP | social enrichment | 1 | WT | Neo+ Arc | NA | NA | social | 48 | 31 | 17 | 4.09722222 | 3.48784335 |
| SAP | social enrichment | 2 | +/- | Neo+ Arc | NA | NA | enriched | 16 | 9 | 7 | 4.22708333 | 1.83049148 |
| SAP | social enrichment | 2 | +/- | Neo+ Arc | NA | NA | social | 8 | 3 | 5 | 4.475 | 1.25391451 |
| SAP | social enrichment | 2 | KO1 | Neo+ Arc | NA | NA | enriched | 12 | 5 | 7 | 6.45277778 | 2.76622699 |
| SAP | social enrichment | 2 | KO1 | Neo+ Arc | NA | NA | non-social | 24 | 12 | 12 | 5.44027778 | 2.56919655 |
| SAP | social enrichment | 2 | WT | Neo+ Arc | NA | NA | enriched | 7 | 5 | 2 | 7.10952381 | 1.90979098 |
| SAP | social enrichment | 2 | WT | Neo+ Arc | NA | NA | social | 43 | 26 | 17 | 4.31472868 | 2.11482925 |
| SAP | social enrichment | 3 | +/- | Neo+ Arc | NA | NA | enriched | 17 | 10 | 7 | 3.9745098 | 2.345342 |
| SAP | social enrichment | 3 | +/- | Neo+ Arc | NA | NA | social | 11 | 6 | 5 | 6.71212121 | 2.31896877 |
| SAP | social enrichment | 3 | KO1 | Neo+ Arc | NA | NA | enriched | 12 | 5 | 7 | 6.81388889 | 3.20083507 |
| SAP | social enrichment | 3 | KO1 | Neo+ Arc | NA | NA | non-social | 24 | 12 | 12 | 5.78472222 | 2.90611514 |
| SAP | social enrichment | 3 | WT | Neo+ Arc | NA | NA | enriched | 7 | 5 | 2 | 7.5 | 1.88345624 |
| SAP | social enrichment | 3 | WT | Neo+ Arc | NA | NA | social | 39 | 24 | 15 | 4.92307692 | 3.24920282 |
| Social approach | OT treatment | 1 | KO1 | Neo+ Arc | KO1 | OT20 | KO1 control | 16 | 10 | 6 | 70.9145833 | 25.5683078 |
| Social approach | OT treatment | 1 | KO1 | Neo+ Arc | KO1 | SAL | KO1 control | 12 | 8 | 4 | 85.1694444 | 15.0840036 |
| Social approach | OT treatment | 1 | WT | Neo+ Arc | WT | OT20 | WT control | 19 | 8 | 11 | 93.2350877 | 21.6056333 |
| Social approach | OT treatment | 1 | WT | Neo+ Arc | WT | SAL | WT control | 15 | 7 | 8 | 100.637778 | 22.4371003 |
| Social approach | OT treatment | 7 | KO1 | Neo+ Arc | KO1 | OT20 | KO1 control | 16 | 10 | 6 | 73.5375 | 23.4386982 |
| Social approach | OT treatment | 7 | KO1 | Neo+ Arc | KO1 | SAL | KO1 control | 12 | 8 | 4 | 85.3194444 | 10.4237148 |
| Social approach | OT treatment | 7 | WT | Neo+ Arc | WT | OT20 | WT control | 19 | 8 | 11 | 91.1912281 | 26.0730215 |
| Social approach | OT treatment | 7 | WT | Neo+ Arc | WT | SAL | WT control | 15 | 7 | 8 | 97.8355556 | 27.938099 |
| Social approach | OT treatment | 14 | KO1 | Neo+ Arc | KO1 | OT20 | KO1 control | 16 | 10 | 6 | 77.3958333 | 17.5007211 |
| Social approach | OT treatment | 14 | KO1 | Neo+ Arc | KO1 | SAL | KO1 control | 12 | 8 | 4 | 86.4305556 | 16.5952132 |
| Social approach | OT treatment | 14 | WT | Neo+ Arc | WT | OT20 | WT control | 19 | 8 | 11 | 95.1736842 | 33.2238007 |
| Social approach | OT treatment | 14 | WT | Neo+ Arc | WT | SAL | WT control | 15 | 7 | 8 | 90.0111111 | 24.0084002 |
| Social approach | OT treatment | 21 | KO1 | Neo+ Arc | KO1 | OT20 | KO1 control | 16 | 10 | 6 | 85.9270833 | 20.5767154 |
| Social approach | OT treatment | 21 | KO1 | Neo+ Arc | KO1 | SAL | KO1 control | 12 | 8 | 4 | 111.580556 | 33.7395714 |
| Social approach | OT treatment | 21 | WT | Neo+ Arc | WT | OT20 | WT control | 19 | 8 | 11 | 101.450877 | 40.0574254 |
| Social approach | OT treatment | 21 | WT | Neo+ Arc | WT | SAL | WT control | 15 | 7 | 8 | 110.828889 | 28.2456567 |
| Social approach | OT treatment | 28 | KO1 | Neo+ Arc | KO1 | OT20 | KO1 control | 16 | 10 | 6 | 94.2479167 | 21.2840958 |
| Social approach | OT treatment | 28 | KO1 | Neo+ Arc | KO1 | SAL | KO1 control | 12 | 8 | 4 | 108.805556 | 26.1735383 |
| Social approach | OT treatment | 28 | WT | Neo+ Arc | WT | OT20 | WT control | 19 | 8 | 11 | 102.984211 | 44.5198144 |
| Social approach | OT treatment | 28 | WT | Neo+ Arc | WT | SAL | WT control | 15 | 7 | 8 | 84.6066667 | 28.7489439 |
| Social approach | genotype differences | 1 | +/- | Neo+ Arc | +/+ | NA | NA | 32 | 16 | 16 | 63.4072917 | 12.7492912 |
| Social approach | genotype differences | 1 | +/- | Neo+ Arc | -/- | NA | NA | 28 | 16 | 12 | 59.2642857 | 11.699284 |
| Social approach | genotype differences | 1 | KO1 | Neo+ Arc | NA | NA | KO1 control | 13 | 6 | 7 | 37.8769231 | 14.9482031 |
| Social approach | genotype differences | 1 | KO1 | Neo- Arc | NA | NA | KO1 control | 31 | 15 | 16 | 60.0505376 | 24.1375455 |
| Social approach | genotype differences | 1 | WT | Neo+ Arc | NA | NA | WT control | 19 | 8 | 11 | 53.2964912 | 24.1228361 |
| Social approach | genotype differences | 1 | WT | Neo- Arc | NA | NA | WT control | 30 | 15 | 15 | 52.9277778 | 17.7210096 |
| Social approach | genotype differences | 2 | +/- | Neo+ Arc | +/+ | NA | NA | 32 | 16 | 16 | 98.9083333 | 20.7320195 |
| Social approach | genotype differences | 2 | +/- | Neo+ Arc | -/- | NA | NA | 27 | 15 | 12 | 97.6012346 | 12.3547475 |
| Social approach | genotype differences | 2 | KO1 | Neo+ Arc | NA | NA | KO1 control | 12 | 6 | 6 | 64.9833333 | 12.584434 |
| Social approach | genotype differences | 2 | KO1 | Neo- Arc | NA | NA | KO1 control | 31 | 15 | 16 | 94.6462366 | 20.6541946 |
| Social approach | genotype differences | 2 | WT | Neo+ Arc | NA | NA | WT control | 18 | 8 | 10 | 78.9203704 | 18.9232061 |
| Social approach | genotype differences | 2 | WT | Neo- Arc | NA | NA | WT control | 30 | 15 | 15 | 107.736667 | 19.2400292 |
| Social approach | genotype differences | 3 | KO1 | Neo- Arc | NA | NA | KO1 control | 31 | 15 | 16 | 125.326882 | 40.8965311 |
| Social approach | genotype differences | 3 | WT | Neo- Arc | NA | NA | WT control | 30 | 15 | 15 | 107.176667 | 21.4387806 |
| Social approach | social enrichment | 1 | +/- | Neo+ Arc | NA | NA | enriched | 18 | 11 | 7 | 37.212963 | 9.26040936 |
| Social approach | social enrichment | 1 | +/- | Neo+ Arc | NA | NA | social | 17 | 11 | 6 | 29.7529412 | 12.301677 |
| Social approach | social enrichment | 1 | KO1 | Neo+ Arc | NA | NA | enriched | 13 | 6 | 7 | 29.3230769 | 10.4554291 |
| Social approach | social enrichment | 1 | KO1 | Neo+ Arc | NA | NA | non-social | 34 | 22 | 12 | 16.9823529 | 4.67744236 |
| Social approach | social enrichment | 1 | WT | Neo+ Arc | NA | NA | enriched | 7 | 5 | 2 | 35.5 | 5.77042267 |
| Social approach | social enrichment | 1 | WT | Neo+ Arc | NA | NA | social | 48 | 31 | 17 | 28.0493056 | 10.6578509 |
| Social approach | social enrichment | 2 | +/- | Neo+ Arc | NA | NA | enriched | 16 | 9 | 7 | 63.3 | 14.5407473 |
| Social approach | social enrichment | 2 | +/- | Neo+ Arc | NA | NA | social | 8 | 3 | 5 | 64.0583333 | 21.7730046 |
| Social approach | social enrichment | 2 | KO1 | Neo+ Arc | NA | NA | enriched | 12 | 5 | 7 | 55.8 | 14.3210236 |
| Social approach | social enrichment | 2 | KO1 | Neo+ Arc | NA | NA | non-social | 24 | 12 | 12 | 56.2486111 | 15.4189543 |
| Social approach | social enrichment | 2 | WT | Neo+ Arc | NA | NA | enriched | 7 | 5 | 2 | 63.2142857 | 17.2406374 |
| Social approach | social enrichment | 2 | WT | Neo+ Arc | NA | NA | social | 43 | 26 | 17 | 61.2224806 | 16.7056429 |
| Social approach | social enrichment | 3 | +/- | Neo+ Arc | NA | NA | enriched | 17 | 10 | 7 | 56.5019608 | 20.2367943 |
| Social approach | social enrichment | 3 | +/- | Neo+ Arc | NA | NA | social | 11 | 6 | 5 | 51.8969697 | 12.1896172 |
| Social approach | social enrichment | 3 | KO1 | Neo+ Arc | NA | NA | enriched | 12 | 5 | 7 | 44.6333333 | 7.4936741 |
| Social approach | social enrichment | 3 | KO1 | Neo+ Arc | NA | NA | non-social | 24 | 12 | 12 | 49.2416667 | 10.1473266 |
| Social approach | social enrichment | 3 | WT | Neo+ Arc | NA | NA | enriched | 7 | 5 | 2 | 56.1380952 | 21.6496466 |
| Social approach | social enrichment | 3 | WT | Neo+ Arc | NA | NA | social | 39 | 24 | 15 | 57.4589744 | 13.9358628 |
| Social escape | OT treatment | 1 | KO1 | Neo+ Arc | KO1 | OT20 | KO1 control | 16 | 10 | 6 | 49.4979167 | 18.2522996 |
| Social escape | OT treatment | 1 | KO1 | Neo+ Arc | KO1 | SAL | KO1 control | 12 | 8 | 4 | 58.5222222 | 13.2813862 |
| Social escape | OT treatment | 1 | WT | Neo+ Arc | WT | OT20 | WT control | 19 | 8 | 11 | 67.7982456 | 19.162249 |
| Social escape | OT treatment | 1 | WT | Neo+ Arc | WT | SAL | WT control | 15 | 7 | 8 | 71.4844444 | 16.7331003 |
| Social escape | OT treatment | 7 | KO1 | Neo+ Arc | KO1 | OT20 | KO1 control | 16 | 10 | 6 | 50.9895833 | 17.3568333 |
| Social escape | OT treatment | 7 | KO1 | Neo+ Arc | KO1 | SAL | KO1 control | 12 | 8 | 4 | 61.6194444 | 8.39596414 |
| Social escape | OT treatment | 7 | WT | Neo+ Arc | WT | OT20 | WT control | 19 | 8 | 11 | 52.1631579 | 22.1148961 |
| Social escape | OT treatment | 7 | WT | Neo+ Arc | WT | SAL | WT control | 15 | 7 | 8 | 58.9866667 | 23.9791172 |
| Social escape | OT treatment | 14 | KO1 | Neo+ Arc | KO1 | OT20 | KO1 control | 16 | 10 | 6 | 54.2479167 | 13.5306181 |
| Social escape | OT treatment | 14 | KO1 | Neo+ Arc | KO1 | SAL | KO1 control | 12 | 8 | 4 | 61.65 | 8.84163132 |
| Social escape | OT treatment | 14 | WT | Neo+ Arc | WT | OT20 | WT control | 19 | 8 | 11 | 51.3982456 | 27.040056 |
| Social escape | OT treatment | 14 | WT | Neo+ Arc | WT | SAL | WT control | 15 | 7 | 8 | 56.4422222 | 25.0146607 |
| Social escape | OT treatment | 21 | KO1 | Neo+ Arc | KO1 | OT20 | KO1 control | 16 | 10 | 6 | 58.1104167 | 17.2415077 |
| Social escape | OT treatment | 21 | KO1 | Neo+ Arc | KO1 | SAL | KO1 control | 12 | 8 | 4 | 75.7972222 | 18.5230812 |
| Social escape | OT treatment | 21 | WT | Neo+ Arc | WT | OT20 | WT control | 19 | 8 | 11 | 56.1578947 | 25.8179066 |
| Social escape | OT treatment | 21 | WT | Neo+ Arc | WT | SAL | WT control | 15 | 7 | 8 | 64.2577778 | 24.1369455 |
| Social escape | OT treatment | 28 | KO1 | Neo+ Arc | KO1 | OT20 | KO1 control | 16 | 10 | 6 | 65.7791667 | 19.236173 |
| Social escape | OT treatment | 28 | KO1 | Neo+ Arc | KO1 | SAL | KO1 control | 12 | 8 | 4 | 76.6111111 | 21.8178841 |
| Social escape | OT treatment | 28 | WT | Neo+ Arc | WT | OT20 | WT control | 19 | 8 | 11 | 62.222807 | 33.6558913 |
| Social escape | OT treatment | 28 | WT | Neo+ Arc | WT | SAL | WT control | 15 | 7 | 8 | 52.0066667 | 20.2596722 |
| Social escape | genotype differences | 1 | KO1 | Neo- Arc | NA | NA | KO1 control | 31 | 15 | 16 | 37.3032258 | 15.4882497 |
| Social escape | genotype differences | 1 | WT | Neo- Arc | NA | NA | WT control | 30 | 15 | 15 | 37.4366667 | 16.2861936 |
| Social escape | genotype differences | 2 | KO1 | Neo- Arc | NA | NA | KO1 control | 31 | 15 | 16 | 58.0849462 | 16.7177214 |
| Social escape | genotype differences | 2 | WT | Neo- Arc | NA | NA | WT control | 30 | 15 | 15 | 66.5255556 | 16.126627 |
| Social escape | genotype differences | 3 | KO1 | Neo- Arc | NA | NA | KO1 control | 31 | 15 | 16 | 64.688172 | 21.168685 |
| Social escape | genotype differences | 3 | WT | Neo- Arc | NA | NA | WT control | 30 | 15 | 15 | 73.2855556 | 16.0828332 |
| Social escape | social enrichment | 1 | +/- | Neo+ Arc | NA | NA | enriched | 18 | 11 | 7 | 27.7833333 | 7.61545771 |
| Social escape | social enrichment | 1 | +/- | Neo+ Arc | NA | NA | social | 17 | 11 | 6 | 25.3137255 | 10.1245443 |
| Social escape | social enrichment | 1 | KO1 | Neo+ Arc | NA | NA | enriched | 13 | 6 | 7 | 24.3128205 | 7.22583506 |
| Social escape | social enrichment | 1 | KO1 | Neo+ Arc | NA | NA | non-social | 34 | 22 | 12 | 12.8696078 | 3.82446816 |
| Social escape | social enrichment | 1 | WT | Neo+ Arc | NA | NA | enriched | 7 | 5 | 2 | 27.5666667 | 7.02245077 |
| Social escape | social enrichment | 1 | WT | Neo+ Arc | NA | NA | social | 48 | 31 | 17 | 22.8118056 | 9.72036805 |
| Social escape | social enrichment | 2 | +/- | Neo+ Arc | NA | NA | enriched | 16 | 9 | 7 | 48.5395833 | 11.0639703 |
| Social escape | social enrichment | 2 | +/- | Neo+ Arc | NA | NA | social | 8 | 3 | 5 | 48.325 | 12.5201393 |
| Social escape | social enrichment | 2 | KO1 | Neo+ Arc | NA | NA | enriched | 12 | 5 | 7 | 39.1527778 | 8.92107321 |
| Social escape | social enrichment | 2 | KO1 | Neo+ Arc | NA | NA | non-social | 24 | 12 | 12 | 40.4319444 | 11.8670676 |
| Social escape | social enrichment | 2 | WT | Neo+ Arc | NA | NA | enriched | 7 | 5 | 2 | 45.9428571 | 11.0423715 |
| Social escape | social enrichment | 2 | WT | Neo+ Arc | NA | NA | social | 43 | 26 | 17 | 44.0581395 | 10.0716904 |
| Social escape | social enrichment | 3 | +/- | Neo+ Arc | NA | NA | enriched | 17 | 10 | 7 | 43.372549 | 12.926476 |
| Social escape | social enrichment | 3 | +/- | Neo+ Arc | NA | NA | social | 11 | 6 | 5 | 39.7878788 | 7.65926125 |
| Social escape | social enrichment | 3 | KO1 | Neo+ Arc | NA | NA | enriched | 12 | 5 | 7 | 30.8194444 | 7.98715457 |
| Social escape | social enrichment | 3 | KO1 | Neo+ Arc | NA | NA | non-social | 24 | 12 | 12 | 34.5541667 | 7.43996138 |
| Social escape | social enrichment | 3 | WT | Neo+ Arc | NA | NA | enriched | 7 | 5 | 2 | 38.7904762 | 10.7552178 |
| Social escape | social enrichment | 3 | WT | Neo+ Arc | NA | NA | social | 39 | 24 | 15 | 42.5641026 | 10.5793144 |
| Stop isolated | OT treatment | 1 | KO1 | Neo+ Arc | KO1 | OT20 | KO1 control | 16 | 10 | 6 | 286.05625 | 77.8183104 |
| Stop isolated | OT treatment | 1 | KO1 | Neo+ Arc | KO1 | SAL | KO1 control | 12 | 8 | 4 | 298.819444 | 42.3693821 |
| Stop isolated | OT treatment | 1 | WT | Neo+ Arc | WT | OT20 | WT control | 19 | 8 | 11 | 224.959649 | 61.1855028 |
| Stop isolated | OT treatment | 1 | WT | Neo+ Arc | WT | SAL | WT control | 15 | 7 | 8 | 217.026667 | 50.986259 |
| Stop isolated | OT treatment | 7 | KO1 | Neo+ Arc | KO1 | OT20 | KO1 control | 16 | 10 | 6 | 302.564583 | 51.8198596 |
| Stop isolated | OT treatment | 7 | KO1 | Neo+ Arc | KO1 | SAL | KO1 control | 12 | 8 | 4 | 280.436111 | 27.3458931 |
| Stop isolated | OT treatment | 7 | WT | Neo+ Arc | WT | OT20 | WT control | 19 | 8 | 11 | 294.964912 | 74.4904721 |
| Stop isolated | OT treatment | 7 | WT | Neo+ Arc | WT | SAL | WT control | 15 | 7 | 8 | 264.437778 | 74.2618574 |
| Stop isolated | OT treatment | 14 | KO1 | Neo+ Arc | KO1 | OT20 | KO1 control | 16 | 10 | 6 | 298.558333 | 54.0342121 |
| Stop isolated | OT treatment | 14 | KO1 | Neo+ Arc | KO1 | SAL | KO1 control | 12 | 8 | 4 | 288.922222 | 39.055998 |
| Stop isolated | OT treatment | 14 | WT | Neo+ Arc | WT | OT20 | WT control | 19 | 8 | 11 | 269.924561 | 105.88563 |
| Stop isolated | OT treatment | 14 | WT | Neo+ Arc | WT | SAL | WT control | 15 | 7 | 8 | 279.117778 | 85.3930314 |
| Stop isolated | OT treatment | 21 | KO1 | Neo+ Arc | KO1 | OT20 | KO1 control | 16 | 10 | 6 | 295.789583 | 58.5811191 |
| Stop isolated | OT treatment | 21 | KO1 | Neo+ Arc | KO1 | SAL | KO1 control | 12 | 8 | 4 | 236.602778 | 57.671239 |
| Stop isolated | OT treatment | 21 | WT | Neo+ Arc | WT | OT20 | WT control | 15 | 8 | 7 | 274.826667 | 97.6900364 |
| Stop isolated | OT treatment | 21 | WT | Neo+ Arc | WT | SAL | WT control | 15 | 7 | 8 | 263.217778 | 73.3562852 |
| Stop isolated | OT treatment | 28 | KO1 | Neo+ Arc | KO1 | OT20 | KO1 control | 16 | 10 | 6 | 254.229167 | 68.3965192 |
| Stop isolated | OT treatment | 28 | KO1 | Neo+ Arc | KO1 | SAL | KO1 control | 12 | 8 | 4 | 257.077778 | 57.9811644 |
| Stop isolated | OT treatment | 28 | WT | Neo+ Arc | WT | OT20 | WT control | 18 | 7 | 11 | 263.537037 | 110.903305 |
| Stop isolated | OT treatment | 28 | WT | Neo+ Arc | WT | SAL | WT control | 14 | 7 | 7 | 218.033333 | 117.860805 |
| Stop isolated | genotype differences | 1 | +/- | Neo+ Arc | +/+ | NA | NA | 32 | 16 | 16 | 279.959375 | 41.773964 |
| Stop isolated | genotype differences | 1 | +/- | Neo+ Arc | -/- | NA | NA | 28 | 16 | 12 | 289.519048 | 39.7208639 |
| Stop isolated | genotype differences | 1 | KO1 | Neo+ Arc | NA | NA | KO1 control | 13 | 6 | 7 | 354.207692 | 35.3554599 |
| Stop isolated | genotype differences | 1 | KO1 | Neo- Arc | NA | NA | KO1 control | 31 | 15 | 16 | 318.587097 | 47.1566402 |
| Stop isolated | genotype differences | 1 | WT | Neo+ Arc | NA | NA | WT control | 19 | 8 | 11 | 294.989474 | 56.2775343 |
| Stop isolated | genotype differences | 1 | WT | Neo- Arc | NA | NA | WT control | 30 | 15 | 15 | 296.844444 | 52.1096981 |
| Stop isolated | genotype differences | 2 | +/- | Neo+ Arc | +/+ | NA | NA | 32 | 16 | 16 | 262.689583 | 47.0205076 |
| Stop isolated | genotype differences | 2 | +/- | Neo+ Arc | -/- | NA | NA | 27 | 15 | 12 | 241.903704 | 37.2604008 |
| Stop isolated | genotype differences | 2 | KO1 | Neo+ Arc | NA | NA | KO1 control | 12 | 6 | 6 | 336.738889 | 38.7256521 |
| Stop isolated | genotype differences | 2 | KO1 | Neo- Arc | NA | NA | KO1 control | 31 | 15 | 16 | 298.576344 | 49.3255739 |
| Stop isolated | genotype differences | 2 | WT | Neo+ Arc | NA | NA | WT control | 18 | 8 | 10 | 270.112963 | 49.5311866 |
| Stop isolated | genotype differences | 2 | WT | Neo- Arc | NA | NA | WT control | 30 | 15 | 15 | 246.372222 | 39.4814677 |
| Stop isolated | genotype differences | 3 | KO1 | Neo- Arc | NA | NA | KO1 control | 31 | 15 | 16 | 267.658065 | 59.6056031 |
| Stop isolated | genotype differences | 3 | WT | Neo- Arc | NA | NA | WT control | 30 | 15 | 15 | 247.381111 | 49.0802843 |
| Stop isolated | social enrichment | 1 | +/- | Neo+ Arc | NA | NA | enriched | 18 | 11 | 7 | 115.524074 | 18.1377871 |
| Stop isolated | social enrichment | 1 | +/- | Neo+ Arc | NA | NA | social | 17 | 11 | 6 | 125.639216 | 20.4151334 |
| Stop isolated | social enrichment | 1 | KO1 | Neo+ Arc | NA | NA | enriched | 13 | 6 | 7 | 140.407692 | 19.8837043 |
| Stop isolated | social enrichment | 1 | KO1 | Neo+ Arc | NA | NA | non-social | 34 | 22 | 12 | 153.194118 | 18.683233 |
| Stop isolated | social enrichment | 1 | WT | Neo+ Arc | NA | NA | enriched | 7 | 5 | 2 | 120.871429 | 12.2339907 |
| Stop isolated | social enrichment | 1 | WT | Neo+ Arc | NA | NA | social | 48 | 31 | 17 | 112.810417 | 28.986906 |
| Stop isolated | social enrichment | 2 | +/- | Neo+ Arc | NA | NA | enriched | 16 | 9 | 7 | 82.3791667 | 23.2573903 |
| Stop isolated | social enrichment | 2 | +/- | Neo+ Arc | NA | NA | social | 8 | 3 | 5 | 92.6541667 | 34.2777078 |
| Stop isolated | social enrichment | 2 | KO1 | Neo+ Arc | NA | NA | enriched | 12 | 5 | 7 | 113.438889 | 25.1063791 |
| Stop isolated | social enrichment | 2 | KO1 | Neo+ Arc | NA | NA | non-social | 24 | 12 | 12 | 107.933333 | 27.2302297 |
| Stop isolated | social enrichment | 2 | WT | Neo+ Arc | NA | NA | enriched | 7 | 5 | 2 | 90.0666667 | 30.923891 |
| Stop isolated | social enrichment | 2 | WT | Neo+ Arc | NA | NA | social | 43 | 26 | 17 | 82.4139535 | 24.6369386 |
| Stop isolated | social enrichment | 3 | +/- | Neo+ Arc | NA | NA | enriched | 17 | 10 | 7 | 93.4392157 | 29.537338 |
| Stop isolated | social enrichment | 3 | +/- | Neo+ Arc | NA | NA | social | 11 | 6 | 5 | 104.721212 | 28.1957888 |
| Stop isolated | social enrichment | 3 | KO1 | Neo+ Arc | NA | NA | enriched | 12 | 5 | 7 | 123.461111 | 17.1948303 |
| Stop isolated | social enrichment | 3 | KO1 | Neo+ Arc | NA | NA | non-social | 24 | 12 | 12 | 119.709722 | 23.943354 |
| Stop isolated | social enrichment | 3 | WT | Neo+ Arc | NA | NA | enriched | 7 | 5 | 2 | 101.157143 | 28.1782009 |
| Stop isolated | social enrichment | 3 | WT | Neo+ Arc | NA | NA | social | 39 | 24 | 15 | 87.8666667 | 28.1370674 |

| experiment | day | event | genotype | KO_postweaning | treatment | group1 | group2 | n1 | n2 | statistic | p | p.adj | p.adj.signif |
|---|---|---|---|---|---|---|---|---|---|---|---|---|---|
| genotype differences | 3 | Cuddling | NA | NA | NA | Neo- Arc KO1 | WT | 31 | 30 | -3.36861728 | 0.00075546 | 0.00075546 | *** |
| genotype differences | 1 | FollowZone | NA | NA | NA | +/-;M:KO | WT | 28 | 40 | -4.54930101 | 5.38E-06 | 3.77E-05 | **** |
| genotype differences | 1 | FollowZone | NA | NA | NA | +/-;M:WT | WT | 32 | 40 | -5.11454288 | 3.15E-07 | 2.52E-06 | **** |
| genotype differences | 1 | FollowZone | NA | NA | NA | Neo- Arc KO1 | WT | 13 | 40 | 2.82680261 | 0.00470153 | 0.01880612 | * |
| genotype differences | 2 | FollowZone | NA | NA | NA | +/-;M:KO | WT | 27 | 42 | -5.54094468 | 3.01E-08 | 2.41E-07 | **** |
| genotype differences | 2 | FollowZone | NA | NA | NA | +/-;M:WT | WT | 32 | 42 | -4.06730064 | 4.76E-05 | 0.00033293 | *** |
| genotype differences | 2 | FollowZone | NA | NA | NA | Neo- Arc KO1 | WT | 20 | 42 | 3.08918952 | 0.00200703 | 0.01003517 | * |
| genotype differences | 3 | FollowZone | NA | NA | NA | Neo- Arc KO1 | WT | 20 | 24 | 2.33435734 | 0.01957702 | 0.01957702 | * |
| genotype differences | 2 | Make contact | NA | NA | NA | Neo- Arc KO1 | WT | 31 | 30 | 2.87811198 | 0.00400063 | 0.00400063 | ** |
| genotype differences | 1 | Move in contact | NA | NA | NA | +/-;M:KO | WT | 28 | 49 | -2.73490228 | 0.00623988 | 0.02495952 | * |
| genotype differences | 1 | Move in contact | NA | NA | NA | +/-;M:WT | WT | 32 | 49 | -3.16114171 | 0.00157152 | 0.01100064 | * |
| genotype differences | 1 | Move in contact | NA | NA | NA | Neo+ Arc KO1 | WT | 13 | 49 | 3.32848759 | 0.00087319 | 0.00698551 | ** |
| genotype differences | 2 | Move in contact | NA | NA | NA | Neo+ Arc KO1 | WT | 12 | 48 | 3.93878405 | 8.19E-05 | 0.00073706 | *** |
| genotype differences | 1 | Nose contact | NA | NA | NA | Neo+ Arc KO1 | WT | 13 | 49 | 2.95605662 | 0.003116 | 0.02181199 | * |
| genotype differences | 2 | Nose contact | NA | NA | NA | Neo+ Arc KO1 | WT | 12 | 48 | 4.08736135 | 4.36E-05 | 0.00034905 | *** |
| genotype differences | 1 | SAP | NA | NA | NA | Neo+ Arc KO1 | WT | 13 | 49 | -3.33078669 | 0.00086601 | 0.00779408 | ** |
| genotype differences | 3 | SAP | NA | NA | NA | Neo- Arc KO1 | WT | 31 | 30 | -4.09005676 | 4.31E-05 | 4.31E-05 | **** |
| genotype differences | 2 | Social approach | NA | NA | NA | Neo+ Arc KO1 | WT | 12 | 48 | 4.48034829 | 7.45E-06 | 7.45E-05 | **** |
| genotype differences | 1 | Stop isolated | NA | NA | NA | Neo+ Arc KO1 | WT | 13 | 49 | -3.70058819 | 0.0002151 | 0.0017208 | ** |
| genotype differences | 2 | Stop isolated | NA | NA | NA | Neo- Arc KO1 | WT | 31 | 48 | -3.51908892 | 0.00043303 | 0.00259819 | ** |
| genotype differences | 2 | Stop isolated | NA | NA | NA | Neo+ Arc KO1 | WT | 12 | 48 | -4.47661794 | 7.58E-06 | 6.83E-05 | **** |
| genotype differences | 1 | Cuddling | WT | NA | NA | F | M | 15 | 15 | -4.12890526 | 3.64E-05 | 3.64E-05 | **** |
| genotype differences | 2 | Cuddling | Neo- Arc KO1 | NA | NA | F | M | 15 | 16 | -3.12274919 | 0.0017917 | 0.0017917 | ** |
| genotype differences | 1 | Move in contact | Neo- Arc KO1 | NA | NA | F | M | 15 | 16 | -1.97642354 | 0.04810683 | 0.04810683 | * |
| genotype differences | 2 | Move in contact | +/-;M:KO | NA | NA | F | M | 15 | 12 | 2.29406554 | 0.02178673 | 0.02178673 | * |
| genotype differences | 3 | Move in contact | Neo- Arc KO1 | NA | NA | F | M | 15 | 16 | 2.76699295 | 0.0056576 | 0.0056576 | ** |
| genotype differences | 3 | Move in contact | WT | NA | NA | F | M | 15 | 15 | 2.42646697 | 0.01524664 | 0.01524664 | * |
| genotype differences | 1 | Nose contact | +/-;M:KO | NA | NA | F | M | 16 | 12 | -2.5533109 | 0.01067042 | 0.01067042 | * |
| genotype differences | 1 | Nose contact | WT | NA | NA | F | M | 28 | 26 | -2.45833016 | 0.01395848 | 0.01395848 | * |
| genotype differences | 2 | Nose contact | +/-;M:WT | NA | NA | F | M | 16 | 16 | 2.18595725 | 0.02881873 | 0.02881873 | * |
| genotype differences | 1 | Social approach | WT | NA | NA | F | M | 28 | 26 | -2.40639361 | 0.0161109 | 0.0161109 | * |
| genotype differences | 2 | Social approach | +/-;M:KO | NA | NA | F | M | 15 | 12 | 2.14698016 | 0.03179486 | 0.03179486 | * |
| genotype differences | 2 | Social approach | +/-;M:WT | NA | NA | F | M | 16 | 16 | 3.16586912 | 0.0015462 | 0.0015462 | ** |
| genotype differences | 3 | Social approach | Neo- Arc KO1 | NA | NA | F | M | 15 | 16 | 2.2926513 | 0.02186809 | 0.02186809 | * |
| genotype differences | 1 | Social escape | WT | NA | NA | F | M | 15 | 15 | -2.30203276 | 0.02133332 | 0.02133332 | * |
| genotype differences | 3 | Social escape | Neo- Arc KO1 | NA | NA | F | M | 15 | 16 | 3.04369225 | 0.00233694 | 0.00233694 | ** |
| genotype differences | 2 | Stop isolated | +/-;M:WT | NA | NA | F | M | 16 | 16 | -2.52515751 | 0.01156464 | 0.01156464 | * |
| genotype differences | 3 | Stop isolated | Neo- Arc KO1 | NA | NA | F | M | 15 | 16 | -2.68793601 | 0.00718952 | 0.00718952 | ** |
| social enrichment | 3 | FollowZone | NA | control | NA | KO | WT | 17 | 33 | 3.26171818 | 0.00110739 | 0.00332218 | ** |
| social enrichment | 1 | Make contact | NA | control | NA | KO | WT | 34 | 48 | 3.76811159 | 0.00016449 | 0.00049346 | *** |
| social enrichment | 1 | Move in contact | NA | control | NA | KO | WT | 34 | 48 | 4.04144368 | 5.31E-05 | 0.00015937 | *** |
| social enrichment | 3 | Move in contact | NA | control | NA | KO | WT | 24 | 39 | 2.90855967 | 0.00363098 | 0.01089294 | * |
| social enrichment | 1 | Nose contact | NA | control | NA | KO | WT | 34 | 48 | 4.49312691 | 7.02E-06 | 2.11E-05 | **** |
| social enrichment | 1 | Social approach | NA | control | NA | KO | WT | 34 | 48 | 5.15757326 | 2.50E-07 | 7.51E-07 | **** |
| social enrichment | 1 | Social escape | NA | control | NA | KO | WT | 34 | 48 | 5.19733237 | 2.02E-07 | 6.07E-07 | **** |
| social enrichment | 3 | Social escape | NA | control | NA | KO | WT | 24 | 39 | 3.33398204 | 0.00085612 | 0.00256837 | ** |
| social enrichment | 1 | Stop isolated | NA | control | NA | KO | WT | 34 | 48 | -6.00817392 | 1.88E-09 | 5.63E-09 | **** |
| social enrichment | 2 | Stop isolated | NA | control | NA | KO | WT | 24 | 43 | -3.43575084 | 0.00059091 | 0.00177274 | ** |
| social enrichment | 3 | Stop isolated | NA | control | NA | KO | WT | 24 | 39 | -4.3487327 | 1.37E-05 | 4.11E-05 | **** |
| social enrichment | 1 | Cuddling | KO | NA | NA | control | enriched | 32 | 13 | 4.11200046 | 3.92E-05 | 3.92E-05 | **** |
| social enrichment | 3 | FollowZone | KO | NA | NA | control | enriched | 17 | 12 | 2.85786832 | 0.00426497 | 0.00426497 | ** |
| social enrichment | 1 | Move in contact | KO | NA | NA | control | enriched | 34 | 13 | 4.16363601 | 3.13E-05 | 3.13E-05 | **** |
| social enrichment | 1 | Nose contact | KO | NA | NA | control | enriched | 34 | 13 | 3.37820002 | 0.00072962 | 0.00072962 | *** |
| social enrichment | 1 | Nose contact | WT | NA | NA | control | enriched | 48 | 7 | 2.23544642 | 0.02538806 | 0.02538806 | * |
| social enrichment | 2 | SAP | WT | NA | NA | control | enriched | 43 | 7 | 2.9779955 | 0.0029014 | 0.0029014 | ** |
| social enrichment | 3 | SAP | WT | NA | NA | control | enriched | 39 | 7 | 2.78352756 | 0.00537713 | 0.00537713 | ** |
| social enrichment | 1 | Social approach | KO | NA | NA | control | enriched | 34 | 13 | 3.37849322 | 0.00072884 | 0.00072884 | *** |
| social enrichment | 1 | Social escape | KO | NA | NA | control | enriched | 34 | 13 | 4.42496623 | 9.65E-06 | 9.65E-06 | **** |
| OT treatment | 1 | Cuddling | NA | NA | SAL | KO | WT | 12 | 15 | 3.02529023 | 0.00248395 | 0.00248395 | ** |
| OT treatment | 1 | Move in contact | NA | NA | OT20 | KO | WT | 16 | 19 | 2.98040652 | 0.00287866 | 0.00287866 | ** |
| OT treatment | 1 | Move in contact | NA | NA | SAL | KO | WT | 12 | 15 | 4.09878031 | 4.15E-05 | 4.15E-05 | **** |
| OT treatment | 1 | Nose contact | NA | NA | OT20 | KO | WT | 16 | 19 | 2.91417527 | 0.0035663 | 0.0035663 | ** |
| OT treatment | 1 | Nose contact | NA | NA | SAL | KO | WT | 12 | 15 | 3.31806025 | 0.00090645 | 0.00090645 | *** |
| OT treatment | 1 | Social approach | NA | NA | OT20 | KO | WT | 16 | 19 | 2.78151795 | 0.00541053 | 0.00541053 | ** |
| OT treatment | 1 | Social escape | NA | NA | OT20 | KO | WT | 16 | 19 | 2.81463126 | 0.00488332 | 0.00488332 | ** |
| OT treatment | 1 | Stop isolated | NA | NA | SAL | KO | WT | 12 | 15 | -3.51324026 | 0.00044268 | 0.00044268 | *** |
| OT treatment | 7 | Nose contact | NA | NA | OT20 | KO | WT | 16 | 19 | 2.84774457 | 0.00440302 | 0.00440302 | ** |
| OT treatment | 7 | Nose contact | NA | NA | SAL | KO | WT | 12 | 15 | 3.02575207 | 0.00248016 | 0.00248016 | ** |
| OT treatment | 14 | Nose contact | NA | NA | OT20 | KO | WT | 16 | 19 | 3.16254248 | 0.00156398 | 0.00156398 | ** |
| OT treatment | 14 | Rearing | NA | NA | OT20 | KO | WT | 16 | 19 | -3.52780284 | 0.00041902 | 0.00041902 | *** |
| OT treatment | 21 | Nose contact | NA | NA | OT20 | KO | WT | 16 | 19 | 3.26188936 | 0.00110672 | 0.00110672 | ** |
| OT treatment | 21 | Periphery Zone | NA | NA | OT20 | KO | WT | 16 | 19 | 4.23850354 | 2.25E-05 | 2.25E-05 | **** |
| OT treatment | 28 | Periphery Zone | NA | NA | OT20 | KO | WT | 16 | 19 | 2.91397119 | 0.00356863 | 0.00356863 | ** |
| OT treatment | 1 | Rearing | WT | NA | NA | OT20 | SAL | 19 | 15 | 2.01184883 | 0.04423587 | 0.04423587 | * |
| OT treatment | 1 | Social approach | KO | NA | NA | OT20 | SAL | 16 | 12 | 2.08907255 | 0.03670119 | 0.03670119 | * |
| OT treatment | 7 | Social escape | KO | NA | NA | OT20 | SAL | 16 | 12 | 2.08907255 | 0.03670119 | 0.03670119 | * |
| OT treatment | 21 | Cuddling | KO | NA | NA | OT20 | SAL | 16 | 12 | 2.04264872 | 0.04108722 | 0.04108722 | * |
| OT treatment | 21 | Move in contact | KO | NA | NA | OT20 | SAL | 16 | 12 | 1.99622489 | 0.04590945 | 0.04590945 | * |
| OT treatment | 21 | Periphery Zone | KO | NA | NA | OT20 | SAL | 16 | 12 | 2.71616602 | 0.00660428 | 0.00660428 | ** |
| OT treatment | 21 | Social approach | KO | NA | NA | OT20 | SAL | 16 | 12 | 2.08907255 | 0.03670119 | 0.03670119 | * |
| OT treatment | 21 | Social escape | KO | NA | NA | OT20 | SAL | 16 | 12 | 2.3911547 | 0.01679547 | 0.01679547 | * |
| OT treatment | 21 | Stop isolated | KO | NA | NA | OT20 | SAL | 16 | 12 | -2.64615857 | 0.00814116 | 0.00814116 | ** |
| OT treatment | 28 | Nose contact | WT | NA | NA | OT20 | SAL | 19 | 15 | -2.34137579 | 0.01921282 | 0.01921282 | * |

**Table S3. All raw data, mean, and statistics from behavior tests performed in pups.**

| var_s3 | pup_genotype | N | mean | sd |
|---|---|---|---|---|
| duration_vocalization | Neo+ Arc +/- | 85 | 49.7351866 | 11.0361411 |
| duration_vocalization | Neo+ Arc KO1 | 63 | 40.9978991 | 6.75324454 |
| duration_vocalization | WT | 70 | 45.6495472 | 9.56964462 |
| latency_nest | Neo+ Arc +/- | 85 | 53.6951429 | 83.8325753 |
| latency_nest | Neo+ Arc KO1 | 63 | 67.6980526 | 80.5066531 |
| latency_nest | WT | 70 | 24.8395 | 56.7165856 |
| latency_vocalize | Neo+ Arc +/- | 85 | 1.69394118 | 3.51701427 |
| latency_vocalize | Neo+ Arc KO1 | 63 | 4.30123529 | 5.58921977 |
| latency_vocalize | WT | 70 | 1.32723913 | 1.56354929 |
| maxamp_mean | Neo+ Arc +/- | 85 | -65.6379183 | 5.11542161 |
| maxamp_mean | Neo+ Arc KO1 | 63 | -66.7588887 | 4.05861211 |
| maxamp_mean | WT | 70 | -68.8572031 | 4.67345564 |
| maxfreq_mean | Neo+ Arc +/- | 85 | 65.125605 | 3.98962759 |
| maxfreq_mean | Neo+ Arc KO1 | 63 | 65.8766498 | 3.30332117 |
| maxfreq_mean | WT | 70 | 68.6352679 | 3.68774047 |
| meanDuration_nestEntries | Neo+ Arc +/- | 85 | 15.1329429 | 28.4441598 |
| meanDuration_nestEntries | Neo+ Arc KO1 | 63 | 12.1270526 | 17.5934688 |
| meanDuration_nestEntries | WT | 70 | 26.9240714 | 29.0272326 |
| meanfreq_mean | Neo+ Arc +/- | 85 | 65.8608447 | 4.34175017 |
| meanfreq_mean | Neo+ Arc KO1 | 63 | 66.4667687 | 4.34813015 |
| meanfreq_mean | WT | 70 | 69.3788383 | 4.49763199 |
| number_immobile | Neo+ Arc +/- | 85 | 2.28571429 | 2.82396705 |
| number_immobile | Neo+ Arc KO1 | 63 | 0.84210526 | 1.21395396 |
| number_immobile | WT | 70 | 1.35714286 | 1.80973265 |
| number_nestEntries | Neo+ Arc +/- | 85 | 3.65714286 | 3.92535392 |
| number_nestEntries | Neo+ Arc KO1 | 63 | 3.10526316 | 3.3647641 |
| number_nestEntries | WT | 70 | 6.64285714 | 3.50887009 |
| number_vocalizations | Neo+ Arc +/- | 85 | 466.823529 | 247.774907 |
| number_vocalizations | Neo+ Arc KO1 | 63 | 272.058824 | 180.636538 |
| number_vocalizations | WT | 70 | 488.391304 | 187.979143 |
| totalTime_immobile | Neo+ Arc +/- | 85 | 36.6044857 | 48.8136556 |
| totalTime_immobile | Neo+ Arc KO1 | 63 | 11.3711579 | 20.680017 |
| totalTime_immobile | WT | 70 | 21.9346071 | 40.0280612 |
| totalTime_nestEntries | Neo+ Arc +/- | 85 | 64.2721429 | 86.3622889 |
| totalTime_nestEntries | Neo+ Arc KO1 | 63 | 60.7066842 | 87.8881921 |
| totalTime_nestEntries | WT | 70 | 162.275429 | 104.320138 |
| totalTime_vocalizing | Neo+ Arc +/- | 85 | 23930.6176 | 12516.2986 |
| totalTime_vocalizing | Neo+ Arc KO1 | 63 | 11669.8529 | 8965.10645 |
| totalTime_vocalizing | WT | 70 | 22876.3261 | 10440.8787 |

| var_s3 | group1 | group2 | n1 | n2 | statistic | p | p.adj | p.adj.signif |
|---|---|---|---|---|---|---|---|---|
| duration_vocalization | Neo+ Arc +/- | Neo+ Arc KO1 | 17 | 17 | -2.6864149 | 0.00722233 | 0.021667 | * |
| latency_nest | Neo+ Arc +/- | WT | 35 | 28 | -2.26774025 | 0.02334504 | 0.04669009 | * |
| latency_nest | Neo+ Arc KO1 | WT | 19 | 28 | -3.38288364 | 0.00071729 | 0.00215187 | ** |
| latency_vocalize | Neo+ Arc +/- | Neo+ Arc KO1 | 17 | 17 | 2.77940619 | 0.00544584 | 0.01633751 | * |
| maxfreq_mean | Neo+ Arc +/- | WT | 17 | 23 | 2.416453 | 0.01567255 | 0.04701765 | * |
| maxfreq_mean | Neo+ Arc KO1 | WT | 17 | 23 | 2.416453 | 0.01567255 | 0.04701765 | * |
| meanDuration_nestEntries | Neo+ Arc +/- | WT | 35 | 28 | 3.26575864 | 0.00109171 | 0.00327514 | ** |
| meanDuration_nestEntries | Neo+ Arc KO1 | WT | 19 | 28 | 2.70358978 | 0.00685949 | 0.01371898 | * |
| meanfreq_mean | Neo+ Arc +/- | WT | 17 | 23 | 2.41597125 | 0.0156933 | 0.0470799 | * |
| number_nestEntries | Neo+ Arc +/- | WT | 35 | 28 | 3.34618762 | 0.00081931 | 0.00245793 | ** |
| number_nestEntries | Neo+ Arc KO1 | WT | 19 | 28 | 3.14020599 | 0.00168829 | 0.00337658 | ** |
| number_vocalizations | Neo+ Arc +/- | Neo+ Arc KO1 | 17 | 17 | -2.37131619 | 0.01772486 | 0.03544972 | * |
| number_vocalizations | Neo+ Arc KO1 | WT | 17 | 23 | 3.07698074 | 0.00209109 | 0.00627326 | ** |
| totalTime_nestEntries | Neo+ Arc +/- | WT | 35 | 28 | 3.78581036 | 0.00015321 | 0.00045963 | *** |
| totalTime_nestEntries | Neo+ Arc KO1 | WT | 19 | 28 | 3.32887904 | 0.00087196 | 0.00174393 | ** |
| totalTime_vocalizing | Neo+ Arc +/- | Neo+ Arc KO1 | 17 | 17 | -2.95505639 | 0.00312612 | 0.00937836 | ** |
| totalTime_vocalizing | Neo+ Arc KO1 | WT | 17 | 23 | 2.84761835 | 0.00440477 | 0.00937836 | ** |

**Table S4. All raw data, mean, and statistics from western blot data in HEK293 cells.**

| condition | plasmid | var_s4_stat | group1 | group2 | n1 | n2 | statistic | p | p.adj | p.adj.signif |
|---|---|---|---|---|---|---|---|---|---|---|
| DMEM | NA | pERK_to_ERK_normalized | pcDNA3.1 | pUC19 | 5 | 5 | -2.78543007 | 0.00534568 | 0.00534568 | ** |
| NA | pUC19 | pS6_to_S6_normalized | DMEM | SVF | 5 | 5 | 2.78543007 | 0.00534568 | 0.00534568 | ** |
| NA | pcDNA3.1 | pS6_to_S6_normalized | DMEM | SVF | 5 | 5 | 1.98448528 | 0.04720177 | 0.04720177 | * |
| NA | pUC19 | phospho_S6 | DMEM | SVF | 5 | 5 | 2.61116484 | 0.00902344 | 0.00902344 | ** |
| NA | pcDNA3.1 | phospho_S6 | DMEM | SVF | 5 | 5 | 2.19337847 | 0.02828012 | 0.02828012 | * |

| var_s4 | plasmid | condition | N | mean | sd |
|---|---|---|---|---|---|
| ERK | pUC19 | DMEM | 10 | 11677.1343 | 5920.8986 |
| ERK | pUC19 | SVF | 10 | 8935.5725 | 4249.83131 |
| ERK | pcDNA3.1 | DMEM | 10 | 7977.4904 | 4935.46868 |
| ERK | pcDNA3.1 | SVF | 10 | 7107.1741 | 4324.5222 |
| ERK_to_GAPDH | pUC19 | DMEM | 10 | 0.92864882 | 0.43109111 |
| ERK_to_GAPDH | pUC19 | SVF | 10 | 0.75356517 | 0.26932804 |
| ERK_to_GAPDH | pcDNA3.1 | DMEM | 10 | 0.70743437 | 0.44016379 |
| ERK_to_GAPDH | pcDNA3.1 | SVF | 10 | 0.62443717 | 0.26269784 |
| GAPDH | pUC19 | DMEM | 18 | 12485.8947 | 2208.07502 |
| GAPDH | pUC19 | SVF | 18 | 11786.6873 | 2303.52084 |
| GAPDH | pcDNA3.1 | DMEM | 18 | 11453.5202 | 2469.76845 |
| GAPDH | pcDNA3.1 | SVF | 18 | 10533.2288 | 2514.73536 |
| S6 | pUC19 | DMEM | 10 | 9768.8152 | 3690.46949 |
| S6 | pUC19 | SVF | 10 | 12051.1651 | 4559.52957 |
| S6 | pcDNA3.1 | DMEM | 10 | 10039.9013 | 3849.10892 |
| S6 | pcDNA3.1 | SVF | 10 | 9834.1258 | 3001.55619 |
| S6_to_GAPDH | pUC19 | DMEM | 10 | 0.76413399 | 0.24477183 |
| S6_to_GAPDH | pUC19 | SVF | 10 | 0.99242539 | 0.39893185 |
| S6_to_GAPDH | pcDNA3.1 | DMEM | 10 | 0.86804233 | 0.38376373 |
| S6_to_GAPDH | pcDNA3.1 | SVF | 10 | 0.95057921 | 0.38441322 |
| pERK_to_ERK | pUC19 | DMEM | 10 | 0.45537314 | 0.2905289 |
| pERK_to_ERK | pUC19 | SVF | 10 | 0.59115968 | 0.4210605 |
| pERK_to_ERK | pcDNA3.1 | DMEM | 10 | 1.00345316 | 0.70494097 |
| pERK_to_ERK | pcDNA3.1 | SVF | 10 | 1.73747818 | 1.27849145 |
| pS6_to_S6 | pUC19 | DMEM | 10 | 1.04827167 | 0.82999265 |
| pS6_to_S6 | pUC19 | SVF | 10 | 1.81026144 | 1.12296576 |
| pS6_to_S6 | pcDNA3.1 | DMEM | 10 | 1.14422676 | 1.05226876 |
| pS6_to_S6 | pcDNA3.1 | SVF | 10 | 1.88235678 | 1.02702494 |
| phospho_ERK | pUC19 | DMEM | 10 | 4873.6277 | 3295.0332 |
| phospho_ERK | pUC19 | SVF | 10 | 4214.9116 | 1824.76088 |
| phospho_ERK | pcDNA3.1 | DMEM | 10 | 7157.6541 | 6235.0469 |
| phospho_ERK | pcDNA3.1 | SVF | 10 | 7688.0423 | 4031.93147 |
| phospho_S6 | pUC19 | DMEM | 10 | 8318.4889 | 3415.60374 |
| phospho_S6 | pUC19 | SVF | 10 | 17998.8278 | 2790.48605 |
| phospho_S6 | pcDNA3.1 | DMEM | 10 | 9039.2244 | 4746.66104 |
| phospho_S6 | pcDNA3.1 | SVF | 10 | 16034.7297 | 3086.84985 |

**Table S5. All raw data, mean, and statistics from qPCR data in dams and naïve mice.**

| var_s5 | structure | condition | genotype | N | n_females | n_males | mean | sd |
|---|---|---|---|---|---|---|---|---|
| Avp | CPU | dams | Neo+ Arc KO1 | 6 | 6 | 0 | -11.3995533 | 0.92875372 |
| Avp | CPU | dams | Neo- Arc KO1 | 7 | 7 | 0 | -9.03944807 | 1.20227005 |
| Avp | CPU | dams | WT | 9 | 9 | 0 | -8.90346914 | 0.86082377 |
| Avp | CPU | naive mice | Neo+ Arc KO1 | 14 | 8 | 6 | -10.2587642 | 1.36608629 |
| Avp | CPU | naive mice | WT | 13 | 8 | 5 | -10.8994067 | 1.43984174 |
| Avp | NAC | dams | Neo+ Arc KO1 | 4 | 4 | 0 | -9.75192243 | 0.68666996 |
| Avp | NAC | dams | Neo- Arc KO1 | 8 | 8 | 0 | -9.43403206 | 0.90986339 |
| Avp | NAC | dams | WT | 10 | 10 | 0 | -9.51094524 | 0.82115445 |
| Avp | NAC | naive mice | Neo+ Arc KO1 | 13 | 7 | 6 | -10.4742006 | 0.8618336 |
| Avp | NAC | naive mice | WT | 13 | 8 | 5 | -9.72474411 | 0.46025262 |
| Avp | PFC | dams | Neo+ Arc KO1 | 5 | 5 | 0 | -10.5710438 | 1.5918167 |
| Avp | PFC | dams | Neo- Arc KO1 | 8 | 8 | 0 | -8.50991344 | 0.4673804 |
| Avp | PFC | dams | WT | 9 | 9 | 0 | -8.9958074 | 0.94226404 |
| Avp | PFC | naive mice | Neo+ Arc KO1 | 13 | 7 | 6 | -9.70008272 | 1.61147124 |
| Avp | PFC | naive mice | WT | 10 | 5 | 5 | -10.3022184 | 0.84905452 |
| Avp | PVN | dams | Neo+ Arc KO1 | 6 | 6 | 0 | -4.47260755 | 1.43903245 |
| Avp | PVN | dams | Neo- Arc KO1 | 6 | 6 | 0 | -3.4030169 | 0.90463363 |
| Avp | PVN | dams | WT | 8 | 8 | 0 | -5.11975786 | 2.20771139 |
| Avp | PVN | naive mice | Neo+ Arc KO1 | 14 | 8 | 6 | -4.2113028 | 1.8566728 |
| Avp | PVN | naive mice | WT | 13 | 8 | 5 | -3.9387352 | 2.97789574 |
| Avp | SON | dams | Neo+ Arc KO1 | 6 | 6 | 0 | -2.58468347 | 2.20741916 |
| Avp | SON | dams | Neo- Arc KO1 | 8 | 8 | 0 | -3.51920056 | 3.08660813 |
| Avp | SON | dams | WT | 8 | 8 | 0 | -4.68145277 | 2.32726397 |
| Avp | SON | naive mice | Neo+ Arc KO1 | 13 | 7 | 6 | -4.44571957 | 2.60010185 |
| Avp | SON | naive mice | WT | 13 | 8 | 5 | -5.2528746 | 4.04824411 |
| Avpr1a | CPU | dams | Neo+ Arc KO1 | 5 | 5 | 0 | -9.86604512 | 1.89471052 |
| Avpr1a | CPU | dams | Neo- Arc KO1 | 7 | 7 | 0 | -11.8292039 | 0.86305708 |
| Avpr1a | CPU | dams | WT | 8 | 8 | 0 | -10.7982412 | 1.52553108 |
| Avpr1a | CPU | naive mice | Neo+ Arc KO1 | 14 | 8 | 6 | -10.3468016 | 1.82711381 |
| Avpr1a | CPU | naive mice | WT | 13 | 8 | 5 | -11.1100912 | 2.15309044 |
| Avpr1a | NAC | dams | Neo+ Arc KO1 | 5 | 5 | 0 | -8.49014158 | 0.79911512 |
| Avpr1a | NAC | dams | Neo- Arc KO1 | 7 | 7 | 0 | -11.7679143 | 1.32073721 |
| Avpr1a | NAC | dams | WT | 10 | 10 | 0 | -10.2316501 | 1.78533384 |
| Avpr1a | NAC | naive mice | Neo+ Arc KO1 | 13 | 7 | 6 | -10.7233884 | 1.77394007 |
| Avpr1a | NAC | naive mice | WT | 13 | 8 | 5 | -10.6727367 | 2.23836224 |
| Avpr1a | PFC | dams | Neo+ Arc KO1 | 5 | 5 | 0 | -12.5268862 | 0.24016134 |
| Avpr1a | PFC | dams | Neo- Arc KO1 | 8 | 8 | 0 | -12.1238799 | 0.5239606 |
| Avpr1a | PFC | dams | WT | 8 | 8 | 0 | -12.5829899 | 0.50004029 |
| Avpr1a | PFC | naive mice | Neo+ Arc KO1 | 13 | 7 | 6 | -8.2440993 | 2.95457489 |
| Avpr1a | PFC | naive mice | WT | 10 | 5 | 5 | -9.46602346 | 2.19865775 |
| Avpr1a | PVN | dams | Neo+ Arc KO1 | 6 | 6 | 0 | -11.0336415 | 0.78112633 |
| Avpr1a | PVN | dams | Neo- Arc KO1 | 7 | 7 | 0 | -8.77622191 | 0.39215041 |
| Avpr1a | PVN | dams | WT | 8 | 8 | 0 | -9.78652945 | 1.15946836 |
| Avpr1a | PVN | naive mice | Neo+ Arc KO1 | 14 | 8 | 6 | -3.9422718 | 2.93727426 |
| Avpr1a | PVN | naive mice | WT | 13 | 8 | 5 | -5.74093197 | 1.57805492 |
| Avpr1a | SON | dams | Neo+ Arc KO1 | 6 | 6 | 0 | -12.1099111 | 0.3069019 |
| Avpr1a | SON | dams | Neo- Arc KO1 | 8 | 8 | 0 | -10.0800112 | 0.52352728 |
| Avpr1a | SON | dams | WT | 8 | 8 | 0 | -10.587507 | 1.65802352 |
| Avpr1a | SON | naive mice | Neo+ Arc KO1 | 13 | 7 | 6 | -5.9299838 | 3.04359541 |
| Avpr1a | SON | naive mice | WT | 13 | 8 | 5 | -6.59773393 | 3.57851112 |
| Cartpt | NAC | dams | Neo+ Arc KO1 | 5 | 5 | 0 | -3.94565129 | 0.50290061 |
| Cartpt | NAC | dams | Neo- Arc KO1 | 8 | 8 | 0 | -4.12917122 | 0.23913007 |
| Cartpt | NAC | dams | WT | 10 | 10 | 0 | -4.12594048 | 0.42376932 |
| Cartpt | NAC | naive mice | Neo+ Arc KO1 | 13 | 7 | 6 | -4.18739476 | 0.54766675 |
| Cartpt | NAC | naive mice | WT | 13 | 8 | 5 | -3.97446214 | 0.57500383 |
| Cartpt | PVN | dams | Neo+ Arc KO1 | 5 | 5 | 0 | -5.39261991 | 0.7393825 |
| Cartpt | PVN | dams | Neo- Arc KO1 | 7 | 7 | 0 | -4.45932109 | 0.94876736 |
| Cartpt | PVN | dams | WT | 8 | 8 | 0 | -4.68808175 | 0.77309241 |
| Cartpt | PVN | naive mice | Neo+ Arc KO1 | 14 | 8 | 6 | -4.93503342 | 1.47848739 |
| Cartpt | PVN | naive mice | WT | 13 | 8 | 5 | -4.93812952 | 0.72236277 |
| Cartpt | SON | dams | Neo+ Arc KO1 | 6 | 6 | 0 | -6.2958651 | 0.28049339 |
| Cartpt | SON | dams | Neo- Arc KO1 | 7 | 7 | 0 | -6.58823635 | 0.32816614 |
| Cartpt | SON | dams | WT | 8 | 8 | 0 | -6.25945711 | 0.24711411 |
| Cartpt | SON | naive mice | Neo+ Arc KO1 | 13 | 7 | 6 | -5.88743084 | 0.66532185 |
| Cartpt | SON | naive mice | WT | 13 | 8 | 5 | -6.37895452 | 0.52893831 |
| Cck | PFC | dams | Neo+ Arc KO1 | 6 | 6 | 0 | -2.72507494 | 0.34316547 |
| Cck | PFC | dams | Neo- Arc KO1 | 7 | 7 | 0 | -3.11255088 | 0.13657823 |
| Cck | PFC | dams | WT | 8 | 8 | 0 | -2.99863842 | 0.32265755 |
| Cck | PFC | naive mice | Neo+ Arc KO1 | 12 | 6 | 6 | -1.79144845 | 1.38878805 |
| Cck | PFC | naive mice | WT | 10 | 5 | 5 | -2.64480846 | 0.52151789 |
| Cck | PVN | dams | Neo+ Arc KO1 | 5 | 5 | 0 | -4.59857359 | 0.49134979 |
| Cck | PVN | dams | Neo- Arc KO1 | 7 | 7 | 0 | -6.35472492 | 0.42762752 |
| Cck | PVN | dams | WT | 8 | 8 | 0 | -6.09556673 | 0.92891153 |
| Cck | PVN | naive mice | Neo+ Arc KO1 | 14 | 8 | 6 | -4.83072859 | 1.39724416 |
| Cck | PVN | naive mice | WT | 13 | 8 | 5 | -5.49729329 | 0.57526513 |
| Cck | SON | dams | Neo+ Arc KO1 | 6 | 6 | 0 | -4.33894364 | 0.44700009 |
| Cck | SON | dams | Neo- Arc KO1 | 8 | 8 | 0 | -3.30716198 | 0.4045821 |
| Cck | SON | dams | WT | 7 | 7 | 0 | -4.07364696 | 0.75394824 |
| Cck | SON | naive mice | Neo+ Arc KO1 | 13 | 7 | 6 | -4.56923549 | 1.45524029 |
| Cck | SON | naive mice | WT | 13 | 8 | 5 | -4.77051642 | 1.21302555 |
| Cntnap2 | CPU | dams | Neo+ Arc KO1 | 6 | 6 | 0 | -6.45049751 | 0.6165757 |
| Cntnap2 | CPU | dams | Neo- Arc KO1 | 7 | 7 | 0 | -6.51675295 | 0.32803721 |
| Cntnap2 | CPU | dams | WT | 8 | 8 | 0 | -6.48159941 | 0.48914543 |
| Cntnap2 | CPU | naive mice | Neo+ Arc KO1 | 14 | 8 | 6 | -5.44430991 | 0.53147481 |
| Cntnap2 | CPU | naive mice | WT | 13 | 8 | 5 | -5.65086667 | 0.46278877 |
| Cntnap2 | NAC | dams | Neo+ Arc KO1 | 5 | 5 | 0 | -5.07445237 | 0.65969718 |
| Cntnap2 | NAC | dams | Neo- Arc KO1 | 7 | 7 | 0 | -5.91415322 | 0.24754277 |
| Cntnap2 | NAC | dams | WT | 9 | 9 | 0 | -5.83536455 | 0.60024635 |
| Cntnap2 | NAC | naive mice | Neo+ Arc KO1 | 13 | 7 | 6 | -5.22984692 | 0.61802584 |
| Cntnap2 | NAC | naive mice | WT | 13 | 8 | 5 | -5.70928596 | 0.49444816 |
| Cntnap2 | PFC | dams | Neo+ Arc KO1 | 5 | 5 | 0 | -7.11631309 | 0.39626381 |
| Cntnap2 | PFC | dams | Neo- Arc KO1 | 7 | 7 | 0 | -7.04306367 | 0.16671974 |
| Cntnap2 | PFC | dams | WT | 9 | 9 | 0 | -7.21993517 | 0.16188082 |
| Cntnap2 | PFC | naive mice | Neo+ Arc KO1 | 13 | 7 | 6 | -5.98803004 | 1.04303152 |
| Cntnap2 | PFC | naive mice | WT | 10 | 5 | 5 | -6.66736565 | 0.33032864 |
| Cntnap2 | PVN | dams | Neo+ Arc KO1 | 6 | 6 | 0 | -6.10106663 | 0.21842681 |
| Cntnap2 | PVN | dams | Neo- Arc KO1 | 7 | 7 | 0 | -5.28837747 | 0.19613384 |
| Cntnap2 | PVN | dams | WT | 8 | 8 | 0 | -5.89461428 | 0.43292432 |
| Cntnap2 | PVN | naive mice | Neo+ Arc KO1 | 14 | 8 | 6 | -3.56424163 | 1.80414061 |
| Cntnap2 | PVN | naive mice | WT | 13 | 8 | 5 | -3.83551437 | 1.49346237 |
| Cntnap2 | SON | dams | Neo+ Arc KO1 | 6 | 6 | 0 | -6.13950779 | 0.08853755 |
| Cntnap2 | SON | dams | Neo- Arc KO1 | 8 | 8 | 0 | -5.97887146 | 0.27122194 |
| Cntnap2 | SON | dams | WT | 7 | 7 | 0 | -5.98533093 | 0.20532104 |
| Cntnap2 | SON | naive mice | Neo+ Arc KO1 | 13 | 7 | 6 | -4.39502716 | 1.79383722 |
| Cntnap2 | SON | naive mice | WT | 13 | 8 | 5 | -5.12282945 | 0.56566659 |
| Cpeb4 | CPU | dams | Neo+ Arc KO1 | 6 | 6 | 0 | -5.10231816 | 0.65474243 |
| Cpeb4 | CPU | dams | Neo- Arc KO1 | 7 | 7 | 0 | -5.10401007 | 0.17659617 |
| Cpeb4 | CPU | dams | WT | 8 | 8 | 0 | -5.05649962 | 0.23644236 |
| Cpeb4 | CPU | naive mice | Neo+ Arc KO1 | 14 | 8 | 6 | -4.97181486 | 0.37709556 |
| Cpeb4 | CPU | naive mice | WT | 13 | 8 | 5 | -4.92945731 | 0.21496826 |
| Cpeb4 | NAC | dams | Neo+ Arc KO1 | 5 | 5 | 0 | -4.73742651 | 0.27720987 |
| Cpeb4 | NAC | dams | Neo- Arc KO1 | 8 | 8 | 0 | -4.86170579 | 0.17925564 |
| Cpeb4 | NAC | dams | WT | 10 | 10 | 0 | -4.87730616 | 0.22749098 |
| Cpeb4 | NAC | naive mice | Neo+ Arc KO1 | 13 | 7 | 6 | -4.56421227 | 0.6352815 |
| Cpeb4 | NAC | naive mice | WT | 13 | 8 | 5 | -4.98345943 | 0.18035351 |
| Cpeb4 | PFC | dams | Neo+ Arc KO1 | 6 | 6 | 0 | -4.91788116 | 0.12011156 |
| Cpeb4 | PFC | dams | Neo- Arc KO1 | 6 | 6 | 0 | -4.94387644 | 0.07719273 |
| Cpeb4 | PFC | dams | WT | 8 | 8 | 0 | -4.98642907 | 0.07318755 |
| Cpeb4 | PFC | naive mice | Neo+ Arc KO1 | 13 | 7 | 6 | -5.11481205 | 0.27279561 |
| Cpeb4 | PFC | naive mice | WT | 10 | 5 | 5 | -5.06264825 | 0.21114229 |
| Cpeb4 | PVN | dams | Neo+ Arc KO1 | 6 | 6 | 0 | -5.10667057 | 0.13049564 |
| Cpeb4 | PVN | dams | Neo- Arc KO1 | 7 | 7 | 0 | -4.69877854 | 0.07641059 |
| Cpeb4 | PVN | dams | WT | 8 | 8 | 0 | -5.05538212 | 0.2074659 |
| Cpeb4 | PVN | naive mice | Neo+ Arc KO1 | 14 | 8 | 6 | -5.05693282 | 1.35022088 |
| Cpeb4 | PVN | naive mice | WT | 13 | 8 | 5 | -4.74512598 | 1.03187263 |
| Cpeb4 | SON | dams | Neo+ Arc KO1 | 5 | 5 | 0 | -5.25947263 | 0.16121716 |
| Cpeb4 | SON | dams | Neo- Arc KO1 | 8 | 8 | 0 | -4.8515583 | 0.17191947 |
| Cpeb4 | SON | dams | WT | 8 | 8 | 0 | -5.14804271 | 0.31234606 |
| Cpeb4 | SON | naive mice | Neo+ Arc KO1 | 13 | 7 | 6 | -5.15549946 | 0.77060688 |
| Cpeb4 | SON | naive mice | WT | 13 | 8 | 5 | -5.67073749 | 0.52903766 |
| Egr1 | CPU | dams | Neo+ Arc KO1 | 6 | 6 | 0 | -3.07059934 | 0.69153582 |
| Egr1 | CPU | dams | Neo- Arc KO1 | 7 | 7 | 0 | -3.55264521 | 0.52711062 |
| Egr1 | CPU | dams | WT | 9 | 9 | 0 | -3.37995883 | 0.79448439 |
| Egr1 | CPU | naive mice | Neo+ Arc KO1 | 14 | 8 | 6 | -1.57591073 | 0.46939155 |
| Egr1 | CPU | naive mice | WT | 13 | 8 | 5 | -1.80235373 | 0.73682102 |
| Egr1 | NAC | dams | Neo+ Arc KO1 | 6 | 6 | 0 | -3.88024473 | 0.6729407 |
| Egr1 | NAC | dams | Neo- Arc KO1 | 8 | 8 | 0 | -4.7561415 | 0.55764483 |
| Egr1 | NAC | dams | WT | 10 | 10 | 0 | -4.50788834 | 0.89182819 |
| Egr1 | NAC | naive mice | Neo+ Arc KO1 | 13 | 7 | 6 | -2.88197723 | 0.52269854 |
| Egr1 | NAC | naive mice | WT | 13 | 8 | 5 | -3.3636896 | 0.58212299 |
| Egr1 | PFC | dams | Neo+ Arc KO1 | 6 | 6 | 0 | -4.0704252 | 0.6665758 |
| Egr1 | PFC | dams | Neo- Arc KO1 | 8 | 8 | 0 | -5.11349506 | 0.65034374 |
| Egr1 | PFC | dams | WT | 9 | 9 | 0 | -4.61391416 | 0.6683017 |
| Egr1 | PFC | naive mice | Neo+ Arc KO1 | 13 | 7 | 6 | -2.21547773 | 0.72330883 |
| Egr1 | PFC | naive mice | WT | 10 | 5 | 5 | -2.91819429 | 0.69160821 |
| Esr1 | CPU | dams | Neo+ Arc KO1 | 5 | 5 | 0 | -10.7257917 | 2.88029368 |
| Esr1 | CPU | dams | Neo- Arc KO1 | 7 | 7 | 0 | -13.138674 | 0.94072594 |
| Esr1 | CPU | dams | WT | 9 | 9 | 0 | -12.3320732 | 1.55881858 |
| Esr1 | CPU | naive mice | Neo+ Arc KO1 | 14 | 8 | 6 | -10.2560741 | 2.83750125 |
| Esr1 | CPU | naive mice | WT | 13 | 8 | 5 | -11.2549257 | 1.60938387 |
| Esr1 | NAC | dams | Neo+ Arc KO1 | 5 | 5 | 0 | -12.6570554 | 0.52577796 |
| Esr1 | NAC | dams | Neo- Arc KO1 | 8 | 8 | 0 | -12.0371007 | 0.7092539 |
| Esr1 | NAC | dams | WT | 10 | 10 | 0 | -11.8242937 | 0.71530351 |
| Esr1 | NAC | naive mice | Neo+ Arc KO1 | 13 | 7 | 6 | -12.2502246 | 0.94116786 |
| Esr1 | NAC | naive mice | WT | 13 | 8 | 5 | -12.2979397 | 0.8219893 |
| Esr1 | PVN | dams | Neo+ Arc KO1 | 5 | 5 | 0 | -8.13086176 | 0.62630639 |
| Esr1 | PVN | dams | Neo- Arc KO1 | 7 | 7 | 0 | -5.59195734 | 0.61115853 |
| Esr1 | PVN | dams | WT | 8 | 8 | 0 | -6.84316718 | 1.81128443 |
| Esr1 | PVN | naive mice | Neo+ Arc KO1 | 14 | 8 | 6 | -5.82118324 | 1.18695634 |
| Esr1 | PVN | naive mice | WT | 13 | 8 | 5 | -6.65487286 | 0.80680811 |
| Esr1 | SON | dams | Neo+ Arc KO1 | 6 | 6 | 0 | -10.0601945 | 0.26772709 |
| Esr1 | SON | dams | Neo- Arc KO1 | 8 | 8 | 0 | -9.28136254 | 0.62078833 |
| Esr1 | SON | dams | WT | 8 | 8 | 0 | -9.74596923 | 0.75766004 |
| Esr1 | SON | naive mice | Neo+ Arc KO1 | 12 | 7 | 5 | -9.08055654 | 1.49326452 |
| Esr1 | SON | naive mice | WT | 13 | 8 | 5 | -9.27430338 | 1.20328214 |
| Fmr1 | CPU | dams | Neo+ Arc KO1 | 5 | 5 | 0 | -6.42073255 | 0.25091976 |
| Fmr1 | CPU | dams | Neo- Arc KO1 | 7 | 7 | 0 | -6.72219087 | 0.14382811 |
| Fmr1 | CPU | dams | WT | 8 | 8 | 0 | -6.4671115 | 0.25176027 |
| Fmr1 | CPU | naive mice | Neo+ Arc KO1 | 14 | 8 | 6 | -5.98330006 | 0.40080097 |
| Fmr1 | CPU | naive mice | WT | 13 | 8 | 5 | -6.03365997 | 0.42366141 |
| Fmr1 | NAC | dams | Neo+ Arc KO1 | 6 | 6 | 0 | -7.01064111 | 0.38822623 |
| Fmr1 | NAC | dams | Neo- Arc KO1 | 7 | 7 | 0 | -6.90271359 | 0.28821277 |
| Fmr1 | NAC | dams | WT | 10 | 10 | 0 | -6.91084793 | 0.31951087 |
| Fmr1 | NAC | naive mice | Neo+ Arc KO1 | 13 | 7 | 6 | -6.71820642 | 0.22572014 |
| Fmr1 | NAC | naive mice | WT | 13 | 8 | 5 | -6.91908806 | 0.2684046 |
| Fmr1 | PFC | dams | Neo+ Arc KO1 | 6 | 6 | 0 | -6.49181137 | 0.30516213 |
| Fmr1 | PFC | dams | Neo- Arc KO1 | 8 | 8 | 0 | -6.62321718 | 0.19106453 |
| Fmr1 | PFC | dams | WT | 9 | 9 | 0 | -6.52837287 | 0.20331304 |
| Fmr1 | PFC | naive mice | Neo+ Arc KO1 | 13 | 7 | 6 | -6.3814362 | 0.31532783 |
| Fmr1 | PFC | naive mice | WT | 10 | 5 | 5 | -6.53111964 | 0.34832414 |
| Fmr1 | PVN | dams | Neo+ Arc KO1 | 6 | 6 | 0 | -6.8038605 | 0.14830837 |
| Fmr1 | PVN | dams | Neo- Arc KO1 | 7 | 7 | 0 | -6.75840821 | 0.12411237 |
| Fmr1 | PVN | dams | WT | 7 | 7 | 0 | -6.85768845 | 0.09373914 |
| Fmr1 | PVN | naive mice | Neo+ Arc KO1 | 14 | 8 | 6 | -6.33383341 | 0.82106095 |
| Fmr1 | PVN | naive mice | WT | 13 | 8 | 5 | -6.38368806 | 0.6342381 |
| Fmr1 | SON | dams | Neo+ Arc KO1 | 5 | 5 | 0 | -6.7873707 | 0.13928162 |
| Fmr1 | SON | dams | Neo- Arc KO1 | 8 | 8 | 0 | -6.48110135 | 0.22366668 |
| Fmr1 | SON | dams | WT | 7 | 7 | 0 | -6.64173589 | 0.23192675 |
| Fmr1 | SON | naive mice | Neo+ Arc KO1 | 13 | 7 | 6 | -6.4813324 | 0.72410518 |
| Fmr1 | SON | naive mice | WT | 13 | 8 | 5 | -6.61579455 | 0.73872307 |
| Fos | CPU | dams | Neo+ Arc KO1 | 6 | 6 | 0 | -8.40527963 | 0.9826842 |
| Fos | CPU | dams | Neo- Arc KO1 | 7 | 7 | 0 | -8.99728203 | 0.91352986 |
| Fos | CPU | dams | WT | 9 | 9 | 0 | -9.27533572 | 1.29824777 |
| Fos | CPU | naive mice | Neo+ Arc KO1 | 14 | 8 | 6 | -7.04644537 | 1.021191 |
| Fos | CPU | naive mice | WT | 13 | 8 | 5 | -7.31315438 | 1.39387235 |
| Fos | NAC | dams | Neo+ Arc KO1 | 6 | 6 | 0 | -8.52814793 | 0.471479 |
| Fos | NAC | dams | Neo- Arc KO1 | 8 | 8 | 0 | -9.93481516 | 0.89268566 |
| Fos | NAC | dams | WT | 10 | 10 | 0 | -9.97048894 | 1.02952305 |
| Fos | NAC | naive mice | Neo+ Arc KO1 | 13 | 7 | 6 | -7.4304126 | 0.59137025 |
| Fos | NAC | naive mice | WT | 13 | 8 | 5 | -7.75170792 | 0.54488165 |
| Fos | PFC | dams | Neo+ Arc KO1 | 6 | 6 | 0 | -8.40137041 | 0.55996607 |
| Fos | PFC | dams | Neo- Arc KO1 | 8 | 8 | 0 | -8.38351421 | 0.57212597 |
| Fos | PFC | dams | WT | 9 | 9 | 0 | -8.6765164 | 0.60469139 |
| Fos | PFC | naive mice | Neo+ Arc KO1 | 13 | 7 | 6 | -5.59600665 | 1.21929457 |
| Fos | PFC | naive mice | WT | 10 | 5 | 5 | -6.78587271 | 1.00392409 |
| Fos | PVN | dams | Neo+ Arc KO1 | 6 | 6 | 0 | -8.57207131 | 0.26551629 |
| Fos | PVN | dams | Neo- Arc KO1 | 5 | 5 | 0 | -9.8034293 | 0.56869532 |
| Fos | PVN | dams | WT | 7 | 7 | 0 | -9.40268414 | 0.66934931 |
| Fos | PVN | naive mice | Neo+ Arc KO1 | 14 | 8 | 6 | -3.92683931 | 1.87438764 |
| Fos | PVN | naive mice | WT | 13 | 8 | 5 | -4.9735965 | 1.00695579 |
| Fos | SON | dams | Neo+ Arc KO1 | 6 | 6 | 0 | -8.72297229 | 0.16633043 |
| Fos | SON | dams | Neo- Arc KO1 | 8 | 8 | 0 | -9.15517939 | 0.90589787 |
| Fos | SON | dams | WT | 7 | 7 | 0 | -9.23431292 | 0.77981416 |
| Fos | SON | naive mice | Neo+ Arc KO1 | 13 | 7 | 6 | -5.54094011 | 1.7322835 |
| Fos | SON | naive mice | WT | 13 | 8 | 5 | -6.35586674 | 0.90213751 |
| Foxp1 | CPU | dams | Neo+ Arc KO1 | 5 | 5 | 0 | -4.86289951 | 0.32466067 |
| Foxp1 | CPU | dams | Neo- Arc KO1 | 7 | 7 | 0 | -5.04047253 | 0.20542957 |
| Foxp1 | CPU | dams | WT | 8 | 8 | 0 | -4.84517978 | 0.40381764 |
| Foxp1 | CPU | naive mice | Neo+ Arc KO1 | 14 | 8 | 6 | -3.87601263 | 0.72733653 |
| Foxp1 | CPU | naive mice | WT | 13 | 8 | 5 | -4.1024409 | 0.33236968 |
| Foxp1 | NAC | dams | Neo+ Arc KO1 | 5 | 5 | 0 | -5.47518439 | 0.29736587 |
| Foxp1 | NAC | dams | Neo- Arc KO1 | 8 | 8 | 0 | -5.79749269 | 0.22843197 |
| Foxp1 | NAC | dams | WT | 9 | 9 | 0 | -5.65027538 | 0.28324505 |
| Foxp1 | NAC | naive mice | Neo+ Arc KO1 | 13 | 7 | 6 | -5.35677269 | 0.67164785 |
| Foxp1 | NAC | naive mice | WT | 13 | 8 | 5 | -5.48464616 | 0.31007817 |
| Foxp1 | PFC | dams | Neo+ Arc KO1 | 6 | 6 | 0 | -7.27543808 | 0.37206762 |
| Foxp1 | PFC | dams | Neo- Arc KO1 | 7 | 7 | 0 | -7.06589495 | 0.29344125 |
| Foxp1 | PFC | dams | WT | 9 | 9 | 0 | -6.83945861 | 0.47608727 |
| Foxp1 | PFC | naive mice | Neo+ Arc KO1 | 13 | 7 | 6 | -6.02858934 | 1.72944442 |
| Foxp1 | PFC | naive mice | WT | 10 | 5 | 5 | -6.86643656 | 0.53664902 |
| Gal | CPU | dams | Neo+ Arc KO1 | 3 | 3 | 0 | -13.2329431 | 1.29467933 |
| Gal | CPU | dams | Neo- Arc KO1 | 7 | 7 | 0 | -13.344635 | 0.79592883 |
| Gal | CPU | dams | WT | 9 | 9 | 0 | -12.9252946 | 1.01434119 |
| Gal | CPU | naive mice | Neo+ Arc KO1 | 13 | 7 | 6 | -10.6208543 | 2.4334399 |
| Gal | CPU | naive mice | WT | 13 | 8 | 5 | -10.1539085 | 3.65784913 |
| Gal | NAC | dams | Neo+ Arc KO1 | 6 | 6 | 0 | -11.2234765 | 2.30505864 |
| Gal | NAC | dams | Neo- Arc KO1 | 8 | 8 | 0 | -11.5851574 | 0.81947217 |
| Gal | NAC | dams | WT | 9 | 9 | 0 | -10.5032585 | 1.71047667 |
| Gal | NAC | naive mice | Neo+ Arc KO1 | 13 | 7 | 6 | -10.0251531 | 2.148622 |
| Gal | NAC | naive mice | WT | 13 | 8 | 5 | -10.3067821 | 1.4121385 |
| Gal | PVN | dams | Neo+ Arc KO1 | 6 | 6 | 0 | -7.26471934 | 1.16555039 |
| Gal | PVN | dams | Neo- Arc KO1 | 7 | 7 | 0 | -2.20424081 | 0.85738229 |
| Gal | PVN | dams | WT | 8 | 8 | 0 | -4.17861739 | 2.96556019 |
| Gal | PVN | naive mice | Neo+ Arc KO1 | 14 | 8 | 6 | -4.81365019 | 1.31827184 |
| Gal | PVN | naive mice | WT | 13 | 8 | 5 | -4.88074105 | 1.7171107 |
| Gal | SON | dams | Neo+ Arc KO1 | 6 | 6 | 0 | -8.4731276 | 0.3760576 |
| Gal | SON | dams | Neo- Arc KO1 | 7 | 7 | 0 | -8.89576818 | 1.58963846 |
| Gal | SON | dams | WT | 8 | 8 | 0 | -8.49753122 | 1.01961013 |
| Gal | SON | naive mice | Neo+ Arc KO1 | 13 | 7 | 6 | -5.44053009 | 3.23030902 |
| Gal | SON | naive mice | WT | 13 | 8 | 5 | -7.58938831 | 1.34783158 |
| Homer1a | CPU | dams | Neo+ Arc KO1 | 5 | 5 | 0 | -5.70005315 | 0.3838561 |
| Homer1a | CPU | dams | Neo- Arc KO1 | 7 | 7 | 0 | -5.68165213 | 0.38052076 |
| Homer1a | CPU | dams | WT | 8 | 8 | 0 | -5.95253599 | 0.52782381 |
| Homer1a | CPU | naive mice | Neo+ Arc KO1 | 14 | 8 | 6 | -4.78552969 | 0.52597004 |
| Homer1a | CPU | naive mice | WT | 13 | 8 | 5 | -4.31781955 | 0.42912706 |
| Homer1a | NAC | dams | Neo+ Arc KO1 | 5 | 5 | 0 | -6.47362639 | 0.26124845 |
| Homer1a | NAC | dams | Neo- Arc KO1 | 7 | 7 | 0 | -6.90940737 | 0.73296794 |
| Homer1a | NAC | dams | WT | 10 | 10 | 0 | -6.74757114 | 0.64968963 |
| Homer1a | NAC | naive mice | Neo+ Arc KO1 | 13 | 7 | 6 | -5.73133696 | 0.56065534 |
| Homer1a | NAC | naive mice | WT | 13 | 8 | 5 | -6.15697807 | 0.42703751 |
| Homer1a | PFC | dams | Neo+ Arc KO1 | 6 | 6 | 0 | -6.77753479 | 0.64995391 |
| Homer1a | PFC | dams | Neo- Arc KO1 | 8 | 8 | 0 | -8.69422382 | 0.57694931 |
| Homer1a | PFC | dams | WT | 9 | 9 | 0 | -7.72701547 | 0.90456193 |
| Homer1a | PFC | naive mice | Neo+ Arc KO1 | 13 | 7 | 6 | -4.31557674 | 1.1674295 |
| Homer1a | PFC | naive mice | WT | 10 | 5 | 5 | -5.30408977 | 0.73993092 |
| Homer1a | PVN | dams | Neo+ Arc KO1 | 6 | 6 | 0 | -9.26708159 | 0.24221423 |
| Homer1a | PVN | dams | Neo- Arc KO1 | 6 | 6 | 0 | -8.64437801 | 0.43039593 |
| Homer1a | PVN | dams | WT | 8 | 8 | 0 | -9.03495377 | 0.28529958 |
| Homer1a | PVN | naive mice | Neo+ Arc KO1 | 14 | 8 | 6 | -4.48688934 | 1.92974212 |
| Homer1a | PVN | naive mice | WT | 13 | 8 | 5 | -5.36378459 | 1.59239201 |
| Homer1a | SON | dams | Neo+ Arc KO1 | 6 | 6 | 0 | -8.7519731 | 0.30541682 |
| Homer1a | SON | dams | Neo- Arc KO1 | 8 | 8 | 0 | -7.48881212 | 0.548243 |
| Homer1a | SON | dams | WT | 8 | 8 | 0 | -7.8973682 | 0.87357386 |
| Homer1a | SON | naive mice | Neo+ Arc KO1 | 13 | 7 | 6 | -6.06726266 | 1.39142242 |
| Homer1a | SON | naive mice | WT | 13 | 8 | 5 | -7.23219045 | 0.92682592 |
| Oxt | CPU | dams | Neo+ Arc KO1 | 4 | 4 | 0 | -7.46926013 | 0.70434055 |
| Oxt | CPU | dams | Neo- Arc KO1 | 7 | 7 | 0 | -7.51299661 | 0.5183049 |
| Oxt | CPU | dams | WT | 9 | 9 | 0 | -7.53977505 | 0.72065568 |
| Oxt | CPU | naive mice | Neo+ Arc KO1 | 14 | 8 | 6 | -6.5416905 | 2.79685246 |
| Oxt | CPU | naive mice | WT | 13 | 8 | 5 | -7.50649227 | 1.51274927 |
| Oxt | NAC | dams | Neo+ Arc KO1 | 5 | 5 | 0 | -6.64006908 | 1.52789111 |
| Oxt | NAC | dams | Neo- Arc KO1 | 8 | 8 | 0 | -8.63282621 | 1.21424903 |
| Oxt | NAC | dams | WT | 10 | 10 | 0 | -8.05597897 | 1.28055553 |
| Oxt | NAC | naive mice | Neo+ Arc KO1 | 13 | 7 | 6 | -7.59044522 | 2.49141749 |
| Oxt | NAC | naive mice | WT | 13 | 8 | 5 | -7.49920654 | 1.40718048 |
| Oxt | PFC | dams | Neo+ Arc KO1 | 5 | 5 | 0 | -8.37899166 | 1.32813957 |
| Oxt | PFC | dams | Neo- Arc KO1 | 8 | 8 | 0 | -7.48085802 | 0.53130304 |
| Oxt | PFC | dams | WT | 8 | 8 | 0 | -7.8895256 | 1.15488698 |
| Oxt | PFC | naive mice | Neo+ Arc KO1 | 13 | 7 | 6 | -5.00395055 | 4.23317646 |
| Oxt | PFC | naive mice | WT | 10 | 5 | 5 | -6.85323883 | 2.53257507 |
| Oxt | PVN | dams | Neo+ Arc KO1 | 6 | 6 | 0 | -1.73016046 | 0.61447183 |
| Oxt | PVN | dams | Neo- Arc KO1 | 7 | 7 | 0 | 1.26879641 | 0.63412391 |
| Oxt | PVN | dams | WT | 8 | 8 | 0 | -0.09056611 | 2.3592462 |
| Oxt | PVN | naive mice | Neo+ Arc KO1 | 14 | 8 | 6 | 0.79575591 | 1.06435022 |
| Oxt | PVN | naive mice | WT | 13 | 8 | 5 | 0.40319247 | 1.12141106 |
| Oxt | SON | dams | Neo+ Arc KO1 | 6 | 6 | 0 | -4.01928736 | 1.52493774 |
| Oxt | SON | dams | Neo- Arc KO1 | 7 | 7 | 0 | -4.48417782 | 3.03330585 |
| Oxt | SON | dams | WT | 8 | 8 | 0 | -5.02675024 | 1.92056433 |
| Oxt | SON | naive mice | Neo+ Arc KO1 | 13 | 7 | 6 | -2.637814 | 3.01399324 |
| Oxt | SON | naive mice | WT | 13 | 8 | 5 | -3.92700841 | 2.63748412 |
| Oxtr | CPU | dams | Neo+ Arc KO1 | 4 | 4 | 0 | -8.2825512 | 1.74610185 |
| Oxtr | CPU | dams | Neo- Arc KO1 | 7 | 7 | 0 | -7.93290805 | 0.77256359 |
| Oxtr | CPU | dams | WT | 9 | 9 | 0 | -7.6547974 | 0.99851111 |
| Oxtr | CPU | naive mice | Neo+ Arc KO1 | 14 | 8 | 6 | -5.73275867 | 3.07229164 |
| Oxtr | CPU | naive mice | WT | 13 | 8 | 5 | -7.78702265 | 2.21208252 |
| Oxtr | NAC | dams | Neo+ Arc KO1 | 6 | 6 | 0 | -6.43331901 | 1.09687908 |
| Oxtr | NAC | dams | Neo- Arc KO1 | 8 | 8 | 0 | -8.37395466 | 0.64772827 |
| Oxtr | NAC | dams | WT | 9 | 9 | 0 | -8.02000121 | 1.27069506 |
| Oxtr | NAC | naive mice | Neo+ Arc KO1 | 13 | 7 | 6 | -6.120188 | 2.83950284 |
| Oxtr | NAC | naive mice | WT | 13 | 8 | 5 | -6.04328477 | 2.83122637 |
| Oxtr | PFC | dams | Neo+ Arc KO1 | 6 | 6 | 0 | -9.04602426 | 0.67945224 |
| Oxtr | PFC | dams | Neo- Arc KO1 | 8 | 8 | 0 | -6.72944992 | 0.27469091 |
| Oxtr | PFC | dams | WT | 9 | 9 | 0 | -7.31967821 | 0.68078239 |
| Oxtr | PFC | naive mice | Neo+ Arc KO1 | 13 | 7 | 6 | -4.60859178 | 3.87849433 |
| Oxtr | PFC | naive mice | WT | 10 | 5 | 5 | -7.14560742 | 1.61167262 |
| Oxtr | PVN | dams | Neo+ Arc KO1 | 5 | 5 | 0 | -10.1301111 | 0.36513476 |
| Oxtr | PVN | dams | Neo- Arc KO1 | 7 | 7 | 0 | -8.05394101 | 0.42250825 |
| Oxtr | PVN | dams | WT | 8 | 8 | 0 | -8.46662573 | 1.21524214 |
| Oxtr | PVN | naive mice | Neo+ Arc KO1 | 14 | 8 | 6 | -1.82961232 | 2.37867369 |
| Oxtr | PVN | naive mice | WT | 13 | 8 | 5 | -3.98521733 | 2.63225821 |
| Oxtr | SON | dams | Neo+ Arc KO1 | 6 | 6 | 0 | -9.11065659 | 1.10447821 |
| Oxtr | SON | dams | Neo- Arc KO1 | 8 | 8 | 0 | -7.28001669 | 0.4098787 |
| Oxtr | SON | dams | WT | 8 | 8 | 0 | -8.21701645 | 0.9653421 |
| Oxtr | SON | naive mice | Neo+ Arc KO1 | 13 | 7 | 6 | -4.16495607 | 3.36214835 |
| Oxtr | SON | naive mice | WT | 13 | 8 | 5 | -5.69322939 | 2.05404762 |
| Shank3 | CPU | dams | Neo+ Arc KO1 | 5 | 5 | 0 | -3.28008563 | 0.18449238 |
| Shank3 | CPU | dams | Neo- Arc KO1 | 7 | 7 | 0 | -3.10417702 | 0.37499371 |
| Shank3 | CPU | dams | WT | 8 | 8 | 0 | -2.91153956 | 0.44150473 |
| Shank3 | CPU | naive mice | Neo+ Arc KO1 | 14 | 8 | 6 | -2.64942565 | 0.50867547 |
| Shank3 | CPU | naive mice | WT | 13 | 8 | 5 | -2.41476982 | 0.3139888 |
| Shank3 | NAC | dams | Neo+ Arc KO1 | 6 | 6 | 0 | -3.85052136 | 0.20315787 |
| Shank3 | NAC | dams | Neo- Arc KO1 | 7 | 7 | 0 | -3.40929131 | 0.19096008 |
| Shank3 | NAC | dams | WT | 10 | 10 | 0 | -3.43799146 | 0.24310901 |
| Shank3 | NAC | naive mice | Neo+ Arc KO1 | 13 | 7 | 6 | -3.43678319 | 0.29852321 |
| Shank3 | NAC | naive mice | WT | 13 | 8 | 5 | -3.40887576 | 0.32511257 |
| Shank3 | PFC | dams | Neo+ Arc KO1 | 6 | 6 | 0 | -4.16508998 | 0.39073504 |
| Shank3 | PFC | dams | Neo- Arc KO1 | 8 | 8 | 0 | -4.39909537 | 0.23794146 |
| Shank3 | PFC | dams | WT | 7 | 7 | 0 | -4.3987431 | 0.15027721 |
| Shank3 | PFC | naive mice | Neo+ Arc KO1 | 13 | 7 | 6 | -3.32301867 | 0.91819765 |
| Shank3 | PFC | naive mice | WT | 10 | 5 | 5 | -3.91541828 | 0.5269285 |
| Shank3 | PVN | dams | Neo+ Arc KO1 | 5 | 5 | 0 | -4.53000538 | 0.19152315 |
| Shank3 | PVN | dams | Neo- Arc KO1 | 7 | 7 | 0 | -4.60826176 | 0.28136207 |
| Shank3 | PVN | dams | WT | 8 | 8 | 0 | -4.73359384 | 0.15398088 |
| Shank3 | PVN | naive mice | Neo+ Arc KO1 | 14 | 8 | 6 | -5.14937245 | 0.80795964 |
| Shank3 | PVN | naive mice | WT | 13 | 8 | 5 | -4.79504035 | 0.28607794 |
| Shank3 | SON | dams | Neo+ Arc KO1 | 6 | 6 | 0 | -5.10922914 | 0.1322453 |
| Shank3 | SON | dams | Neo- Arc KO1 | 8 | 8 | 0 | -4.11918035 | 0.27592264 |
| Shank3 | SON | dams | WT | 8 | 8 | 0 | -4.45135857 | 0.42595823 |
| Shank3 | SON | naive mice | Neo+ Arc KO1 | 13 | 7 | 6 | -4.52357922 | 0.37200809 |
| Shank3 | SON | naive mice | WT | 13 | 8 | 5 | -4.61796312 | 0.34517717 |

| structure | var_s5 | condition | group1 | group2 | n1 | n2 | statistic | p | p.adj | p.adj.signif |
|---|---|---|---|---|---|---|---|---|---|---|
| CPU | Avp | dams | Neo+ Arc KO1 | WT | 6 | 9 | 3.01930547 | 0.00253355 | 0.00760065 | ** |
| PVN | Cck | dams | Neo+ Arc KO1 | WT | 5 | 8 | -2.56472264 | 0.01032583 | 0.02065166 | * |
| PVN | Cntnap2 | dams | Neo- Arc KO1 | WT | 7 | 8 | -2.70806418 | 0.00676769 | 0.01353539 | * |
| PVN | Cpeb4 | dams | Neo- Arc KO1 | WT | 7 | 8 | -2.74142842 | 0.00611727 | 0.01223454 | * |
| NAC | Fos | dams | Neo+ Arc KO1 | WT | 6 | 10 | -2.75687021 | 0.00583575 | 0.01750725 | * |
| PVN | Fos | dams | Neo+ Arc KO1 | WT | 6 | 7 | -2.2526181 | 0.02428324 | 0.04856648 | * |
| NAC | Oxtr | dams | Neo+ Arc KO1 | WT | 6 | 9 | -2.31572027 | 0.02057355 | 0.04114709 | * |
| PFC | Oxtr | dams | Neo+ Arc KO1 | WT | 6 | 9 | 2.39342901 | 0.01669171 | 0.03338342 | * |
| PVN | Oxtr | dams | Neo+ Arc KO1 | WT | 5 | 8 | 2.63884757 | 0.00831884 | 0.01761918 | * |
| NAC | Shank3 | dams | Neo+ Arc KO1 | WT | 6 | 10 | 2.61726777 | 0.00886368 | 0.01772736 | * |
| SON | Shank3 | dams | Neo+ Arc KO1 | WT | 6 | 8 | 2.55446045 | 0.01063525 | 0.0212705 | * |
| SON | Homer1a | naive mice | Neo+ Arc KO1 | WT | 13 | 13 | -2.02564103 | 0.04280158 | 0.04280158 | * |
| PFC | Oxtr | naive mice | Neo+ Arc KO1 | WT | 13 | 10 | -1.98455575 | 0.04719392 | 0.04719392 | * |
| PVN | Oxtr | naive mice | Neo+ Arc KO1 | WT | 14 | 13 | -2.03809866 | 0.04154007 | 0.04154007 | * |

**Table S6. List of all primers used for genotyping and qPCR.**

| ID | accession ID | gene | forward | Reverse | cDNA tissue | R square |
|---|---|---|---|---|---|---|
| *Oxt* | NM_011025 | oxytocin | ACCATCACCTACAGCGGATCT | [illegible]CGAGGTCAGAGCCAGTAAG | olfactory bulb | 0.9982 |
| *Oxtr* | NM_001081147 | oxytocin receptor | CTTAGGGCCAAAAGGTGTCA | GCAGGTTTCTATGCCCTCTG | olfactory bulb | 0.8923 |
| *Avp* | NM_009732 | Arginine vasopressine | ACACTACGCTCTTCCGCTTGT | CACTGTCTCAGCTCCATGTCA | olfactory bulb | 0.9861 |
| *Avpr1a 2* | NM_016847.2 | Arginine vasopressine receptor 1A | CCTCTGCTGGACACCTTTCTT | AAGGGTTTTCGGAATCGGTCC | olfactory bulb | 0.902 |
| *Homer1a* | NM_011982.4 | homer scaffolding protein 1 (Homer1), transcript variant S (Homer1a) | TGAAAAATCTCAGGTCAGACTCCT | GCTCAATGCTCCTTTTTCCACA | olfactory bulb | 0.9935 |
| *Fos* | NM_010234 | FBJ osteosarcoma oncogene (c-Fos) | GAAGGGAACGGAATAAGATG | CATCTTCAAGTTGATCTGTCTC | olfactory bulb | 0.9716 |
| *Egr1* | NM_007913 | early growth response 1 (Zif268) | CACCTGACCACAGAGTCCTTT | CGGCCAGTATAGGTGATGGG | olfactory bulb | 0.9698 |
| *Foxp1* | NM_053202 | forkhead box P1 | CCACTGTAACTCGAAGCGGT | GCGGCAGCACAGATACAAAG | olfactory bulb | 0.9917 |
| *Shank3* | NM_021423 | SH3/ankyrin domain gene 3 | CTCACCAACAGAGCAGGTCA | AGCGGAACTGACCCTGTAAA | olfactory bulb | 0.9691 |
| *Fmr1* | NM_008031 | Fragile X mental retardation protein (FMRP) translational regulator 1 | CCACGTAATCCAAGAGAGGCT | TTACGATCTTTACCCGTGCG | olfactory bulb | 0.9607 |
| *Cpeb4* | NM_026252 | cytoplasmic polyadenylation element binding protein 4 | AGAAGGGAACGTTATTTAAGTGCG | ACTACAACCTCCCCATACACA | olfactory bulb | 0.9433 |
| *Cntnap2* | NM_001004357 | contactin associated protein-like 2 | GTCTTCAGCCACTGACCCTT | TATAGCTTGGCCTTGTCCTGG | olfactory bulb | 0.9985 |
| *Cck* | NM_001284508 | cholecystokinin | TCCCCATCCAAAGCCATGAA | AGCTTCTGCAGGGACTACCG | olfactory bulb | 0.9812 |
| *Gal* | NM_010253.4 | galanin and GMAP prepropeptide | TAGGCTGGCTCCTGTTGGTT | TCTCTTCTCCTTTGCAGGCATC | Paraventricular nucleus of hypothalamus in Oprm1 KO | 0.9971 |
| *Cartpt* | NM_013732 | CART prepropeptide | AAGAAGTACGGCCAAGTCCC | CAGTCACACAGCTTCCCGAT | Striatum | 0.9948 |
| *Esr1* | NM_007956.5 | estrogen receptor 1 alpha | AAGAGAGTGCCAGGCTTTGG | CGCCAGACGAGACCAATCAT | Paraventricular nucleus of hypothalamus in Oprm1 KO | 0.9627 |
| *Gapdh* | NM_008084 | glyceraldehyde-3-phosphate dehydrogenase | ATGGCCTTCCGTGTTCCTAC | TCAGATGCCTGCTTCACCAC | olfactory bulb | 0.9737 |

## SI References